\documentclass[a4paper,11pt]{article}
\pdfoutput=1
\usepackage{jheppub}
\usepackage[mathscr]{eucal}
\usepackage{epsfig,amsfonts}
\usepackage{graphicx}
\usepackage{amsmath}
\usepackage{amsthm,amssymb}
\usepackage{graphicx}
\usepackage{hhline}
\usepackage{psfrag}
\usepackage{placeins}
\usepackage{relsize}
\usepackage{enumerate}
\usepackage{tikz}
\usepackage{textcomp}
\usepackage{hyperref}
\usepackage{mathrsfs} 
\usepackage{enumitem}
\usepackage{multirow}
\usepackage{comment}
\DeclareMathAlphabet{\mathpzc}{OT1}{pzc}{m}{it}

\usetikzlibrary{patterns}
\tikzset{>=latex}
\makeatletter
\@addtoreset{equation}{section}
\makeatother

\def\be{\begin{equation}}
\def\ee{\end{equation}}
\def\bdm{\begin{displaymath}}
\def\edm{\end{displaymath}}
\def\bea{\begin{eqnarray}}
\def\eea{\end{eqnarray}}

\def\XXint#1#2#3{{\setbox0=\hbox{$#1{#2#3}{\int}$}
    \vcenter{\hbox{$#2#3$}}\kern-.5\wd0}}

\def\be{\begin{equation}}
\def\ee{\end{equation}}

\def\beq{\begin{equation}}
\def\eeq{\end{equation}}
\def\half{{\textstyle \frac{1}{2}}}

\newcommand{\rd}{\mbox{d}}
\newcommand{\ri}{{\rm i}}
\newcommand{\re}{{\rm e}}

\title{\bf Quantum transfer-matrices for the sausage model}

\author[a]{Vladimir V. Bazhanov,}
\author[a,b]{Gleb A.  Kotousov}
\author[b,c]{and Sergei  L. Lukyanov}

\affiliation[a]{Department of Theoretical Physics,
         Research School of Physics and Engineering,\\
    Australian National University, Canberra, ACT 2601, Australia}
\affiliation[b]{NHETC, Department of Physics and Astronomy,
     Rutgers University,\\
     Piscataway, NJ 08855-0849, USA}
\affiliation[c]{L.D. Landau Institute for Theoretical Physics,\\
Chernogolovka 142432, Russia}

\abstract{In this work we revisit the problem of the quantization of the two-dimensional $O(3)$  non-linear sigma model and
its one-parameter integrable deformation -- the sausage model.
Our consideration is based on the so-called ODE/IQFT  correspondence, a variant of the Quantum Inverse Scattering Method.
The approach allowed  us to explore the integrable structures underlying the quantum $O(3)$/sausage model.
Among the obtained results is a system of non-linear integral  equations for the computation of the vacuum eigenvalues of the 
quantum transfer-matrices.}

\arxivnumber{1706.09941}

\emailAdd{vladimir.bazhanov@anu.edu.au}
\emailAdd{kotoousov@physics.rutgers.edu}
\emailAdd{sergei@physics.rutgers.edu}

\dedicated{To the memory of Ludwig Dmitrievich Faddeev}

\begin{document}
\maketitle
\flushbottom
\tikzset{
    partial ellipse/.style args={#1:#2:#3}{
        insert path={+ (#1:#3) arc (#1:#2:#3)}
    }
}

\section{Introduction}

\begin{figure}
\centering
\scalebox{.9}{
\begin{tikzpicture}
\draw[thick] (0,0) -- (0,5);
\draw[thick] (2.5,0) -- (2.5,5);
\fill[fill=yellow!80!white] (1.25,0) [partial ellipse = 180:360:1.25cm and 0.3cm];
\fill[fill=yellow!80!white] (0.02,0) rectangle (2.48,5);
\fill[fill=white] (1.25,5) [partial ellipse = 180:360:1.25cm and 0.3cm];
\draw [dashed] (1.25,5) [partial ellipse = 180:360:1.25cm and 0.3cm];
\draw [dashed] (1.25,5) [partial ellipse = 0:180:1.25cm and 0.3cm];
\draw [dashed] (1.25,0) [partial ellipse = 180:360:1.25cm and 0.3cm];
\draw [thick] (1.25,2) [partial ellipse = 180:360:1.25cm and 0.3cm];
\draw [thick] (1.25,2) [partial ellipse = 0:180:1.25cm and 0.3cm];
\draw [thick] (1.25,3.5) [partial ellipse = 180:360:1.25cm and 0.3cm];
\draw [thick] (1.25,3.5) [partial ellipse = 0:180:1.25cm and 0.3cm];
\node [right] at (3.2,2.5) {\Large $x\sim x+R$};
\draw [thick, ->] (-1,2.5) -- (-1,4.5) ;
\node [left] at (-1.3, 3.5) {\Large $t$};
\end{tikzpicture}}
\caption{The integration contour for the Wilson loop
can be moved freely along the space-time cylinder.}\label{fig0t}
\end{figure}
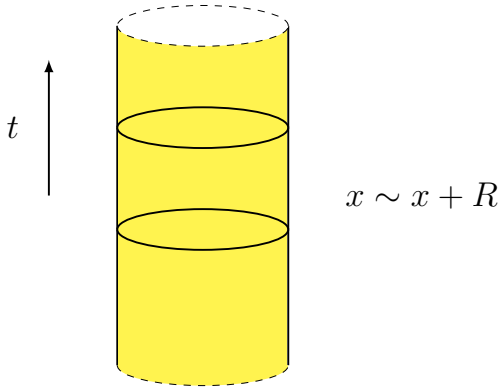

Integrability is a traditional area of mathematical physics having a long history.
The notion  is relatively well understood in finite
dimensional systems:  it requires the system to have exactly $n$
isolating and commuting integrals of motion  where  $n$ is the number of
degrees of freedom.
In the context of 1+1 dimensional field theory,
 where the continuous number
of  degrees of freedom makes the traditional definition insufficient,
 a suitable paradigm of integrability was also discovered.
The  key  ingredient in  this case
is  a Lie algebra-valued  world sheet connection involving an analytic
spectral parameter,
whose   flatness condition is equivalent to  the classical equations of motion.
Since the Wilson loops
\bea\label{aoispspa}
T={\rm Tr}\,{\overset{\leftarrow}{\cal P}}\exp\int_C{\boldsymbol { A}}
\eea
\noindent
remain unchanged under continuous  deformations of the integration contour (see fig.\,\ref{fig0t}), they
%(since the connection  depends on   the spectral  parameter)
generate  an infinite   family
of  conserved quantities which can be used to solve the field theory
within the framework of  the inverse scattering method.
At  the end of the   seventies the  Quantum Inverse Scattering Method (QISM)
was  proposed\, \cite{Faddeev:1979gh}.
%,Sklyanin:1980ij
The approach was   inspired by the  pioneering works  of Baxter on lattice statistical systems \cite{Baxter:1972wg,Baxter:1982zz} and
based on the 
 study of the  common  spectrum  
 of  the transfer-matrices ($T$-operators) -- the quantum counterpart of the classical Wilson loops.
 The original formulation of the QISM was   restricted to  the   so-called ``ultralocal'' models
 where  the elementary transport matrices  ${\boldsymbol M}_n={\overset{\leftarrow}{\cal P}}
  \exp\int_{x_n}^{x_{n+1}}{\boldsymbol  {A}}$
 commute for different segments  of the discretized path.
 The study of the algebra of  the ultralocal operators ${ {\boldsymbol M}}_n$
 led to the
discovery of  new remarkable mathematical structures collectively known
today as the Yang-Baxter algebras.
%, in turn stimulating areas of pure
%mathematics, most notably quantum groups and representations.
The most studied  class of  integrable
models is the one  where  the  Yang-Baxter algebra of the ultralocal operators ${{\boldsymbol M}}_n$  admits a finite-dimensional representation.
In this case  the  discretized quantum system can be interpreted  as an  exactly soluble
statistical model 
%whose local Boltzmann weights take  a finite
%number of admissible values,  and 
whose solution can be obtained by means of
the  Bethe ansatz method. The  solution of  the  continuous  QFT  is achieved  by taking a proper scaling limit.
An archetype  of this scenario is the   sine-Gordon model, while  the corresponding  statistical system
 is known as the  inhomogeneous  6-vertex model \cite{Destri:1987ze}.

However  the QISM
% known as  Quantum Inverse Scattering Method (QISM),
fails
 when it is applied to  many interesting systems including   classically  integrable Nonlinear Sigma Models (NLSM).
%In the most interesting  cases the elementary transport matrices
%do not satisfy the ultralocality condition.
%\footnote{More precisely,
%the ultralocality is not a gauge invariant condition and
 %it is not generally  known  whether
%the  ``ultralocal'' gauge for   the flat connection   exists.}
The origin of the problem can be illustrated  by  the $O(3)$
NLSM  governed  by the Lagrangian 
\bea\label{ajsiasui}
{\cal L}=\frac{1}{2  }\ \big( \, (\partial_t{\boldsymbol n})^2- (\partial_x{\boldsymbol n})^2\, \big)\ ,
\eea
where   the $3$-dimensional vector ${\boldsymbol n}$ is  subject to  the  constraint 
${\boldsymbol n}\cdot{\boldsymbol n}\,=\,1$.
% Geometrically, the constraint
%means  that ${ \boldsymbol n}$ lies on the round sphere so that the model possesses all of its isometries, i.e., $O(3)$ invariance.
%Geometrically, it means that ${\boldsymbol n}$ lies on the round sphere so that the model possesses all of its isometries, i.e., $O(3)$ %invariance.
In this  case
the light-cone components of the   flat connection, ${\boldsymbol { A}}_\pm\equiv \frac{1}{2}\, ( {\boldsymbol { A}}_t\pm  
{\boldsymbol { A}}_x)$,   are  given by the relations \cite{Pohlmeyer:1975nb,Zakharov:1973pp}
 \bea\label{jassais}
{\boldsymbol { A}}_\pm=\frac{2}{1\pm\zeta}\  [\partial_\pm {\check {\boldsymbol n}},\, {\check {\boldsymbol n}}]\ ,\ \ \ \ \ \ \ \ \ 
 {\check {\boldsymbol n}}=n_a{\tt t}_a \in\mathfrak{su}(2)\ ,
\eea
where $\partial_\pm=\frac{1}{2}(\partial_t\pm \partial_x)$ while   $ {\check {\boldsymbol n}}$  is
built  from
the components  of the unit vector  ${\boldsymbol n}$ and 
${\tt t}_a $:
\bea\label{iasosaisa}
[{\tt t}_a,{\tt t}_b]=\ri\,\varepsilon_{abc}\, {\tt t}_c\ .
\eea 
With  periodic boundary conditions, ${\boldsymbol n}(t,x+R)={\boldsymbol n}(t,x)$, the  flat connection turns out to be
 a single-valued 1-form
 on  the space-time cylinder and
 the Wilson loops
\eqref{aoispspa} are conserved charges    for any value of the  spectral parameter $\zeta$.
The
equal time  Poisson  brackets   of  \eqref{jassais} respect   the  general 
structure
\bea\label{masssah}
\{ { \boldsymbol {A}}_\mu(x)
\begin{array}{ccc} \\[-0.4cm] \otimes \\[-0.35cm] , \end{array}
%\underset{{}^,}{\otimes}\, 
{ \boldsymbol { A}}_\nu(y)\}={ \boldsymbol  C}^{(0)}_{\mu\nu}(x)\ \delta(x-y)+
{ \boldsymbol  C}^{(1)}_{\mu\nu}(x)\ \delta'(x-y)\ ,
\eea
where ${ \boldsymbol  C}^{(i)}_{\mu\nu}(x)$ are   local fields built from $n_a$ and $\partial_t n_a$ and 
take
values in the tensor product $ \mathfrak{su}(2)\otimes  \mathfrak{su}(2)$.
Because  of the    derivative of the $\delta$-function,  relations of this  type are usually  referred  to as 
non-ultralocal.  The presence of  such terms creates  serious  problems  with the rigorous proof of 
Poisson commutativity of the conserved charges  \eqref{aoispspa}  for different values of the
spectral parameter. This, in turn,  hampers  the  first-principles  quantization of the model.

\bigskip

The problem with non-ultralocality   is a general  issue  
for  integrable NLSM and   is a major  obstacle which  prevents  the  incorporation of these models  into the framework of   the  QISM.
This    was first  observed    in attempts to quantize  the principal chiral field and $O(N)$ models.
In the case of the $O(4)$-model (i.e., the model \eqref{ajsiasui} with ${\boldsymbol n}\in {\mathbb S}^3$),
Polyakov and Wiegmann \cite{Polyakov:1983tt} and later 
 Faddeev  and  Reshetikhin \cite{Faddeev:1985qu} 
managed to bypass  the problem with  non-ultralocality.
  In  both works,
the   NLSM  was replaced
by  a different  model   satisfying the ultralocality condition;  Polyakov  and Wiegmann  considered a
model with $N_f$  fermion flavors, whereas  Faddeev  and Reshetikhin focused on   a certain spin-$S$ chain.
They studied the 
 thermodynamics  using  the Bethe ansatz technique and gained 
valuable results for the $O(4)$ sigma model        through  the  large     $N_f$ and   $S\to \infty$ limit, respectively.
Both limiting procedures yielded   the same system of
 Thermodynamic Bethe Ansatz Equations (TBA),  which  was then  justified by a  comparison
with perturbative calculations and the exact results  from the $S$-matrix bootstrap \cite{ZZ79}.
In the next 30 years   an enormous amount  of  TBA systems  were  discovered.
However, in spite of significant achievements in their  study, the original  problem of
the construction of  quantum transfer-matrices (including the  calculation of  their spectrum) has remained  unsolved
even  for the   $O(3)$ and $O(4)$ NLSM.

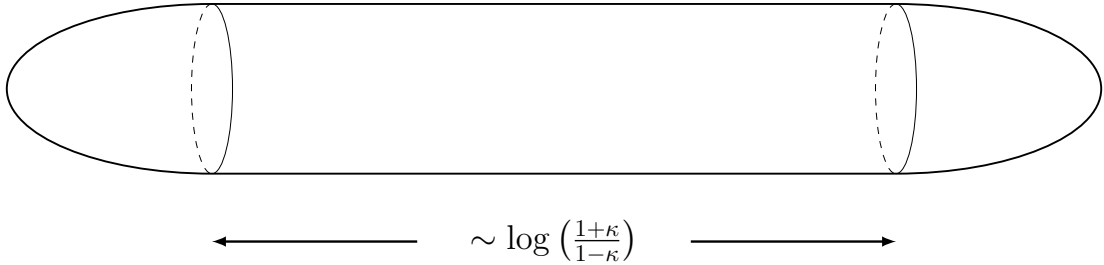
\begin{figure}
\centering
\scalebox{.9}{
\begin{tikzpicture}
\draw[thick] (2,0) -- (12,0);
\draw[thick] (2,2.5) -- (12,2.5);
\draw[thick] (2,1.25) [partial ellipse =90:270:3cm and 1.25cm];
\draw[thick] (12,1.25) [partial ellipse =270:450:3cm and 1.25cm];
\draw[line width = 0.35 mm,->] (9,-1) --(12,-1); 
\draw[line width = 0.35 mm,->] (5,-1) --(2,-1); 
\node at (7,-1) {\Large $\sim \log\big(\frac{1+\kappa}{1-\kappa}\big)$};
\draw (2,1.25) [partial ellipse =270:450:0.3cm and 1.25cm];
\draw[dashed] (2,1.25) [partial ellipse =90:270:0.3cm and 1.25cm];
\draw (12,1.25) [partial ellipse =270:450:0.3cm and 1.25cm];
\draw[dashed] (12,1.25) [partial ellipse =90:270:0.3cm and 1.25cm];
\end{tikzpicture}
}
\caption{A depiction of the target manifold of the NLSM \eqref{iassaosaio} for $1-\kappa\ll 1$.}
\label{fig1t}
\end{figure}

Is the non-ultralocal structure \eqref{masssah} an immanent property
of an integrable NLSM, or is it just the result of an awkward choice
of gauge for the flat connection?  Unfortunately, we do not know the
answer for the most interesting models, however in the case of the
$O(3)$ sigma model, it turns out that the ``ultralocal'' gauge does
exist \cite{Bytsko:1994ae}.\footnote{% 
  It is worth noting that a transition to an
  ultralocal gauge has previously been utilised for some other important but
  apparently non-ultralocal systems, such as the (quantum) KdV theory
  \cite{Bazhanov:1994ft} and its discretized analogs
  \cite{Fioravanti:2001bx,Fioravanti:2002sq}.}
Further, we found that such a gauge also exists   for 
% the one parameter deformation
%of
 the theory described by the Lagrangian
\bea\label{iassaosaio}
{\cal L}=\frac{1}{2}\  \frac{\partial_\mu {\boldsymbol n}\partial^\mu {\boldsymbol n}}
{\kappa^{-1}-\kappa\, n_3^2}\ .
\eea
This  theory  also  belongs  to the class of NLSM,  and, assuming $0<\kappa<1$,   the corresponding target space
 is topologically  the  two-sphere. 
As $\kappa\to 0$ one can neglect the second term in the denominator  which  results  in the  $O(3)$ Lagrangian
multiplied by $\kappa$. Hence,  the model \eqref{iassaosaio} is  the  one parameter deformation
of the $O(3)$ sigma model. It is colloquially  known as  the ``sausage model'' since
for $\kappa\to 1^-$ the target manifold can be
pictured  as a  long sausage with  length $\propto \log(\frac{1+\kappa}{1-\kappa})$ (see fig.\ref{fig1t}).  
The sausage model  was introduced in the work \cite{Fateev:1992tk}  where  strong evidence for its quantum integrability was   presented. Its
zero curvature representation    was found in ref.\cite{Lukyanov:2012zt}.
Remarkably,     with the proper gauge transformation, 
 the flat connection can be chosen to 
satisfy   the ultralocal  Poisson bracket  relations
\bea\label{hsasausa}
&&\big\{{\boldsymbol {A}}_\pm(x|\mu) \begin{array}{ccc} \\[-0.4cm] \otimes \\[-0.35cm] , \end{array}  {\boldsymbol { A}}_\pm (x'|\mu')\big\}
=
\pm\, \big[
{\boldsymbol { A}}_\pm(x|\mu)\otimes {\boldsymbol  1}+
 {\boldsymbol 1}\otimes {\boldsymbol {A}}_\pm(x'|\mu'), {\boldsymbol r}(\mu/\mu')\big]\, \delta(x-x')\nonumber\\ 
&&\big\{{\boldsymbol {A}}_\pm(x|\mu)  \begin{array}{ccc}
 \\[-0.4cm] \otimes \\[-0.35cm] , \end{array} {\boldsymbol {A}}_\mp(x'|\mu')\big\}=0\, .
\eea
Here
\bea\label{oieur}
{\boldsymbol r}(\mu)=\frac{1}{\mu-\mu^{-1}}\ \Big(\,
2\ {\tt t}_1\otimes{\tt t}_1+2\ {\tt t}_2\otimes{\tt t}_2+ \big(\mu+\mu^{-1}\big)\ {\tt t}_3\otimes{\tt t}_3\,\Big)\ ,
\eea
and the dependence 
 on  the  multiplicative spectral  parameter $\mu$ is made explicit ($\mu$ substitutes the auxiliary  parameter   $\zeta$  in eq.\eqref{jassais}).

To present  explicit formulae for the ultralocal flat connection, we first
 resolve  the  constraint  ${\boldsymbol n}^2=1$  by means of the
stereographic projection.  It allows 
the triple  ${\boldsymbol n}=(n_1,n_2,n_3)$ to be substituted by 
the complex  unconstrained field $Q$,
\bea\label{aisosa}
\re^Q=\sqrt{\frac{1-n_3}{1+n_3}\  \frac{n_1+\ri n_2}{n_1-\ri n_2}}\ ,\ \ \ \ \ \ 
\re^{ Q^*}=\sqrt{\frac{1-n_3}{1+n_3}\  \frac{n_1-\ri n_2}{n_1+\ri n_2}}\ .
\eea
%In term of $Q$ and ${Q}^*$ the Lagrangian , \eqref{iassaosaio} takes the form
%\bea\label{kasissauw}
%{\cal L}=
%\frac{\kappa (1-z^2)}{4\,(1-\kappa^2 z^2)}\ \eta^{\mu\nu}\ \big(\partial_\mu  Q{\partial_\nu}{ Q}^*+
%\partial_\nu Q{\partial_\mu}{ Q}^*\big)\ ,
%\eea
The pair of  momenta $(\Pi, \Pi^*)$ canonically conjugate to $(Q, Q^*)$
are given by the relations
\bea
\Pi= \frac{1-n_3^2}{2(\kappa^{-1}-\kappa n_3^2)}\ \  \partial_t{ Q}^*\ ,\  \ \  \  \  \ \ 
 \Pi^*= \frac{1-n_3^2}{2(\kappa^{-1}-\kappa n_3^2)}\ \ \partial_t { Q}\ ,
\eea
where  $n_3=-\tanh\big(\frac{1}{2}(Q+Q^*)\big)$.
Let us also introduce the shortcut notations
\bea
\Pi_+=\Pi+
\frac{1-n_3^2}{2\,(\kappa^{-1}-\kappa\, n_3^2)}\ \partial_x  {Q}^*\ ,\ \ \ \ \ \ \ \ 
{ \Pi}_-=\Pi^*-
\frac{1-n_3^2}{2\,(\kappa^{-1}-\kappa\, n_3^2)}\ \partial_x  Q \ ,
\eea
and the field independent element of ${\mathfrak {sl}}(2,{\mathbb C})$
\bea\label{iasoisosaoias}
{\boldsymbol a}(\mu)=f(\mu)\, {\tt t}_3+\ri\, g(\mu)\ {\tt  t}_1\ ,
\eea
where
\bea\label{iasosao}
f(\mu)&=&
%\frac{1-\kappa+(1+\kappa)\mu^2}{1-\kappa-(1+\kappa)\mu^2}=
\frac{(1-\kappa)\,\mu+(1+\kappa)\, \mu^{-1}}{(1-\kappa)\,\mu-(1+\kappa)\,\mu^{-1}}\\
g(\mu)&=& \frac{ 2\sqrt{1-\kappa^2}}{(1-\kappa)\,\mu-(1+\kappa)\, \mu^{-1}}\ .\nonumber
\eea
It is straightforward to check  now that the formulae\footnote{\label{foot1}Notice that   the   connection in the ultralocal gauge
takes  values in the complexification of the Lie algebra ${\mathfrak {su}}(2)$. This should not worry us because
only  reality conditions imposed on  gauge invariant quantities are  a subject of interest and importance.
In what follows we'll prefer to use 
the Cartan-Weyl generators ${\tt h}=2{\tt t}_3,\ {\tt e}_\pm={\tt t}_1\pm\ri {\tt t}_2$: 
\bea
[{\tt h},\, {\tt e}_\pm]=\pm 2\,  {\tt e}_\pm\ ,\ \ \ \ \ [{\tt e}_+,{\tt e}_-]={\tt h}\nonumber
\eea
instead of  ${\tt t}_a$ \eqref{iasosaisa}.
}
\bea\label{kassau}
{\boldsymbol { A}}_+&=&+{ \Pi}_+ \   
 \re^{ Q{\tt t}_3}\ 
{\boldsymbol a}(\mu)\ 
  \re^{- Q{\tt t}_3}
\\ [5.pt]
{\boldsymbol {A}}_-&=&-{ \Pi}_- \   \re^{\ri\pi {\tt t}_3 }\,  \re^{- Q^* {\tt t}_3}\ 
{\boldsymbol a}(\mu^{-1})\ 
 \re^{ Q^*{\tt t}_3}\   \re^{-\ri\pi{\tt t}_3 }\nonumber
\eea
define the flat connection, i.e.,
\bea
\big[\partial_+-{\boldsymbol {A}}_+,\partial_--{\boldsymbol {A}}_-]=0\ ,
\eea
satisfying  the ultralocality  conditions \eqref{hsasausa}.

Having an explicit formula for the ultralocal flat connection we may turn to the construction of the 
time-independent Wilson loops.
In fact their  construction requires some assumptions to be made.
First of all in writing eq.\eqref{aoispspa} it is  assumed that  the connection is  single valued on the space-time cylinder.  
In the case 
under consideration this can be achieved by imposing  periodic boundary conditions  on the unit vector  ${\boldsymbol n}$.
Such boundary conditions  look   natural for  the $O(3)$ NLSM, since they preserve the global invariance of the Lagrangian.
In the case of the sausage model, the $O(3)$  symmetry is broken down to   $U(1)$ and  we will
consider more general  boundary conditions
\beq
n_3(t,x+R)=n_3(t,x)\,,\quad n_\pm(t,x+R)=\re^{\pm 2\pi\ri k}\,  n_\pm(t,x)\qquad (n_\pm\equiv n_1\pm\ri  n_2)
\eeq
which depend  on the twist parameter $k\sim k+1$. In terms of the real and imaginary part of the
complex field $Q$,
\bea\label{jsausa}
Q=\phi+\ri\, \alpha \ ,
\eea
the twisted boundary conditions  are equivalently given by
\bea\label{aisosaio}
\phi(t, x+R)=\phi(t, x)\ ,\ \ \ \  \alpha(t, x+R)=\alpha(t, x)+2\pi k\ .
\eea
As it follows from eq.\eqref{kassau} 
the connection is not single valued on the space-time cylinder now,
\bea
 {\boldsymbol  {A}}_\mu(t, x+R)=\re^{\ri\pi k {\tt h}}\ 
 {\boldsymbol  {A}}_\mu(t, x)\, \re^{-\ri\pi k {\tt h}}\ ,
\eea
however with a  little modification one can still introduce the time independent conserved charges
\bea\label{oisasasp}
T_j(\mu)={\rm Tr}\Big[\re^{-\ri \pi k {\tt h}}\  {\boldsymbol M}_j(\mu)\,\Big]\ ,\ \ \  \
{\boldsymbol M}_j=\pi_j\bigg[\,{\overset{\leftarrow} {\cal P}}\exp\int_C\rd x\, {\boldsymbol  A}_x
\bigg]\ .
\eea
Notice that the definition of the Wilson  loop requires a choice of the representation of the Lie algebra
and  the subscript $j$ in \eqref{oisasasp}  labels the representations $\pi_j$   of  $\mathfrak{sl}(2)$.
In our case we will focus on the finite dimensional irreps, so that in  the standard convention $j=\frac{1}{2},\,1,\ldots$\ .

The ultralocal Poisson structure \eqref{hsasausa}  implies that the monodromy matrix
generates  the classical
Yang-Baxter  Poisson algebra:
\bea\label{aissaoas}
\big\{{\boldsymbol M}_j(\mu) \begin{array}{ccc} \\[-0.4cm] \otimes \\[-0.35cm] , \end{array}{\boldsymbol M}_{j'}(\mu')\,\big\}=
\big[{\boldsymbol M}_j(\mu)\,{\otimes}\, {\boldsymbol M}_{j'}(\mu')\,, {\boldsymbol r}_{jj'}(\mu/\mu')\,\big]\ ,
\eea
where ${\boldsymbol r}_{jj'}=\pi_j\otimes\pi_{j'}[\,{\boldsymbol r}\,].$
This, supplemented  by the   easily established property
\bea
\big[\, \re^{-\ri \pi k {\tt h}}\otimes \re^{-\ri \pi k {\tt h}}\, ,\, {\boldsymbol r}(\mu/\mu')\,\big]=0\ ,
\eea
implies  the infinite set of relations
\bea\label{asuiasas}
\big\{T_j(\mu), \,T_{j'}(\mu')\big\}=0\ .
\eea
%In fact  the   $T_j$ corresponding to the different values of $j$ are functionally dependent -- they all can be 
%expressed in terms of   $T\equiv T_{\frac{1}{2}}$ corresponding to the fundamental  representation of   $\mathfrak{sl}(2,C)$
Thus, we have arrived at formulae \eqref{aissaoas} and \eqref{asuiasas} which 
are key ingredients in  the Hamiltonian approach to the inverse scattering method \cite{Sklyanin:1979gh,Faddeev:1987ph}.
 %which can be used to exploit the infinite family of involutive, conserved charges $T_j(\mu)$  to solve the classical theory.

%When attempting the quantization of the sausage NLSM, the first thing that comes to mind is to discretize the path-ordered integral  \eqref{oisasasp} onto $N$ small %%segments.
When faced with the problem of quantizing the sausage NLSM, a simple idea that may come to mind 
is to discretize the path-ordered integral onto $N$ small segments.
Due to ultralocality,  the $N$ elementary transport matrices  $\pi_j\Big[{\overset{\leftarrow}{\cal P}}
  \exp\int_{x_n}^{x_{n+1}}\rd x\, {\boldsymbol  {A}}_x\Big]$ satisfy the same type of  Poisson bracket relation as \eqref{aissaoas} and Poisson commute for different segments. These relations can be formally quantized leading to a certain 
  quantum Yang-Baxter algebra. The major problem now is to construct a suitable representation of this abstract algebraic structure.
In the case under consideration, the representation is, in all likelihood, infinite dimensional even for finite $N$. 
At this moment, it is  not clear for us how to construct and handle such representations, let alone take the scaling limit with 
$N\to \infty$.

In this work, we will try to avoid discretization as much as we can and mostly
 follow the so-called BLZ approach -- the variant of the QISM developed in the series of works \cite{Bazhanov:1994ft,Bazhanov:1996dr,Bazhanov:1998dq}. For  integrable Conformal Field
Theories (CFT), it  was demonstrated that the $T$-operators can be constructed   without
any
discretization procedure.
Later  it was  observed   that many deep properties of
representations of Yang-Baxter algebras in integrable CFT
can be encoded in the monodromies of certain linear Ordinary
Differential Equations (ODE) \cite{Voros:1992,Voros:1999,Dorey:1998pt,Bazhanov:1998wj,
Suzuki:1999hu,Bazhanov:2003ni,AlZ}.
These  results  were extended to massive
Integrable Quantum Field Theories (IQFT) \cite{Lukyanov:2010rn}
 (for recent developments, see also
refs.\cite{Dorey:2007zx,Dorey:2012bx,
Adamopoulou:2014fca,Ito:2015nla,Masoero:2015rcz1,Masoero:2015rcz2,Ito:2016qzt,Babenko:2017fmu}).
The general
relation of this type will be referred to in the paper as the ODE/IQFT correspondence.
% (see
%refs.\cite{Dorey:2007zx,Dorey:2012bx,
%Adamopoulou:2014fca,Ito:2015nla,Masoero:2015rcz,Ito:2016qzt,Babenko:2017fmu}   for recent developments).
%For further  developments of the approach, see refs.\cite{Dorey:2007zx,Dorey:2012bx,
%Adamopoulou:2014fca,Ito:2015nla,Masoero:2015rcz,Ito:2016qzt,Babenko:2017fmu}.

Broadly speaking, the ODE/IQFT correspondence means that for a given IQFT   the eigenvalues of the quantum   $T$-operators
are identified  with  certain connection coefficients  for the  system of equations,
\bea
{\boldsymbol { D}}(\theta)\, {\boldsymbol \Psi}=0\ ,\ \ \ \ \ {\overline{\boldsymbol { D}}}(\theta)\, {\boldsymbol \Psi}=0\ ,
\eea
where ${\boldsymbol  {D}}(\theta)$  and ${\overline {\boldsymbol { D}}}(\theta)$ stand for   (singular) differential operators depending on the
auxiliary parameter $\theta$  which is found to be a  function of the original  spectral parameter from  the quantum theory.
The system  of ODE can be then  interpreted  as  an auxiliary linear problem, whose compatibility condition,
$[{\boldsymbol {D}}(\theta),\, { \overline{\boldsymbol { D}}}(\theta)]=0$, coincides with the zero-curvature representation for   some 
{\it   classically} integrable
field theory.
Thus the  ODE/IQFT correspondence
reduces   the   calculation   of  the spectrum of  quantum
transfer-matrices  to a certain problem in the theory of classical  integrable equations.
The latter  can be effectively  treated by  the inverse scattering transform method.
This makes the  ODE/IQFT correspondence  a  very powerful  tool.
In particular,  it gives a practical way to make progress in the conceptual
long standing problem of the quantization of  integrable NLSM. The ultimate goal of this work is to demonstrate 
this for  the case
of the quantum sausage model.

\section{Chiral transfer-matrices for the cigar\label{sec20}}

The BLZ approach  \cite{Bazhanov:1994ft,Bazhanov:1996dr,Bazhanov:1998dq} begins with  an analysis of the 
RG fixed point which controls the  ultraviolet  behaviour of  the integrable QFT. 
With this in mind, let's  take a quick  look at    the sausage NLSM  \eqref{iassaosaio}.
In the traditional path-integral quantization, the model  should be equipped with a  UV cutoff $\Lambda$.
A consistent removal of the UV  divergences requires  that  the ``bare'' coupling   in the Lagrangian   \eqref{iassaosaio}
be given a certain dependence on the cutoff momentum, i.e., $\kappa=\kappa(\Lambda)$. To the first  perturbative order
the RG flow equation is given by \cite{Fateev:1992tk}
\bea
\Lambda\ \frac{\partial \kappa}{\partial \Lambda}=\frac{\hbar }{2\pi}\ (1-\kappa^2)+O(\hbar^2)\  ,
%\ \ \   \ \  \hbar=\frac{2\pi}{n}\ .
\eea
where $\hbar$ stands for the  (dimensionless) Planck constant.
Integrating this equation leads to
 \bea\label{jsyyd}
\frac{1-\kappa}{1+\kappa}= (E_*/\Lambda)^\nu\ ,
\eea
where $\nu=\frac{\hbar }{\pi}+O(\hbar^2)$.
The  energy scale  $E_*$   is an RG invariant (i.e., it's kept fixed  with changing $\Lambda$), 
so that 
  $\kappa\to 1$   as  $\Lambda\to \infty$. Having in mind the quantization of the model, 
  this  simple analysis   shows  that   the classical field theory   \eqref{iassaosaio} deserves special  attention 
  when
  $\kappa$ is  close to one.

\subsection{Wilson loops for the long sausage}

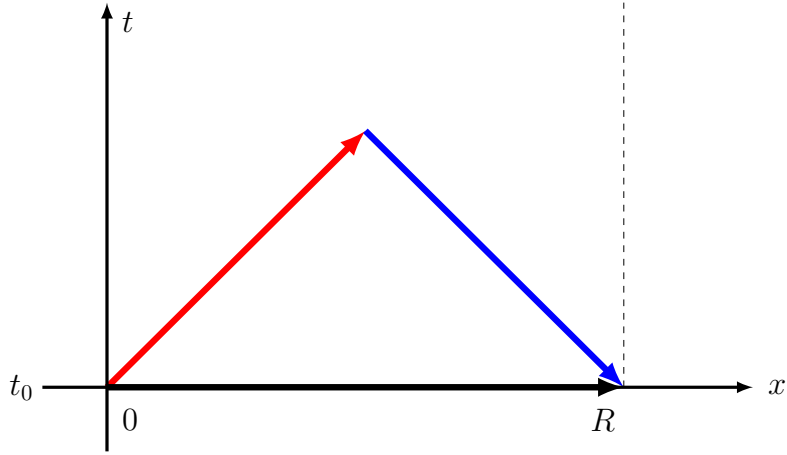
\begin{figure}[t]
\centering
\scalebox{0.85}{
\begin{tikzpicture}
\draw [line width = 1mm,->,red] (0,0) -> (4,4);
\draw [line width = 1mm,->,blue] (4,4) -> (8,0);
\draw [ line width = 1mm,->] (0,0) -> (8,0);
\draw [line width = 0.5mm, ->] (0,-1) -- (0,6);
\draw [line width = 0.5mm,->] (-1,0)-- (10,0);
\draw [dashed] (8,0) -- (8,6);
\node [below right] at (0.1,-0.2) {\Large $0$};
\node [below left] at (8,-0.2) {\Large $R$};
\node [left] at (-1,0) {\Large $t_0$};
\node [right] at (0.1,5.7) {\Large $t$};
\node [right] at (10.1,0) {\Large $x$};
\end{tikzpicture}
}
\caption{The integration contour along the time slice $t\,=\,t_0$ (black arrow) in eq\,(1.20) can be replaced by an integration contour 
along the characteristics: $x_-\,=\,t_0$ with $0<x_+<t_0+R$ (red arrow) and $x_+\,=\,t_0+R$ with $t_0<x_-<t_0-R$ (blue arrow).}
\label{fig03}
\end{figure}

 Consider  the 
 %mutually Poisson commuting
 conserved charges
 $T_j(\mu)$ \eqref{oisasasp} for $1-\kappa\ll 1$.
 If the integration contour  appearing  in the definition \eqref{oisasasp}
  is chosen at the time slice $t=t_0$,  $T_j(\mu)$ are expressed in terms of  the pair of
  real variables
  $(\phi,\alpha)$ \eqref{jsausa}   and the
  corresponding canonically conjugate momenta.
  We can  use the magic of  the zero-curvature representation
  to re-express $T_j(\mu)$ in terms of the values of the fields $\phi(x,t)$ and  $\alpha(x,t)$
  on the characteristics  $x_+\equiv t+x=t_0+R$ and $x_-\equiv t-x=t_0$, 
  \bea\label{apsosapas}
  &&\phi_+(x_+)=\phi|_{x_-=t_0}\ ,\ \ \ \ \ \ \ \ \   \ \ \ \   \alpha_+(x_+)= \alpha|_{x_-=t_0}\nonumber\\
  &&\phi_-(x_-)=\phi|_{x_+=t_0+R}\ ,\ \ \ \ \ \ \ \ \ \    \alpha_-(x_-)= \alpha|_{x_+=t_0+R}\ .
  \eea
Indeed, the original integration  along the time slice $t=t_0$ in \eqref{oisasasp} can be replaced by
the path-ordered integral  over the  contour glued from two light-cone segments as shown in fig.\,\ref{fig03}.
Denoting the light-cone values of the connection as 
\beq
{\boldsymbol A}_+(x_+)={\boldsymbol A}_+(t,x)|_{x_-=t_0}\, ,\qquad
{\boldsymbol A}_-(x_-)={\boldsymbol A}_-(t,x)|_{x_+=t_0+R}\, ,
\eeq
and 
taking into account the boundary conditions \eqref{aisosaio}, one can rewrite eq.\eqref{oisasasp} in the form
\bea\label{aoaspsaoas}
&&T_j\equiv 
{\rm Tr}\big[\, \re^{-\ri \pi k {\tt h}}\ {\boldsymbol M}_j\, \big]
={\rm Tr}\big[\,    \pi_j( {\boldsymbol{\cal M}})\,\big]\ ,
\eea
where
\beq\label{lightcone}
{\boldsymbol{\cal M}}=\re^{-\frac{1}{2} \ri\, \alpha(t_0,R)\,{\tt h}} 
\  
{\overset{\leftarrow} {\cal P}}\exp\bigg(\int_{t_0}^{t_0-R}
{\boldsymbol A}_-(x_-)\, \rd x_- \bigg)\,
{\overset{\leftarrow} {\cal P}}
\exp\bigg(\int_{t_0}^{t_0+R} {\boldsymbol A}_+(x_+)\,\rd x_+\,\bigg)\ 
\re^{\frac{1}{2} \ri \,\alpha(t_0,0)\,{\tt h}}. 
\eeq
This formula  
is a convenient starting point for the quantization procedure, though 
one should have in mind that the light-cone fields 
 are complicated  functionals  of the canonical set at $t=t_0$. 
In general,  they are not independent variables and their Poisson
brackets are not known explicitly. Fortunately, we can overcome this
problem when considering the limit $\kappa\to 1$. 
In this limit the flat connection \eqref{kassau} simplifies
considerably, since the quantities 
 $\Pi_+\approx \Pi+\half\,\partial_x Q^*$, $\Pi_-\approx
\Pi^*-\half\,\partial_x Q$ are then linearly expressed through 
the light-cone fields $\phi_\pm$ and $\alpha_\pm$.   
It is convenient to write the constant Lie algebra elements 
 ${\boldsymbol a}(\mu^{\pm 1})$ \eqref{iasoisosaoias}
in the form
\bea
 {\boldsymbol a}\big(\mu^{\pm 1}\big)=
\frac{1}{2(\lambda_\pm-\lambda_\pm^{-1})}
\Big[(\lambda_\pm+\lambda_\pm^{-1})\,{\tt h}-2\ri\,\big({\tt e}_++{\tt e}_-\big)\Big]\,,
\eea
where the variables  
$\lambda_\pm$ are defined by the relations 
 \bea\label{lambdapm}
\lambda_+\lambda_-=\frac{1-\kappa}{1+\kappa}\ ,\qquad 
\lambda_+/\lambda_-=\mu^2
\eea
and $({\tt h},{\tt e}_+,{\tt e}_-)$ are the Cartan-Weyl generators 
(see footnote \ref{foot1} on page \pageref{foot1}). 
Then the connection component ${\boldsymbol A}_+(x_+)$
only depends on the parameter $\lambda_+$, while ${\boldsymbol
  A}_-(x_-)$ only depends on $\lambda_-$.
According to \eqref{lambdapm} at least  one of the
$(\lambda_+,\lambda_-)$ vanishes as $\kappa\to1$. Let us assume
that $\lambda_-\to0$ and $\lambda_+$ is kept fixed. Then, it is easy to see that
 \bea\label{asasaisao}
 \tau_j(\lambda_+)\equiv \lim_{\kappa\to1\atop\lambda_+-{\rm fixed}}T_j={\rm Tr}\Big[\, {\boldsymbol L}_j(\lambda_+)\ 
 \re^{-\pi P_1 {\tt h}}\, \Big]\ .
 \eea
Here  we use the notations
 \begin{eqnarray}\label{oapasosap}
{\boldsymbol L}_j(\lambda_+)=\pi_j\bigg[
{\overset{\leftarrow} {\cal P}}
\exp\bigg\{\!\frac{\ri\lambda_+}{1-\lambda_+^{2}}\!\!
\int\limits_{t_0}^{t_0+R}\!\! \rd x_+\Big({  V}^+ {\tt e}_++V^- {\tt e}_-\!+\ri
\lambda_+(\partial_+\phi_+-\ri\,\partial_+\alpha_+)\,{\tt h}\Big)\! \bigg\}
\re^{-\pi P_1{\tt h}}\bigg]
\nonumber
\\[0.2cm]
\end{eqnarray}
where
\beq\label{oapsasap}
V^\pm=(\partial_+\phi_+-\ri\, \partial_+\alpha_+)\ \re^{\pm 2\phi_+}
%{\bar V}^{\pm}=(\partial_-\phi_-+\ri\,\partial_-\alpha_-)\ \re^{\mp 2\phi_-}\
\eeq
and 
\bea\label{P1-def}
P_1=\frac{1}{2\pi}\ \big(\, \phi_+(t_0+R)-\phi_+(t_0)\,\big)\ .
%=\frac{1}{2\pi}\ \big(\, \phi_-(t_0)-\phi_-(t_0-R)\,\big)\ .
\eea
Notice that the $r$-matrix in the classical Yang-Baxter Poisson algebra \eqref{aissaoas} depends only on the ratio of the spectral parameters and is independent of $\kappa$. Hence, we can conclude that
\bea\label{aopsasap}
\big\{{\boldsymbol L}_j(\lambda_+)
 \begin{array}{ccc} \\[-0.4cm] \otimes \\[-0.35cm] , \end{array}
 {\boldsymbol L}_{j'}(\lambda'_+)\,\big\}=
\big[{\boldsymbol L}_j(\lambda_+)\,{\otimes}\, {\boldsymbol L}_{j'}(\lambda'_+)\,, {\boldsymbol r}_{jj'}(\lambda_+/\lambda'_+)\,\big]\, ,
\eea 
and also that the
$\tau_j$ are in involution:
 \bea
 \{\tau_j(\lambda_+),\,\tau_{j'}(\lambda'_+)\}=0\ .
 \eea

For  $1-\kappa\ll  1$ the target space of the sausage model  consists of two cigars
glued  together, whose  
tips are separated by a   distance  $\propto \log(\frac{1+\kappa}{1-\kappa})$  (see fig.\,\ref{fig1t}).
In  the  $\kappa\to 1$ limit,  the sausage is broken down into two disjoint  half-infinite  Hamilton's cigars\,\cite{Hamilton}.
 Let us focus  on  one of the cigars, say  the left one.
 The Lagrangian of the  NLSM which has this  target space follows from the sausage Lagrangian  \eqref{iassaosaio}:
 one should
 express the sausage Lagrangian in terms of the fields $\phi$ and $\alpha$, perform the constant shift $\phi\to
 \phi+\half  \log(\frac{1-\kappa}{1+\kappa})$, and finally take the limit $\kappa\to 1$. As a result one 
 finds \cite{Elitzur:1991cb,Witten:1991yr}
 \bea\label{iisaoos}
{\cal L}=
\frac{\re^{2\phi}}{2(1+\re^{2\phi})}\,
\big(\, (\partial_\mu\phi)^2+(\partial_\mu\alpha)^2\, \big)\ .
\eea
This suggests that the $\tau_j$ introduced through the limiting
procedure \eqref{asasaisao}, can be interpreted as conserved
quantities in the cigar NLSM \eqref{iisaoos}.  The fields $\phi_\pm$
and $\alpha_\pm$ which appear in the construction of ${\boldsymbol
  L}_j$ (see eqs.\,\eqref{oapsasap},\,\eqref{oapasosap}), should be
now understood as the light-cone values of the fields in the cigar
NLSM, in the case when the field $\phi$ becomes large, i.e., in the
asymptotically flat domain of the target manifold. The cigar equations
of motion in this domain asymptotically approach the pair of
D'Alembert equations $\partial_\mu\partial^\mu\phi\approx 0$ and
$\partial_\mu\partial^\mu\alpha\approx 0$, whose general solution can
be written as $\phi\approx \phi_+(x_+)+ \phi_-(x_-)$ and
$\alpha\approx \alpha_+(x_+)+ \alpha_-(x_-)$.  Taking the time slice
$t=t_0$ where the field $\phi$ becomes large, one finds
\bea\label{siossiaoas}
&&\{\phi_\pm(x), \phi'_\pm(y)\}= \half\,  
\delta(x-y)\ ,\ \ \ \{\alpha'_\pm(x),\alpha'_\pm(y)\}=\half\, \ \delta'(x-y)\\[0.2cm]
&&\{\phi'_\pm(x), \phi'_\mp(y)\}=\{\alpha'_\pm(x),\alpha'_\mp(y)\}=
\{\phi_\pm(x),\alpha'_\pm(y)\}=\{\phi_\pm(x),\alpha'_\mp(y)\}=0\ ,
\nonumber 
\eea
where the prime stands for the derivative w.r.t. the argument. Recall that  the $\tau_j$ are time independent charges,
and their value does not depend on the choice of $t_0$. Hence, we come to the conclusion that the matrix ${\boldsymbol L}_j$
\eqref{oapasosap}, built from the fields $\phi_+(x)$,
$\alpha_+(x)$ satisfying the Poisson bracket relations
\eqref{siossiaoas}, will obey the ``$r$-matrix'' Poisson bracket
algebra \eqref{aopsasap}.

\bigskip
The following comment is in order here.  It is not difficult to see
that the matrix ${\boldsymbol L}_j$ does not change under the constant
shifts $\phi_+\to\phi_++const$ so that the Poisson bracket relations
\eqref{siossiaoas} can be lifted to
\bea
\{\phi_+(x), \phi_+(y)\}=   
-{\textstyle\frac{1}{4}}\, \epsilon(x-y)\ ,\ \ \ \ %{\alpha_+(x),\alpha_+(y)\}=-{\textstyle\frac{1}{4}}\, \epsilon(x-y)\ ,
\eea
where $\epsilon(x)=2n+1$ for $ n R<x<  (n+1) R;\ n\in{\mathbb Z}$. 
The later are consistent with the quasiperiodicity  condition
\eqref{P1-def}. 

\subsection{Definition  and   basic  properties of  the  chiral transfer-matrices\label{sec22}}
 
We may now turn to the problem of quantization. The quantum
counterpart of the conserved charges \eqref{asasaisao} will be
referred to as the chiral transfer-matrices and in what follows, will
be denoted
%(with some abuse of notation)
by the  same symbol ${\tau}_j$.
 Their construction almost identically follows the steps elaborated in refs.\,\cite{Bazhanov:1994ft,Bazhanov:1998dq}
 in the context   of the quantum KdV theory.
 Here we present them very briefly, referring the reader to those  works for detailed explanations.

\bigskip

 First of all we  should ``quantize'' the Lie algebra  ${\mathfrak {sl}}(2)$, so that 
  ${\tt h}$, ${\tt e}_\pm$ are understood now as the generators  of the quantum universal enveloping
  algebra $U_q\big({\mathfrak {sl}}(2)\big)$:
  \bea
[{\tt h},\, {\tt e}_\pm]=\pm 2\,  {\tt e}_\pm\ ,\ \ \ \ \ [{\tt e}_+,{\tt e}_-]=\frac{q^{\tt h}-q^{-\tt h}}{q-q^{-1}}\ ,
\eea
 where $q=\re^{\frac{\ri\hbar}{2}}$.  
 Consequently the symbol $\pi_j$ will stand for the $(2j+1)$-dimensional representation of  the quantum algebra.
 % $U_q\big({\mathfrak {sl}}(2)\big)$.
 Instead of the Planck constant $\hbar$,  for convenience we will 
use the parameter  $n$:
 \bea
 \hbar\equiv{\textstyle  \frac{2\pi}{n}}\ ,\ \ \ \ \ \ q=\re^{\frac{\ri\pi}{n}}\ .
 \eea

  The quantum counterpart to ${\boldsymbol L}_j$ is the following $(2j+1)\times(2j+1)$ operator valued matrix
 \bea\label{hasuassaasu}
{\boldsymbol L}_j(\lambda_+)=\pi_j\bigg[
{\overset{\leftarrow} {\cal P}}
\exp\bigg( \ri \lambda_+ \int_{t_0}^{t_0+R} \rd x\,\ \big({ V}^{+}\ 
 q^{\frac{\tt h}{2}}\, {\tt e}_++V^-\   q^{-\frac{\tt h}{2}}\,{\tt e}_-\big) \bigg)\ 
\re^{-\pi { P}_1{\tt h}}\,\bigg]\ .
\eea
 The vertex operators $V^{\pm}$ are defined by the set of relations:
 \bea\label{aisaiasoas}
V^{\pm}(x)= \big(\, \half\, c^{\pm}\,\,\partial_x-\ri\,  
{\textstyle\frac{\sqrt{n+2}}{n}}\ 
 \vartheta'_+(x\,)\big)
 %c^{\pm}\, \varphi'_+-\ri\,\, \vartheta'_+)\ 
\  \re^{\pm
\frac{2\varphi_+}{\sqrt{n}}}(x)\  ,
\eea
 where $c^\pm$ are some constants and
\bea\label{iaisia}
&&\varphi_+(x)=Q_1+\frac{2\pi x}{R}\  \sqrt{n}\,{ P}_1+\ri\,
 \sum_{m\not= 0}\frac{a_m}{m}\ \re^{-\frac{2\pi\ri m}{R} x}
 \\
&&\vartheta_+(x)=Q_2+\frac{2\pi x}{R}\  \sqrt{n+2}\,{  P}_2+
\ri\, \sum_{m\not= 0}\frac{b_m}{m}\ \re^{-\frac{2\pi\ri m}{R} x}\ ,\nonumber
\eea
 with
\bea\label{isaosasao}
[\,a_m,a_l\,]=[\,b_m,b_l\,]={\textstyle \frac{m}{2}}\  \delta_{m+l,0}\ ,\ \ \  \ \ 
\big[\, Q_1, \sqrt{n}\, { P}_1\,\big]=\big[\, Q_2, 
\sqrt{n+2}\,{ P}_2\, \big]={\textstyle \frac{\ri}{2}}\ . 
\eea
Let ${\cal F}_{p_1,p_2}\equiv{\cal F}_{\bf p} $ (``Fock space'') be the highest weight module of the Heisenberg algebra
 \eqref{isaosasao} with the highest weight vector $|\,{\bf p}\,\rangle $  defined by the equations
 \bea\label{kasasusa}
 { P}_1\,|\,{\bf p}\,\rangle =\frac{p_1}{n}\, |\,{\bf p}\,\rangle\ ,\ \ \ \ \ 
 { P}_2\,|\,{\bf p}\,\rangle =\frac{ p_2}{n+2}\, \ |\,{\bf p}\,\rangle\ .
 \eea
 It is easy to see that
 \bea
 V^{\pm}(x)\ :\ \ \ {\cal F}_{p_1,p_2}\mapsto  {\cal F}_{p_1\mp \ri ,p_2}\ ,
 \eea
 and therefore the matrix elements of ${\boldsymbol L}_j(\lambda)$ are operators
 in $\oplus_{m=-\infty}^{\infty}  {\cal F}_{p_1+\ri m ,p_2}$. The 
 expression \eqref{hasuassaasu} contains the ordered exponential 
which can be formally written  in terms of a power series
 in $\lambda$ as
 \beq\label{aoaspsap}
 {\boldsymbol L}_j(\lambda_+)=\pi_j\bigg[\,\sum_{m=0}^\infty(\ri \lambda_+)^m\int_{t_0+R>x_m>\ldots x_1>t_0}\rd x_m\cdots \rd x_1\ 
 {\boldsymbol  K}(x_m)\cdots {\boldsymbol  K}(x_1)\, \re^{-\pi { P}_1{\tt h}}\,\bigg]\, ,
\eeq
where
 \bea
{\boldsymbol  K}(x)={ V}^{+}\ 
 q^{\frac{\tt h}{2}}\, {\tt e}_++V^-\   q^{-\frac{\tt h}{2}}\,{\tt e}_-\ .
\eea
However,  since
\bea\label{sasossiaoas}
V^\pm(x_2)V^\mp(x_1)\big|_{ x_2\to x_1+0}
\sim\ \frac{a}{2n^2}\ q^{-1}\ \big(x_2-x_1\big)^{-2 (1- \frac{1}{n})}\ ,
\eea
where $a=n+2 + (n-2)\ c^+ c^-$, 
the integrals in \eqref{aoaspsap} diverge. As  explained in 
\cite{Bazhanov:1998dq},  the commutation relations
\bea\label{comm-relations}
V^{\sigma_1}(x_1)V^{\sigma_2}(x_2)=q^{2\sigma_1\sigma_2}\ V^{\sigma_2}(x_2)V^{\sigma_1}(x_1)\ ,\  \ \ \ \ \ x_2>x_1\ \ \ \ 
(\sigma_{1,2}=\pm)
\eea
%where $\sigma_{1,2}=\pm$,
allow one to re-express the integrals in \eqref{aoaspsap} in terms of  two basic
contour integrals
\bea\label{asiopsap}
{\cal  X}_0=\frac{1}{q-q^{-1}}\ \int_{t_0}^{t_0+R}\rd x\ V^-(x)\ ,\ \ \  \ {\cal X}_1=\frac{1}{q-q^{-1}}\ \int_{t_0}^{t_0+R}\rd x\ V^+(x)\ .
\eea
This procedure yields an unambiguous definition of the ordered exponential in \eqref {hasuassaasu}  for
$n\not=2,\,4,\,6\ldots$\ . The case of even $n$ needs some special attention and we will return to it later.

Notice that the above analytical regularization of the ${\cal P}$-ordered exponential
\eqref{hasuassaasu} is not applicable at $n\to\infty$. 
This limit can be studied within the renormalized perturbation theory in $\hbar=\frac{2\pi}{n}$.
As usual, the perturbative expansion will involve the counterterms depending on the UV regulator.
At $\hbar=0$ the counterterms give rise to anomalous contributions which 
remain finite  in the limit where the UV regulator goes away. 
This makes the classical limit not entirely straightforward, however
if the anomalous terms are properly taken into account, the formula
\eqref{hasuassaasu} precisely reduces to its classical version \eqref{oapasosap} in the limit $\hbar\to 0$.
The relevant calculations 
will be presented elsewhere.

The operator valued  matrices  ${\boldsymbol L}_j$ \eqref{hasuassaasu}  are designed
in  such a way that,  for arbitrary chosen constants $c^\pm$ and $t_0$, they obey the quantum Yang-Baxter algebra
\beq\label{kusysiiua}
{\boldsymbol R}_{jj'}(\lambda'_+/\lambda_+)\, \big({\boldsymbol  L}(\lambda_+)\otimes 1\big)\, 
(1\otimes {\boldsymbol  L}(\lambda'_+)\big)= \big(1\otimes{\boldsymbol  L}(\lambda'_+)\big)\,\big(
{\boldsymbol  L}(\lambda_+)\otimes 1\big)\,  {\boldsymbol  R}_{jj'}(\lambda'_+/\lambda_+)\, ,
\eeq
where the matrix  ${\boldsymbol R}_{jj'}(\lambda)$ is the trigonometric solution to the Yang-Baxter
equation which acts in the space $\pi_j\otimes\pi_{j'}$. In particular
\bea\label{kasusau}
{\boldsymbol  R}_{\text{\textonehalf\textonehalf}}(\lambda)=
\begin{pmatrix}
{q^{-1}\,\lambda-q\,\lambda^{-1}}&0&0&0\\
0&{\lambda-\lambda^{-1}}& {q^{-1}-q}&0\\
0&{q^{-1}-q}&{\lambda-\lambda^{-1}}&0\\
0&0&0&{q^{-1}\,\lambda-q\,\lambda^{-1}}
\end{pmatrix}\ .
\eea
The proof of eq.\eqref{kusysiiua} follows that  from the work \cite{Bazhanov:1998dq}.

\bigskip
The chiral transfer-matrices, defined similar to \eqref{asasaisao},
\bea\label{aospas}
{ \tau}_j(\lambda_+)={\rm Tr}\Big[\, {\boldsymbol L}_j(\lambda_+)\ 
 \re^{-\pi { P}_1 {\tt h}}\, \Big]\, ,
 \eea
satisfy the commutativity condition
 \bea 
 [\, { \tau}_j(\lambda_+),\,{ \tau}_{j'}(\lambda'_+)\,]=0\, ,
 \eea
as a simple consequence of \eqref{kusysiiua}.
Notice that  the chiral transfer-matrices act inside a
single Fock space, 
whereas the same is not true  for an arbitrary  element  of ${\boldsymbol  L}_j(\lambda)$. Furthermore, 
the  Fock space ${\cal F}_{\bf p}$ naturally splits into the  finite dimensional ``level subspaces''
\bea
{\cal F}_{\bf p}=\oplus_{{{L}}=0}^\infty\,{\cal F}^{{({L})}}_{\bf p} :\ \ \ {\mathbb L}\, {\cal F}^{({L})}_{\bf p}=
 L\, {\cal F}^{({L})}_{\bf p}\ ,
\eea
where the grading operator is given by
\bea\label{jsayasasy}
{\mathbb   L}=2\, \sum_{m=1}^\infty\big(\, a_{-m}a_m+b_{-m}b_m\, \big)\,  .
\eea
Using the relation,
\bea
V^\pm(x+R)=q^2\ \re^{\pm 4\pi { P}_1}\  V^\pm(x)
\eea
one can show (see Appendix C from  \cite{Bazhanov:1998dq})  that the ${ \tau}_j(\lambda_+)$  commute with the
grading operator, and therefore,
act invariantly in each  finite-dimensional level subspace:
\bea
 { \tau}_j(\lambda_+)\ :\ \ \ {\cal F}^{{({L})}}_{\bf p}\mapsto  {\cal F}^{{({L})}}_{\bf p}\, .
 \eea
 The Fock space  ${\cal F}_{\bf p}$ can be equipped with  an inner product consistent with the Hermiticity
conditions $a^\dagger_m=a_{-m}\ ,\  b^\dagger_m=b_{-m}$ imposed
on the Heisenberg operators \eqref{isaosasao}. 
It is not difficult to  show  that  for  real $p_1^2,\ p_2^2$ and $\lambda^2_+$,  ${  \tau}(\lambda_+)$ 
is a  Hermitian operator and
\bea
\big[{ \tau}(\lambda_+)\big]^\dagger={ \tau}(\pm \lambda^*_+)\, .
 \eea

Notice that  the commutativity with the grading operators can be interpreted as the independence of the chiral transfer-matrix on
 the arbitrary chosen constant $t_0$. It turns out that  they further  do not depend on the constants $c^\pm$ appearing in
 the definition  of the vertex operators $V^{\pm}$\,\eqref{aisaiasoas}.
Also, a simple dimensional analysis  shows that the spectral parameter $\lambda_+$ and $R$  occur in the chiral transfer-matrix through 
 the combination  $\lambda^2_+R^{\frac{2}{n}}$ only. It is convenient to introduce a
  dimensionless spectral parameter
 $\lambda$ by means of the relation
  \bea\label{isaiasias}
 \lambda^2= \Gamma^2\big(1+{\textstyle \frac{1}{n}}\big)\ \big({\textstyle \frac{nR}{2\pi}}\big)^{\frac{2}{n}}\ \lambda^2_+
 \eea
 and treat  the chiral transfer-matrices as  functions of this variable rather than the dimensionful 
 $\lambda_+$.

\bigskip

 The chiral transfer-matrices are not independent operators for different values of $j=\half,\,1,\,\ldots\ .$ They can be expressed
 through the ``fundamental'' transfer-matrix  ${ \tau}_{\frac{1}{2}}(\lambda)$ by the so-called  fusion relation \cite{Str79,BP82,KR87a}
 \bea\label{asusuas}
 { \tau}_j\big(\lambda\,q^{j+\frac{1}{2}}\big)\ 
{ \tau}_{\frac{1}{2}}(\lambda)=
 { \tau}_{j+\frac{1}{2}}\big(\lambda\, q^{j}\big)
 + { \tau}_{j-\frac{1}{2}}\big(\lambda\, q^{j+1}\big)\, ,
 \eea
supplemented by the condition ${ \tau}_0={1}$. 
In what follows, we will mostly focus on the fundamental transfer-matrix and 
use the notation
${\tau}\equiv { \tau}_{\frac{1}{2}}$.
The integrable structures associated with the commuting family of operators ${\tau}_j(\lambda)$
 were already  studied  in the context of the
 so-called paperclip model -- an integrable model with boundary interaction \cite{Lukyanov:2005nr}. 
 %\cite{Lukyanov:2003nj,Lukyanov:2005nr}.
 % \cite{Lukyanov:2005nr}. 
 Here for convenience we make a short summary   of some  basic properties of the operator $ { \tau}(\lambda)$.
 %acting in the Fock space ${\cal F}_{\bf p}$.
 
 \bigskip
 For arbitrary complex ${\bf p}=(p_1,p_2)$,
 the operator $ {  \tau}(\lambda)\in {\rm End}\big({\cal F}_{\bf p}\big)$ is an entire function of $\lambda^2$ in the sense that all 
its matrix elements and eigenvalues  are entire functions of this variable.
Thus the power   series 
  \bea\label{saosapassap}
  {  \tau}(\lambda)=
  2\,\cosh\big({\textstyle \frac{2\pi p_1}{n}}\big)
+\sum_{m=1}^\infty{{\mathfrak t}}_m\ \lambda^{2m}
  \eea 
 converges  in the whole complex plane of $\lambda^2$ and defines
  an entire function  with an essential
 singularity at $\lambda^2= \infty$.
 The  asymptotic expansion near the essential singularity
 is of primary
 interest. It can be written as
 \bea\label{asasisao}
{ \tau}(\lambda)=   \exp\Big(\, {\textstyle \frac{2\pi }{\sin(\frac{\pi n}{2})}}\, (-\lambda^2)^{\frac{n}{2}}\, \Big)\ 
{ {\tilde \tau}}\Big(\,\ri\, (-\lambda^2)^{-\frac{n}{2(n+2)}}\Big)\, ,
\eea
where ${ {\tilde \tau}}$ is a formal power series of the form
\bea\label{jasusau}
{ {\tilde \tau}}({\tilde \lambda})\asymp 2\, \cos\big(
{\textstyle\frac{2\pi p_2}{n+2}}\big)+
\sum_{m=1}^\infty {  {\tilde {\mathfrak t}}}_{m}\ {\tilde\lambda}^{2m}\ .
\eea
 This  asymptotic expansion  can be applied for arbitrary complex ${\bf p}=(p_1,p_2)$ and $n\not=2,\,4,\,6\ldots\ $.
 Furthermore, in the case  $n \geq 1$ it holds true  for $|\arg(-\lambda^2)|<\pi$.

 The expansion coefficients in \eqref{saosapassap} and   \eqref{jasusau}
form  two infinite sets of mutually commuting operators.
Using the terminology of the work \cite{Bazhanov:1996dr},  we will refer to  $\big\{ {{ {\mathfrak t}}}_{m}\big\}_{m=1}^\infty$  and
 $\big\{ {{ \tilde {\mathfrak t}}}_{m}\big\}_{m=1}^\infty$
 as the  nonlocal  and dual nonlocal Integrals of Motions (IM), respectively.
 Remarkably,  the formal power series ${{\tilde \tau}}({\tilde \lambda})$  can be  written in a  form
   similar to \eqref{aospas}. Namely \cite{Lukyanov:2005nr},
 \bea\label{aossapsap}
{ {\tilde \tau}}({\tilde \lambda})={\rm Tr}\bigg[\,
{\overset{\leftarrow} {\cal P}}
\exp\bigg( \ri {\tilde  \lambda}_+ \int_{t_0}^{t_0+R} \rd x\ \big(\,{ \Psi}^{+}\ 
  {\sigma}_++\Psi^-\  {\sigma}_-\big) \bigg)\ 
\re^{-2\pi \ri {P}_2{\sigma_3}}\,\bigg]\ ,
\eea
where $\sigma_3,\ \sigma_\pm=\half\,(\sigma_1\pm \ri \sigma_2)$ are the conventional Pauli matrices
and
the vertex operators $\Psi^{\pm}$ are given by
 \bea\label{aisaiasoasaaa}
\Psi^{\pm}(x)= \big( 
{\textstyle\frac{\sqrt{n}}{n+2}}\ 
 \varphi'_+(x)+\half\, \,\,\partial_x\big)
 %c^{\pm}\, \varphi'_+-\ri\,\, \vartheta'_+)\ 
\  \re^{\pm
\frac{2\ri \theta_+}{\sqrt{n+2}}}(x)\  .\nonumber
\eea
The scale  dimension of  $\Psi^{\pm}$ is equal to $1+\frac{1}{n+2}$ and we assume here  that they are normalized in such 
a way that
\bea
\Psi^\pm(x_2)\,\Psi^\mp(x_1)\big|_{ x_2\to x_1+0}
\sim\ {\textstyle \frac{2}{(n+2)^2}}\ \ \big(x_2-x_1\big)^{{-2 (1+ \frac{1}{n+2})}}\ .
\eea
Because of the  divergencies, the path ordered exponential in \eqref{aossapsap} should be understood   
in the same manner  as \eqref{hasuassaasu}, i.e.,  the formal  expansion in a  power series of ${\tilde \lambda}_+$ should be
rewritten in terms of the basic contour integrals similar to \eqref{asiopsap}. With this analytical
regularization the   r.h.s. of eq.\,\eqref{aossapsap} becomes a  formal power series   in ${\tilde \lambda}^2_+R^{-\frac{2}{n+2}}$  with
unambiguously defined
expansion coefficients.
Up to a factor similar to that in  \eqref{isaiasias}, this combination can be identified with ${\tilde \lambda}^2$ in
eq.\,\eqref{jasusau}:
 \bea
 {\tilde \lambda}^2= \Gamma^2\big(1-{\textstyle \frac{1}{n+2}}\big)\ 
 \big({\textstyle \frac{(n+2)R}{2\pi}}\big)^{-\frac{2}{n+2}}\ {\tilde \lambda}^2_+\ .
 \eea
  For future reference we present here explicit formulae for the  ``vacuum'' eigenvalues of the
  operators  ${{  {\mathfrak t}}}_{1}$ and ${{ \tilde {\mathfrak t}}}_{1}$ corresponding to the highest  weight vector 
  $|\,{\bf p}\,\rangle\in {\cal F}_{\bf p}$  \eqref{kasasusa}:
\beq\label{jssusau}
{t}_1(p_1,p_2)=\Big(\frac{2}{  n}\Big)^{\frac{2}{ n}}\ 
{\Gamma({1\over 2}+{1\over n})\over \sqrt{\pi}
\Gamma(1+{1\over n})}
\ \Big(\frac{n+2}{ n-2}+\frac{4p_2^2}{1+4p_1^2}\, \Big)\ 
{\pi^2\over \Gamma({1+2\ri p_1\over n})\Gamma({1-2\ri p_1\over n})}
\eeq
and
\beq\label{asuausai}
{\tilde {t}}_{1}(p_1,p_2)=\Big({ n+2\over 2}\Big)^{{2\over n+2}}\ 
{\Gamma({1\over 2}-{1\over n+2})\over \sqrt{\pi}
\Gamma(1-{1\over n+2})}
\ \Big(\frac{n}{ n+4}-\frac{4p_1^2}{1-4 p_2^2}\, \Big)\ 
{\pi^2\over \Gamma(-\frac{1-2 p_2}{ n+2})\Gamma(-\frac{1+2p_2}{ n+2})}\, .
\eeq
 An   efficient integral representation
 for calculating the vacuum eigenvalue  ${\tilde {t}}_{2}(p_1,p_2)$ can be found in 
 Appendix A of ref.\cite{Lukyanov:2005nr}.

 For even  $n$,  the chiral transfer-matrices require some careful handling.
 In this case  ${\tau}(\lambda)$ can be  defined through the limiting procedure
 \bea\label{iuczvx}
 { \tau}(\lambda)|_{n=2 l}=\lim_{\epsilon \to 0}\ 
  \exp\big(\, -{\textstyle \frac{4}{\epsilon }}\,\, \lambda^{2l}\, \big)\ 
 { \tau}(\lambda)\big|_{n=2l+\epsilon}\ \ \ \ \ \ \ \  (l=1,\,2,\,3\ldots)\ ,
  \eea
  so that   the asymptotic formula \eqref{asasisao}  should be substituted by
\bea\label{asasi}
{ \tau}(\lambda)|_{n=2 l}=   \exp\Big(\, 2\, \lambda^n\ 
\log (-\lambda^2)\, \Big)\ 
{{\tilde \tau}}\Big(\,\ri\, (-\lambda^2)^{-\frac{l}{2(l+1)}}\Big)\Big|_{n=2 l}\ .
\eea
%Here $C$ is  some non-universal constant  which can be chosen at will.

\bigskip  
The  formulae  \eqref{asasisao},\,\eqref{jasusau} are  not valid  for positive real $\lambda^2$.
In order to describe the asymptotic behaviour for $\lambda^2\to+\infty$, it is convenient to substitute
the set of  dual nonlocal IM \eqref{jasusau}
by the set $\big\{  {\tilde {\mathfrak g}}_{m}\big\}_{m=1}^\infty$
 which are algebraically expressed in terms of the former 
 through the relation
\bea\label{jsasuausa}
%{\hat {\tilde \tau}}({\tilde \lambda})=
%{\boldsymbol {\tilde \tau}}({\tilde \lambda})\asymp 
2\, \cos(2\pi P_2)+
\sum_{m=1}^\infty {  {\tilde {\mathfrak t}}}_{m}\ z^m=
2\, \cos(2\pi P_2)\ 
\exp\Big(\sum_{m=1}^\infty { {\tilde {\mathfrak g}}}_{m}\   {z}^{m}\,
\Big)\ .
\eea
Then, 
for arbitrary complex ${\bf p}=(p_1,p_2)$, $n\not=2,\,4,\,6\ldots\,$, $n\geq 1$,
\bea\label{tseriega}
{ { \tau}}(\lambda)=4\, \cos\big({\textstyle \frac{2\pi p_2}{n+2}}\big)\ \re^{{H}(\lambda^2)}\ 
\cos\big({G}(\lambda^2)\big)\ \ \ \ \ \ {\rm as}\ \ \ \ \lambda^2\to+\infty\ ,
 \eea
 where
 \bea\label{ieurweiukj}
  {H}(z)&\asymp&2\pi\cot\big({\textstyle\frac{\pi n}{2}}\big)\  z^{\frac{n}{2}}+
 \sum_{m=1}^\infty \, {\tilde {\mathfrak g}}_{m}\ \cos\big({\textstyle\frac{2\pi  m}{n+2}}\big)\ z^{-\frac{n m}{n +2}}+O(z^{-\infty})\nonumber\\
 {G}(z)&\asymp&2\pi\, z^{\frac{n}{2}}+\sum_{m=1}^\infty\, {\tilde  {\mathfrak g}}_{m}\ 
 \sin\big({\textstyle\frac{2\pi  m}{n+2}}\big)\ z^{-\frac{nm}{n +2}}+O(z^{-\infty})\ .
\eea
For even $n$, the first term in the formal power  series  $H(z)$ should be replaced by
$2 z^\frac{n}{2}\ \log(z)$.

\subsection{Basic facts about  the quantum  cigar\label{sec2.3}}

In  the previous subsection, we described the formal algebraic construction
of the chiral transfer-matrices. Here we briefly discuss how $\tau_j(\lambda)$ can be understood as 
operators  in the quantum cigar NLSM (for more details on the  quantum cigar see, e.g., ref.\cite{Lukyan}).

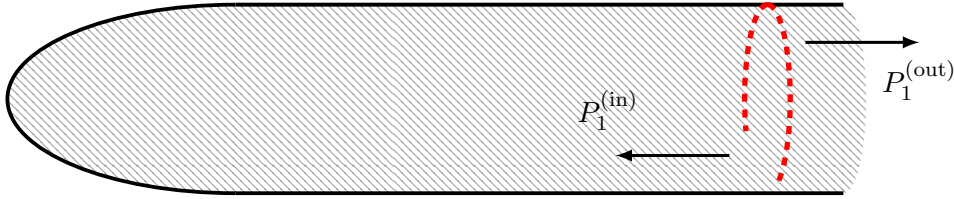
\begin{figure}
\centering
\scalebox{1}{
\begin{tikzpicture}
\draw[line width = 0.5mm] (2,0) -- (10,0);
\draw[line width = 0.52mm] (2,2.5) -- (10,2.5);
\draw[line width = 0.5mm] (2,1.25) [partial ellipse =90:270:3cm and 1.25cm];
\fill[pattern=north west lines, opacity = 0.6] (2,0) rectangle (10,2.5);
\fill[pattern = north west lines,opacity = 0.6] (2,1.25) [partial ellipse =90:270:3cm and 1.25cm];
\fill[pattern = north west lines,opacity = 0.6] (10,1.25) [partial ellipse =-90:90:0.3cm and 1.25cm];
\draw[line width = 0.7mm,dashed,red] (9,1.25) [partial ellipse = 90:200:0.3cm and 1.25cm];
\draw[line width = 0.7mm,dashed,red] (9,1.25) [partial ellipse = -60:90:0.3cm and 1.25cm];
\draw[->,line width = 0.4mm] (8.5,0.5) -- (7,0.5);
\draw[->,line width = 0.4mm] (9.5,2) -- (11,2);
\node at (6.9,1.1) {$P_1^{\rm (in)}$ };
\node at (11,1.5) {$P_1^{\rm (out)}$};
\end{tikzpicture}
}
\caption{The classical scattering problem in the cigar NLSM.
 From the asymptotically flat domain the string approaches the tip, scatters and then escapes back to the flat region. 
 After the scattering process the zero mode momentum changes sign.}
\label{fig4}
\end{figure}

 The cigar NLSM was introduced before at the classical level by means of the Lagrangian \eqref{iisaoos}.
In  the classical field theory, it is natural to consider  the following scattering problem.
 Suppose that at $t\to -\infty$  we are given the field configuration within 
the asymptotically flat domain of the target manifold, i.e., 
\bea\label{hssysay}
\phi(t,x)|_{t\to -\infty}&\asymp&  \phi^{(\rm in)}_0+
\frac{4\pi }{R}\ {  P}^{(\rm in)}_1\, t+
\sum_{m\not= 0}\frac{\ri }{m}\ \big(\,a^{\rm (in)}_m\ \re^{-\frac{2\pi\ri m}{R} (t+x)}
+{\bar a}^{\rm (in)}_m\ \re^{-\frac{2\pi \ri m}{R} (t-x)}\,\big)\\
\alpha(t,x)|_{t\to -\infty}&\asymp& \alpha^{(\rm in)}_0+
\frac{2\pi }{R}\  (k x+{\tilde k} t)+
\sum_{m\not= 0}\frac{\ri}{m}\ \big(\,b^{\rm (in)}_m\ \re^{-\frac{2\pi\ri m}{R} (t+x)}
+{\bar b}^{\rm (in)}_m\ \re^{-\frac{2\pi\ri m}{R} (t-x)}\,\big)\ . \nonumber
\eea
In writing this equation, we took into account the boundary conditions \eqref{aisosaio}. Also,
 the constant ${\tilde k}$ is the  conserved  charge for the Noether $U(1)$-current associated with the Lagrangian \eqref{iisaoos}.
The set,
${\cal A}^{{\rm (in)}}=\{ \phi^{(\rm in)}_0,\,P_1^{\rm(in)},\,\alpha^{(\rm in)}_0,\,\tilde{k},\, a^{\rm (in)}_m, \,b^{\rm (in)}_m\}$,
  can be interpreted as  a classical  ``in-state''  for a
string propagating
on the target manifold (see fig.\,\ref{fig4}).  
The nontrivial interaction occurs at some finite time  when the fields  take
values in the vicinity of the tip of the cigar. After  scattering at the tip, as $t\to+\infty$,  the field configuration
returns to the asymptotically flat domain and  takes the same form as in the  r.h.s. of \eqref{hssysay} with  the  in-state  ${\cal A}^{{\rm (in)}}$
  replaced by  the  out-state
 ${\cal A}^{{\rm (out)}}=\{ \phi^{(\rm out)}_0,\,P_1^{\rm(out)},$
 $\,\alpha^{(\rm out)}_0,\,\tilde{k},\, a^{\rm (out)}_m, \,b^{\rm (out)}_m\}$.\footnote{Strictly speaking, the winding number
 $k$ is only conserved modulo an  integer, i.e., $k^{\rm ( out)}-k^{\rm (in)}\in {\mathbb Z}$. 
 Here  we ignore this and assume that $k^{\rm ( out)}=k^{\rm (in)}\in (-\half,+\half]$.}
 %\bea
%\phi(t,x)|_{t\to +\infty}&\asymp&  \phi^{(\rm out)}_0+
%\frac{4\pi }{R}\ {  P}^{(\rm out)}_1\, t+
%\sum_{m\not= 0}\frac{\ri }{m}\ \big(\,a^{\rm (out)}_m\ \re^{-\frac{2\pi\ri m}{R} (t+x)}
%+{\bar a}^{\rm (out)}_m\ \re^{-\frac{2\pi \ri m}{R} (t-x)}\,\big)\\
%\alpha(t,x)|_{t\to -\infty}&\asymp& \alpha^{(\rm out)}_0+
%\frac{2\pi }{R}\  (k\, x+{\tilde k} t)+
%\sum_{m\not= 0}\frac{\ri }{m}\ \big(\,b^{\rm (out)}_m\ \re^{-\frac{2\pi\ri m}{R} (t+x)}
%+{\bar b}^{\rm (out)}_m\ \re^{-\frac{2\pi\ri m}{R} (t-x)}\,\big)\ .\nonumber
%\eea
The classical scattering problem can be formulated  as the problem of  finding the canonical transformation
 which maps ${\cal A}^{{\rm (in)}}$ to 
${\cal A}^{{\rm (out)}}$.
It turns out that the theory possesses two infinite sets of  left- and right-currents \cite{Bakas:1991fs}, i.e.,
\bea\label{aisiosas}
\partial_- W_s=0\ ,\ \ \ \ \ \ \ \partial_+\overline {W}_s=0\,,\ \ \ \ \  \ (s=2,\,3,\ldots)\ \, ,
\eea
so that the classical dynamics of the fields are strongly constrained. In particular,  the magnitude of the zero-mode momentum
remains unchanged after the scattering (see fig.\,\ref{fig4}),
\bea
{  P}^{(\rm out)}_1=-{  P}^{(\rm in)}_1\ .
 \eea

\bigskip

Consider now  the quantum theory.
First of all we note that  the  value of the $U(1)$ charge is quantized so that $(n+2)\,{\tilde k}=m\in \mathbb{Z}$.
Thus the space of  states  of the quantum theory  splits into orthogonal subspaces 
${\cal H}_{k,m}$ labeled  by the twist parameter $k$ and the integer $m$.
The quantum theory still possesses the chiral currents satisfying eqs.\,\eqref{aisiosas}. As a result, 
${\cal H}_{k,m}$  can be decomposed into 
the highest weight  irreps
of the  $W$-algebra, ${\mathfrak W}\otimes {\bar {\mathfrak W}}$
generated by the fields $W_s$ and ${\bar W}_s$ \cite{Bakas:1991fs}.
 % having the Lorentz spin $(+s)$ and $(-s)$, respectively.
Let  ${\cal V}_h$ $({\bar {\cal V}}_{\bar h})$  
be   the  highest weight representation of the chiral  
$W$-algebra ${\mathfrak W}$  $({\bar {\mathfrak W}})$  labeled by the
highest weight $h$ $({\bar h})$. Then, schematically,
\begin{equation}\label{aisasisao}
{\cal H}_{k,m}\,=\,\begin{array}{c} \\[-0.1cm]  \oplus \\[-0.15cm] \scriptstyle{\{h,\bar{h}\}} 
\end{array}{\cal V}_h\,\otimes \, \overline{{\cal V}}_{\bar{h}} \, .
\end{equation}
The  highest weight $h$ can be chosen  to be a   pair of  numbers $(\Delta, w)$, where $\Delta$ coincides with
the conformal dimensions of the highest weight   vector, while $w$ is the  eigenvalue of the  dimensionless
conserved charge $R^2\int_0^R\rd x\, W_3(x)$, and similar for ${\bar h}$.
Let us first  focus on the ``left'' component  ${\cal V}_h$ in the tensor product
 ${\cal V}_h\otimes \overline{  {\cal V}}_{\bar h}$.

It should be clear that the quantum counterpart  to the left components of the in-asymptotic fields \eqref{hssysay}
can be identified with the fields $\varphi_+$ and $\vartheta_+$ given by \eqref{iaisia}. 
Since the quantum   fields  $W_s $  are chiral currents, i.e. $W_s(t,x)=W_s(t+x)$, they can be    expressed in terms 
of  the asymptotic fields $\varphi_+$ and  $\vartheta_+$. Indeed, for given $s$,  $W_s$ is
a  certain  order-$s$ {  homogeneous} polynomial with constant coefficients  w.r.t.  the fields
 $\varphi'_+$, $\vartheta'_+$ and their higher derivatives
(in other words, any monomials appearing within  $W_s$  contains exactly $s$ derivative symbols).
This implies that 
the Fock space ${\cal F}_{\bf p}$, which is the space  of representation for the fields $\varphi'_+$, $\vartheta'_+$,
possesses the structure of the highest weight representation of the chiral $W$-algebra.
 It turns out   that for real ${\bf p}$,   the Fock space ${\cal F}_{\bf p}$
 coincides with irrep ${\cal V}_{h}$  as a linear space, provided that $h=(\Delta, w)$ is related to ${\bf p}=(p_1,p_2)$ as follows
\bea\label{jssa}
&&\Delta(p_1,p_2)=\frac{p_1^2}{n}+\frac{p^2_2}{n+2}+\frac{1}{4n}\\
&&w(p_1,p_2)=p_2\ \bigg(  p_1^2+\frac{3n+2}{3(n+2)}\ p_2^2-\frac{2n+1}{12}\bigg)\ .\nonumber
\eea
In fact, one can use these formulae to conveniently parameterize the highest weight $h$  by the pair
$(p_1,p_2)$: ${\cal V}_{h}\equiv {\cal V}_{p_1,\,p_2}$. With this notation,  a more accurate version  of  eq.\,\eqref{aisasisao} reads as
\bea\label{aiossao}
{\cal H}_{k,m}={\int\limits_{p_1<0}}^{\!\!\oplus}\  {\cal V}_{p_1,\,p_2}\otimes \overline{{\cal V}}_{p_1,\,{\bar p}_2}\ ,
\eea
where
\bea
p_2=\half\, \big( \,m+  (n+2)\,k\, \big)\ , \ \ \ \ \ \ {\bar p}_2=\half\, \big(\, m-  (n+2)\,k\, \big)\ .
\eea
The direct integral in \eqref{aiossao} does not include the domain with positive $p_1$, since, as follows from eqs.\,\eqref{jssa}, 
${\cal V}_{p_1,\,p_2}\equiv {\cal V}_{-p_1,\,p_2}$.

A basis of   in-asymptotic states in ${\cal H}_{k,m}$  is formed  by
\bea
a^{(\rm in)}_{-m_1}\ldots a^{(\rm in)}_{-m_N}\, {\bar a}^{(\rm in)}_{-{\bar m}_1}\ldots {\bar a}^{(\rm in)}_{-{\bar m}_{\bar N}}\, 
{b}^{(\rm in)}_{-m_1}\ldots b^{(\rm in)}_{-m_M}\, {\bar b}^{(\rm in)}_{-{\bar m}_1}\ldots {\bar b}^{(\rm in)}_{-{\bar m}_{\bar M}}\, |\,{\rm vac}\,\rangle
\eea
and can be identified with the states from the tensor product of the Fock space ${\cal F}_{p_1,\,p_2}\otimes \bar{ { \cal F}}_{p_1,\,{\bar p}_2}$:
\bea\label{aiosaosao}
a_{-m_1}\ldots a_{-m_N}\, {\bar a}_{-{\bar m}_1}\ldots {\bar a}_{-{\bar m}_{\bar N}}\,
b_{-m_1}\ldots b_{-m_M}\, {\bar b}_{-{\bar m}_1}\ldots {\bar b}_{-{\bar m}_{\bar M}}\,  |\,p_1,\,p_2\,\rangle\otimes |\,p_1,\,{\bar p}_2\,\rangle\ .
\eea
Similarly for the out-states, one has
\bea
&&a^{(\rm out)}_{-m_1}\ldots a^{(\rm out)}_{-m_N}\,{\bar  a}^{(\rm out)}_{-{\bar m}_1}\ldots {\bar a}^{(\rm out)}_{-{\bar m}_{\bar N}}\,
b^{(\rm out)}_{-m_1}\ldots b^{(\rm out)}_{-m_M}\, {\bar b}^{(\rm out)}_{-{\bar m}_1}
\ldots {\bar b}^{(\rm out)}_{-{\bar m}_{\bar M}}\, |\,{\rm vac}\,\rangle\sim\\
&&a_{-m_1}\ldots a_{-m_N}\, {\bar a}_{-{\bar m}_1}\ldots {\bar a}_{-{\bar m}_{\bar N}}\,
b_{-m_1}\ldots b_{-m_M}\, {\bar b}_{-{\bar m}_1}\ldots {\bar b}_{-{\bar m}_{\bar M}}\,  |\,-p_1,\,p_2\,\rangle\otimes |\,-p_1,\,{\bar p}_2\,\rangle\ .\nonumber
\eea
Usually, the $S$-matrix  is  introduced as a  unitary operator which relates  the in- and  out- asymptotic bases.
In the case under consideration, the $S$-matrix can be interpreted as the intertwiner acting between the Fock spaces:
\bea
{\hat S}\ : \ \ {\cal F}_{p_1,\, p_2}\otimes {\bar { \cal F}}_{p_1,\,{\bar p}_2}\mapsto  {\cal F}_{-p_1,\,p_2}\otimes {\bar { \cal F}}_{-p_1,\,{\bar p}_2}\ .
\eea
It turns out that  the operator ${\hat S}$ has the following  structure
\bea\label{ahsassauasu}
{\hat S}=S_0({\bf p})\ {\hat S}_L\otimes  {\hat S}_R\ ,
\eea
where ${\hat S}_L$
  intertwines the level subspaces, 
${\hat S}_L\, :\ {\cal F}^{{({L})}}_{p_1,\,p_2}\mapsto {\cal F}^{{({L})}}_{-p_1,\,p_2}$, and is  normalized  by the condition
${\hat S}_L\,|\,p_1,p_2\,\rangle=|-p_1,\,p_2\,\rangle$, and  similarly for ${\hat S}_R$.
For a given level $\ell$, the construction of the operators ${\hat S}_{L,R}$ is a straightforward algebraic
task.  The  more delicate problem is finding  the overall  scalar factor $S_0({\bf p})$.  
It was obtained in the minisuperspace approximation  in ref.  \cite{Dijkgraaf:1991ba}.
The exact form of $S_0({\bf p})$ has been known since the unpublished work   of  the Zamolodchikov brothers  \cite{ZAM}.
%This problem was originally solved in the
%unpublished work of  the Zamolodchikov brothers. 

Returning to the chiral transfer-matrices,  let us note that these operators should act in the Hilbert space of the quantum cigar, and therefore
their action should commute with the intertwiner ${\hat S}$:\footnote{The intertwiner
${\hat S}$ should  not to be confused with the so called ``reflection'' operator ${\hat R}:\, 
{\cal F}_{p_1,\, p_2}\otimes {\bar {\cal F}}_{p_1,\,{\bar p}_2}\mapsto  {\cal F}_{p_1,\,p_2}\otimes
 {\bar { \cal F}}_{p_1,\,{\bar p}_2}$,
and  $[{\hat R},\tau(\lambda)]=0$. Note that $ {\hat  R}={\hat \sigma}\circ \,{\hat S}$ where 
${\hat \sigma}={\hat \sigma}_L\otimes{\hat \sigma}_R$ and the chiral intertwiners 
${\hat \sigma}_L:\,  {\cal F}^{({L})}_{p_1,\, p_2}\mapsto {\cal F}^{{(L})}_{-p_1,\, p_2}$ are defined by the conditions
${\hat \sigma}_L a_m=- a_m{\hat \sigma}_L,\, { \hat \sigma}_L b_m=+ b_m{\hat \sigma}_L,\ 
{\hat \sigma}_L\, |\,p_1,p_2\,\rangle=|-p_1,p_2\,\rangle$, and similar for ${ \hat \sigma}_R$ (see, e.g.,\,\cite{Lukyan}).}
\bea
{\hat S}\, \tau(\lambda)= \tau(\lambda)\ {\hat S}\ .
\eea
In practice, this condition implies that all matrix elements of the (dual) nonlocal IM in the basis of Fock states \eqref{aiosaosao}
are even functions of $p_1$ (for illustration see eqs.\,\eqref{jssusau},\,\eqref{asuausai}). 

\bigskip
 The quantum cigar  also possesses an infinite set of the so-called local IM acting in ${\cal H}_{k,m}$.
To get some feeling for these operators,  we need to remind ourselves of an important feature of the model. Namely,
it admits an  equivalent ``dual'' description in terms of the so-called sine-Liouville model. The dual Lagrangian
is given by \cite{ZAM}
\bea\label{sasasa}
{\cal L}^{\rm (dual)}={\textstyle \frac{1}{4\pi}}\ \big(\, (\partial_\sigma\varphi)^2+ (\partial_\sigma\vartheta)^2\big)+2 \mathpzc{M}\, 
\re^{-\sqrt{n}\varphi}\ \cos\big(\sqrt{n+2}\, \vartheta\big)\,,
\eea
with the sine-Liouville fields satisfying  the boundary conditions
\bea\label{aisaisoa}
\varphi(t, x+R)=\varphi(t, x)\ ,\ \ \ \ \ 
\vartheta(t,x+R)=\vartheta(t, x)+\frac{2\pi m}{\sqrt{n+2}}\ .
\eea
Notice that the ``coupling'' $\mathpzc{M}$ is  a somewhat fake
parameter of the Lagrangian --  by an additive shift $\varphi\mapsto \varphi+const$ the value of    $\mathpzc{M}$ can be
chosen to be any real number. Nevertheless, it is convenient to keep it unspecified.

To understand the relation between the fields in the NLSM and its dual description,
let us take the ``zero-mode'' of the
field $\varphi$
\bea\label{hasysat}
\varphi_0=\int_0^{R}\frac{\rd x}{R}\ \varphi(x)\ ,
\eea
and consider the region $\varphi_0\to+\infty$ in configuration space.
In this asymptotic domain, the potential term in the action \eqref{sasasa} can be neglected and
$\frac{\varphi}{\sqrt{n}}\asymp \phi+const$, while    $ \frac{\vartheta}{\sqrt{n+2}}$
can   be identified  with  ${\tilde \alpha}$ -- the field from the cigar NLSM defined by the 
 relation $J_\mu=\epsilon_{\mu\nu}\partial_\nu{\tilde \alpha}$, where $J_\mu$  stands for  the 
Noether $U(1)$-current.

The twist parameter $k$  has   a natural interpretation in the dual description -- it can be identified with the
so-called quasimomentum.
The sine-Liouville Lagrangian  is invariant under the transformation $\vartheta\mapsto \vartheta+\frac{2\pi}{\sqrt{n+2}}$.
Due to this periodicity,  the space of states of the theory 
with  the boundary conditions \eqref{aisaisoa}, splits on the orthogonal  subspaces ${\cal H}_{k,m}$
such that for any  state $|\, A \,\rangle\in {\cal H}_{k,m}$, the corresponding wave functional 
$\Psi_{A}[\,\varphi(x),\vartheta(x)\,]$
transforms
as
\bea
\Psi_{A}\big[\,\varphi(x),\vartheta(x)+{\textstyle \frac{2\pi}{\sqrt{n+2}}}\,\big]=
 \re^{2\pi\ri k}\ \Psi_{A}[\,\varphi(x),\vartheta(x)\,]\ .
\eea
%Notice also that, 
%in the dual description, the $S$-matrix \eqref{ahsassauasu} is a  complete analog of the  reflection $S$-matrix in the quantum Liouville theory [].

\bigskip
Let $P_{s}(\partial _+{ \varphi},\ \partial _+{ \vartheta},\ldots)$
 be a local
field of spin $s$, and
a polynomial of
$\partial_+ { \varphi}, \,\partial_+{\vartheta}$
and their higher derivatives.
All  such fields are periodic in $x$, so that
one can introduce the integral,
\bea\label{ospssapo}
{\mathfrak i}_{s-1}=\Big(\frac{R}{2\pi}\Big)^{s-1}\ \int_0^{R}\frac{\rd x}{2\pi}\ P_{s}(\partial _+{ \varphi},\ \partial _+{ \vartheta},\ldots)\ .
\eea
It turns out that for any even $s=2 j$ there exists a local density (defined modulo the addition of a  total derivative and an overall
multiplicative constant)   such  that ${\mathfrak i}_{2j-1}$ is an  integral  of  motion and satisfies   the commutativity conditions
 \bea\label{hssaast}
 [\,{\mathfrak i}_{2j-1},\,\tau(\lambda)\,]= [\,{\mathfrak i}_{2j-1},\, {\mathfrak i}_{2j'-1}\,]=0\ .
\eea
These operators are referred to as the  (chiral)  local IM.
They were studied in ref.\cite{Lukyanov:2003nj}, where the explicit form for the
 first local IM  and their  vacuum eigenvalues, $i_{2j-1}(p_1,p_2)$ for $j=1,2,3$, can be found.
Here we only note that for any $j=1,\,2,\,\ldots$
\bea\label{jasssay}
P_{2j}= \sum_{l+m=j } C^{(j)}_{lm}\ (\partial_+\varphi)^{2 l}(\partial_+\vartheta)^{2m }+\ldots\ ,
\eea
where the dots stand for monomials which include higher derivatives  of
$\partial _+\varphi$ and  $\partial_+ \vartheta$ and the numerical coefficients
 $ C^{(s)}_{lm}$ can be written as
\beq\label{sausuas}
 C^{(j)}_{lm}=C_{2j-1}\  \frac{(-2)^{j+1} (2j-2)!}{(j+1)!}\ 
\frac{\big( (n+2) (\frac{1}{2}-j)\big)_l\,\big( (-n) (\frac{1}{2}-j)\big)_{m}}{l!\, m!}\ (-n)^{l-1}\,(n+2)^{m-1}\ .
\eeq
Here   $(a)_m=\prod_{i=0}^{m-1}(a+i)$ is the Pochhammer symbol.
 The overall normalization constant  $C_{2j-1}$ is  usually 
set to
\bea\label{kasausau}
C_{2j-1}=\frac{ 2^{-3j}\ (j+1)!\  n (n+2)}{\big( (n+2) (\frac{1}{2}-j)\big)_j\,\big( (-n) (\frac{1}{2}-j)\big)_{j}}\ .
\eea

\section{Chiral transfer-matrix  for ${\mathbb Z}_n$ parafermions}

While quantizing the sausage model within the BLZ approach, we have run into the  problem
of finding the spectrum of the chiral transfer-matrices  for the cigar  NLSM. As it has been  explained,
we can consider $\tau(\lambda)$ as
an operator acting in the Fock space ${\cal F}_{\bf p}$ with real ${\bf p}=(p_1,p_2)$.  
From the formal point of view, the same  spectral 
problem can be posed for any complex values of  ${\bf p}$. Notice, that for
real $p_2$, $\lambda^2$ and 
 pure imaginary  $p_1$, 
the operator $\tau(\lambda)$ is Hermitian.
The spectral problem  in this case (except  for $p_1=0$)
is not directly related to the  quantization of the sausage model,
 however for $n=2,\,3,\ldots$  and  a certain  discrete  set of  $p_1$ and $p_2$,
  it gives a better understanding of  the interplay between  the BLZ approach
and that based on the discretization of the quantum system.
It will be the subject of our study here.

\subsection{Bosonization of  ${\mathbb Z}_n$ parafermions}

Let us take a closer look at the vertex operators  $V^{(\pm)}$\ \eqref{aisaiasoas}, which appear  in the construction
of the chiral transfer-matrices  $\tau_j$. As it was already mentioned, the  constants $c^\pm$ can be arbitrarily chosen.
Let us set $c^\pm=1$ and assume that  $n\geq 2$ is a positive integer.
Then, eq.\,\eqref{aisaiasoas} can be recognized as the bosonization relations for the 
Fateev-Zamolodchikov  ${\mathbb Z}_n$ parafermions \cite{Gerasimov:1989mz}.
More precisely, as follows from the normalization condition \eqref{sasossiaoas},
the chiral nonlocal fields
\bea\label{asiosasaoia}
\psi^\pm=\sqrt{n}\  q^{-\frac{1}{2}}\ V^{\pm}
\eea
can be understood as canonically normalized  parafermion currents,
\bea
\psi^\pm(x_2)\,\psi^\mp(x_1)\big|_{ x_2\to x_1+0}\,\sim 1\times \big(x_2-x_1\big)^{-2\Delta_\psi}
\eea
of the conformal dimension $\Delta_\psi=1-\frac{1}{n}$.

The chiral algebra of parafermion currents was introduced by Fateev and Zamolodchikov in ref.\cite{Fateev:1985mm},
in the construction of the ${\mathbb Z}_n$ CFT models with  central charge
 \bea\label{sisias}
 c_n=\frac{2(n-1)}{n+2}\ ,
 \eea
 describing the multicritical points of the  ${\mathbb Z}_n$ statistical systems (certain
 generalizations
of the ${\mathbb Z}_2$  invariant Ising model) \cite{Fateev:1982wi}.
The chiral component of the Hilbert space
 of the ${\mathbb Z}_n$ CFT
 can be decomposed into  irreps ${\cal V}_{\mathfrak j}$  of the chiral algebra.  Here,  the subscript ${\mathfrak j}$  stands 
 for the highest weight of the irrep with highest weight vector $|\,\sigma_{\mathfrak j}\,\rangle$
 having  
 conformal dimension
 \bea\label{hssaysasy}
 \Delta_{\mathfrak j}=\frac{{\mathfrak j}\,(n-2 {\mathfrak j})}{n\,(n+2)}\ .
\eea 
 The admissible values of ${\mathfrak j}$ are given by non-negative integers and half-integers  restricted  by the condition
 \bea
 {\mathfrak j}=0,\,\half,\,1,{\textstyle\frac{3}{2}},\ldots,\, \half\, \big[{\textstyle \frac{n}{2}}\big]\ ,
 \eea
where $\big[{\textstyle \frac{n}{2}}\big]$ is the integer part of $n/2$. 
 The fundamental parafermion currents $\psi^+$ and $\psi^-$ act in   ${\cal V}_{\mathfrak j}$ and carry  the 
 ${\mathbb Z}_n$-charges $+2$ and $-2$ respectively:
 \bea
 \Omega\, \psi^{\pm}\,  \Omega^{-1}=\omega^{\pm 2}\,  \psi^{\pm}\ ,\ \ \ \  \ \ \ \ \ {\rm where}\ \ \ \ \ \ \ \omega=\re^{-\frac{2\pi\ri}{ n}}\ .
 \eea
Note that ${2\mathfrak j}$ can be identified with the  ${\mathbb Z}_n$-charge of the
 highest weight vector:\footnote{To be more precise, 
  the  chiral component of the Hilbert space
 of the ${\mathbb Z}_n$ CFT   contains, together with the irrep ${\cal V}^{(+)}_{\mathfrak j}\equiv  {\cal V}_{\mathfrak j}$,
 the   irrep ${\cal V}^{(-)}_{\mathfrak j}$ whose
  highest weight vector  has the same conformal dimension  \eqref{hssaysasy} but carries
   the ${\mathbb Z}_n$-charge $\omega^{-2{\mathfrak j}}$.  For even $n$, ${\cal V}^{(-)}_{\frac{n}{4}}={\cal V}^{(+)}_{\frac{n}{4}}$.
   The chiral transfer-matrix \eqref{aoss} is a ${\mathbb Z}_2$-invariant 
   operator which does not distinguish between the irreps ${\cal V}^{(+)}_{\mathfrak j}$ and  ${\cal V}^{(-)}_{\mathfrak j}$. }
 \bea
 \Omega\ |\,\sigma_{\mathfrak j}\,\rangle=\omega^{2{\mathfrak j}}\, |\,\sigma_{\mathfrak j}\,\rangle\ .
 \eea
The irrep  ${\cal V}_{\mathfrak j}$  naturally splits   on the subspaces ${\cal V}_{\mathfrak j}^{(\mathfrak m)}$ characterized
by a  definite value of the ${\mathbb Z}_n$-charge,
\bea
{\cal V}_{\mathfrak j}=\Big[\oplus_{s=0}^{2{\mathfrak j}}\, {\cal V}^{(2{\mathfrak j}-2s)}_{\mathfrak j}\Big]\oplus
\Big[\oplus_{s=1}^{n-2{\mathfrak j}-1}\, {\cal V}^{(2{\mathfrak j}+2s)}_{\mathfrak j}\Big]
\ :\ \ \ 
 \Omega\, {\cal V}_{\mathfrak j}^{({\mathfrak m})}=
 \omega^{{\mathfrak m}}\  {\cal V}_{\mathfrak j}^{({\mathfrak m})}\ .
 \eea
 The lowest possible  conformal dimension in  the subspace ${\cal V}_{\mathfrak j}^{({\mathfrak m})}$ is given
 by $\Delta_{{\mathfrak j}, {\mathfrak m}}$ for  ${\mathfrak m}=-2{\mathfrak j},\, -2{\mathfrak j}+2,\,\ldots, 2{\mathfrak j}$, and
 $\Delta_{{\mathfrak j}, {\mathfrak m}}+\half\, ({\mathfrak m}-2{\mathfrak j})$ for  ${\mathfrak m}=2{\mathfrak j}+2,\ldots,2 n-2{\mathfrak j}-2$.
 Here we use the notation
 \bea\label{aoosaoas}
\Delta_{{\mathfrak j}, {\mathfrak m}}=\frac{{\mathfrak j}({\mathfrak j}+1)}{n+2}-\frac{{\mathfrak m} ^2}{4n}\ .
\eea
In what follows  $|\,\sigma_{{\mathfrak j},{\mathfrak m}} \,\rangle$  will  denote the state from  the subspace
${\cal V}_{\mathfrak j}^{({\mathfrak m})}$  with ${\mathfrak m}=2{\mathfrak j},\, 2{\mathfrak j}-2,\,\ldots,- 2{\mathfrak j}$
of the lowest  conformal dimension $\Delta_{{\mathfrak j}, {\mathfrak m}}$.

From the mathematical point of view the bosonization   of the algebra of parafermion currents implies that
the subspaces ${\cal V}_{\mathfrak j}^{({\mathfrak m})}$   with ${\mathfrak m}=2{\mathfrak j},\, 2{\mathfrak j}-2,\,\ldots,- 2{\mathfrak j}$
can be understood  as a cohomology of the Fock space
${\cal F}_{p_1,p_2}$ where 
\bea\label{asjsasai}
p_1={\textstyle\frac{\ri}{2}}\ {\mathfrak m} \, ,\ \ \ \ \  \ \ p_2={\mathfrak j}+\half\ ,
\eea
with  respect to a certain BRST complex ${\grave{\rm a}}$ la the Felder  complex \cite{Felder:1988zp} involved  in the  bosonization
of  the highly reducible
Verma modules  over  the Virasoro algebra.
Among other  things, the bosonization  formula \eqref{asiosasaoia} leads to the following  relation for the matrix elements of the
parafermion currents:
\beq
\langle\, \sigma_{{\mathfrak j},{\mathfrak m}}\,|\, \prod_{m=1}^M\psi^{\varepsilon_m}(x_m)\,|\,  \sigma_{{\mathfrak j},{\mathfrak m}}
\,\rangle= (n q^{-1})^{\frac{M}{2}}\ \langle\, p_1,\,p_2\,|\,  \prod_{m=1}^M V^{\varepsilon_m}(x_m)\,|\, p_1,\,p_2\,\rangle\ \ \ \ 
(\varepsilon_m=\pm)\ ,
\eeq
provided $\sum_{m=1}^L\varepsilon_m=0$ and the pairs $({\mathfrak j},\,{\mathfrak m})$ are related
to $(p_1,p_2)$ as in eq.\,\eqref{asjsasai}.
It is not difficult to see  that the ${\mathbb Z}_n$-charge operator  is bosonized by the relation
\bea
\Omega=\re^{4\pi P_1}
\eea
and  the operator $\tau(\lambda)$ can be written in the form
 \bea\label{aoss}
{ { \tau}}({ \lambda})={\rm Tr}\bigg[\,
{\overset{\leftarrow} {\cal P}}
\exp\bigg( \ri\, {\textstyle \frac{ {  \lambda}_+}{\sqrt{n}}}\, \int_{t_0}^{t_0+R} \rd x\ \big(\,{ \psi}^{+}\ 
  {\sigma}_++\psi^-\  {\sigma}_-\big) \bigg)\ 
\Omega^{-{\small\frac{1}{2}}\,\sigma_3}\,\bigg]\ .
\eea
Therefore, $\tau(\lambda)$ can be understood  as an operator which  invariantly acts in the 
subspaces ${\cal V}_{\mathfrak j}^{({\mathfrak m})}$ of the irrep ${\cal V}_{\mathfrak j}$ of the 
algebra of parafermion currents.  We can now address the problem of the diagonalization of this operator.
Notice that  it is sufficient to consider  ${\mathfrak m}\geq 0$, and in what follows we will always assume that
\bea\label{siaisaiais}
{\mathfrak j}&=&0,\,\half,\,1,{\textstyle\frac{3}{2}},\ldots,\, \half\, \big[{\textstyle \frac{n}{2}}\big]\nonumber\\[0.2cm]
{\mathfrak m}&=&2\, {\mathfrak j},\, 2\,{\mathfrak j}-2,\, \ldots , 2\, {\mathfrak j}-2\, [{\mathfrak j}]\ .
\eea

\subsection{Discretization of the chiral transfer-matrix}
The goal of this section is to propose a lattice version of the parafermionic  chiral transfer-matrix \eqref{aoss}. For this purpose
we   return back to   the formula \eqref{oisasasp} for the classical 
conserved charges in  the sausage model and follow the approach based on discretization that was mentioned in the introduction.
%\bea
%%%T_j(\mu)={\rm Tr}\Big[\re^{-\ri \pi k {\tt h}}\  {\boldsymbol M}_j(\mu)\,\Big]\ ,\ \ \  \
%%%{\boldsymbol M}_j=\pi_j\bigg[{\overset{\leftarrow} {\cal P}}\exp\int_C\rd x\, {\boldsymbol  A}_x
%\bigg]\ .\nonumber
%\eea

\begin{figure}
\centering
\begin{tikzpicture}
\draw [line width = 0.5mm,->] (-1,0) -- (12,0);
\draw [line width = 0.5mm,->] (0,-1) -- (0,3);
\draw [dashed] (10,0) -- (10,3);
\draw[thick] (0,0) -- (1,1);
\draw[line width=0.9mm,->] (0.6,0.6)--(0.7,0.7);
\draw[thick] (1,1) -- (2,0);
\draw[line width=0.9mm,->] (1.6,0.4)--(1.7,0.3);
\draw[thick] (2,0) -- (3,1);
\draw[line width=0.9mm,->] (2.6,0.6)--(2.7,0.7);
\draw[thick] (3,1) -- (4,0);
\draw[line width=0.9mm,->] (3.6,0.4)--(3.7,0.3);
\draw[thick] (4,0) -- (5,1);
\draw[line width=0.9mm,->] (4.6,0.6)--(4.7,0.7);
\draw[thick] (5,1) -- (6,0);
\draw[line width=0.9mm,->] (5.6,0.4)--(5.7,0.3);
\draw[thick] (6,0) -- (7,1);
\draw[line width=0.9mm,->] (6.6,0.6)--(6.7,0.7);
\draw[thick] (7,1) -- (8,0);
\draw[line width=0.9mm,->] (7.6,0.4)--(7.7,0.3);
\draw[thick] (8,0) -- (9,1);
\draw[line width=0.9mm,->] (8.6,0.6)--(8.7,0.7);
\draw[thick] (9,1) -- (10,0);
\draw[line width=0.9mm,->] (9.6,0.4)--(9.7,0.3);
\draw[line width=0.9mm,->] (5.2,0)--(5.25,0);
\node [left]  at (-1,0) {\Large $t_0$};
\node [below] at (10,0) {\Large $ R$};
\node [below right] at (0,0) {\Large $0$};
\node [right = 0.2cm] at (0,2.8) {\Large $t$};
\node [above] at (11.3,0) {\Large $x$};
\node at (4.3,1.3) {\Large $\boldsymbol{L}^{\rm ( s)}$};
\node at (6,1.3) {\Large $\bar{\boldsymbol{L}}^{\rm ( s)}$};
\node at (5,-0.8) {\Large $\boldsymbol{M}$};
\end{tikzpicture}
\caption{By replacing the integration over the segment   at the time slice $t=t_0$ by integration over the light-cone 
pieces, the monodromy matrix can be expressed as a product of the elementary ``light-cone'' transport matrices:
${\boldsymbol M}(\mu)={\overset{\leftarrow}\prod}_{s=1}^{N}{\bar {\boldsymbol L}}^{(\rm s)}(\mu)\,
 {\boldsymbol L}^{(\rm s)}(\mu)$. }
\label{fig05}
\end{figure}
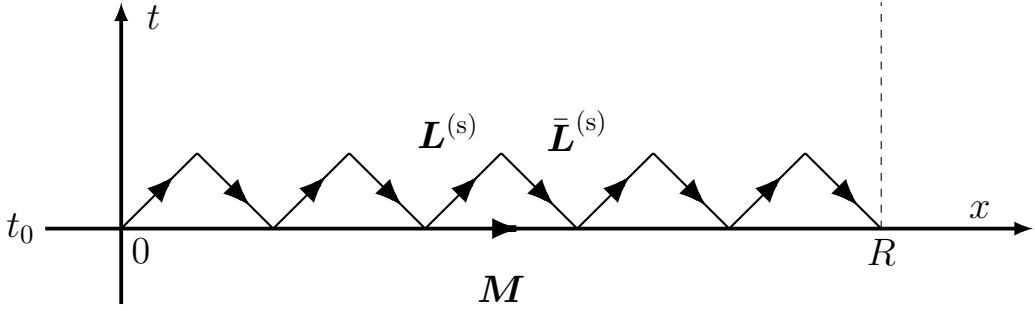

Split the integration contour onto  $N$ small segments of size $\delta$ and consider 
the elementary transport matrices in the fundamental representation:
\bea
{\boldsymbol M}^{(\rm s)}(\mu)=\pi_{\text{\textonehalf}}\bigg[\,
{\overset{\leftarrow} {\cal P}}\exp\int_{x_{\rm s}-\scriptscriptstyle{{\delta}/{2}}}^{x_{\rm s}+\scriptscriptstyle{{\delta}/{2}}} \rd x\, {\boldsymbol A}_x\,\bigg]
\ \ \ \ \ \ \ \ \ \ \ \ 
({\rm s}=1,\ldots, N)\ .
\eea
These  can be expressed in terms of   the  elementary  ``light-cone'' transport matrices
% ${\boldsymbol L}^{(a)}$ and
%${\bar {\boldsymbol L}}^{(a)}$
$
 {\boldsymbol M}^{(\rm s)}(\mu)={\bar {\boldsymbol L}}^{(\rm s)}(\mu)\,
 {\boldsymbol L}^{(\rm s)}(\mu)$
  (see fig.\,\ref{fig05})
 and,
 as it follows from eq.\,\eqref{hsasausa},
 \bea\label{oiwehsfd}
\big\{{\boldsymbol L}^{(\rm s)}(\mu)\begin{array}{ccc} \\[-0.3cm] \otimes \\[-0.35cm] , \end{array} 
{\boldsymbol L}^{(\rm s')}(\mu')\,\big\}&=&
\big[{\boldsymbol L}^{(\rm s)}(\mu)\,{\otimes}\,{\boldsymbol L}^{(\rm s')}(\mu')\,, 
{\boldsymbol r}_{\text{\textonehalf\textonehalf}}(\mu/\mu')\big]\ \delta_{\rm s s'}
\\
\big\{{\bar {\boldsymbol L}}^{(\rm s)}(\mu) \begin{array}{ccc} \\[-0.5cm] \otimes \\[-0.35cm] , \end{array} 
{\bar {\boldsymbol L}}^{(\rm s')}(\mu')\,\big\}&=&
\big[{\bar {\boldsymbol L}}^{(\rm s)}(\mu)\,\otimes\,{\bar  {\boldsymbol L}}^{(\rm s')}(\mu')\,, 
{\boldsymbol r}_{\text{\textonehalf\textonehalf}}(\mu/\mu')\big]\ \delta_{\rm s s'}\nonumber\\
 \big\{{\boldsymbol L}^{(\rm s)}(\mu)\begin{array}{ccc} \\[-0.5cm] \otimes \\[-0.35cm] , \end{array} {\bar  {\boldsymbol L}}^{(\rm s')}(\mu')\,\big\}&=&0\ ,\nonumber
\eea
where
\bea
{\boldsymbol r}_{\text{\textonehalf\textonehalf}}(\mu)=
\begin{pmatrix}
a(\mu)& 0&0&0\\
0&0& c(\mu)&0\\
0&c(\mu)&0&0\\
0&0&0&a(\mu)
\end{pmatrix}\ \  \qquad \text{with} \qquad \begin{array}{l} a(\mu)=\frac{1}{2}\ \frac{\mu+\mu^{-1}}{\mu-\mu^{-1}}\ \\[0.5cm] c(\mu)=\frac{1}{\mu-\mu^{-1}} \end{array}\ .
\eea

Consider  the structure of  ${\boldsymbol L}^{(\rm s)}(\mu)=
\pi_{\text{\textonehalf}}\Big[\,
{\overset{\leftarrow} {\cal P}}\exp\int\limits_{x_{\rm s}-\scriptscriptstyle{{\delta}/{2}}}^{x_{\rm s}+\scriptscriptstyle{{\delta}/{2}}} \rd x\, {\boldsymbol A}_+\,\Big]$.
From the explicit form of ${\boldsymbol A}_+$\,\eqref{kassau}, one has
\bea\label{asusaisa}
{\boldsymbol L}(\mu)=
%\pi_{\frac{1}{2}}\bigg[
%{\overset{\leftarrow} {\cal P}}\exp\bigg(\int_{x_a-\frac{\delta}{2}}^{x_a+\frac{\delta}{2}} \rd x\, {\boldsymbol A}_+\bigg)
%\bigg]=
\begin{pmatrix}
1+\frac{f(\mu)}{4}\ H &\frac{\ri}{2}\, g(\mu)\, E_{-}\\
\frac{\ri}{2}\, g(\mu)\, E_{+} & 1-\frac{f(\mu)}{4}\ H
\end{pmatrix}+O(\delta^2)\ ,
\eea
with
\bea\label{aosaoaas}
E_{\pm}=
\ \int_{x_{\rm s}-\scriptscriptstyle{{\delta}/{2}}}^{x_{\rm s}+\scriptscriptstyle{{\delta}/{2}}}\rd x \, \Pi_+(x)\,\re^{\mp Q(x)}\ ,\ \  \ \ \ \ \ 
H=2 \int_{x_{\rm s}-\scriptscriptstyle{{\delta}/{2}}}^{x_{\rm s}+\scriptscriptstyle{{\delta}/{2}}}\rd x\,\Pi_+(x)\ ,
\eea
and, as it follows from the canonical commutation relations,
\bea\label{kajasu}
\{E_{+},E_{-}\}=
-H\,, \ \ \ \ \ \ \  \{H, E_{\pm}\}=\pm  2 E_{\pm}\  \,.
\eea
Here, to simplify the notation, we have temporarily dropped the superscript ``${\rm s}$'' and are focusing on a single site.
Let's look at the above  formulae from a slightly different angle. Suppose we are given the matrices ${\boldsymbol L}$ of the form \eqref{asusaisa} with arbitrary functions $g(\mu)$ and $f(\mu)$ where $H$, $E_{\pm}$ satisfy the Poisson bracket relations \eqref{kajasu}. The requirement that  ${\boldsymbol L}$ obeys the Yang-Baxter Poisson algebra \eqref{oiwehsfd} leads  to two equations imposed on the functions $f$ and $g$:
\bea\label{jashsay}
&&a(\mu/\lambda)\, g(\lambda)-c(\mu/\lambda)\, g(\mu)=\frac{1}{2}\ g(\lambda)\, f(\mu)\\
&& c(\mu/\lambda)=\frac{1}{2}\ \frac{g(\lambda) g(\mu)}{f(\lambda)-f(\mu)}\ .\nonumber
\eea
One can show that, modulo the rescaling $\mu\mapsto const\,\mu$,  the most general solution to these equations is given by formula \eqref{iasosao} with $\kappa$ an arbitrary parameter.

This simple calculation hints as to how we should proceed with the deduction of the quantum counterpart of the above formulae. Strictly speaking, there is no canonical prescription for the quantization of the Poisson brackets \eqref{kajasu}, however, it seems natural to substitute them by the defining relations of the  $U_q(\mathfrak{sl}(2))$ quantum algebra with $q=\re^{\frac{\ri\hbar}{2}}$:
\bea\label{sosoasaspo}
\big[\,{\tt E}_+\,,{\tt E}_-\,\big]=\frac{q^{{\tt H}}-q^{-{\tt H}}}{q-q^{-1}}\ ,\ \ \qquad \quad
\big[\,{\tt H}\,,\, {\tt  E}_\pm\,\big]
=\pm  2\,  \ {\tt  E}_\pm\, .
\eea
For the quantum version of eq.\,\eqref{asusaisa}, we put forward the following ansatz
\bea\label{iasssaoasi}
{\boldsymbol L}(\mu)=
\begin{pmatrix}
F_-(\mu)\, q^{-\frac{ 1}{2}{\tt H}}+F_+(\mu)\, q^{+\frac{1}{2}{\tt H}} &(q-q^{-1})\, G(\mu)\ {\tt E}_-\\[0.3cm]
(q-q^{-1})\, G(\mu)\ {\tt E}_+&F_-(\mu)\, q^{+\frac{ {1}}{2}{\tt H}}+F_+(\mu)\, q^{-\frac{ {1}}{2}{\tt H}}
\end{pmatrix}\ ,
\eea
where $F_\pm(\mu)$ and $G(\mu)$ are some functions. The classical matrix will be  recovered if we  assume that  as $\hbar \to 0$,
\bea
q^{ \pm \frac{1}{2}{\tt H}}=1\pm {\textstyle  \frac{1}{4}\, \ri\hbar\   {\tt  H}+o(\delta)\ ,}
\eea
while
\bea
{\tt H}=-
\ri\,\hbar^{-1}\  H+O(\hbar^0)\ ,
\qquad\qquad
{\tt E}^{(\pm)}=
%\frac{1}{\hbar}
\hbar^{-1}\ E^{(\pm)}+O(\hbar^0)\, ,
\eea
and also
\bea\label{pasij}
\lim_{\hbar\to 0}F_\pm(\mu)=\frac{1}{2}\ \big(1\pm f(\mu)\big)\ ,\ \ \ \ \lim_{\hbar\to 0} G(\mu)=
\frac{1}{2}\ g(\mu)\ .
\eea
It is clear that the operator valued matrix ${\boldsymbol L}$ must satisfy the Yang-Baxter algebra
\bea\label{kusys}
{\boldsymbol R}_{\text{\textonehalf\textonehalf}}(\mu'/\mu)\, \big({\boldsymbol  L}(\mu)\otimes 1\big)\, 
(1\otimes {\boldsymbol  L}(\mu')\big)= \big(1\otimes{\boldsymbol  L}(\mu')\big)\big(
{\boldsymbol  L}(\mu)\otimes 1\big)\,  {\boldsymbol  R}_{\text{\textonehalf\textonehalf}}(\mu'/\mu)\ ,
\eea
where ${\boldsymbol R}_{\text{\textonehalf\textonehalf}}$ is given by eq.\,\eqref{kasusau}. The ansatz  \eqref{iasssaoasi}, combined with this relation, yields
\bea
F_+(\mu)=+{a}{\mu}\ G(\mu)\ , \ \ \ \ \  \ F_-(\mu)=-(a\mu)^{-1}\ G(\mu)\ ,
\eea
where $a$ is an arbitrary constant. Consistency with eq.\,\eqref{pasij} and the explicit form for $f$ and $g$ \eqref{iasosao} requires that
\bea
\lim_{\hbar\to 0}( a^2)=\frac{1-\kappa}{1+\kappa}\ .
\eea

We may now use the well known fact that 
 the $U_q(\mathfrak{sl}(2))$ algebra with defining relations \eqref{sosoasaspo} and $q=\re^{\frac{\ri\hbar}{2}}$
 admits  a formal realization in terms of the Heisenberg algebra
 \bea
 [\,{\tt Q}\,, {\tt P}\,]=\ri\, \hbar\ .
 \eea
 Namely \cite{Izergin:1981mc},
 \bea
{\tt  E}_{+}&=&\re^{-\frac{1}{2}{\tt Q}}\ 
\ \frac{\sinh(\, \half\, {\tt P }+
 %\frac{\tt P}{2}-
\hbar C\,)}{\sin(\frac{1}{2}\hbar)}\ \  \re^{-\frac{1}{2}{\tt Q}}\nonumber\\
{\tt  E}_{-}&=&\re^{+\frac{1}{2}{\tt Q}}\ \ 
 \frac{\sinh(\, \half\, {\tt P }
 -\hbar C\,)}{\sin(\frac{1 }{2}\hbar)}\ \ \re^{+\frac{1}{2}{\tt Q}}\\
 {\tt H}\,\,\,&=&-2\ri\ \hbar^{-1}\ 
 %{\textstyle \frac{2\ri}{\hbar}}\ 
 {\tt P}\ .\nonumber
\eea
It is not difficult to see that this can be thought of as the quantum counterpart of eqs.\,\eqref{aosaoaas}.
The constant $C$ is arbitrary and is related to the value of the quantum Casimir. In fact, it is convenient to substitute it by $\ell$: $C= \ri \,
 (2\ell+1)/4$, then
\bea
\frac{1}{2}\, \bigg[(q+q^{-1})\, \big(q^{{\tt  H}}+q^{-{\tt H}}\big)
+(q-q^{-1})^2\,
\ \big({\tt  E}_{-} {\tt  E}_{+}+ {\tt E}_{+}{\tt E}_{-}\big)\,\bigg]=
%2+4\, \sinh^2\Big(\frac{\hbar}{2}\, C\Big).
q^{2\ell+1}+q^{-2\ell-1}\ .
\eea

Let us introduce  the Heisenberg group generators, subject to the Weyl commutation relations
\bea
{\tt V}=\exp\big(\,\half\, {\tt P}\,\big)
%\re^{\frac{1}{2}{\tt P}}\, 
,\ \ \ \ {\tt U}=\exp(\,{\tt Q}\,)
%\re^{\tt Q}
\ :\ \ \ \ \  
{\tt U}{\tt  V}=q\, {\tt V}{\tt  U}\ .
\eea
Our analysis  suggests  that the $2\times 2$  operator valued matrix  ${\boldsymbol L}(\mu)=
{\boldsymbol {\cal L}}^{(\ell)}(\mu|{\tt U}, {\tt  V})$, where  \cite{Sklyanin:1983ig,Bazhanov:1989nc}
\bea\label{kassysa1}
{\boldsymbol {\cal L}}^{(\ell)}(\mu|{\tt U}, {\tt  V})=
\begin{pmatrix}
 \big(\mu\  {\tt V} -\mu^{-1}  \, {\tt V}^{-1}\big) &
\ri\, \big(q^{-\ell}\ {\tt V}-q^{+\ell}\ {\tt V}^{-1}\big)\, {\tt U}\\
\ri\, \big(q^{+\ell}\ {\tt V}-
q^{-\ell}\ {\tt V}^{-1}\big)\, {\tt U}^{-1}
&\big(\mu\  {\tt V}^{-1}-\mu^{-1}\  {\tt V}\,\big)
\end{pmatrix}\ 
\eea
satisfies the Yang-Baxter relation \eqref{kusys}. Furthermore,
it is easy to see that the same properties still hold for the matrix which depends on a set of  six parameters
$\{\mathpzc{a,b,c,g, r, \ell}\}$:
\begin{equation}\label{kassysa}
{\boldsymbol {\cal L}}
{\scriptsize{\big[\,\,\begin{array}{c}
\!\!\!\mathpzc{g\, r\, \ell}\\[-0.05cm]
\!\!\! \mathpzc{a\,b\, c}\,
\end{array}\big]}}(\mu\,|\,{\tt U}, {\tt  V})
=\mathpzc{ g}\, \mathpzc{ r}^{\frac{\sigma_3}{2}}\ {\boldsymbol {\cal L}}^{(\ell )}( \mathpzc{ a}\mu\, |\,  \mathpzc{b}\,{\tt U},  \mathpzc{c}\,{\tt V})\  \mathpzc{ r}^{\frac{\sigma_3}{2}}\ .
\end{equation}
Most of the parameters, except maybe $\mathpzc{ r}$ and $\ell$,  look trivial when we consider only a single site. However, for the discretized
system of $N$ sites, the possibility that the parameters may be different at different sites should be considered.
In any case, one can expect that for a properly adjusted
set $\{\mathpzc{a_{\rm s},b_{\rm s},c_{\rm s},g_{\rm s}, r_{\rm s}, \ell_{\rm s}}\}_{\rm s=1}^N$
the discretized quantum version of the formulae \eqref{asasaisao},\,\eqref{oapasosap} in the fundamental representation
is given by
\bea\label{aioasio}
{ {\cal T}}^{{ (N)}}(\mu)=(-\mu)^{N}\ {\rm Tr}\bigg[\,
{\overset{\leftarrow} {\cal P}} \Big(\prod_{{\rm s}=1}^{N}
{\boldsymbol {\cal L}}
{\scriptsize{\big[\,\,\begin{array}{c}
\!\!\!\mathpzc{g_{\rm s}\, r_{\rm s}\, \ell_{\rm s}}\\[-0.05cm]
\!\!\! \mathpzc{a_{\rm s}\,b_{\rm s}\, c_{\rm s}}\,
\end{array}\big]}}(\mu\,|\,{\tt U}_{\rm s}, {\tt  V}_{\rm s})
\Big)\,(q^\mathpzc{d}\,{\tt  V})^{-\sigma_3}\,\bigg]\ ,
\eea
where ${\tt V}= \prod_{{\rm s}=1}^N {\tt V}_{\rm s}$
is the discretized  counterpart  to the exponential $\re^{\pi P_1}$ and
$\mathpzc{d}$ is  some constant.
The overall factor $(-\mu)^N$ is inserted to ensure that the transfer-matrix is a polynomial  in $\mu^2$ of order $N$. Notice that
\bea 
{ {\cal T}}^{{ (N)}}(0)={\tt V}^{-2}\, q^{-d}\ \prod_{s=1}^Ng_{\rm s}\, r_{\rm s} c^{-1}_{\rm s}+{\tt V}^{+2}\, q^{d}\
\prod_{s=1}^Ng_{\rm s}\, r^{-1}_{\rm s} c_{\rm s}\ 
\eea
%$\mathpzc{abcdefghijklmnopqrstuvwxyz}$  
%$\mathscr{AB}$
is expressed in terms of integer powers of 
\bea
{\tt X}_{\rm s}\equiv
{\tt V}^{2}_{\rm s}
\eea
rather than ${\tt V}_{\rm s}$. In fact, this is true for any $\mu$, and it
can be made explicit by rewriting \eqref{aioasio} in the equivalent form:
\bea\label{jasasysa}
%{\tau}^{\scriptscriptstyle{ (N)}}(\mu)=
\!\!\!{{\cal  T}}^{( N)}(\mu)=
\mathscr{C} \,{\rm Tr}\bigg[
{\overset{\leftarrow} {\cal P}} \Big(\prod_{j=1}^N
\big({\boldsymbol {\cal L}}^\mathpzc{ (r_{\rm s}\, \ell_{\rm s})}_-
\big(  \mathpzc{b}_{\rm s}{\tt  U}_{\tt s},   \mathpzc{c}^2_{\rm s}\,{\tt X}_{\rm s} \big)\!-\!\mathpzc{a_{\rm s}}^2
\mu^2\, {\boldsymbol {\cal L}}^\mathpzc{ (r_{\rm s}\, \ell_{\rm s})}_+\big(  \mathpzc{b}_{\rm s}{\tt  U}_{\tt s},   \mathpzc{c}^2_{\rm s}\,{\tt X}_{\rm s} \big)\big)\Big) \!\!
\begin{pmatrix}
\, q^{- \mathpzc{d}}\,{\tt Z}^{-1}&0\\
0&q^\mathpzc{d}
\end{pmatrix}\!\nonumber
\bigg] \\[0.2cm]
\eea
where $\mathscr{C}$ is a constant, ${\boldsymbol {\cal L}}_\pm$ are triangular matrices:
\bea
{\boldsymbol {\cal L}}^{(\mathpzc{r} \ell)}_-({\tt U}, {\tt X})&=&
 \begin{pmatrix}
\mathpzc{r}  &
0\\
-\ri\, \big(q^{-\ell-1}-q^{+\ell-1}\ {\tt X}
\big)\, {\tt U}^{-1}
&\mathpzc{r}^{-1}\, {\tt X}
\end{pmatrix}\nonumber
\\[0.2cm]
{\boldsymbol {\cal L}}^{( \mathpzc{r}\ell)}_+({\tt U}, {\tt X})&=&
\begin{pmatrix}
\mathpzc{r}\, {\tt X} &
\ri\, \big(q^{1+\ell}-q^{1-\ell}\, {\tt X}\big)\, {\tt U}\\
0
&\mathpzc{r}^{-1}
\end{pmatrix}\ ,
\eea
and
\bea\label{jasasasuy}
{\tt Z}= \prod_{{\rm s}=1}^N {\tt X}_{\rm s}\ .
\eea
Finally, let us note that
the set of formal operators $\big\{{\tt U}_{\rm s}, {\tt X}_{\rm s}\big\}_{{\rm s}=1}^N$ satisfy the commutation relations
\bea\label{aopaasasp}
\big[{\tt U}_{\rm s},\,{\tt U}_{\rm s'}\big]=\big[{\tt X}_{\rm s},\,{\tt X}_{\rm s'}\big]=\big[{\tt U}_{\rm s},\,{\tt X}_{\rm s'}\big]=0\ \ \ ({\rm s}\not={\rm s}')\ ,\ \ \ \ \ \ 
{\tt U}_{\rm s}\,{\tt X}_{\rm s}=q^2\, {\tt X}_{\rm s}\,{\tt U}_{\rm s}\ ,
\eea
and also that
${\tt Z}$ commutes with ${\cal T}^{{ (N)}}(\mu)$ for any values of the parameters in   \eqref{jasasysa}.

\bigskip
We are now faced with the task of specifying  the parameters  in   \eqref{jasasysa}. Our analysis outlined below shows that in all  likelihood it is enough
to consider the case with
\bea\label{jsaussai}
\mathscr{C}=\mathpzc{a}_{\rm s}= \mathpzc{b}_{\rm s}= \mathpzc{c}_{\rm s}= \mathpzc{r}_{\rm s}=1\,,\ \ \ \ \ \ \ 
  \ \ell=-\half\, , \ \ \  \ \ \ \mathpzc{d}=2p_2-1\ .
\eea
In this case, we expect
that, with a  properly chosen representation of
the algebra \eqref{aopaasasp} and with a  properly understood scaling limit, 
the operator  ${\cal T}^{{ (N)}}(\mu)$
can be identified with the chiral transfer-matrix $\tau(\lambda)$
 defined in eq.\,\eqref{aospas} with $j=\half$.
The discretized operator should be restricted to the sector with
\bea\label{asiissiaias}
{\tt Z}=q^{1-2(p_2+\ri p_1)}\ ,
\eea
where, 
perhaps,  some constraints need to be  imposed on $(p_1,p_2)$.
 Recall that  the pair $(p_1,p_2)$ label the Fock space ${\cal F}_{p_1,p_2}$ in which $\tau(\lambda)$ acts.

 \bigskip

We came to the  above conjecture   through  the analysis  of the case of integer $n\equiv\frac{2\pi}{\hbar}$.
At $n=2,\,3,\ldots$, the formal algebra \eqref{aopaasasp} admits an  $n^N$ dimensional representation where
the operators associated with each site are given by the $n\times n$ matrices
\bea\label{jasasauyas}
{\tt X}^\alpha_\beta=\delta_{\alpha+1,\beta}\ ,\ \ \ \ \ 
{\tt U}^\alpha_\beta=\omega^{\alpha}\, \delta_{\alpha,\beta}\ ,\ \ \ \ \omega=\re^{-\frac{2\pi\ri}{n}}\ .
\eea
Here $\alpha,\beta=0,\,\ldots\,, n-1$ and
\bea
\delta_{\alpha\beta}=
\begin{cases}
1\  ,\ \ \ \ \ \alpha=\beta\ \ \ \ ({\rm mod}\  n)\\
0\  ,\ \ \ \ \ \alpha\not=\beta\ \ \ \ ({\rm mod}\  n)
\end{cases}\ .
\eea
Now ${\cal T}^{(N)}(\mu)$ is an  $n^N\times n^N$ matrix.  We plan to discuss the diagonalization of the operator \eqref{jasasysa}, with the parameters $\{\mathpzc{a_{\rm s},b_{\rm s},c_{\rm s}, r_{\rm s}, \ell_{\rm s}}\}$ not depending on ``${\rm s}$'',
 in  a separate publication. Here we only note that such an operator  but without the diagonal matrix
$\begin{pmatrix}
\ q^{- \mathpzc{d}}\,{\tt Z}^{-1}&0\\
0&q^\mathpzc{d}
\end{pmatrix}$, was studied in the works \cite{Bazhanov:1989nc,Baxter:1999mn} in the  context of the chiral Potts model. The extra diagonal matrix does not significantly change the diagonalization procedure.

%TTo be more precise, the operator considered in these works was  of the form \eqref{jasasysa}
%with  site-independent values of the parameters and
   %without the insertion of the 
% diagonal matrix $\begin{pmatrix}
%\ q^{- \mathpzc{d}}\,{\tt Z}^{-1}&0\\
%0&q^\mathpzc{d}
%%\%end{pmatrix}$.
%The presence of the extra diagonal matrix does not significantly change the procedure  in ref.[];
%%%%we plan to discuss  the  diagonalization of the operator \eqref{jasasysa}
%with  site-independent  parameters
%%%in a separate publication. 
 
 %Let ${\cal H}^{(N)}_M$ be the subspace in the  tensor product $\big({\mathbb C}^n\big)^{\otimes N}$ with
%${\tt Z}=\omega^{-M}$. 

 In the case  \eqref{jsaussai}, the  diagonalization problem
simplifies dramatically and can be solved within the standard Bethe ansatz framework (see next section for some details).
This allows a thorough investigation of the scaling limit.
An important point is that the scaling procedure requires
a choice of  some reference state  (``vacuum'') and only states whose
``energy'' measured from  the vacuum energy remains  finite as $N\to\infty$, should be taken into account. Let ${\cal H}^{(N)}_M$ be the subspace in the  tensor product ${\cal H}^{(N)}\equiv \big({\mathbb C}^n\big)^{\otimes N}$ with
${\tt Z}=\omega^{M}$.
The operator ${\cal T}^{(N)}(\mu)$,  with parameters as in eq.\,\eqref{jsaussai}, restricted to
the
 ${\cal H}^{(N)}_{{\mathfrak j}-\frac{\mathfrak m}{2}}$ subspace,
with ${\mathfrak j}$ and ${\mathfrak m}$ 
satisfying the conditions \eqref{siaisaiais}, commutes with the Hamiltonian of the 
 Fateev-Zamolodchikov   ${\mathbb Z}_n$
spin chain  
\begin{equation}\label{sisisai}
{\mathbb H}^{(N)}=\,\left. - \frac{1}{n}\ \sum\limits_{{\tt s}=1}^N\sum\limits_{l=1}^{n-1}\, \frac{({\tt X}_{\tt s})^l+\big({\tt U}_{\tt s}
{\tt U}_{{\tt s}+1}^\dag\big)^l}{\sin\big(\frac{\pi l}{n}\big)}\ \right|_{{\tt Z}
=\omega^{{\mathfrak j}-\frac{\mathfrak m}{2}}}
\end{equation}
with twisted  boundary conditions
\bea\label{sisisai2}
{\tt U}_{N+1}= \omega^{{\mathfrak j}+\frac{{\mathfrak m}}{2}}\ {\tt U}_1\ .
\eea
It  also commutes with the lattice shift operator 
\bea\label{isisaai}
{ {\mathbb  P}}^{(N)}=\,
\delta_{\beta_2}^{\alpha_1}\delta_{\beta_3}^{\alpha_2}\ldots\delta_{\beta_1+{\mathfrak j}+\frac{{\mathfrak m}}{2}}^{\alpha_N}
\ \Big|_{{\tt Z}
=\omega^{{\mathfrak j}-\frac{\mathfrak m}{2}}}\  .
\eea
Our numerical work for the vacuum state of the Hamiltonian $ \mathbb{H}^{(N)}$ in the sector ${\cal H}^{(N)}_{{\mathfrak j}-\frac{\mathfrak m}{2}}$ 
%in  ${\cal H}^{(N)}_{{\mathfrak j}-\frac{\mathfrak m}{2}}$
%for the   Fateev-Zamolodchikov  ${\mathbb Z}_n$ 
%%%%spin chain in  ${\cal H}^{(N)}_{{\mathfrak j}-\frac{\mathfrak m}{2}}$ 
for different admissible values
of $n$, ${\mathfrak j}$ and ${\mathfrak m}$  gives strong support to
the following relations (see Appendix \ref{askjhdaA})
\bea\label{jashyas}
\tau(\lambda)={\rm s}\!\!\!\lim\limits_{N\to \infty}\ F^{(N)}(\lambda)\,{\cal T}^{{ (N)}}\Big(\big({\textstyle\frac{\pi}{N}}\big)^{\frac{1}{n}}\,\lambda\,\Big)\ ,
\eea
where
\bea
{F}^{(N)}(\lambda)=
 \begin{cases}
\exp\Big(\sum_{l=1}^{[\frac{n}{2}]}\  \frac{ \pi^{\frac{2l}{n}} }{l\cos(\frac{\pi l}{n})}\    N^{1-\frac{2l}{n}}\  \lambda^{2l}\, \Big)\
\ \ \ \ \ \ \ \ \ \ \ \ \ \ \ 
&(n\not=2,\,4,\,\ldots)\\[0.4cm]
 \big(\frac{N\re}{\pi}\big)^{\frac{4}{ n}\lambda^n}\ \ 
 \exp\Big(\sum_{l=1}^{[\frac{n}{2}]-1} \frac{\pi^{\frac{2l}{n}}}{l\cos(\frac{\pi l}{n})}\    N^{1-\frac{2l}{n}}\ \lambda^{2l}\, \Big)\ \ \ \ 
&(n=2,\,4,\ldots)\\
\end{cases}\ ,
\eea
and the symbol ``slim'' in  \eqref{jashyas} stands for the scaling limit
which assumes that only the low-energy states  are taken into account.
The operator $\tau(\lambda)$ in \eqref{jashyas} should be understood as the chiral parafermionic transfer-matrix
\eqref{aoss} acting in the space ${\cal V}_{\mathfrak j}^{(\mathfrak m)}$ discussed in the previous subsection.

Similarly to \eqref{jashyas}, one can consider the scaling limit
$$
{\rm s}\!\!\!\lim\limits_{N\to \infty}\  F^{(N)}(\lambda^{-1})\,
(-1)^N\  \mu^{-2 N}\ {\cal T}^{{ (N)}}(\mu)\,\Big|_{\mu=\small{({\pi}/{N})^{\frac{1}{n}}}\,\lambda^{-1}}\ .
$$
This  can be identified with the anti-chiral  parafermionic transfer-matrix ${\bar \tau}(\lambda^{-1})$ acting in the space
$\overline{ {\cal V}}_{\mathfrak j}^{(2\mathfrak j)}$.
Therefore,  in the scaling limit, (at least some of) the low energy states of  
${\cal H}^{(N)}_{{\mathfrak j}-\frac{\mathfrak m}{2}}$ organize into  the sector ${\cal V}_{\mathfrak j}^{(\mathfrak m)}\otimes
\overline{ {\cal V}}_{\mathfrak j}^{(2\mathfrak j)}$ of the $\mathbb{Z}_n$ CFT Hilbert space.  In this sector
the low energy spectrum of the
Hamiltonian \eqref{sisisai} and the corresponding  eigenvalues of the  lattice shift operator 
have  the form
\bea\label{hsaasyas}
E^{(N)}&=&e_0\, N+{\textstyle \frac{2\pi}{N}}\ \big(\Delta_{{\mathfrak j},\,{\mathfrak m}}+\Delta_{{\mathfrak j},\,2{\mathfrak j}}-
{\textstyle \frac{c_n}{12}}+{L}+\bar {L}\,\,\big)+o\big(N^{-1}\big)\nonumber\\[0.2cm]
P^{(N)}&=&\exp\big(\, {\textstyle\frac{2\pi \ri }{N}}\, 
(\,\Delta_{{\mathfrak j},\,{\mathfrak m}}-\Delta_{{\mathfrak j},\,2{\mathfrak j}}\,+
{ {L}}-\bar {{L}}
\,\,)
\,\big)\ ,\\[-0.4cm]
\nonumber
\eea
where $e_0$ is some constant and 
${L}$ and $\bar {{L}}$ are  integers.
The central charge and  conformal dimensions are given by eqs.\,\eqref{sisias} and  \eqref{aoosaoas}, respectively.

\bigskip

Returning to the formal operator ${\cal  T}^{(N)}(\mu)$     \eqref{jasasysa}-\eqref{asiissiaias}
for arbitrary values of $n$,  we note that 
the case of real $(p_1,p_2)$ is of prime interest to the cigar NLSM.
Perhaps the most 
promising approach to the construction of a suitable representation of the algebra \eqref{aopaasasp} and the diagonalization of 
${\cal  T}^{(N)}(\mu)$ is based on the  method of separation of variables \cite{Sklyanin:1995bm}.
% This
%was successfully applied to a similar problem which appears in the quantization of the   sinh-Gordon model [].
%Another interesting possibility is related to the work [], where some  spectral properties of
%the cigar NLSM were observed to appear in the scaling limit of a certain inhomogeneous version of the 6-vertex model.

 \section{Spectrum  of the chiral transfer-matrix}
 
 In the previous sections the construction of the 
 chiral transfer-matrices has  been discussed. We are now ready to tackle   the  calculation of their  spectrum.
  A powerful approach  to this problem is the
 ODE/IQFT correspondence and here, we'll illustrate the method  by calculating the
 vacuum eigenvalues of $\tau(\lambda)$.

 \subsection{Operators $\zeta_\pm(\theta)$\label{sec4.1}}

 As it was already mentioned,  a comprehensive discussion of the diagonalization
 procedure of the  discretized  chiral transfer-matrix ${\cal T}^{{ (N)}}(\mu)$ \eqref{jasasysa}-\eqref{asiissiaias} will be dealt with in a separate publication. However,
 it would be useful  here to make a short summary of the important integrable structures which play a crucial r$\rm\hat{o}$le
 in the procedure. Namely,
it is possible to  explicitly  construct  two matrices  ${{\cal Z}}_\pm(\mu)$
satisfying the following set of conditions:\footnote{All  matrices  appearing  below are understood as operators  acting
invariantly
 in  ${\cal H}_{{\mathfrak j}-\frac{\mathfrak m}{2}}^{(N)}$ -- the eigenspace  of ${\tt Z}$ in 
 $\big({\mathbb C}^n\big)^{\otimes N}$ corresponding to  the eigenvalue 
$\omega^{{\mathfrak j}-\frac{\mathfrak m}{2}}$, where $\mathfrak{j}$ and $\mathfrak{m}$ are restricted as in \eqref{siaisaiais}.}

\begin{enumerate}[label=(\roman*)]

\item \label{reli} {\bf Commutativity}
\bea
&&\big[{ {\cal Z}}_\pm(\mu),{ {\cal Z}}_\pm(\mu')\big]=
\big[{ {\cal Z}}_+(\mu),{ {\cal Z}}_-(\mu')\big]=0\nonumber
\\[6.pt]
&&
\big[{{\cal Z}}_\pm(\mu),{{\cal T}}^{{ (N)}}(\mu')\big]=
\big[{{\cal Z}}_\pm(\mu), { {\mathbb H}}^{(N)}\,\big]=
\big[{ {\cal Z}}_\pm(\mu), {{\mathbb P}}^{(N)}\,\big]=0\nonumber
\eea

\item \label{relii}  {\bf ``Quantum Wronskian'' type  relations}

Odd $n$:
\bea\label{jsasaasu}
 (1+\mu)^{2N}\ {\cal Z}_+( q^{-\frac{1}{2}}\mu)\,  {\cal Z  }_+(q^{+\frac{1}{2}}\mu)-
(1-\mu)^{2N}\, {\cal Z}_-( q^{-\frac{1}{2}}\mu)\,  {\cal Z}_-(q^{+\frac{1}{2}}\mu)= W(\mu)\ {{\mathbb  P}}^{(N)} \nonumber
\eea
Even $n$:
\bea\label{jassayas}
{\cal Z}_+(  q^{-\frac{1}{2}}\mu)\, {\cal Z  }_+( q^{+\frac{1}{2}}\mu)-
(1-\mu^2)^{2N}\  {\cal Z}_-(  q^{-\frac{1}{2}}\mu)\,  {\cal Z}_-( q^{+\frac{1}{2}}\mu)=  W(\mu)\ {{\mathbb P}}^{(N)}\nonumber
%\lambda^n\ {G}_{N-1}(\lambda^{2n})
\eea
with
\bea\label{asiosiasi}
W(\mu)=(1+\mu^n\big)^{2N}- (1-\mu^n\big)^{2N}\nonumber
\eea

 \item\label{reliii} {\bf ``$T-Q$'' type  relations}
 
 Odd $n$:
 \bea\label{kasusausa}
{{\cal T}}^{(N)}(\mu)\, { {{\cal  Z}}}_\pm(\mu)&=&(1\mp q^{-\frac{1}{2}}\,\mu)^{2N}\ 
\,  {{\cal Z}}_\mp ( q^{-1}\mu)+(1\mp q^{+\frac{1}{2}}\,\mu)^{2N}\ 
\,  {{\cal Z}}_\mp(q^{+1}\mu )\nonumber
\eea
Even $n$:
\bea\label{kasksjkasjsaaks}
{{\cal T}}^{(N)}(\mu)\, {{\cal  Z}}_-(\mu)&=&
 { {{\cal  Z}}}_+(q^{-1}\mu)+
{ { {\cal Z}}}_+(q^{+1}\mu) \nonumber
\\[0.2cm]
{{\cal T}}^{(N)}(\mu)\, { {{\cal Z}}}_+(\mu)&=&(1-q^{-1}\,\mu^2)^{2N}
\,  {{\cal  Z}}_-(q^{-1}\mu)+(1-q^{+1}\,\mu^2)^{2N}
\,  {{\cal  Z}}_-(q^{+1}\mu) \nonumber
\eea

\item \label{reliv}{\bf  Analytical conditions}

Odd $n$:
\bea \label{sdfjkdOD}
{{\cal Z}}_\pm(\mu)\ \ = \ \mu^{{\mathfrak m}}\ \times\big(\ {\rm  polynomial\ \  in}\ \ \mu\ \  {\rm of \ degree}\ \ 
(n-1)\, N-2{\mathfrak j}-{\mathfrak m}\ \big) \nonumber
\eea
Even $n$:
\bea\label{sdfjkdEV}
{{\cal Z}}_+(\mu)&=&\mu^{{\mathfrak m}}\ \times\big(\ {\rm  polynomial\ \ in}\ \ \mu^2\ \   {\rm of \ degree}\ \ \ \ 
\half\, nN-{\mathfrak j}-\half\, {\mathfrak m}\ \big)\nonumber\\[0.2cm]
{ {\cal Z}}_-(\mu)&=&\mu^{{\mathfrak m}}\ \times\big(\ {\rm  polynomial\  \ in}\ \ \mu^2\ \   {\rm of \ degree}\ \ \ \ 
\half\,(n-2)\, N-{\mathfrak j}-\half\,{\mathfrak m}\ \big) \nonumber
\eea

\item \label{relv} {\bf  $\mu\to-\mu$ symmetry }
\bea\label{asisiaasi}
{\rm Odd}\ \ n\ :\ \ \ \ {{\cal Z}}_\pm(-\mu)&=&(-1)^{{\mathfrak m}}\ {{\cal Z}}_\mp(\mu)
\nonumber\\[0.2cm]
{\rm Even}\ n\ :\ \ \ \ {{\cal Z}}_\pm(-\mu)&=&(-1)^{{\mathfrak m}}\ {{\cal Z}}_\pm(\mu) \nonumber
\eea

\end{enumerate}

 \bigskip
 
 \noindent
 The scaling limit of the  operators ${{\cal Z}}_\pm(\mu)$ is of special interest. 
 In Appendix \,\ref{p_b} we  present evidence that the  following scaling limits, similar to \eqref{jashyas}, do exist:
  \begin{equation}\label{ZOD}
  \arraycolsep=0.1cm
\begin{array}{llll}\quad\ \ \, {\rm Odd}\ \   n\ : & {\zeta}_\pm =\lambda^{\mp 2\,\lambda^n}  
 & {\rm s}\!\!\!\lim\limits_{N\to \infty}\, G^{(N)}(\pm \lambda)\ {{\cal  Z}}_\pm
 \Big(\big({\textstyle \frac{\pi}{ N}}\big)^{\frac{1}{n}}\,\lambda\Big)\ 
 \end{array}
 \end{equation}
 \begin{equation}\label{ZEV}
  \arraycolsep=0.1cm
 \begin{array}{llll}
{\rm Even}\ \ n\ : & 
{\zeta}_+=  
&{\rm s}\!\!\!\lim\limits_{N\to \infty}\, G^{(N)}_+(\lambda)\  {\cal Z}_+\Big(\big({\textstyle \frac{\pi}{ N}}\big)^{\frac{1}{n}}
\,\lambda\Big)  \\[0.4cm]
 &{ \zeta}_-=\  \lambda^{4\lambda^n}
&{\rm s}\!\!\!\lim\limits_{N\to \infty}\, G^{(N)}_-(\lambda)\ {\cal Z}_-\Big(\big({\textstyle \frac{\pi}{ N}}\big)^{\frac{1}{n}}\, \lambda\Big)\\
\end{array}
\end{equation}
with
\bea
G^{(N)}(\lambda)&=&
\Big({\frac{\re N}{\pi}}\Big)^{-\frac{2(n-1)}{n}\, \lambda^n}\ \exp\bigg(\sum_{l=1}^{n-1}(-1)^{l+1} \frac{\pi^{\frac{l}{n}}}{l\cos(\frac{\pi l}{2n})}\    N^{1-\frac{l}{n}}\ \lambda^l\, \bigg)\ \nonumber\\
G^{(N)}_+(\lambda)&=&
\Big({ \frac{\re N}{\pi}}\Big)^{-2\lambda^n} \\
G^{(N)}_-(\lambda)&=&
\Big({\frac{\re N}{\pi}}\Big)^{\frac{2(n-2)}{n}\, \lambda^n}\ \
\exp\bigg(-\sum_{l=1}^{\frac{n}{2}-1}
\frac{\pi^{\frac{2l}{n}}}{l\cos(\frac{\pi l}{n})}\    N^{1-\frac{2l}{n}}\ \lambda^{2l}\bigg)\, .\nonumber
\eea
Notice that  for odd  $n$, we include the ``strange'' extra factor  $\lambda^{\mp 2\lambda^n}$
 in the formula for
${ \zeta}_\pm$. A similar factor $\lambda^{4\lambda^n}$ appears for $\zeta_-$ with even $n$. 
At  first glance they
look artificial and, furthermore, make  ${ \zeta}_\pm$  multivalued functions of $\lambda$.
However, these ``strange'' factors allow one to write the scaling version of the quantum Wronskian type relations  \ref{relii}
 in a form which is applicable for both odd and even $n$:
\bea\label{aksasus}
{\zeta}_+\big(\theta+{\textstyle\frac{\ri\pi}{2}}\big)\,{\zeta}_+\big(\theta-{\textstyle\frac{\ri\pi}{2}}\big)-
{\zeta}_-\big(\theta+{\textstyle\frac{\ri\pi}{2}}\big)\,{\zeta}_-\big(\theta-{\textstyle\frac{\ri\pi}{2}}\big)
=2\, \sinh\big(2\pi\re^\theta\big)\ .
\eea
Here we have introduced $\theta$,
\bea\label{kjshdfueqw}
\lambda=\re^{\frac{\theta}{n}}
\eea
and later we'll argue that ${\zeta}_\pm$ are single valued functions of this variable.
The $T-Q$ type  relations  \ref{reliii}  in the scaling limit also have the same form for odd and even $n$,
\bea\label{jasusauu}
\tau(\lambda)\ {\zeta}_\pm(\theta)={\zeta}_\mp(\theta-\ri\pi)+{\zeta}_\mp(\theta+\ri\pi)\ .
\eea
Since the operator $\tau(\lambda)$ acts in the parafermionic space ${\cal V}_{\mathfrak j}^{({\mathfrak m})}$
it is natural to expect that the same holds true for ${\zeta}_\pm(\theta)$.
\bigskip

The relations \eqref{aksasus} and \eqref{jasusauu} work for $n=2,\,3,\,4\ldots$ ,
but is it possible to extend them to non-integer $n$? We conjecture, that for any $n>0$ and general values of $(p_{1},p_{2})$,
 there  exists a pair of operators which act  invariantly  in the Fock  level subspaces
\bea\label{osossa}
\zeta_\pm(\theta)\ :\ \ \ {\cal F}^{(L)}_{p_1,p_2}\mapsto{\cal F}^{(L)}_{p_1,p_2}
\eea
such that
\bea
[\zeta_\pm(\theta),\,\zeta_\pm(\theta')]=[\zeta_+(\theta),\,\zeta_-(\theta')]=[\zeta_\pm(\theta),\,\tau(\lambda')]=0\ ,
\eea
 satisfying the relations \eqref{aksasus} and \eqref{jasusauu}.
 \bigskip

 Unfortunately, at this moment we don't know  how to explicitly construct the operators \eqref{osossa}.
 Nevertheless, there are strong physical arguments that they do exist.  In the works \cite{Lukyanov:2003nj} and  \cite{Lukyanov:2005nr},
 $\zeta_+$ and $\zeta_-$ were introduced and studied
for  real $(p_{1}, p_2)$, as  the boundary state operators in the paperclip model
with topological angle equal to $0$ and $\pi$, respectively.\footnote{In the notations of ref. \cite{Lukyanov:2003nj}:
 ${\mathbb B}(\kappa)=2^{-\frac{1}{2}}\ \exp(2\theta\,\re^\theta)\ \zeta_+(\theta)$, provided $\kappa=\re^\theta$.}
Among the results of those works is the large-$\theta$  behaviour. It was proposed  that
the  operators $\zeta_\pm$ possess  the following asymptotic at $\theta\to+\infty$:
\bea\label{finew}
\zeta_+(\theta)&\asymp& \exp
\Big(\,- \big(\,2\theta+ \pi \cot\big({\textstyle \frac{\pi n}{2}}\big)-C\,\big)\,\re^\theta-\sum_{j=1}^{\infty}\, {\mathfrak i}_{2j-1}\,
\re^{-\theta (2j-1)}\, \Big)\\
\zeta_-(\theta)&\asymp&
{\tilde\tau}\Big(\re^{-\frac{\theta}{n+2}}\Big)\ 
\ \exp\Big(\,  \big(\,2\theta-\pi\cot\big({\textstyle \frac{\pi n}{2}}\big)-C\,\big)\,\re^\theta+\sum_{j=1}^{\infty} \, {\mathfrak i}_{2j-1}\
\re^{-\theta (2j-1)} \,  \Big)\ .\nonumber
\eea
Here ${\tilde \tau}({\tilde \lambda})$ is the formal series \eqref{jasusau} generating the set of dual nonlocal IM while
$\{{\mathfrak i}_{2j-1}\}_{j=1}^\infty$ is the infinite  set of local IM \eqref{ospssapo}-\eqref{kasausau}.
The real  constant $C$ is non-universal and can
be chosen at will.  In ref.\cite{Lukyanov:2003nj}, it was set to $\pi\cot\big({\textstyle {\frac{\pi n}{2}}}\big)$. 
For our purposes it is convenient to set it to zero,
\bea
C=0\ .
\eea
%where $\psi(z)=\partial_z\log \Gamma(z)$ and $\gamma_E$ stands for the Euler constant.
With this choice  it turns out that for odd $n$ and $(p_{1},p_2)$ restricted by the conditions \eqref{asjsasai},\eqref{siaisaiais},
$\zeta_\pm$ are the same functions that appear in the scaling limit \eqref{ZOD}.
However, one can see from eq.\eqref{finew} that the operators $\zeta_\pm$ become singular for even $n$. They can be analytically regularized
similar to as in eq.\,\eqref{iuczvx},
\bea\label{asoisaiasi}
\zeta^{(\rm reg)}_\pm(\theta)\big|_{n=2l}=\lim_{\epsilon \to 0}\ 
  \exp\big(\, {\textstyle \frac{2}{\epsilon }}\,\, \re^\theta\, \big)\ 
 { \zeta}_\pm(\theta)\big|_{n=2l+\epsilon}\ \ \ \ \ \ \ \  (l=1,2,3\ldots)\ .
  \eea
When the  regularized operators are restricted to the parafermionic space  ${\cal V}_{\mathfrak j}^{({\mathfrak m})}$, they are the 
same as the ones on the left hand side of eq.\,\eqref{ZEV}.\footnote{Here, we have abused notation because the 
$\zeta_\pm$ in \eqref{ZEV} denote the continuous operators obtained by means of the lattice regularization procedure. 
%we would like
%the  T-Q relation to have the same form for any $n$. The T-Q relation for even $n$ should be understood as
%$\tau^{(\rm reg)}(\lambda)\ {\zeta}^{(\rm reg)}_\pm(\theta)={\zeta}^{(\rm reg)}_\mp(\theta-\ri\pi)+{\zeta}^{(\rm reg)}_\mp(\theta+\ri\pi)$
%%where $\tau^{(\rm reg)}$ is defined by eq.\,\eqref{iuczvx}.
%The operators $\zeta_\pm$ in \eqref{ZEV} denote the continuous operators obtained by means of the lattice regularization procedure.
}

It is expected that the operators $\zeta_\pm$ are entire functions of $\theta$ (in the sense that all 
their  matrix elements and eigenvalues  are entire functions of this variable) satisfying, for real $(p_1,p_2)$,  the  Hermiticity condition
\bea\label{sassao}
 \big[\zeta_\pm(\theta)\big]^\dagger= \zeta_\pm(\theta^*)\ .
\eea
As it was pointed out, the asymptotic formulae \eqref{finew} are written for large positive $\theta$, however  in all likelihood,  they hold true for complex values,
 at least in the strip
 $|\Im m (\theta)|<\pi$ with $\Re e(\theta)\to+\infty$.
 \bigskip
 
Also, it deserves mentioning that the chiral transfer-matrices $\tau_j(\lambda)$ with $j=0,\half,1,\ldots\,$,  can be expressed in terms of the operators $\zeta_\pm(\theta)$. Namely,
 for $j=\frac{1}{2},\,\frac{3}{2},\ldots$
\bea\label{jkdfghsdf1}
\tau_{j}(\lambda)&=&\frac{(-1)^{j+\frac{1}{2}}}{2 \sinh(2\pi\ri\, \re^\theta)}\\
&\times& \Big[
{\zeta}_+\big(\theta-\ri\pi (j+{\textstyle \frac{1}{2}})\big){\zeta}_-\big(\theta+\ri\pi (j+{\textstyle \frac{1}{2}})\big)-
{\zeta}_-\big(\theta-\ri\pi (j+{\textstyle \frac{1}{2}})\big){\zeta}_+\big(\theta+\ri\pi (j+{\textstyle \frac{1}{2}})\big)\Big]\nonumber
\eea
whereas  for $j=0,\, 1,\ldots$
\bea\label{jkdfghsdf2}
\tau_{j}(\lambda)&=&\frac{(-1)^{j}}{2\sinh(2\pi \re^\theta)}\\
&\times& \Big[
{\zeta}_+\big(\theta-\ri\pi (j+{\textstyle \frac{1}{2}})\big){\zeta}_+\big(\theta+\ri\pi (j+{\textstyle \frac{1}{2}})\big)-
{\zeta}_-\big(\theta-\ri\pi (j+{\textstyle \frac{1}{2}})\big){\zeta}_-\big(\theta+\ri\pi (j+{\textstyle \frac{1}{2}})\big)\Big].\nonumber
\eea
 where $\lambda$ and $\theta$ are related as in eq.\,\eqref{kjshdfueqw}.
 With these  formulae  the  fusion relation \eqref{asusuas}, as well as the $T-Q$ type relations  \eqref{jasusauu}, are  satisfied identically.

 \bigskip 
 Finally, let us consider the  ``$\mu\to-\mu$ symmetry'' relations  \ref{relv}. It is easy to see that  in the scaling limit
 they become
 
%\bea
%{\rm Odd}\ n\ \ : \  \: {\zeta}_\pm(\theta+\ri\pi n)&=&(-1)^{{\mathfrak m}}\ \re^{\pm 2\pi\ri\re^\theta}\ {\zeta}_\mp(\theta)\nonumber\\[10.pt]
%{\rm Even}\ n\ \ : \  \:
%{ \zeta}_+(\theta+\ri\pi n)&=&(-1)^{{\mathfrak m}}\  {\zeta}_+(\theta)\\
%{\zeta}_-(\theta+\ri\pi n)&=&(-1)^{{\mathfrak m}}\ \re^{4\pi\ri \re^\theta}\ {\zeta}_-(\theta)\ .\nonumber
%\eea
\bea
{\zeta}_\pm(\theta+\ri\pi n)&=&(-1)^{{\mathfrak m}}\ \re^{\pm 2\pi\ri\re^\theta}\ {\zeta}_\mp(\theta)\ \ \ \ \ \ (n-{\rm odd})
\eea
 and
\bea
{ \zeta}_+(\theta+\ri\pi n)&=&(-1)^{{\mathfrak m}}\  {\zeta}_+(\theta)\ \ \ \ \ \ \ \  \ \ \ \ \ \ \ \ (n-{\rm even})\\
{\zeta}_-(\theta+\ri\pi n)&=&(-1)^{{\mathfrak m}}\ \re^{4\pi\ri \re^\theta}\ {\zeta}_-(\theta)\ .\nonumber
\eea
As we will see below, these equations are not satisfied for non-integer $n$ and only hold for
the admissible values of ${\mathfrak j}$  and ${\mathfrak m}$ \eqref{siaisaiais}. In these particular cases, they can be used to truncate
the so called $Y$-system -- the chain of relations for $Y_j(\lambda)\equiv\tau_{j-\frac{1}{2}}(\lambda)\,\tau_{j+\frac{1}{2}}(\lambda)$ 
(see, e.g., refs.\cite{Bazhanov:1994ft,Bazhanov:1998dq}).
In turn, the truncated $Y$-system allows one to derive the set of TBA equations from which the eigenvalues of the chiral transfer-matrices can be computed.
For the vacuum eigenvalues with ${\mathfrak j}={\mathfrak m}=0$, the TBA system can be found in ref.\cite{Lukyanov:2005nr}.
 
  \subsection{ODE/IQFT correspondence for the vacuum eigenvalues}

  Let us first consider  the  vacuum eigenvalues $\zeta^{(\rm vac)}_{\pm}(\theta)$ of the operators  $\zeta_{\pm}(\theta)$ corresponding to the real pair $(p_1,p_2)$.
 The main ingredient in the ODE/IQFT correspondence for this case, is  
the ordinary differential equation
\bea\label{diffa} \bigg[ -{{\rd^2}\over{\rd x^2}}   -
p^2_1\   { {\re^x}\over{1+\re^x}} +
\big({\textstyle\frac{1}{4}}-p_2^2\,\big) \ {{\re^x}\over{(1+\re^x)^2}} +
\re^{2\theta}\ \big(1+ \re^x\big)^n \bigg]\, \Psi(x) = 0\, .
 \eea 
 We
assume for the moment that $\theta$ is real.
Eq.\,\eqref{diffa} has the form of a stationary zero energy Schr${\rm \ddot
o}$dinger equation with the potential $V(x)$ given by the
last three terms in \eqref{diffa}. The potential
$V(x)$ is positive and grows fast at large positive $x$ so that
\eqref{diffa} has a solution $\Xi (x)$ decaying  at
$x\to+\infty$; this condition specifies $\Xi(x)$ uniquely up to
normalization. To fix
the normalization, we assume that 
\bea\label{ksidef} \Xi(x) \to
\re^{-\frac{\theta}{2}}\ 
 \exp\bigg( -\big({\textstyle{n\over 4}}+\re^\theta\,\big)\, x- \re^\theta\,
\int_0^{\re^x} {\rd u\over u}\, \big((1+u)^{n\over 2}-1\big)\, \bigg)
\eea as $ x\to+\infty$. On the other hand, $V(x)$ approaches
the positive constant $\re^{2\theta}$ at large negative $x$. Hence
eq.\,\eqref{diffa} has a solution which decays for large negative
$x$; we denote this solution $\Psi_{+}(x)$. The condition,
\bea\label{psiplusdef} \Psi_{+}(x) \to\, \sqrt{\pi}\  \frac{\exp\big(\re^\theta \,x\big)}{
\Gamma(1+2\,\re^\theta)} \quad \ \ \ \ \ {\rm as} \quad\ \ \ \  x\to-\infty\ ,
\eea
specifies the solution $\Psi_{+}(x)$ uniquely, including its
normalization. Then 
\bea\label{zet0} 
\zeta_+^{(\rm vac)}(\theta)
=   \exp\big( C_+\,\re^{\theta}\, \big) \ W\big[\Xi, \Psi_{+}\big]\ ,
\eea
where
\begin{equation}\label{kljasdfd}
C_+=-\pi\, \cot\big({\textstyle \frac{\pi n}{2}}\big) +2\log(2)-2
\end{equation}
and $W[f,g]$ denotes the Wronskian $f(x)g'(x)-g(x)f'(x)$.
Eq.\,\eqref{zet0} was proposed in the work\,\cite{Lukyanov:2003nj}.

\bigskip

 In the paper   \cite{Lukyanov:2005nr} this was extended to the vacuum eigenvalue $\zeta_-^{(\rm vac)}(\theta)$.
 The starting point was the same differential
equation\ \eqref{diffa}, but instead of $\Psi_{+}$ in
\eqref{zet0}, another solution which grows as
$\exp(-\re^\theta x)$ at large negative $x$ was taken. Of course, this condition
alone does not define the solution uniquely since, besides the overall
normalization, one can always add any amount of $\Psi_{+}(x)$.
It is usually difficult to define a
growing solution unambiguously, but in our case the following property of\
\eqref{diffa} helps. Let us consider $x$ as a complex variable.
The potential $V(x)$ is an analytic function of $x$ with
branch-point singularities at all points where $\re^x$ turns to
$-1$. Let us make branch cuts starting at each of the points $x=\ri\pi\, (2
N+1)$, $N=0,\pm 1, \pm 2, \ldots$ and going to $+\infty$ parallel to the
real axis, and choose the branch of $V(x)$ for which $(1+\re^x)^n$
is real and positive on the real axis of the $x$-plane. 
Restricting attention to the domain $\Re e( x) < 0$, one finds that
the potential $V(x)$ has the periodicity property
\bea\label{vperiod} V(x+2\pi\ri) = V(x) \qquad\ \ \  \big(\,\Re e( x) < 0\,\big)\,.
\eea Consequently,  equation\ \eqref{diffa} has two Bloch-wave
solutions ($2\re^{\theta}\not\in {\mathbb  Z}$): \bea\label{blochwaves}
\Psi_{\pm}(x+2\pi \ri) = \re^{\pm 2\pi{\rm
i}\,\re^\theta}\,\Psi_{\pm}(x) \ \ \qquad \big(\Re e( x )< 0\big)\,, 
\eea 
where
the Bloch factors can be found by taking the limit $\Re e(x) \to
-\infty$. At this point we assume that $2\re^\theta$ is not an
integer, so that the conditions\ \eqref{blochwaves} specify 
two independent solutions $\Psi_{\pm}(x)$ uniquely, up to their
normalizations. Of course, the solution $\Psi_{+}(x)$ defined this
way decays as $\exp(\re^\theta x)$ at $\Re e(x) \to -\infty$, and the
asymptotic condition\  \eqref{psiplusdef} also fixes its
normalization. The solution $\Psi_{-}(x)$ grows at large negative
$\Re e(  x)$, and its normalization can be fixed by specifying the
leading asymptotic in this domain. Thus we define $\Psi_{-}(x)$ by
the conditions \bea\label{psiminusdef} &&\Psi_{-}(x + 2\pi\ri) =
\re^{-2\pi{\rm i}\,\re^\theta}\,\Psi_{-}(x) \qquad\quad \big(\Re e( x )< 0\big)\,\\
&& \Psi_{-}(x) \to\ \sqrt{\pi}\ \ \frac{\exp\big(-\re^\theta \,x\big)}{
\Gamma(1-2\,\re^\theta)}
 \  \
\ \  \quad {\rm as} \quad \ \ \ \ \ \Re e(x) \to -\infty\,.
\nonumber 
\eea 
{}It is possible to show that both $\Psi_{+}(x)$
and $\Psi_{-}(x)$ defined by \ \eqref{psiplusdef} and \
\eqref{psiminusdef} are entire functions of $\re^\theta$, and
\bea\label{psiana}\Psi_{-}\big(x\,|\,\re^\theta\big) = \Psi_{+}\big(x\,|-\re^\theta\big)\,,
\eea
where we temporarily exhibited the dependence of $\Psi_{\pm}$ on
the parameter $\re^\theta$. From the definitions \eqref{psiplusdef}
and 
\eqref{psiminusdef} we have 
\bea\label{wpm}
W\big[\Psi_{-},\Psi_{+}\big] ={\sin (2\pi\re^\theta)}\,.
\eea 
The  proposal in \cite{Lukyanov:2005nr} was that %for $\zeta_-^{(\rm vac)}(\theta)$ was
 \bea\label{zetpi}
\zeta_-^{(\rm vac)}(\theta)= 
\exp\big( C_-\,\re^{\theta} \,\big)\ 
W\big[\Xi, \Psi_{-}\big]\,, \eea
where 
\begin{equation}\label{gkjdas}
C_-=-\pi\, \cot\big({\textstyle \frac{\pi n}{2}}\big) -2\log(2)+2-2\gamma_E-2\, \psi(1+{\textstyle\frac{n}{2}}\big)
\end{equation}
with $\psi(z)=\partial_z\log\Gamma(z)$ and $\gamma_E$ stands for the Euler constant.

\bigskip
There is much evidence to support  the remarkable relations \eqref{zet0} and \eqref{zetpi}.
Some of them are based on WKB analysis of the differential equation \eqref{diffa}. Using the  method of semiclassical expansion, one
can  systematically study  the    Wronskians in\ \eqref{zet0},\,\eqref{zetpi} at  $\theta\to+\infty$. This  yields  asymptotic expansions whose structures turn
out to
be  identical to the one for $\zeta_\pm^{(\rm vac)}$ following from eq.\,\eqref{finew}. Furthermore the WKB calculations give  non-trivial  predictions 
for the vacuum eigenvalues  of the local and dual nonlocal IM. On the other hand, these vacuum eigenvalues 
can be directly calculated from their definitions. For example, the vacuum eigenvalue of
the first dual nonlocal IM is given  by eq.\,\eqref{asuausai}, while  it is a simple exercise to find the  vacuum eigenvalue of the 
first local IM ${\mathfrak i}_1$:
\bea
i_1(p_1,p_2)=-\frac{1}{12}+\frac{p_1^2}{n}+\frac{p^2_2}{n+2}\ .
\eea
It turns out that the results of the WKB analysis are in full agreement with these direct calculations.

\bigskip
Let us discuss  now the  $\theta\to -\infty$ form of the Wronskians in\ \eqref{zet0} and \eqref{zetpi}. The perturbative evaluation of the solutions
$\Xi(x)$ and $\Psi_{\pm}(x)$
leads to the expansions \cite{Lukyanov:2003nj,Lukyanov:2005nr}
\beq\label{asjssau}
\re^{\pm\pi \kappa\cot(\frac{\pi n}{2})} \ \zeta^{(\rm vac)}_\pm(\theta)=B_{p_1,p_2}\ \re^{\frac{2\ri p_1\theta}{n}}\, F_\pm(\theta\,|\,p_1,p_2)+
B_{-p_1,p_2}\  \re^{-\frac{2\ri p_1\theta}{n}}\, F_\pm(\theta\,|-p_1,p_2)\,,
\eeq
where
\bea\label{jsasasaus}
B_{p_1,p_2}=\sqrt{n}\  n^{-\frac{2\ri p_1}{n}}\ 
\frac{\Gamma(-2\ri p_1) \Gamma(1-\frac{2\ri p_1}{n})}{\Gamma(\frac{1}{2}-p_2-\ri p_1) \Gamma(\frac{1}{2}+p_2-\ri p_1) }
\eea
and
\bea\label{fpiexpan}
F_\pm (\theta\,|p_1,p_2)=\sum_{i,j=0}^\infty f_{i,j}(p_1,p_2)\ (\pm 1)^{i}\ \re^{(i+\frac{2j}{n})\theta}
\eea
with
\bea\label{jsasusau}
f_{0,0}&=&1\\
f_{0,1}&=&
-\frac{\Gamma({1\over 2}+{1\over n})\Gamma(1-{1\over n})}{ 2\sqrt{\pi}}\ 
\Big({2\over n}\Big)^{{2\over n}-1}\, 
\bigg(\frac{n+2}{n-2}+\frac{4p_2^2}{1+4p_1^2}\, \bigg)\, 
{\Gamma(1-\frac{1}{n}-\frac{2\ri p_1}{ n}) \over \Gamma(\frac{1-2\ri p_1}{ n})}
 \nonumber\\[.2cm]
 f_{1,0}&=&-\psi\big({\textstyle\frac{1}{2}}-p_2-\ri p_1\big)- \psi\big({\textstyle\frac{1}{2}}+p_2-\ri p_1\big)+
\psi\big(-{\textstyle\frac{n}{2}}\big) +\gamma_E+2\,\log(2)-2\ . \nonumber
\eea
Note that the integer powers of $\re^\theta$ in\ \eqref{fpiexpan} come
from the perturbative expansion of the solution $\Psi_{\pm}(x)$. In view
of\ \eqref{psiana}, these powers in $F_+$ are related to the corresponding 
powers in $F_-$ by a change of sign, $\re^\theta\to
-\re^{\theta}$. At the same time, the powers of $\re^{\frac{2\theta}{ n}}$
are the result of the expansion of $\Xi(x)$, and hence they
are the same in $F_-$ and
$F_+$.

One of the important properties of the Wronskians on the right hand side of eqs.\eqref{zet0}, \eqref{zetpi}, is that for a given $\theta$,
they are entire functions in both complex variables $p_1^2$ and $p_2^2$.
Thus it is perfectly fine to consider them for pure imaginary values of $p_1$. Let's set $p_1=\frac{\ri}{2}\, {\mathfrak m}$, $p_2=
{\mathfrak j}+\half$ and assume that
${\mathfrak m}\geq 0$. In this notation, the coefficient $B_{p_1,p_2}$ \eqref{jsasasaus}
contains the factor $\Gamma(-{\mathfrak j}+\frac{{\mathfrak m}}{2})$ in the denominator, and therefore vanishes when
 ${\mathfrak j}-\frac{{\mathfrak m}}{2}$ is a non-negative integer.  At the same time,
 the 
coefficient $B_{-p_1,p_2}$ 
 takes the form $B_{-p_1,p_2}=B_s(\mathfrak{m})$ with
 \beq\label{iusdfkjasd}
B_s(\mathfrak{m})=\sqrt{n}\  n^{-\frac{\mathfrak{m}}{n}}\ \Gamma\Big(1-\frac{{\mathfrak m}}{n}\Big)\  \frac{(-1)^s}{s!}\ 
\frac{\Gamma( 1+ \mathfrak{m}+s)}{\Gamma(1+{\mathfrak m})  }
%\prod_{i=1}^s({\mathfrak m}+i)
\ \ 
%\ \Gamma\Big(1-\frac{{\mathfrak m}}{n}\Big)
\quad
\big(s={\mathfrak j}-\half\, {\mathfrak m}=0,1,2,\ldots \big)\quad
\eeq
 which remains finite for  the discrete set of  ${\mathfrak j}$ and   ${\mathfrak m}$ \eqref{siaisaiais} corresponding
 to the bosonization of the parafermionic spaces ${\cal V}_{\mathfrak j}^{(\mathfrak m)}$.
 The  vanishing  of  the coefficient $B_{p_1,p_2}$ does not actually mean that we can neglect the first term in the sum \eqref{asjssau} --
 the expansion coefficients  in $F_\pm(\theta|p_1,p_2)$ may become singular  for $(p_1,p_2)$  given by eqs.\,\eqref{asjsasai},\,\eqref{siaisaiais}.
 Nevertheless, analysis shows that  such ``resonances''  occur only for  the terms $\propto \re^{(-\frac{\mathfrak m}{n}+i+\frac{2j}{n})\theta}$
 with $i\geq 0$ and  $ j\geq {\mathfrak m}$, i.e., having  the same form as monomials in  the double sum  
 $\re^{\frac{\mathfrak m}{n}\theta}\, F_{\pm}(\theta\,|-p_1,p_2)$.
 Notice that for integer $n$, the double  summations in $F_{\pm}$  can be replaced by a single one. This  yields the formulae
 for the regularized (see eq.\,\eqref {asoisaiasi}) vacuum eigenvalues   in the parafermionic spaces ${\cal V}_{\mathfrak j}^{(\mathfrak m)}$:
  \bea\label{hasasasy}
{\tt \zeta}^{(\rm  vac)}_\pm(\theta)=
\begin{cases} 
{B}_{s}({\mathfrak m})\  \lambda^{\mp 2\lambda^{n}}\ \ \ \ \lambda^{\mathfrak m}\ 
 \bigg(1+\
\sum_{j=1}^\infty f _{j} \  (\pm  \lambda)^{j}\bigg)\bigg|_{\lambda=\re^{\frac{\theta}{n}}}\ \ \ \  &(n-{\rm odd})\\[10.pt]
{B}_{s}({\mathfrak m})\   \lambda^{(2\mp 2)\lambda^{n}}\ \lambda^{\mathfrak m}\ 
 \bigg(1+
\sum_{j=1}^\infty f^{(\pm)} _{j} \   \lambda^{2 j}\bigg)\bigg|_{\lambda=\re^{\frac{\theta}{n}}}\ \ \ \  &(n-{\rm even})\,.
\end{cases}
\eea
The appearance of the ``strange''
 factors, $\lambda^{\mp 2\lambda^{n}},\, \lambda^{(2\mp 2)\lambda^{n}}$ here can be understood as follows. The Wronskians in the l.h.s. of
  eqs.\,\eqref{zet0} and  \eqref{zetpi} are, of course,  non-singular functions of $(p_1,p_2)$ and $n>0.
  \footnote{ The only singularities at $n=2,\,4,\ldots$ are produced by
  the cotangent which shows up in  the expression for the constants $C_\pm$ \eqref{kljasdfd}, \eqref{gkjdas}. 
 The reason  that this term was  included  in the definition of
 $\zeta_\pm$ is that  we would like
the  $T-Q$  type relations to have the simple canonical form \eqref{jasusauu}.}$  However, the individual coefficients $f_{i,j}$ in \eqref{fpiexpan}  become
  singular for some  values of the parameters, as it can be seen from the explicit formulae \eqref{jsasusau}. When  the  values of the parameters
  are restricted to
  the  parafermionic case,  the singularities from the different coefficients  must cancel each other  giving rise to
   terms $\propto\theta^m\re^{l\theta}$ which sum up to yield the ``strange'' factors in \eqref{hasasasy}.
Notice also that some of the coefficients in these series expansions are known explicitly. In Appendix \ref{p_b} their  
values are compared with the corresponding results
obtained from the vacuum eigenvalues of the finite 
matrices  ${\cal Z}_\pm(\mu)$.
  
 \bigskip
We now return to the differential equation and  consider   the solution $\Xi$  in more detail. 
For complex $\theta$, the asymptotic condition   \eqref{ksidef} unambiguously defines    $ \Xi(x\,|\,\theta)$ in the strip
 $\big|\Im m(\theta)\big|\leq \frac{\pi}{2}$ including its boundary.  
 The two 
 functions  $ \Xi(x\,|\,\theta+{\textstyle\frac{\ri\pi}{2}})$  and $ \Xi(x\,|\,\theta-{\textstyle\frac{\ri\pi}{2}})$, with real $\theta$, form a linear 
 basis in the space of solutions of  \eqref{diffa} with $\re^{2\theta}$ substituted by $(-\re^{2\theta})$, since as it follows from
 \eqref{ksidef}, 
  \begin{equation}\label{rwfkjsdf}
  W\big[\,\Xi(x\,|\,\theta+{\textstyle\frac{\ri\pi}{2}}),\,\Xi(x\,|\,\theta-{\textstyle\frac{\ri\pi}{2}})\,\big]=2\ri\,.
  \end{equation}
 %%Therefore the two 
 %functions  $ \Xi(x\,|\,\theta+{\textstyle\frac{\ri\pi}{2}})$  and $ \Xi(x\,|\,\theta-{\textstyle\frac{\ri\pi}{2}})$ form a linear 
%% basis in the space of solutions of the differential equation  \eqref{diffa} with $\re^{2\theta}$ substituted by $(-\re^{2\theta})$.
  On the other hand, formulae \eqref{zet0},\,\eqref{zetpi}
 and \eqref{wpm} imply
 \beq\label{aosisisa}
  \Xi(x\,|\,\theta)=\frac{\exp(\xi\,\re^\theta)}{\sin(2\pi\re^\theta)}\, \Big(\, 
  \zeta^{(\rm vac)}_{+}(\theta)\, 
   \re^{-c\, \re^\theta}\,\Psi_+\big(\,x\,|-\re^{\theta}\,\big)-\zeta^{(\rm vac)}_{-}(\theta)\, 
 \re^{+c\,\re^\theta}\,  \Psi_+\big(\,x\,|+\re^{\theta}\,\big)\,\Big)
 \eeq
 which can be used to express $ \Xi(x\,|\,\theta\pm{\textstyle\frac{\ri\pi}{2}})$
   in terms of $\Psi_+(x|\pm \ri\re^\theta)$ and then calculate the l.h.s. of \eqref{rwfkjsdf}.
This yields the quantum Wronskian type  relation \eqref{aksasus} specified at the vacuum eigenvalues of $\zeta_\pm(\theta)$.
Notice that  in eq.\,\eqref{aosisisa} we use the constants $\xi =-\half\,(C_++C_-)$ and $c=\half\, (C_+-C_-)$. 
Of course, the factors
$\exp(\pm c\,\re^{\theta})$ which appear in this formula can be included  in the definition of the solutions $\Psi_\pm$. With the value of $c$ determined by
eqs.\,\eqref{kljasdfd},\,\eqref{gkjdas},
 the
constant $C$,   appearing in the asymptotic expansions  \eqref{finew} for $\zeta_\pm(\theta)$, vanishes.
The constant 
\bea
\xi=-\half\, (C_++C_-)=\pi\, \cot\big({\textstyle \frac{\pi n}{2}}\big) +\gamma_E+ \psi(1+{\textstyle\frac{n}{2}}\big)
\eea
can also be  absorbed  into  the definition of the solution $\Xi(x)$.
% However it turns out to have an important physical interpretation, which will be discussed later.

Let us assume that the solution  $ \Xi(x\,|\,\theta)$ can be unambiguously continued to the whole complex plane from the 
strip $\big|\Im m(\theta)\big|\leq \frac{\pi}{2}\,.$\footnote{It is expected that $ \Xi(x\,|\,\theta)$  is an entire function of $\theta$. However  at this moment, we don't
have a rigorous proof of this statement.} Then the  function $\Xi(x\,|\,\theta+\ri\pi(2j+\half)\big)$ with $j=0,\,\half,\,1,\ldots$ solves the differential
equation \eqref{diffa} with $\re^{2\theta} \mapsto (-\re^{2\theta})$  and can be linearly expressed in terms of  the two
 basic solutions $\Xi(x\,|\,\theta\pm{\textstyle\frac{\ri\pi}{2}})$
\bea\label{aksasuusa}
\Xi(x\,|\,\theta+\ri\pi(2j+\half )\big)=a_j(\theta)\  \Xi\big(x\,|\,\theta-{\textstyle\frac{\ri\pi}{2}}\big)+b_j(\theta)\  
\Xi\big(x\,|\,\theta+{\textstyle\frac{\ri\pi}{2}}\big)\ .
\eea
With manipulations similar to those which lead to the quantum Wronskian type relations, it is straightforward to show that 
\begin{equation}
a_j(\theta)=-b_{j-\frac{1}{2}}(\theta+\ri\pi)
\end{equation}
and the $b_j\big(\theta-\ri\pi j\big)$ are given by the r.h.s. of \eqref{jkdfghsdf1}, \eqref{jkdfghsdf2} where the operators $\zeta_\pm$ are substituted by 
their vacuum eigenvalues. Hence we conclude that
\begin{equation}\label{aksasuusa2}
\tau^{(\rm vac)}_j(\theta)=b_j(\theta-\ri\pi j)\, .
\end{equation}

  \subsection{ODE/IQFT correspondence for the full spectrum\label{sec4.3}}
  
 In the previous subsection, our analysis  was restricted to the vacuum eigenvalues.  As a matter of fact, 
 along the line of ref.\,\cite{Bazhanov:2003ni},
  it can be extended to the whole spectrum of the commuting families of operators.
  This was already done for a more general case in the work\,\cite{Bazhanov:2013cua}. Here we give a sketch of the construction.

Let  $z$ be the the complex coordinate on $\mathbb{CP}^1\backslash\{z_1,z_2, z_3\}$,
the Riemann sphere with three punctures.
Consider the  second order
Fuchsian differential operator  $-\partial_z^2+T_0(z)$,
with $T_0(z)$  given by
\bea\label{ssapsapo}
T_0(z)=-\sum_{i=1}^3\Big(\,\frac{\delta_i}{(z-z_i)^2}+\frac{c_i}{z-z_i}\,
\Big)\ 
\eea
and   the $\delta_i$ are  regarded as independent  parameters, whereas the $c_i$ are unambiguously defined by the
 constraints
\bea\label{ytosaospa}
\sum_{i=1}^3c_i=0\ ,\ \ \ \ \ \sum_{i=1}^3(z_i\,c_i+\delta_i)=0\ ,\ \ \  \
\sum_{i=1}^3(z_i^2\,c_i+2\,z_i\delta_i)=0\ .
\eea
The equation
\bea\label{sopspa}
\big(-\partial_z^2+T_0(z)\,\big)\,\psi=0
\eea
is a second-order differential equation with three regular
singular points.  For a generic choice of the parameters $\delta_i$,
it can be brought to the 
standard hypergeometric form by a change of variables.
In the papers \cite{Lukyanov:2013wra}, a deformation of \eqref{sopspa} was introduced, of the form 
\bea\label{opspspsao}
{\cal D}_{0}(\theta)\,\psi=0\ ,\ \ \ \ \ \ \ \ \ \  \ \ \ \ \ \ \ \ \ 
{\cal D}_{0}(\theta)=
 -\partial^2_z+T_0(z)+\re^{2\theta}\,{\cal P}(z)\ ,
\eea
where $\theta$ stands for an arbitrary complex parameter
and
\bea\label{oasioaq}
{\cal P}(z)=\frac{(z_3-z_2)^{a_1}\,(z_3-z_1)^{a_2}\,(z_2-z_1)^{a_3}}
{(z-z_1)^{2-a_1}(z-z_2)^{2-a_2}(z-z_3)^{2-a_3}}
\eea
with
\bea
a_1+a_2+a_3=0\, .
\eea
Notice that, because of the last relation,  ${\cal P}(z)\,(\rd z)^2$ transforms as a
quadratic differential under the $\mathbb{PSL}(2,\mathbb{C})$ group, so that  the punctures  $z_1,z_2,z_3$ on the Riemann sphere
can  be sent to any desirable positions.

The immediate object of our interest is a particular  case of the differential equation \eqref{opspspsao} with
\bea
a_1=-n\ ,\ \ \ \ \ \ a_2=n+2\ ,\ \ \  \ \ a_3=0
\eea
and
\bea\label{kassa}
\delta_1=\textstyle\frac{1}{4}+p_1^2\ ,\ \ \ \  \ \ \ \delta_2=\textstyle\frac{1}{4}-p_2^2\ ,\ \ \ \  \ \ \ \delta_3=\textstyle\frac{1}{4}\ .
\eea
 Indeed,  the change of  variables
 \bea
  \re^x=\frac{z-z_3}{z-z_1}\ \frac{z_2-z_1}{z_3-z_2}\ ,\ \ \ \ \ \ \  \Psi (x)\ (\rd x)^{-\frac{1}{2}}= \psi (z)\, (\rd z)^{-\frac{1}{2}}\ ,
  \eea
  brings  equation  \eqref{diffa} to the form \eqref{opspspsao} specialized to this set of  parameters.

 As it was explained in \cite{Bazhanov:2013cua}, the description of the spectrum of the commuting family of transfer-matrices in the
Fock  level subspaces ${\cal F}_{p_1,p_2}^{(L)}$ is based on a differential equation of the form similar to \eqref{opspspsao}, where
the differential operator ${\cal D}_0(\theta)$ is substituted by
\bea\label{aoisaisa}
{\cal D}(\theta)=
 -\partial^2_z+T_L(z)+\re^{2\theta}\,{\cal P}(z)\ ,
\eea
which  has  $3+L$ singular points at
$z=z_1,z_2,z_3$ and also $z=x_1,\ldots, x_L$.
The form of $T_L(z)$, including the positions of the extra singularities, $z=x_1,\ldots, x_L$, is determined by the requirement that the monodromy properties of  the solutions to 
the equation ${\cal D}(\theta)\,\psi(z)=0$
 are identical to those for ${\cal D}_0(\theta)\,\psi(z)=0$.
It turns out that this requirement can be fulfilled  if the set of complex numbers  $\{x_i\}_{i=1}^L$ obey a system of $L$ algebraic 
equations 
similar  to that from \cite{
Bazhanov:2003ni,Fioravanti:2004cz,
  Feigin:2007mr}.
  
  As we saw in the previous subsection, the vacuum eigenvalues of the operators $\zeta_\pm(\theta)$ and  $\tau(\theta)$ can
  be identified with certain connection coefficients for the differential equation\ \eqref{diffa}, or equivalently for
  \eqref{opspspsao}-\eqref{kassa}.\footnote{By  ``connection coefficients''  we understand the
  $\theta$-dependent functions which  allow one to relate different bases in the linear space of solutions of the ordinary differential equation,
  see eqs.\eqref{aosisisa} and \eqref{aksasuusa}.} Of course, similar connection coefficients can be associated to 
  the more general differential operator \eqref{aoisaisa}. Since the singularities of the potential $T_L(z)$ at   $z=x_1,\ldots, x_L$
  do not affect
  the monodromy properties of the solutions of ${\cal D}(\theta)\,\psi(z)=0$,  all the relations between the connection coefficients
  remain unchanged. This allows one to identify them with specializations of the operator relations
  like \eqref{aksasus}, \eqref{jasusauu},\,\eqref{jkdfghsdf1},\,\eqref{jkdfghsdf2} and \eqref{finew},
  to the eigenvalues of the commuting families of operators.

   \subsection{Operators $\beta_\pm(\theta)$ and $\alpha_\pm(\theta)$}
   
With the philosophy  of the ODE/IQFT correspondence in mind, let us return to the differential equation \eqref{diffa}.
It has three singular points at $\re^{x}=0,\,\infty,\,-1$, and we have already discussed the  canonical   bases in the space
 of solutions in the neighbourhood of two of them, $\re^{x}=0$ and $\infty$.
% \bea
%{\boldsymbol\Psi}(x)= \Big(\Psi_+\big(x\,|\, -\re^\theta\,\big),\, \Psi_+\big(x\,|\, \re^\theta\,\big)\Big)\ ,\ \ \ \ \ \ \ \ \ 
%{\boldsymbol\Xi}(x)=  \Big(\Xi_+\big(x\,|\, \theta),\, \Xi_+(x\,|\, \theta+\ri\pi)\Big)\ .
 %\eea
 We   now consider the  basis 
% \bea
% {\boldsymbol\Theta}= \Big(\,\Theta_-(x\,|\, \theta),\,\Theta_+(x\,|\, \theta)\,\big)
% \eea
which is canonically defined in the vicinity of $\re^x=-1$. For this purpose
it is convenient to perform  a change of  variables
\bea
\label{transform}
\re^{-x} =-1-\re^{-y}\ ,\ \ \ \ \ \ \ \ 
\Psi(x)=(1+\re^y)^{-{1\over 2}}\ {\tilde  \Psi}(y)\, , 
\eea
which brings\ \eqref{diffa}\ to the form
\bea\label{NODE}
\bigg[\, -{{\rd^2}\over{\rd y^2}}   +
p_2^2\ \ { {\re^y}\over{1+\re^y}} +
\big({\textstyle{1\over 4}}+p_1^2\, \big)\ 
 {{\re^y}\over{(1+\re^y)^2}} + \re^{2\theta}\ \big(1+
\re^y\big)^{-n-2}\, \bigg]\ {\tilde \Psi}(y) = 0 \ .
\eea
 For $p_2>0$
the differential equation\ \eqref{NODE}\ admits a unique solution such that
\bea
{\tilde \Theta}_{+}(y)\to \sqrt{\frac{\pi}{n+2}}\  (n+2)^{-\frac{2 p_2}{n+2}}\ \ \frac{\re^{-p_2 y}}{\Gamma(1+\frac{2p_2}{n+2})}\ \ \ \ {\rm as}\ \ \ 
 y\to +\infty\  .
\eea
For complex $p_2$, one can show that the solution is a meromorphic function,  analytic
in the half plane
$\Re e(p_2)\geq  0$. Since the equation \eqref{NODE}\ is invariant w.r.t.
the
substitution $p_2\mapsto-p_2$, for generic values of $p_2$
one can  define the second linear independent  solution uniquely by
 analytic continuation $\mathscr{A}_{p_2\mapsto \re^{\pm\ri\pi}p_2}$ to the half plane $\Re e(p_2)< 0$:
\bea\label{aiaasisa}
{\tilde \Theta}_{-}(y\,|\,\re^{2\theta},\, p_2)=\mathscr{A}_{p_2\mapsto \re^{\pm\ri\pi}p_2}\Big[{\tilde \Theta}_{+}(y\,|\,\re^{2\theta},p_2)\Big]\ ,
\eea
and it is easy to see that
\bea\label{Owron}
W\big[\,{\tilde { \Theta}}_{+},\,{\tilde { \Theta}}_{-}\,\big]=\sin\big({\textstyle\frac{2\pi p_2}{n+2}}\big)\  .
\eea
Notice that both these solutions are entire functions of the variable $\re^{2\theta}$ whose dependence is emphasized in the  formula\ \eqref{aiaasisa}. 

As $y\to-\infty$, eq.\,\eqref{NODE} admits the two linearly independent solutions
\bea\label{psipfkdd} 
{\tilde \Psi}_{\pm }(y) \to\, \sqrt{\pi}\  \ \frac{\exp\big(\pm \re^\theta \,y\big)}{
\Gamma(1\pm 2\,\re^\theta)} \quad \ \ \ \ \ {\rm as} \quad\ \ \ \  y\to-\infty\ .
\eea
Of course, ${\tilde \Psi}_\pm(y)$ are obtained from $ \Psi_\pm(x)$ \eqref{psiplusdef},\,\eqref{psiminusdef}  by means of the coordinate transformation \eqref{transform}:
\bea\label{hassaasy}
{ \Psi}_\pm(x)\ (\rd x)^{-\frac{1}{2}}=\exp\big(\mp \ri\pi\,\re^\theta\,\big)\  \ {\tilde \Psi}_\pm(y)\ (\rd y)^{-\frac{1}{2}}\ .
\eea
Here the extra phase factor appears because of slightly different normalizations: ${ \Psi}_\pm(x)$ were defined in such a way that they  stay real  for real $x$
 and real  values of the parameters, whereas  ${\tilde  \Psi}_\pm(y)$ \eqref{psipfkdd}  are  real for real $y$ which is related to $x$ as $\re^{-y}=-1-\re^{-x}$.
It is important to keep in mind  the presence of such phase factors, because they affect the reality conditions for the connection coefficients.
 
 Following the philosophy of the ODE/IQFT correspondence, we consider the connection coefficients
 for the bases $\big\{{\tilde  \Psi}_\pm(y)\big\}$ and  $\big\{{\tilde  \Theta}_\pm(y)\big\}$ and interpret them as vacuum eigenvalues of
  certain operators $\beta_\pm (\theta)$:
 \bea\label{assasay}
 \re^{+c\,\re^\theta}\ {\tilde \Psi}_+(y\,|\,\re^{\theta}\big)=\frac{1}{2\sin(\frac{2\pi p_2}{n+2})}\ \Big(\,
  \beta^{(\rm vac)}_+(\theta)\ {\tilde \Theta}_-\big(y\,|\,\re^{2\theta}\big)-
  \beta^{(\rm vac)}_-(\theta)\ {\tilde \Theta}_+\big(y\,|\,\re^{2\theta}\big)
 \,\Big)\,.
 \eea
 %A similar relation can be written  for $ {\tilde \Psi}_-$ with $\beta^{(\rm vac)}_\pm(\theta)$ substituted by $\beta^{(\rm vac)}_\pm(\theta+\ri\pi)$.
 The relation for $ \re^{-c\,\re^\theta}\,{\tilde \Psi}_-$ is similar with $\beta^{(\rm vac)}_\pm(\theta)$ substituted by $\beta^{(\rm vac)}_\pm(\theta+\ri\pi)$.
Notice that  it is expected that
\bea\label{ashsasya}
\beta_\pm(\theta+\ri\pi)=\beta_\pm(\theta-\ri\pi)\ .
 \eea
 Further,  the operators  $\beta_\pm$ are entire functions of the variable $\re^{\theta}$ and can be written in the form of a  convergent series
 \bea\label{jasaasy}
 \beta_\pm(\theta)= b_\pm\ \bigg(1+\sum_{m=1}^\infty {\mathfrak b}^{(\pm)}_m\, \re^{m\theta}\, \bigg)\ .
 \eea
The 
reality condition for the connection coefficients  $\beta^{(\rm vac)}_\pm(\theta)$ suggests the Hermiticity 
 \bea\label{iuywqesadj}
 \big[\beta_\pm(\theta)\big]^\dagger=\beta_\pm(\theta^*)\ .
 \eea
 Using the definition \eqref{assasay}, one can
  express  the Wronskian $W\big[\,{\tilde { \Psi}}_{-},\,{\tilde { \Psi}}_{+}\,\big]$ in terms
of the connection coefficients  $\beta^{(\rm vac)}_\pm$. On the other hand, as follows from \eqref{psipfkdd},  it is equal to $\sin(2\pi\re^\theta)$.
This yields the quantum Wronskian relation
\bea\label{ajsasy}
\beta_+\big(\theta+{\textstyle\frac{\ri\pi}{2}}\big)\,\beta_-\big(\theta-{\textstyle\frac{\ri\pi}{2}}\big)-
\beta_-\big(\theta+{\textstyle\frac{\ri\pi}{2}}\big)\,\beta_+\big(\theta-{\textstyle\frac{\ri\pi}{2}}\big)=-4\ri\, \sinh\big(2\pi\re^\theta\big)\ 
 \sin\big({\textstyle\frac{2\pi p_2}{n+2}}\big)\ ,
 \eea
 where  the superscript ``{\footnotesize{(vac)}}'' is omitted, since we expect that it holds true for all eigenvalues  of the operators $\beta_\pm$.

 It turns out that the operators \eqref{jasaasy}, acting in the  Fock space ${\cal F}_{p_1,p_2}$, 
 and
 satisfying the commutativity conditions
 \bea
 \big[\beta_\sigma(\theta),\,\beta_{\sigma'}(\theta')\,\big]= 
  \big[\beta_\sigma(\theta),\,  \tau(\theta')\,\big]=0\ \ \ \ \ \ \big(\sigma,\,\sigma'=\pm\big)\ ,
 \eea
 can be defined explicitly. Their construction lies beyond the scope of this work. Here we just mention that
% it is based on the affine quantum  superalgebra $U_{q}\big(\hat{\mathfrak{sl}}(2|1)\big)$ and 
it is similar to the construction
  of the $U_q(\hat{\mathfrak{sl}}_2)$  $Q$-operators from refs.\cite{Bazhanov:1996dr,Bazhanov:1998dq}.
  % Also note that the vacuum eigenvalues of the operators
  %${\mathfrak b}^{(\pm)}_1$ coincides with
   %$f_{1,0}(p_1,\pm p_2)-\psi\big(-{\textstyle\frac{n}{2}}\big) -\gamma_E-2\,\log(2)+2$,
   %$f_{1,0}(p_1,\pm p_2)-\pi\cot(\frac{\pi n}{2})$,
  % where $f_{1,0}$ is given by eq.\,\eqref{jsasusau}.

 \bigskip
 
The last set of connection coefficients  relates the basis in the space of solutions  canonically defined at the singular point
$\re^{x}=\infty$ with that defined at $\re^x=-1$, or,  equivalently $\re^y=-1$ with $\re^y=0$.
Let ${\tilde \Xi}(y)\equiv(1+\re^{y})^{\frac{1}{2}}\  \Xi(x)|_{x(y)}$. Then it can be written in the form similar to
\eqref{aosisisa} and \eqref{assasay}:
\bea\label{aassuas}
{\tilde \Xi}(y\,|\,\theta)=\frac{\exp(\xi\,\re^\theta)}{\sin(\frac{2\pi p_2}{n+2})}\ \ \Big(a_+(\theta)\ {\tilde\Theta}_- \big(y\,|\,\re^{2\theta}\big)-
a_-(\theta)\ {\tilde \Theta}_+ \big(y\,|\,\re^{2\theta}\big)\,\Big)\ .
\eea
%with $a_\pm=W[{\tilde \Xi},{\tilde \Theta}_\pm]$.
Since for real $y$,  the solution  ${\tilde \Xi}(y)$ is complex, the reality condition looks simpler if we introduce $\alpha^{(\rm vac)}_\pm(\theta)$, such that
\bea\label{hasta}
%\alpha^{(\rm vac)}_\pm \big(\theta-{\textstyle \frac{\ri\pi n}{2}}\big)=\ri\ \re^{\pm \ri\pi p_2}\  
%\re^{\frac{1}{2}\,(C_++C_-)
%\, \re^\theta}\ 
%a_\pm(\theta)\,,
a_\pm(\theta)=\ri\  \re^{\mp \ri\pi p_2}\ \alpha^{(\rm vac)}_\pm \big(\theta-{\textstyle \frac{\ri\pi n}{2}}\big)\ .
\eea
 The latter  turn out to be real functions for real $\theta$ and $(p_1,p_2)$. Again, we  interpret them as the vacuum eigenvalues  of the
Hermitian operators $\alpha_\pm(\theta)$,
\bea\label{jssasu}
\big[\alpha_{\pm}(\theta)\big]^\dagger=\alpha_{\pm}(\theta^*)\ ,
\eea
which are  also  expected to be entire functions of $\theta$.
Similar to the operators $\zeta_\pm(\theta)$,  one  can    obtain   the quantum Wronskian relation
\bea\label{hsassat}
{\alpha}_+ \big(\theta-{\textstyle\frac{\ri\pi}{2}}\big)\, {\alpha}_-  \big(\theta+{\textstyle\frac{\ri\pi}{2}}\big)-
{\alpha}_-  \big(\theta-{\textstyle\frac{\ri\pi}{2}}\big)\, {\alpha}_+ \big(\theta+{\textstyle\frac{\ri\pi}{2}}\big) =2\ri\, \sin\big({\textstyle\frac{2\pi p_2}{n+2}}\big)\
\eea
and
the $T-Q$ relations, which now have the canonical  form
\bea\label{kasasuas}
\tau(\ri\lambda)\ \alpha_\pm(\theta)=\alpha_\pm(\theta+\ri\pi )+\alpha_\pm(\theta-\ri\pi )
\eea
$\big($recall that $\tau(\ri\lambda)=\tau(-\ri\lambda)$ and $\lambda=\re^{\frac{\theta}{n}}\,\big)$.

\bigskip
Using the WKB approximation, one can explore the asymptotic behaviour of the connection coefficients for $\theta\to+\infty$.
This leads to the asymptotic expansions similar  to formulae\,\eqref{finew} for $\zeta_\pm(\theta)$. In particular, it is expected
that for $\Re e(\theta)\to+\infty$ and $|\Im m(\theta)|<\frac{\pi}{2}\,(n+2)$,
\bea\label{jassysay}
\alpha_\pm(\theta)\,\asymp\,\re^{\mp\ri \pi p_2}\ 
 {\tilde \alpha}_\pm\Big(\re^{\frac{\ri\pi}{2}}\, \re^{-\frac{\theta}{n+2}}\,\Big)\ \ \exp\Big( -{\textstyle \frac{\pi}{\sin(\frac{\pi n}{2})}}\,  \re^{\theta}\, \Big)\ ,
 \eea
where ${\tilde\alpha}_\pm({\tilde\lambda})$ stand  for the formal power series of the form
\bea\label{ashsaysa}
 {\tilde\alpha}_\pm({\tilde\lambda})=
 \big({\tilde \lambda}\big)^{\pm 2 p_2}\ \exp\bigg(-\sum_{m=1}^\infty  {\tilde {\mathfrak s}}^{(\pm)}_m\ {\tilde \lambda}^{2m}\bigg)\ .
\eea
The coefficients $\{ {\tilde {\mathfrak s}}^{(\pm)}_m\}_{m=1}^\infty$ are   dual nonlocal IM, which are algebraically  expressed through
the set  $\{ {\tilde {\mathfrak t}}^{(\pm)}_m\}_{m=1}^\infty$ by means of
the  formal $T-Q$  relation (to be compared with eq.\,\eqref{kasasuas}):
\bea\label{aasasyas}
{\tilde \tau}({\tilde \lambda})\,  {\tilde\alpha}_\pm({\tilde \lambda})={\tilde\alpha}_\pm\big({\tilde q}\,{\tilde \lambda}\big)+
 {\tilde\alpha}_\pm\big({\tilde q}^{-1} {\tilde \lambda}\big)\ ,
\eea
where ${\tilde q}=\re^{-\frac{\ri\pi}{n+2}}$. Substituting the formal power series   \eqref{jasusau} and \eqref{ashsaysa} in \eqref{aasasyas},
one can easily derive the relations between these two sets of dual nonlocal  IM.
For example
%\bea
%{\tilde {\mathfrak s}}^{(\pm)}_1&=&-\frac{{\tilde {\mathfrak t}}_1}{4\sin(\frac{\pi}{n+2})\sin(\frac{\pi (1\pm 2 p_2)}{n+2})}\\
%{\tilde {\mathfrak s}}^{(\pm)}_2&=&-\frac{{\tilde {\mathfrak t}}_2}{4\sin(\frac{2\pi}{n+2})\sin(\frac{2\pi (1\pm  p_2)}{n+2})}+
%\frac{{\tilde {\mathfrak t}}^2_1}{16\sin(\frac{\pi}{n+2})\sin(\frac{2\pi}{n+2})\sin(\frac{\pi (1\pm 2 p_2)}{n+2})\sin(\frac{2\pi (1\pm  p_2)}{n+2})}\nonumber
%\eea
\beq\label{sahsayas}
{\tilde {\mathfrak s}}^{(\pm)}_1=\frac{{\tilde {\mathfrak t}}_1}{[1][1\pm 2p_2]}\ ,\ \ \
{\tilde {\mathfrak s}}^{(\pm)}_2=\frac{{\tilde {\mathfrak t}}_2}{[2][2\pm 2 p_2]}-
\frac{{\tilde {\mathfrak t}}^2_1}{[1][2][1\pm 2p_2][2\pm 2p_2]}+\frac{{\tilde {\mathfrak t}}^2_1}{2[1]^2[1\pm 2p_2]^2}\ ,
\eeq
where we use the shortcut  notation $[x]=2\,\sin(\frac{\pi x}{n+2})$.

Unlike for $\alpha_\pm$, the large-$\theta$ asymptotics of the  operators $\beta_\pm$  include a contribution from the local IM;
for $\Re e(\theta)\to+\infty$ and $|\Im m(\theta)|<\pi$, they read as
\bea\label{qpwoeier}
\beta_\pm(\theta)\,\asymp\,{\tilde \alpha}_\pm\Big( \re^{-\frac{\theta}{n+2}}\,\Big)\ \exp\Big( 
  -2\theta\,\re^\theta-\sum_{j=1}^{\infty}\, {\mathfrak i}_{2j-1}\,
\re^{-\theta (2j-1)}\, \Big)\ .
\eea

 Finally,  using the formulae \eqref{aosisisa},\,\eqref{hassaasy},\,\eqref{assasay},\,\eqref{aassuas},\,\eqref{hasta}, it is 
 straightforward to show that the three sets of  connection coefficients are not functionally 
 independent,  they satisfy the relations
 \bea\label{exna}
\zeta_+(\theta)&=&\frac{\ri\, \re^{-\ri\pi \re^{\theta}}}{2\sin(\frac{2\pi p_2}{n+2})}\  \Big[\,\re^{-\ri\pi p_2}\ \beta_-(\theta)\, 
{\alpha}_{+}\big(\theta-{\textstyle \frac{\ri\pi n}{2}}\big)-
\,\re^{+\ri\pi p_2}\,  \beta_+(\theta)\ {\alpha}_{-}\big(\theta-{\textstyle \frac{\ri\pi n}{2}}\big)\, \Big]\\[0.2cm]
\zeta_-(\theta)&=&\frac{\ri\, \re^{+\ri\pi \re^{\theta}}}{2\sin(\frac{2\pi p_2}{n+2})}\   \Big[\, \re^{-\ri\pi p_2}\ \beta_-(\theta+\ri\pi)\,  
{\alpha}_{+}\big(\theta-{\textstyle \frac{\ri\pi n}{2}}\big)-\re^{+\ri\pi p_2}\ 
\beta_+(\theta+\ri\pi)\,  {\alpha}_{-}\big(\theta-{\textstyle \frac{\ri\pi n}{2}}\big)\, \Big]\, .\nonumber
\eea
It turns out these  formulae  and the quantum Wronskian relations, supplemented by the Hermiticity 
\eqref{sassao},\,\eqref{iuywqesadj},\ \eqref{jssasu}
and  the analyticity  of the operators  $\alpha_\pm(\theta),\beta_\pm(\theta)$,  $\zeta_\pm(\theta)$
constitute a very restrictive set of conditions.
In particular, it leads to the important relation (see Appendix\,\ref{appC1})
\beq\label{jassay}
\alpha_{\pm}\big(\theta\!-\!{\textstyle \frac{\ri\pi (n+1)}{2}}\big)\alpha_{\pm}\big(\theta\!+\!{\textstyle \frac{\ri\pi (n+1)}{2}}\big)\!
-\!\alpha_{\pm}\big(\theta\!-\!{\textstyle \frac{\ri\pi (n-1)}{2}}\big) \alpha_{\pm}\big(\theta\!+\!{\textstyle \frac{\ri\pi (n-1)}{2}}\big)\!=\!
 \beta_\pm\big(\theta\!-\!{\textstyle \frac{\ri\pi}{2}}\big)\beta_\pm\big(\theta\!+\!{\textstyle \frac{\ri\pi}{2}}\big)\,.
\eeq

  \subsection{NLIE for the vacuum eigenvalues\label{sec45}}
The ODE/IQFT correspondence allows one, in principle, to find the spectrum  of the commuting family of operators by 
numerically  solving  differential  equations. Such a numerical procedure is especially convenient for the calculation of the eigenvalues of
the operators $\beta_\pm(\theta)$.
However, it  is a highly non-trivial task to, say, extract the eigenvalues of the chiral transfer-matrix $\tau(\lambda)$ directly from the differential equations.
Here we demonstrate how the functional relations and analytic conditions for the connection coefficients can be used to derive a system of
Non-Linear Integral Equations (NLIE) which prove to be highly efficient in numerical work  \cite{Klummpe:1991vs,Destri:1994bv1,Destri:1994bv2}. 
We will mostly focus on the 
vacuum eigenvalues.
\bigskip

For our purposes it is useful  to rewrite eq.\,\eqref{jassay}  using the notation
\bea\label{hasssay}
\re^{\ri\varepsilon(\theta)}=\exp\big(2\pi\ri\, (\re^\theta-p_2)\big)\ 
\ \frac{\alpha_+\big(\theta-\frac{\ri\pi n}{2}\big)}{\alpha_+\big(\theta+\frac{\ri\pi n}{2}\big)}\ .
\eea
Focusing on the case where   the subscript in \eqref{jassay} is ``+'',    one has
\bea\label{jashsatg}
1-\re^{4\pi\,\re^\theta+\ri\, \varepsilon\big(\theta+{\textstyle\frac{\ri\pi}{2}}\big)-\ri\,\varepsilon\big(\theta-{\textstyle\frac{\ri\pi}{2}}\big)}=
\frac{ \beta_+(\theta-\frac{\ri\pi}{2})\, \beta_+(\theta+\frac{\ri\pi}{2})}
{\alpha_+\big(\theta-\frac{\ri\pi(n+1)}{2}\big)\,\alpha_+\big(\theta+\frac{\ri\pi(n+1)}{2}\big)}\ .
\eea
In order to define the operator $\varepsilon(\theta)$ itself, one should specify the branch of the logarithm $\frac{1}{\ri} \log(\re^{\ri\varepsilon(\theta)})$.
This can be done by supplementing  \eqref{hasssay} with the leading asymptotic behaviour
\bea\label{ahsassat}
\varepsilon(\theta)\to 4\pi\re^\theta -{\textstyle \frac{4\pi p_2}{n+2}}+o(1)\ \ \ \ \ \ \ \ \ \ {\rm as}\ \ \ \ \theta\to+\infty\ ,
\eea
which is chosen   to be consistent with the asymptotic formula  \eqref{jassysay}.  As follows from eq.\,\eqref{jssasu}, the operator $\varepsilon(\theta)$, thus defined
and 
acting  in the Fork space ${\cal F}_{p_1,p_2}$,  satisfies the 
Hermiticity condition
\bea\label{dsfkjhasdjqk}
\big[\varepsilon(\theta)\big]^\dagger=\varepsilon(\theta^*)\ .
\eea
Let us emphasize that \eqref{jashsatg}-\eqref{dsfkjhasdjqk} are operator relations for the commuting family, and therefore the
same relations hold true for  the eigenvalues corresponding to any common eigenvector $|\,\psi\,\rangle$ in the Fock space.
Another important general property (which justifies the ``strange'' first  factor in the definition \eqref{hasssay})
concerns  the zeroes of the eigenvalue $\beta^{(\psi)}_+(\theta)$. Namely it is easy to show   that (see Appendix\,\ref{appC1})
\bea\label{jassaya}
{\rm if} \  \ \theta_j:\ \ \beta^{(\psi)}_+(\theta_j)=0\,, \  \ {\rm then} \ \ \ \ \exp\big(\ri\varepsilon^{(\psi)}(\theta_j-\ri\pi)\big)=-1\ .
\eea
Notice that since $\beta^{(\psi)}_+(\theta)$ is a real analytic and  periodic  function in $\theta$ with period $2\pi\ri$, it is sufficient to consider
its zeroes in the strip $0\leq \Im m(\theta)\leq \pi$, only.

\bigskip
In the case of the vacuum eigenvalues, our numerical work suggests that
for $\frac{2 p_2}{n+2}>-\half$ all the zeroes of $\beta^{({\rm vac})}_+(\theta)$ in the strip 
$|\Im  m(\theta)\big|\leq \pi$ are simple, located on the boundary  $\Im m(\theta)=\pi$, accumulate toward  $\Re e(\theta)\to+\infty$ and
satisfy the condition
\bea\label{aisisi}
\varepsilon^{\rm (vac)}(\theta_j-\ri\pi )=\pi \, (2j-1)\ \ \ \ \ \ \ \ \ (j=1,\,2,\,\ldots)\,.
\eea
This  ``quantization condition'' supplemented by the asymptotic formula \eqref{jassysay}, leads  to an  equation
determining the vacuum roots of  $\beta^{({\rm vac})}_+(\theta)$ which is asymptotically exact as $j\to +\infty$:
\beq\label{hsassatas}
\exp\big(\theta_j^{(\rm vac)}\big)\equiv-\rho_j\  :\ \ \ \rho_j\asymp \half\ \big( j-\half+  {\textstyle \frac{2 p_2}{n+2}}\,\big)-{\textstyle \frac{1}{2\pi}}\ 
\sum_{m=1}^\infty \tilde{s}_m(p_1,p_2)\ \sin\big({\textstyle \frac{2\pi m}{n+2}}\big)\ (\rho_j)^{-\frac{2m}{n+2}}
\eeq
where $\tilde{s}_m(p_1,p_2)$ stands for the vacuum eigenvalues of the dual  nonlocal IM $\tilde{\mathfrak{s}}^+_m$ in \eqref{ashsaysa}, \eqref{sahsayas}.
 In particular
\begin{equation}\label{oqwekjaxfd}
\tilde{s}_1(p_1,p_2)=-\Big({n+2\over 2}\Big)^{{2\over n+2}}\ 
{ \Gamma({1\over n+2})\Gamma({1\over 2}-{1\over n+2})\over
4\, \sqrt{\pi} 
}\ 
\ \bigg[\, \frac{n}{ n+4}-
\frac{4p^2_1}{1-4p_2^2}\, 
\bigg]\
\frac{\Gamma(1+\frac{2p_2+1}{  n+2})}{\Gamma(\frac{2p_2-1}{  n+2})}\ .
\end{equation}
%As for the vacuum eigenvalues $\alpha^{({\rm vac})}_+(\theta)$, it is expected that
%this function does not have any  zeroes in  the strip  $|\Im m(\theta)|<\frac{\pi}{2}\, (n+2)$. In principle, this property should be
%rigorously derived from the differential equation \eqref{NODE}. Unfortunately, we don't have a proof at this moment and we formulate 
%the statement as a conjecture.
%Notice that the domain of applicability of the large-$\theta$ asymptotic expansion \eqref{jassysay}  is also restricted to the same strip.
 
Additional analytical input required for the derivation of the NLIE, is  the behaviour at $\Re e(\theta)\to -\infty$.
It can be  studied  along  the following line:  using the formulae \eqref{exna} and the quantum Wronskian relations
one can express $\alpha_+^{\rm(vac)}$ in terms of $\zeta^{\rm(vac)}_\pm$ and $\beta^{\rm(vac)}_+$.
The asymptotics of   $\zeta^{\rm(vac)}_\pm(\theta)$  at $\theta\to-\infty$ are given by  eqs.\,\eqref{asjssau}. The general  structure
of the $\theta\to-\infty$ behaviour of $\beta^{\rm(vac)}_+$ is dictated by the operator valued series expansion \eqref{jasaasy}.
Using the differential equation \eqref{NODE}, it is not difficult to find the following explicit expressions for
the vacuum eigenvalues of the first expansion coefficients:
% in \eqref{jasaasy}:
 \bea\label{aashsat}
 b_\pm^{(\rm vac)}(p_1,p_2)= (n+2)^{-\frac{1}{2}\mp\frac{2p_2}{n+2}}\ \ \ 
 \frac{2\pi\ \Gamma(1\pm 2 p_2)}{ \Gamma({1\over 2}-\ri p_1\pm p_2)\
\Gamma({1\over 2}+\ri p_1\pm p_2)\,\Gamma(1\pm\frac{ 2 p_2}{n+2})}
 \eea
and  \big(to be compared with the coefficient $f_{1,0}$ in eq.\,\eqref {jsasusau}\big)
\beq
b_1^{(\pm,{\rm vac)}}(p_1,p_2)=-\psi\big(\half-\ri p_1\pm p_2\big)-\psi\big(\half+\ri p_1\pm p_2\big)+\psi\big(1+{\textstyle\frac{n}{2}}\big)+\gamma_E
+2\,\log 2-2\ .
\eeq
Now it is straightforward to derive  the leading asymptotic behaviour of the vacuum eigenvalue of the operator $\varepsilon(\theta)$ 
\eqref{hasssay},\,\eqref{ahsassat}. Below we  list  explicit formulae for $\varepsilon^{(\rm vac)}(\theta)$ and for the vacuum eigenvalues of the operator
\beq\label{ajshsahyt}
\omega(\theta)\equiv \log\Big(\re^{\ri\varepsilon(\theta-\frac{\ri\pi}{2})-\ri\varepsilon(\theta+\frac{\ri\pi}{2})-4\pi\re^\theta}-1\Big)\ ,\ \ \ \  \ \omega(\theta)\to 4\pi\,\re^{\theta}+o(1)\ \ \ \ {\rm as}\ \  \ \theta\to+\infty\ .
\eeq

\bigskip
\begin{enumerate}[label=(\alph*)]
\item  For real $p_1\not=0$ and $ p_2> -\half $: \label{sdkjfhewqc} 
\bea\label{dksjfhsdf1}
\varepsilon^{(\rm vac)}(\theta)&=&-2\pi p_2+\frac{1}{\ri}\ \log\bigg[\frac{\sin\big( \frac{2 p_1 }{n} (\theta_0-\theta+\frac{\ri\pi n}{2})\big)}
{\sin\big( \frac{2 p_1 }{n} (\theta_0-\theta-\frac{\ri\pi n}{2})\big)}\bigg]+o\big(\theta^{-\infty}\big)
\\[0.3cm]
\exp\big(-\omega^{({\rm vac)}}(\theta)\,\big)&=& \frac{\sinh^2(\frac{\pi (n-1) p_1}{n})}{\sinh(2\pi p_1)\sinh(\frac{2\pi p_1}{n})}+
\frac{\sin^2(\frac{2 p_1}{n} (\theta_0-\theta)\big)}{\sinh(2\pi p_1)\sinh(\frac{2\pi p_1}{n})}+
 o\big(\theta^{-\infty}\big)\nonumber
\eea
where
\bea
\theta_0=\log(n)+\frac{n}{4\ri p_1}\ \log\bigg[\,
\frac{\Gamma(1+2\ri p_1) \Gamma(1+\frac{2\ri p_1}{n})}{\Gamma(1-2\ri p_1) \Gamma(1-\frac{2\ri p_1}{n})}
\ 
\frac{\Gamma^2(\frac{1}{2}+p_2-\ri p_1)  }
{\Gamma^2(\frac{1}{2}+p_2+\ri p_1) }\, \bigg]\ .\nonumber
\eea
\item For $p_1=0$ and $p_2>-\half$:\label{sdkjfhewqa} 
\bea\label{aisisosa}
\varepsilon^{(\rm vac)}(\theta)&=&-2\pi p_2+\frac{1}{\ri}\ \log\bigg[\frac{\theta_0-\theta+\frac{\ri\pi n}{2}}{\theta_0-\theta-
\frac{\ri\pi n}{2}}\bigg]+o\big(\theta^{-\infty}\big)\nonumber\\[0.2cm]
\exp\big(-\omega^{({\rm vac)}}(\theta)\,\big)&=& { \frac{(\theta_0-\theta)^2}{n\pi^2}}+{\frac{1}{4 n}\, (n-1)^2}+o\big(\theta^{-\infty}\big) \ ,
\eea
where
\bea\label{assjsaias}
\theta_0=\log(n)-(1+n)\,\gamma_E-n\, \psi\big(\half+p_2\big)\ .\nonumber
\eea

\item For pure imaginary $p_1\equiv\frac{\ri}{2}\ {\mathfrak m}$ with $0< {\mathfrak m}< \frac{n}{2}$  and real 
$p_2\equiv {\mathfrak j}+\half$
such that ${\mathfrak j}-\half\, {\mathfrak m}>-1$:\label{sdkjfhewqb}
\bea\label{hsssaty}
\varepsilon^{(\rm vac)}(\theta)&=& -\pi\, (2{\mathfrak j}+1-{\mathfrak m})+
\frac{2\pi {\mathfrak m}\, n^{-\frac{2{\mathfrak m}}{n}}}{\Gamma^2(1+{\mathfrak m})}\ 
\ 
\frac{\Gamma(1-\frac{{\mathfrak m}}{n})}
{\Gamma(1+\frac{{\mathfrak m}}{n})}\ \frac{\Gamma^2(1+{\mathfrak j}+\frac{{\mathfrak m}}{2})}
{\Gamma^2(1+{\mathfrak j}-\frac{{\mathfrak m}}{2})}\
 \re^{\frac{2{\mathfrak m} \theta}{n}}+
o\big(\re^{\frac{2{\mathfrak m }\theta}{n}}\big)\nonumber\\[0.3cm]
\omega^{(\rm vac)}(\theta)&=&\frac{2{\mathfrak m}}{n}\,\theta-2\, \log\bigg[n^{\frac{1}{2}+\frac{{\mathfrak m}}{n}}
\ \frac{\Gamma(1+{\mathfrak j}-\frac{{\mathfrak m}}{2})
\Gamma({\mathfrak m}) \Gamma(1+\frac{{\mathfrak m}}{n})}
{2\pi \Gamma(1+{\mathfrak j}+\frac{{\mathfrak m}}{2})}\bigg]+O\big(\re^{\frac{2\theta}{n}},\re^\theta\big)\ .
\eea

\end{enumerate}

\noindent
Notice that eqs.\,\eqref{aisisosa} with $p_2={\mathfrak j}+\half$ 
and
\bea
 {\mathfrak j}=
 {\textstyle \frac{1}{2}},\, 1,\ldots, {\textstyle \frac{1}{2}}\,\big[ {\textstyle \frac{n}{2}}\big]\ \ \ \ \ \ \ (n=2,\,3,\,4,\ldots)
  \eea
  can be applied to the parafermion case with ${\mathfrak m}=0$.
For  positive integer ${\mathfrak m}$, restricted to
\bea
 1\leq {\mathfrak m}\leq\big[{\textstyle \frac{n-1}{2}}\big],\ \ \ \ \ \ {\mathfrak j}-\half\,  {\mathfrak m}\geq 0\ ,
 \eea
eqs.\,\eqref{hsssaty} should be used.
 \smallskip

It is expected that for the cases \ref{sdkjfhewqa} and \ref{sdkjfhewqb},
the function $\alpha^{(\rm vac)}(\theta)$ does not have any  zeroes in  the strip  $|\Im m(\theta)|<\frac{\pi}{2}\, (n+2)$.
However, as it follows from 
the asymptotic behaviour \eqref{dksjfhsdf1} and formula \eqref{hasssay}, for $p_1\ne 0$ and $p_2>-\frac{1}{2}$ it
has a sequence of zeroes, $\{\theta_m^{(\alpha)}\}_{m=1}^{\infty}$,
 extending towards $-\infty$ along the real axis such that
\bea\label{oiuqwmnaspx}
\theta_m^{(\alpha)}\,=\,\theta_0\,-\,\frac{\pi n}{2 p_1}\,m\,+\,o\big((m/p_1)^{-\infty}\big)\qquad\qquad (m=1,2,\ldots )\,.
\eea
Again, we expect that there are no other zeroes apart from $\{\theta_m^{(\alpha)}\}_{m=1}^{\infty}$ in the strip 
$|\Im m(\theta)|<\frac{\pi}{2}\, (n+2)$.
In principle, these properties should be
rigorously derived from the differential equation \eqref{NODE}. Unfortunately, we don't have a proof at this moment and we formulate 
the statements as a conjecture.
Notice that the domain of applicability of the large-$\theta$ asymptotic expansion \eqref{jassysay}  is also restricted to the same strip.
 
The outlined analytic properties fully determine all of the vacuum eigenvalues of the commuting  operators 
acting in the Fock space ${\cal F}_{p_1,p_2}$. 
Practically, they can be used to derive  a closed system of integral equations which involve
the vacuum eigenvalues of  $\varepsilon^{(\rm vac)}(\theta)$ and $\omega^{(\rm vac)}(\theta)$.
A few useful formulae appearing in the intermediate steps of the derivation are given in appendices \ref{appC1} and \ref{apDD}.
With the parameters $p_1$ and $p_2$ as in \ref{sdkjfhewqa} and \ref{sdkjfhewqb} above, the final result reads as follows:
\vspace{-0.1cm}
\bea\label{hasatast}
\varepsilon(\theta-\ri\gamma)&=&4\pi\,\re^{\theta-\ri\gamma}-2\pi k+\int_{-\infty}^\infty\frac{\rd \theta'}{2\pi\ri }\
\Big[G(\theta-\theta'-2\ri\gamma)\  \big(L(\theta'-\ri \gamma)\big)^*\nonumber\\[0.2cm]
&-&   G(\theta-\theta')\  L(\theta'-\ri \gamma)\Big]
+
\int_{-\infty}^\infty\frac{\rd \theta'}{2\pi}\ G_1(\theta-\theta'-\ri\gamma)\  \log\big(1+\re^{-\omega(\theta')}\big)\nonumber\\[0.4cm]
\omega(\theta)&=&4\pi\, \re^{\theta}+\Im m\bigg[\int_{-\infty}^\infty\frac{\rd\theta'}{\pi}\,G_1(\theta-\theta'+\ri\gamma)\  L(\theta'-\ri \gamma)
\,\bigg]\\[0.2cm]
&-&
\int_{-\infty}^\infty\frac{\rd\theta'}{\pi}\, G_2(\theta-\theta')\, \log\big(1+\re^{-\omega(\theta')}\big)\nonumber
\\[0.4cm]
%\varepsilon(\theta-\ri\gamma)+2\pi k=
%-\int_{-\infty}^\infty\frac{\rd \theta'}{\pi}\ 
%&\frac{ \log\big(1+\re^{b(\theta')}\big)-4\pi\,\re^{\theta'}}{1+\re^{2\theta-2\theta'-2\ri\gamma}}\\
 L(\theta)&=&\log\big(1+\re^{-\ri \varepsilon(\theta)}\,\big)\ .\nonumber
\eea
\vspace{-0.7cm}

\noindent
Here we drop the superscript ``{\footnotesize{(vac)}}'' in the notation for the vacuum eigenvalues, and the star ``$*$'' as usual, stands for
complex conjugation.
The constant $\gamma$ is an arbitrary number belonging to the open segment $0<\gamma<\frac{\pi}{2}$. We also swap $p_2$ for
\bea
k=\frac{2p_2}{n+2}\ ,
\eea
and the kernels read explicitly as
 \bea\label{hasatast2}
G(\theta)&=&
\frac{\sin(\frac{2\pi}{n+2})}{(n+2)\sinh(\frac{\theta+\ri\pi}{n+2})\sinh(\frac{\theta-\ri\pi}{n+2})}\nonumber\\[0.2cm]
G_1(\theta)&=&
\frac{\sin(\frac{2\pi}{n+2}) \sin(\frac{\pi}{n+2})\ \sinh(\frac{2\theta}{n+2})}
{(n+2)\sinh(\frac{\theta+\frac{\ri\pi}{2}}{n+2})\sinh(\frac{\theta-\frac{\ri\pi}{2}}{n+2})
\sinh(\frac{\theta+\frac{3\ri\pi}{2}}{n+2})\sinh(\frac{\theta-\frac{3\ri\pi}{2}}{n+2})}\\[0.2cm]
G_2(\theta)&=&G(\theta)-
\frac{\sin(\frac{4\pi}{n+2})}{2(n+2)\sinh(\frac{\theta+2\ri\pi}{n+2})\sinh(\frac{\theta-2\ri\pi}{n+2})}\ .\nonumber
\eea
It is important to keep in mind that the integral equations should be supplemented by the asymptotics for the vacuum eigenvalues
given by eqs.\,\eqref{aisisosa}-\eqref{hsssaty}. 
For real $p_1\ne 0$ and $p_2>-\frac{1}{2}$ (as in  case \ref{sdkjfhewqc} above)
the integral equations must be modified by adding extra source terms to the r.h.s. The corresponding NLIE is given by formulae
\eqref{eq-with-cdd0},\,\eqref{eq-with-cdd}
in appendix \ref{apDD} along with some explanations.

 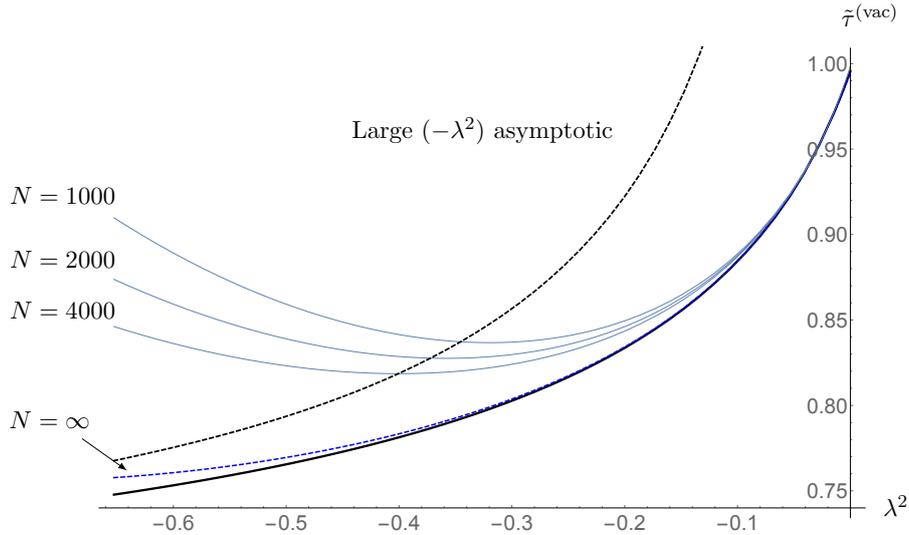
\begin{figure}
\centering
\scalebox{0.85}{
\begin{tikzpicture}
\node at (0,0) {\includegraphics[width=11.9cm]{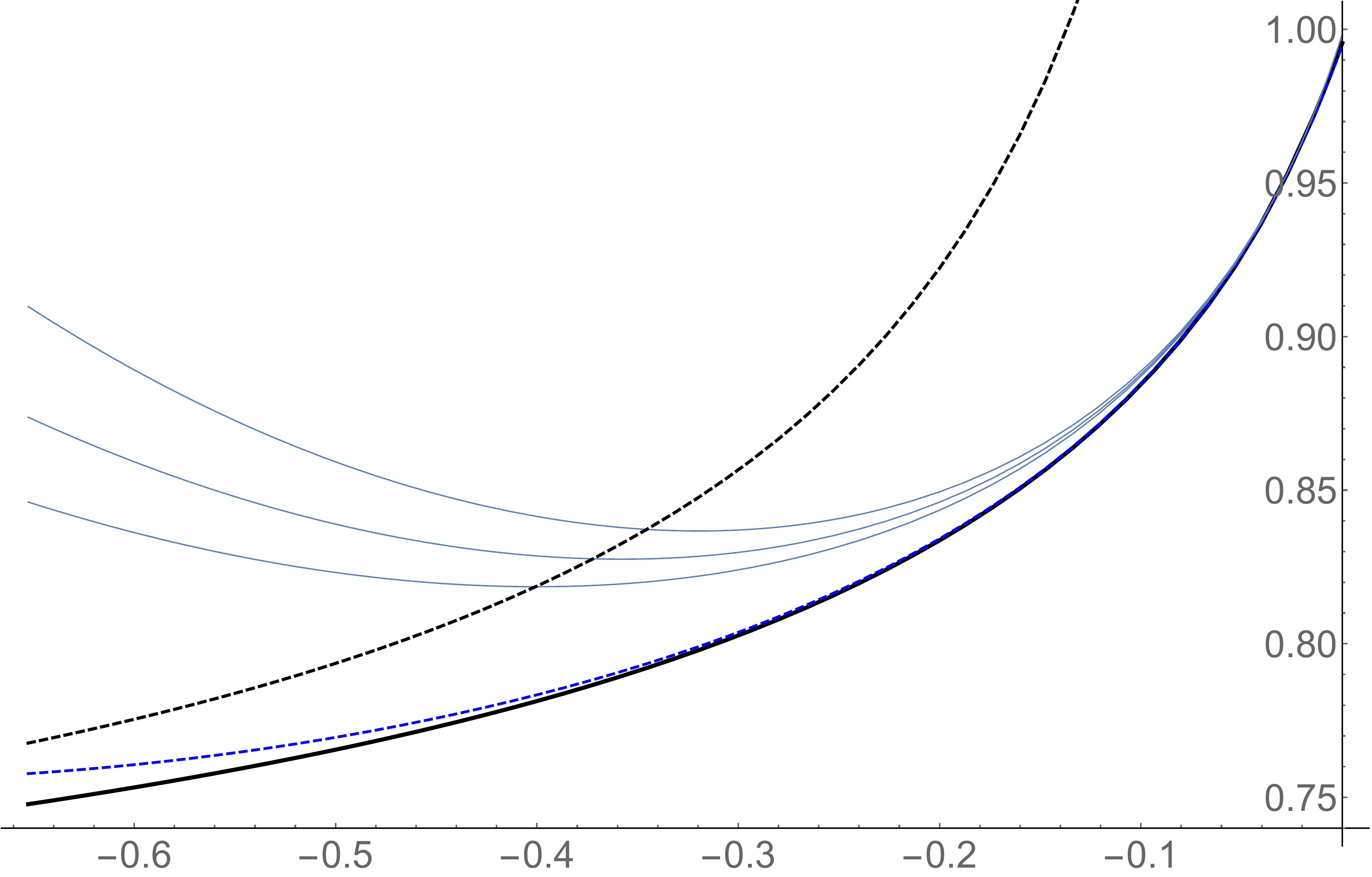}};
\node at (6.4,-3.35) { $\lambda^2$};
\node at (6,4.3) { $\tilde{\tau}^{(\rm vac)}$};
\node at (-6.5,1.5) { $N=1000$};
\node at (-6.5,0.5) { $N=2000$};
\node at (-6.5,-0.3) { $N=4000$};
\node at (-6.7,-2) {$N=\infty$};
\draw [->] (-6.2,-2.3) -- (-5.5,-2.8);
\node at (0,2.5) {Large $(-\lambda^2)$ asymptotic};
\end{tikzpicture}
}
\caption{A plot of the scaling function $\tilde{\tau}^{(\rm vac)}= \tau^{(\rm vac)}\times\exp\big(2\pi\,(-\lambda^2)^{\frac{3}{2}}\big)$ for the parafermion vacuum with
$n=3$, and $2\mathfrak{j}\,=\,\mathfrak{m}\,=\,1$. The thick black line comes from the numerical solution of the NLIE system \eqref{hasatast}. The blue curves 
are the same as in the plot appearing in the right panel of  fig.\,\ref{figA1} from Appendix \ref{askjhdaA}. They
were obtained from the numerical solution of the Bethe ansatz equations for finite $N$ and subsequently interpolating the
data to $N=\infty$.
 The large $(-\lambda^2)$ asymptotic is given by eqs.\,\eqref{asasisao},\,\eqref{jasusau}.\label{figasd}}
\end{figure}

Once the system of NLIE is solved, the numerical data can be used to reconstruct  the vacuum eigenvalues of $\alpha^{({\rm vac})}(\theta)$, 
$\beta^{({\rm vac})}(\theta)$ and $\tau^{({\rm vac})}(\ri\lambda)$. The corresponding formulae are given by \eqref{skdjfhsdfhk1}-\eqref{skdjfhsdfhk2}
from Appendix\,\ref{appC1}. Expressions for the vacuum eigenvalues of the local and dual nonlocal integrals of motion are present there as well. 
In fig.\,\ref{figasd} numerical results for $\tau^{({\rm vac})}$ in the parafermionic vacuum with $n=3$ and
$2\mathfrak{j}=\mathfrak{m}=1$ ($p_1=\frac{\ri}{2}$, $p_2=1$) are shown alongside the results from the Bethe ansatz. 
Also, for numerous cases, we 
compared the numerical results for the connection coefficients computed from the NLIE with those 
obtained by direct integration of the ordinary differential equations \eqref{NODE}.
The agreement we found in all cases  justifies the assumptions made within the derivation of the NLIE system.

\section{Integrable structures in the sausage}

We now turn to the main subject of interest  -- the sausage model. First of all, we recall some
basic facts concerning this quantum field theory. For more details, see, e.g.,\,\cite{Lukyan}.

\subsection{Basic facts about  the  quantum sausage}

\bigskip
At the beginning of sec.\,\ref{sec20}, we briefly touched on the one loop renormalization in  the sausage  model. 
Let us introduce the renormalized coupling $\kappa_r$ which  substitutes  the bare coupling $\kappa$ \eqref{jsyyd}: 
 \bea
\frac{1-\kappa_r}{1+\kappa_r}=(E_*/E)^{\frac{2}{n}}\ .
\eea
Here  $E$  stands for a  typical energy scale, which, in the case under consideration,  can be identified with the  inverse of the circumference of the space-time cylinder, $R^{-1}$.
Recall also  that $n\equiv\frac{2\pi}{\hbar}$ and $E_*$ is a RG invariant energy scale appearing  in the theory through the
mechanism of dimensional transmutation.  
Within the one-loop approximation the bare coupling is replaced  by $\kappa_r$, so that
 the renormalized sausage metric is given by
\bea\label{asa}
G^{({\rm ren)}}_{ab}\, \rd X^a\rd X^b=\frac{n }{2\pi} \ \frac{(\rd\phi)^2+(\rd\alpha)^2}
{\half\,  (\kappa_r^{-1}+\kappa_r)+\half\,  (\kappa_r^{-1}-\kappa_r)\,\cosh(2\phi)}\ ,
\eea
where we use the pair  of real coordinates  $X^a=(\phi,\alpha)$ \eqref{jsausa}.
Consider the ultraviolet  regime where   $E\gg  E^*$, i.e., $1-\kappa_r\ll 1$. In this case,
 the central region of the sausage  depicted in  fig.\,\ref{fig1t} looks like a  long  cylinder equipped with the flat
metric. If  one formally sets  $\kappa_r=1$ in \eqref {asa},  ignoring  the presence of the  two infinitely separated tips, then
\bea\label{jasah}
G^{({\rm ren)}}_{ab}\, \rd X^a\rd X^b\approx\frac{n}{2\pi}\ \big(\, (\rd\phi)^2+(\rd\alpha)^2\,\big)\ .
\eea
The  NLSM  action corresponding to this metric  has the  form of  the massless Gaussian model\footnote{In what follows the Planck constant $\hbar=\frac{2\pi}{n}$  will be  always included in the action.}
\bea\label{jssaasu}
{\cal A}_0
%=\frac{1}{\hbar}\ \int\rd^2 x \,{\cal L}_0
=\frac{n}{4\pi}\ \int\rd^2 x\ \big(\,(\partial_\mu\phi)^2+(\partial_\mu\alpha)^2\, \big)\ .
\eea
We  now apply  the $T$-duality transformation to  the field $\alpha$, i.e.,  we replace $\alpha$ by  the $T$-dual field $\vartheta$ such that
 $\partial_{\mu}{ \alpha}=\frac{1}{\sqrt{n+2}}\  \epsilon_{\mu\nu}\partial_\nu\vartheta$ (the difference between   $n$ and $n+2$ is ignored here,
since  $n$ is assumed to be large).
The substitution  of $(\phi,\alpha)$ by the pair $(\varphi,\vartheta)$, with
$\phi=\frac{1}{\sqrt{n}}\, \varphi$, brings the action to the form
\bea\label{kasausa}
{\tilde {\cal A}}_0=\frac{1}{4\pi}\ \int\rd^2 x\ \big(\,(\partial_\mu\varphi)^2+(\partial_\mu\vartheta)^2\,\big)\ .
\eea
The advantage of  ${\tilde {\cal A}}_0$ compared  to the action \eqref{jssaasu}, is that it allows one to easily
incorporate  the effects  of the   sausage tips. The central region with the left tip of the sausage form the cigar,
and the corresponding NLSM, as it was mentioned in sec.\,\ref{sec2.3}, admits the dual description in terms of   the sine-Liouville action
\bea\label{asusauas}
{\tilde {\cal A}}^{(\rm left)}={\tilde {\cal A}}_{0}+2 \mathpzc{M}\ \int\rd^2 x\ \re^{-\sqrt{n}\varphi}\ \cos\big(\sqrt{n+2}\,\vartheta\,\big)\ .
\eea
Clearly, the cigar NLSM whose target space is glued from the right tip and the central region,  is governed  by the action  which is related to \eqref{asusauas} by the  flip
$\varphi\mapsto -\varphi$, i.e.,
\bea\label{asuas}
{\tilde {\cal A}}^{(\rm right)}={\tilde {\cal A}}_{0}+2 \mathpzc{M}\ \int\rd^2 x\ \re^{+\sqrt{n}\varphi}\ \cos\big(\sqrt{n+2}\,\vartheta\,\big)\ .
\eea
At this point one can guess that the sausage NLSM admits  the dual description by means of   the 
renormalized  action
\bea\label{sasa}
{\tilde {\cal A}}^{\rm (saus)}=\int\rd^2 x\ \Big(\,{\textstyle\frac{1}{4\pi}}\ \big(\, (\partial_\mu\varphi)^2+ (\partial_\mu\vartheta)^2\,\big)+
4 \mathpzc{M}\, 
\cosh\big(\sqrt{n}\varphi\big) \cos\big(\sqrt{n+2}\, \vartheta\,\big)\,\Big)\ .
\eea
Remarkably,  the naive guess turns out to be correct! The dual form of the sausage model was originally proposed
by Aleosha Zamolodchikov and there are many arguments to support its validity, a few of which
 will be mentioned later in the text. 
Contrary to the sine-Liouville model  where  the dimensionless  $ \mathpzc{M}$ is  a somewhat fake parameter, the coupling 
$\mathpzc{M}$ in \eqref{sasa} is an important, dimensionful
characteristic  of the theory. Notice that the field $\cosh\big(\sqrt{n}\varphi\big) \cos\big(\sqrt{n+2}\, \vartheta\big)$ has the scale dimensions equal to one w.r.t.
the conventional energy momentum tensor  of the ``unperturbed''  free  theory  \eqref{kasausa}. Therefore  $\mathpzc{M}$ has  dimensions of energy, i.e., $ \mathpzc{M}\propto  E_*$ up to some  dimensionless constant.

\bigskip
Consider now  the general structure of the Hilbert space of the sausage NLSM in   finite volume equipped with the boundary conditions of the form  \eqref{aisaisoa}. 
The quantum number $m$ 
in that  formula  must  take  integer values only, it is  a conserved charge associated with the $U(1)$-isometry  
of the sausage metric. In what follows  we will focus on the neutral sector of the theory with ${m}=0$, and therefore
\bea\label{aaisoa}
\varphi(t, x+R)=\varphi(t, x)\ ,\ \ \ \ \ 
\vartheta(t, x+R)=\vartheta(t,x)\ .
\eea
Since  the action \eqref{sasa} and the boundary conditions  are both  invariant under the transformation $\vartheta\mapsto \vartheta+\frac{2\pi}{\sqrt{n+2}}$,
the
 space of  the neutral  states   of the sausage model
 is somewhat similar to the Hilbert space of a quantum  particle  in  a periodic potential:
 it is split on the orthogonal subspaces ${\cal H}_{k}^{(K)}$ 
 characterized by the quasimomentum  restricted to the first Brillouin zone, $  -\frac{1}{2}<  k< \frac{1}{2}$, and 
   a positive  integer  $K$ -- the band number.
% In the cigar NLSM/sine-Liouville model, the
%%forbidden bands have  zero width,  and therefore,   the splitting into  bands appears to be somewhat artificial  in that case.  In the sausage NLSM the
%widths of the forbidden bands are non-zero and  depend on  the dimensionless parameter $E_*R$.
Our considerations below will be mostly restricted to
the $k$-vacuum,  $|{\rm vac}\rangle_k\in {\cal H}^{(1)}_k$  -- the lowest energy neutral state  in  the first band.

In sec.\,\ref{sec2.3} it was  mentioned  that 
the Hilbert space of the  cigar/sine-Liouville theory is classified w.r.t.  the action of a certain 
${\mathfrak W}\otimes \overline{ {\mathfrak W}}$-algebra.  The $W$-algebras related by the   reflection $\varphi\to-\varphi$  are algebraically isomorphic (for details see, e.g., \cite{Lukyanov:2003nj,Lukyan}).
This  property  allows one to identify   the  spaces of states  for the ``left'' and  ``right'' sine-Liouville models \eqref{asusauas} and \eqref{asuas} which can be then interpreted  as
  an extended   Hilbert space of the sausage NLSM.
  %Therefore, in view of eq.\,\eqref{aiossao}, we can expect that
%\bea\label{kasauaysstsu}
%{\cal H}^{(1)}_{k}\in \begin{array}{c} \\[-0.1cm]  \oplus \\[-0.15cm] \scriptstyle{p_1<0\atop w\in{\mathbb Z}} 
%\end{array}{\cal V}_{p_1, p_2(w)}\otimes {\bar {\cal V}}_{p_1, -p_2(w)}\ \ \ \ \ \ \ {\rm with}\ \ \ \ \ p_2(w)=\half\, (n+2)\, (k+w)\ .
%\eea
However, contrary to the case of the cigar NLSM,  
%the  individual components  in the direct sum  are not invariant subspaces of the Hamiltonian anymore.
%Also, 
 instead of the  continuous summation  as   in \eqref{aiossao},   the zero-mode momentum $p_1$  in the sausage 
takes a certain discrete set of admissible values which depends on  $R$.
As $\mathpzc{M}R\to 0$ the mechanism of the  quantization of $p_1$ is similar to that in the sinh-Gordon model considered in 
ref.\,\cite{Zamolodchikov:1995aa}. 
The discussions from this work can be easily adopted
 to our problem (see, e.g.,\,\cite{Lukyanov:2003nj}). 

When  $\mathpzc{M}R\ll 1$,   
the quantization condition  which determines the value of $p_1=p_1(R)$  for  the $k$-vacuum$ , |{\rm vac}\rangle_k\in {\cal H}^{(1)}_k$,
reads as follows
\bea\label{sasaa}
-\frac{8p_1}{n}\ \log\Big(\frac{ \mathpzc{M}R}{2}\Big)+2\, \delta(p_1,p_2)=2\pi\ .
\eea
Here $p_2\equiv\half (n+2)\, k$,  $\delta(p_1,p_2)=-\ri\, \log S_0({\bf p})$ and $S_0({\bf p})$ is the overall scalar factor for the $S$-matrix \eqref{ahsassauasu}. 
This factor was first derived by A. and Al. Zamolodchikov \cite{ZAM}
and, using the notation of $B_{p_1,p_2}$ from \eqref{jsasasaus}, it is given by $S_0(p_1,p_2)=-B_{-p_1,p_2}/B_{p_1,p_2}$.  Thus,
\begin{eqnarray}\label{ksjshs}
\delta(p_1,p_2)={ \frac{4p_1}{n}}\,\log(n)-\ri\,
\log\Bigg[\frac{\Gamma(\frac{1}{2}-p_2-\ri p_1)\, \Gamma(\frac{1}{2}+p_2-\ri p_1)}
{\Gamma(\frac{1}{2}-p_2+\ri p_1) \Gamma(\frac{1}{2}+p_2+\ri p_1)}
 \frac{\Gamma(1+2\ri p_1)}{\Gamma(1-2\ri p_1)}
\frac{\Gamma(1+\frac{2\ri p_1}{n})}{\Gamma(1-\frac{2\ri p_1}{n})}\Bigg]  \nonumber \\[0.2cm]
\end{eqnarray}
where  the  branch of the logarithm is chosen  so that $\delta(0,p_2)=0$.  
The vacuum energy in the sector ${\cal H}^{(1)}_k$ is approximately the corresponding vacuum energy 
  of the free theory \eqref{kasausa}, 
  with the zero mode momentum for the field $\varphi$ determined by the quantization condition \eqref{sasaa}:
  \bea\label{hassaysa}
E^{(\rm vac)}_k\approx\frac{\pi}{ R}\ \Big(-\frac{1}{3}+ \frac{4}{n}\ \big(p_1(R)\big)^2+ (n+2)\, k^2\,\Big)\ .
\eea
This result can be obtained  both from the theory  described by the action \eqref{sasa} and directly  from the sausage NLSM within the so-called minisuperspace approximation, which,
in fact, is a strong  argument for the validity of the dual description of the quantum sausage.

A few comments to the formula \eqref{hassaysa} are in order. First of all a brief inspection of  the 
scattering phase $\delta(p_1,p_2)$\,\eqref{ksjshs}, shows that  the quantization condition  can be  applied literally only for $|p_2|<\half$, or, equivalently,
for $|k|<\frac{1}{n+2}$. For $p_2=\half$, notice that  eq.\,\eqref{sasaa} has an $R$-independent solution $p_1=0$, whereas  for  $\frac{1}{n+2}<|k|<\half $,
the precise  form of  the quantization  condition, to the best of our knowledge,  is currently not known. The next comment deals  with  corrections to the approximate formula \eqref{hassaysa}.
A  superficial analysis shows that for $n>2$ the main correction is of order $R^{\frac{4}{n}-1}$, while for  $0<n<2$,
 a term  $\propto  R\,\log(R)$  dominates.
The latter    can be understood as the  one-instanton contribution  and has the following explicit  form
\bea
\delta E^{(\rm 1- inst)}=-
\pi \mathpzc{M}^2 R\ \Big(4\,\log\big(R\Lambda^{\rm (inst)} \big)+e(k)+ O\big( p^2_1(R)\big)\Big)\ .
\eea
Here $\Lambda^{\rm (inst)}$ is  a  cut-off energy  scale which regularizes the  contribution of the small-size instantons \cite{Fateev:1979dc,Bukhvostov:1980sn,Luscher:1981tq,Fateev:1992tk} and
\bea\label{jsasasay}
e(k)=-2-2\,\psi\,\big(\half -{\textstyle\frac{(n+2) k}{2}}\big)-2\psi\,\big(\half +{\textstyle\frac{(n+2) k}{2}}\big)\ .
\eea

\bigskip
Now, let us turn to the large-$R$ limit. In this limit, $E^{(\rm vac)}_k$ contains an extensive part which is proportional to the spatial size of the system and does not depend on $k$.
As follows from the results of ref.\cite{Fateev:1992tk}, 
the specific bulk energy ${\cal E}\equiv\lim\limits_{R\to\infty}E^{(\rm vac)}_k/R$ is given by
\bea
{\cal E}=\pi  \mathpzc{M}^2 \ \Big(\, 4\, \log\big(  \mathpzc{M}/ \Lambda^{\rm (inst)}\big)+
\pi\cot\big({\textstyle\frac{\pi n}{2}}\big)+2\,\gamma_E+2\,\psi\big(1+{\textstyle\frac{n}{2}}\big)\,\Big)\ .
 \eea
The large-$R$ behaviour of  the difference $E^{(\rm vac)}_k-R\,{\cal E}$ is dictated by the factorized scattering theory associated with the model. In  the original work \cite{Fateev:1992tk}, it was proposed that
the spectrum of the sausage model in  infinite volume consists of a triplet of particles of the same mass $m$. The mass scale 
is simply related to the dimensionful coupling in the
renormalized action\ \eqref{sasa}:
\bea\label{jassasaytast}
m=4\pi\,  \mathpzc{M}\ .
\eea
Two of the  particles $A_+ $ and $A_-$ carry the $U(1)$ charge $+1$ and $-1$, respectively, whereas the
 third particle $A_0$  is neutral. The factorized scattering  is completely determined by the two-particle $S$-matrix   which can be interpreted as a structure constant in the
formal Zamolodchikov-Faddeev associative algebra:
\bea
A_a(\theta_1)\,A_b(\theta_2)=S_{ab}^{cd}(\theta_1-\theta_2)\ A_d(\theta_2)\, A_c(\theta_1)\ .
 \eea
 For convenience, we collect in Appendix\,\ref{appD} explicit expressions  for  the   scattering amplitude  $S_{ab}^{cd}$  proposed
   in \cite{Fateev:1992tk}.
 Taking a closer look at these amplitudes, one can  observe that they are trivialized  for $n=0$.
This is consistent with the fact that the dual  theory  \eqref{sasa} can be understood, by  use of the
 Coleman-Mandelstam bosonization, as an interacting theory of a massive scalar  and Dirac fermion.
At $n=0$, the interaction disappears and we end up with a theory of three non-interacting particles. In fact, this was 
the starting point of Al. Zamolodchikov's proposal for the dual description of the sausage. He performed perturbative calculations for small $n$ and found that the perturbative amplitudes match the small-$n$ expansion of
the exact two-particle $S$-matrix.

As usual for a massive  quantum field theory,  
the large-$R$ expansion of $E^{(\rm vac)}_k-R\,{\cal E}$ can be  represented   in the form
 \bea\label{hsgst}
 E^{(\rm vac)}_k-R\,{\cal E}=\Delta E^{1-{\rm particle}}+\Delta E^{2-{\rm particle}}+\ldots\ ,
 \eea
 where  the individual terms  correspond to the   virtual contributions  of  $N$-particle states.
 The one-particle contribution does not depend on the details of the interaction -- it is the same as for the non-interacting theory and hence is
 given by
 \bea\label{ajsasys}
 \Delta E^{1-{\rm particle}}=
  -{\rm Tr}_1\big({\boldsymbol K}^{(1)}\big)\ \int_{-\infty}^\infty\frac{\rd\theta}{2\pi}\ \ m\cosh(\theta)\, \re^{-mR\cosh(\theta)}\ .
 \eea
Here the trace is taken over the  isotopic component of the one-particle sector of the theory in the infinite volume and   ${\boldsymbol K}^{(1)}$ is a $3\times 3$  diagonal  matrix
 of the (complex) fugacities:
 \bea\label{nssg}
 {\boldsymbol K}^{(1)}=\begin{pmatrix}
 \re^{+2\pi\ri k} &0&0\\
 0&1&0\\
 0&0& \re^{-2\pi\ri k}
 \end{pmatrix}\,.
  \eea
 Following    L${\rm \ddot u}$scher  \cite{Luscher:1985dn1,Luscher:1985dn2} (see also ref.\,\cite{BalogN}), 
 one can derive the explicit formula for the two-particle contribution in eq.\,\eqref{hsgst}:
 \bea\label{hssst}
 \Delta E^{2-{\rm particle}}&=&
  m \int_{-\infty}^\infty\frac{\rd\theta\rd\theta'}{(2\pi)^2}
   \,\cosh(\theta)\ \re^{-mR \cosh(\theta)-mR\cosh(\theta')}\\[0.2cm]
   &\times &
  {\rm Tr}_2\Big[\,{\boldsymbol K}^{(2)}\
  \Big( \,\pi{\boldsymbol I}^{(2)}\,  \delta(\theta-\theta')+
   \ri\,  \partial_\theta \log{ \boldsymbol  S}^{\rm (2\mapsto 2)}(\theta-\theta')\, \Big)\, \Big]\ ,\nonumber
  \eea
where ${ \boldsymbol  S}^{\rm (2\mapsto 2)}$ is the $9\times 9$ matrix acting in the isotopic component of the two-particle sector, ${\boldsymbol K}^{(2)}={\boldsymbol K}^{(1)}\otimes {\boldsymbol K}^{(1)}$
and ${\boldsymbol I}^{(2)}$ is the identity matrix.

\bigskip

For future reference let us make a short summary of the properties of the $k$-vacuum energy  discussed above.
For this purpose it is convenient to introduce  the scaling variable
\bea
r=mR
\eea
and dimensionless scaling function
\bea\label{kjsdfoiqwec}
{\mathfrak  F}(r,k)=\frac{R}{\pi}\ \big(E^{(\rm vac)}_k-R\,{\cal E}\big)\ .
 \eea
 Notice that, $c_{\rm eff}\equiv -6\, {\mathfrak  F}(r,k)$ is sometimes interpreted as the effective central charge for the off-critical theory.
 As $r\to 0$ our discussion  suggests that for $|k|<\frac{1}{n+2}$ and fixed $n$
 \bea\label{jashsaasy}
 {\mathfrak  F}(r,k)=\begin{cases}
 -\frac{1}{3}+ \frac{4}{n}\ p^2(r)+ (n+2)\, k^2+O(r^{\frac{4}{n}})\ \ \ \ \ &{\rm for }\ \ \ n>2\\[0.5cm]
 -\frac{1}{3}+ \frac{4}{n}\ p^2(r)+ (n+2)\, k^2+\delta {\mathfrak  F}^{(\rm 1- inst)}+O\big(r^2\, p^2(r)\big)\ \ \ \ \ &{\rm for }\ \ \ 0<n<2
 \end{cases}
 \eea
 Here $p(r)$ is defined as the solution of the quantization condition 
 \bea\label{jsasasa}
 -{\textstyle \frac{8p}{n}}\ \log\big({\textstyle \frac{r}{8\pi}}\big)+2\, \delta(p,p_2)=2\pi\ \ \ \ \ \ \ \ \ \ \  \big(p_2<\half\big)
 \eea
 where $\delta(p_1,p_2)$ is given by \eqref{ksjshs},  and
 \bea
 \delta {\mathfrak  F}^{(\rm 1- inst)}=
-
{\textstyle \frac{r^2}{16\pi^2}}\ \Big(4\,\log\big({\textstyle {\frac{ r }{4\pi }}}\big)+\pi\,\cot\big({\textstyle \frac{\pi n}{2}}\big)
+2\,\gamma_E+2\,\psi\big(1+{\textstyle\frac{n}{2}}\big)+e_1(k)\Big)
 \eea
 with $e_1(k)$  defined in \eqref{jsasasay}.

 The large $r$-behaviour of the  scaling function ${\mathfrak  F}(r,k)$ is determined by eqs.\eqref{hsgst}-\eqref{hssst}. It can be equivalently described by the following formula which is convenient
 for numerical calculations:
 \bea\label{aosiaosioa}
&& {\mathfrak F}(r,k)=-\frac{r}{\pi^2 }\, \big(\, 2\, c(2k)+1\big)\ K_1(r)+\frac{r}{2\pi^2}\ \big(\,2\,c(2k)+1\,\big)^2\   K_1(2r)-\\
&&\frac{2r}{\pi^3}\, \big(\,1+c(2k)\,\big)\ \int_{-\infty}^\infty\rd\nu\ \frac{K_{1-\ri\nu}(r)K_{\ri\nu}(r)}{\sinh\big(\frac{\pi (n+2)\nu}{2}\big)}\ 
\Big[\,2 c(2k) \sinh\big({\textstyle\frac{\pi n\nu}{2}}\big)-  \sinh\big({\textstyle\frac{\pi(n-2)\nu}{2}}\big)\Big]+
O\big(\re^{-3r}\big)\nonumber
\eea
where $c(x)\equiv \cos(\pi x)$, $K_s(z)$ denotes the modified Bessel function of the second order,
 \bea
  K_s(z)=\frac{1}{2}\ \int_{-\infty}^\infty\rd\theta\ \re^{s\theta-z\cosh(\theta)}\ ,
 \eea
 and the symbol $O(\re^{-3 r })$ stands for terms which decay  faster than $\re^{-(3-\epsilon)r}$ as $r\to+\infty$, for any  small $\epsilon>0$.

Finally, for $n=0$, the scaling function ${\mathfrak  F}(r,k)$ is given explicitly by:
  \bea\label{hssysys}
 {\mathfrak F}(r,k)=-\frac{r}{2\pi^2 }\ \int_{-\infty}^{\infty}\rd\theta\ \re^{\pm \theta}\,
 \log\bigg(\frac{\big(1+\re^{2\pi\ri k-r\cosh(\theta)}\big)\big(1+\re^{-2\pi\ri k-r\cosh(\theta)}\big)}
 {1-\re^{-r\cosh(\theta)}}\bigg)\ .
 \eea
 Notice that the small $r$-asymptotic \eqref{jashsaasy} can not be applied to this exact formula because of the noncommutativity of the limits $r\to 0$ and $n\to 0$.
\bigskip

   \subsection{NLIE  for the $k$-vacuum eigenvalues in the sausage model }
   \bigskip
   
  With some experience in working  with nonlinear integral equations in integrable QFT,
  one expects that the generalization of the massless equations   to  the massive ones requires little
  effort.
  For this reason, before exploring  the general integrable structures, we make a simple-minded shortcut
 and   guess the NLIE describing the $k$-vacuum eigenvalues  in the sausage model.
  Of course, this route  requires  careful consistency checks which will be the main subject of our discussion here.

  \bigskip

  As usual, the major modification required to get the massive NLIE is related to the source terms. In the case under consideration
  it is not difficult to guess that the system  \eqref{hasatast} should be modified to the following
\bea\label{aatast}
\varepsilon(\theta-\ri\gamma)&=&r\,\sinh(\theta-\ri\gamma)-2\pi k+\int_{-\infty}^\infty\frac{\rd \theta'}{2\pi\ri }\
\Big[G(\theta-\theta'-2\ri\gamma)\  \big(L(\theta'-\ri \gamma)\big)^*\nonumber\\[0.2cm]
&-&   G(\theta-\theta')\  L(\theta'-\ri \gamma)\Big]
+
\int_{-\infty}^\infty\frac{\rd \theta'}{2\pi}\ G_1(\theta-\theta'-\ri\gamma)\  \log\big(1+\re^{-\omega(\theta')}\big)\nonumber\\[0.2cm]
\omega(\theta)&=&r\, \cosh(\theta)+\Im m\bigg[\int_{-\infty}^\infty\frac{\rd\theta'}{\pi}\,G_1(\theta-\theta'+\ri\gamma)\  L(\theta'-\ri \gamma)
\,\bigg]\\[0.2cm]
&-&
\int_{-\infty}^\infty\frac{\rd\theta'}{\pi}\, G_2(\theta-\theta')\, \log\big(1+\re^{-\omega(\theta')}\big)\nonumber
\\[0.2cm]
%\varepsilon(\theta-\ri\gamma)+2\pi k=
%-\int_{-\infty}^\infty\frac{\rd \theta'}{\pi}\ 
%&\frac{ \log\big(1+\re^{b(\theta')}\big)-4\pi\,\re^{\theta'}}{1+\re^{2\theta-2\theta'-2\ri\gamma}}\\
 L(\theta)&=&\log\big(1+\re^{-\ri \varepsilon(\theta)}\,\big)\ .\nonumber
\eea
Unlike  the massless  case, there is no need to supplement these  equations by the asymptotic conditions at $\theta\to-\infty$ -- the source
terms in \eqref{aatast} control
the solution's  behaviour  both  at $\theta\to\pm \infty$.
 In this subsection we will discuss the $k$-vacuum energy only. Having at hand the formula \eqref{jasshaa} for the vacuum eigenvalue of the 
 conformal local IM ${\mathfrak i}_1$, one expects that for the massive case,
\bea\label{hasstsa}
{\mathfrak F}(r,k)=\frac{r}{2\pi^2}\ \int_{-\infty}^{\infty}\rd\theta \, 
 \bigg(\pm 2\, \Im m\Big[  \re^{\pm (\theta-\ri\gamma)}\, L(\theta-\ri \gamma)\,\Big]
 -\re^{\pm \theta}\, \log\big(1+\re^{-\omega(\theta)}\big)\bigg)
\eea 
 and this should be valid for both choices of the sign $\pm$.

Some superficial observations can been made at this point. First we note that the kernels  in \eqref{aatast} which are given by eqs.\,\eqref{hasatast2},
can be expressed through the two-particle scattering amplitudes for the sausage model. Indeed, using the explicit  formulae  from Appendix\,\ref{appD}, it is straightforward to check that
\bea
&&G(\theta)=-\ri\ \partial_\theta \log S(\theta)\,, \hspace{2.5cm}
G_1(\theta)= \partial_\theta  \log t\big(\theta+{\textstyle\frac{\ri\pi}{2}}\big)\nonumber\\[0.2cm]
&&G_2(\theta)=-{\textstyle \frac{\ri}{8}}\ \partial_\theta \log \det\Big( {\boldsymbol  S}^{(2\mapsto 2)}(\theta)\Big)\ .
\eea
The next observation is that the system \eqref{aatast} admits a simple solution for $n=0$. In this case the kernels $G(\theta)$ and $G_1(\theta)$ vanish, whereas 
$G_2(\theta)$ turns to be $\pi\, \delta(\theta)$. This brings the NLIE to the form
\bea
\varepsilon(\theta)=r\,\sinh(\theta)-2\pi k\ ,\ \ \  \ \omega(\theta)=r\, \cosh(\theta)-\log\big(1+\re^{-\omega(\theta)}\big)\ ,
\eea
and using eq.\,\eqref{hasstsa}, one arrives at \eqref{hssysys}. Furthermore, one can perturbatively solve the NLIE for small $n$, and compare the results to those from 
the weak coupling expansion based on the dual action
\eqref{sasa} for the sausage model. We found complete agreement to the first non-trivial order in the expansion.

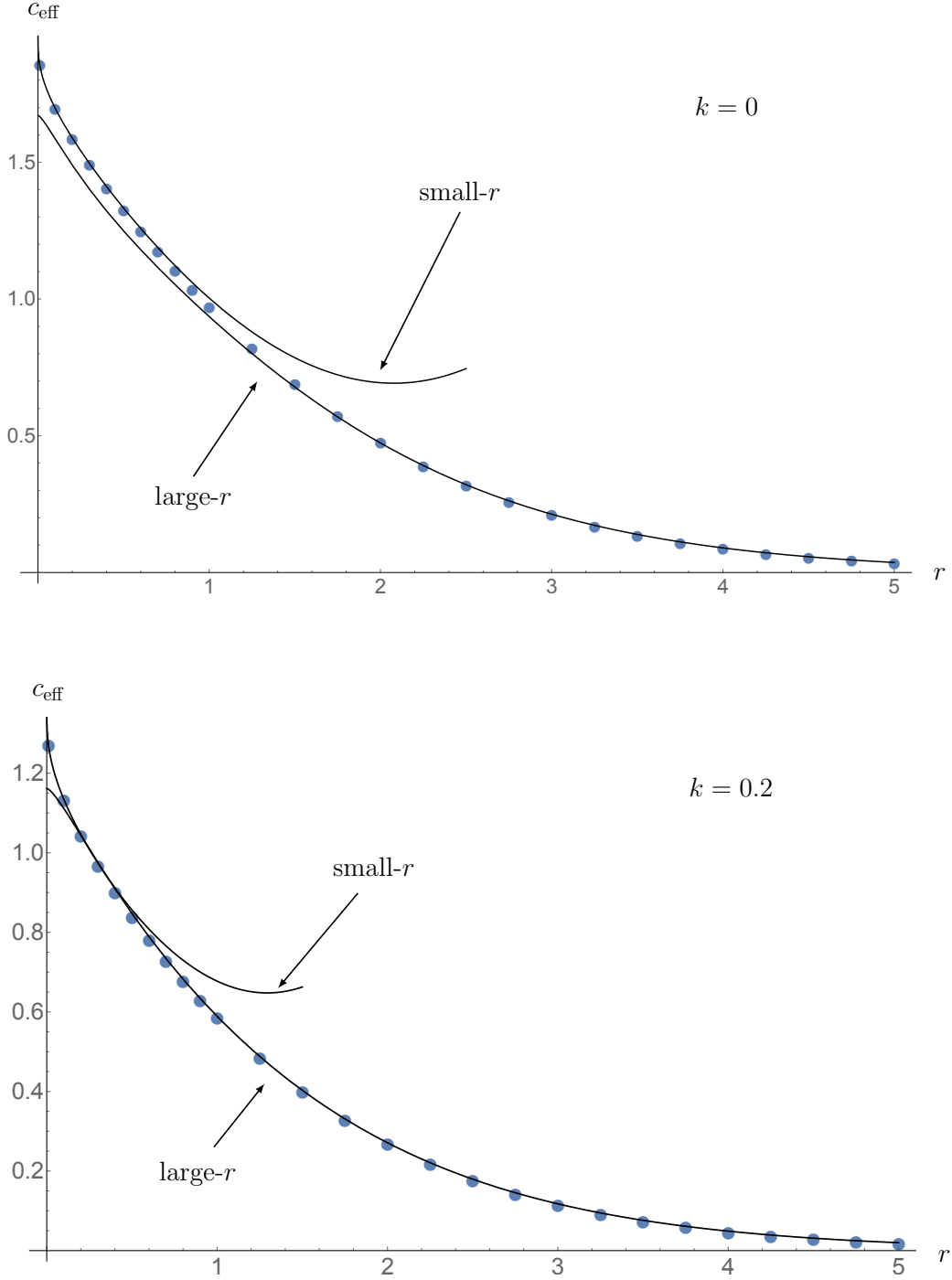
\begin{figure}
\begin{center}
\resizebox{0.92\textwidth}{!}{
\begin{tikzpicture}
%\draw[help lines] (0,0) grid (16,10);
\node at (8,5) {\includegraphics[width=17cm]{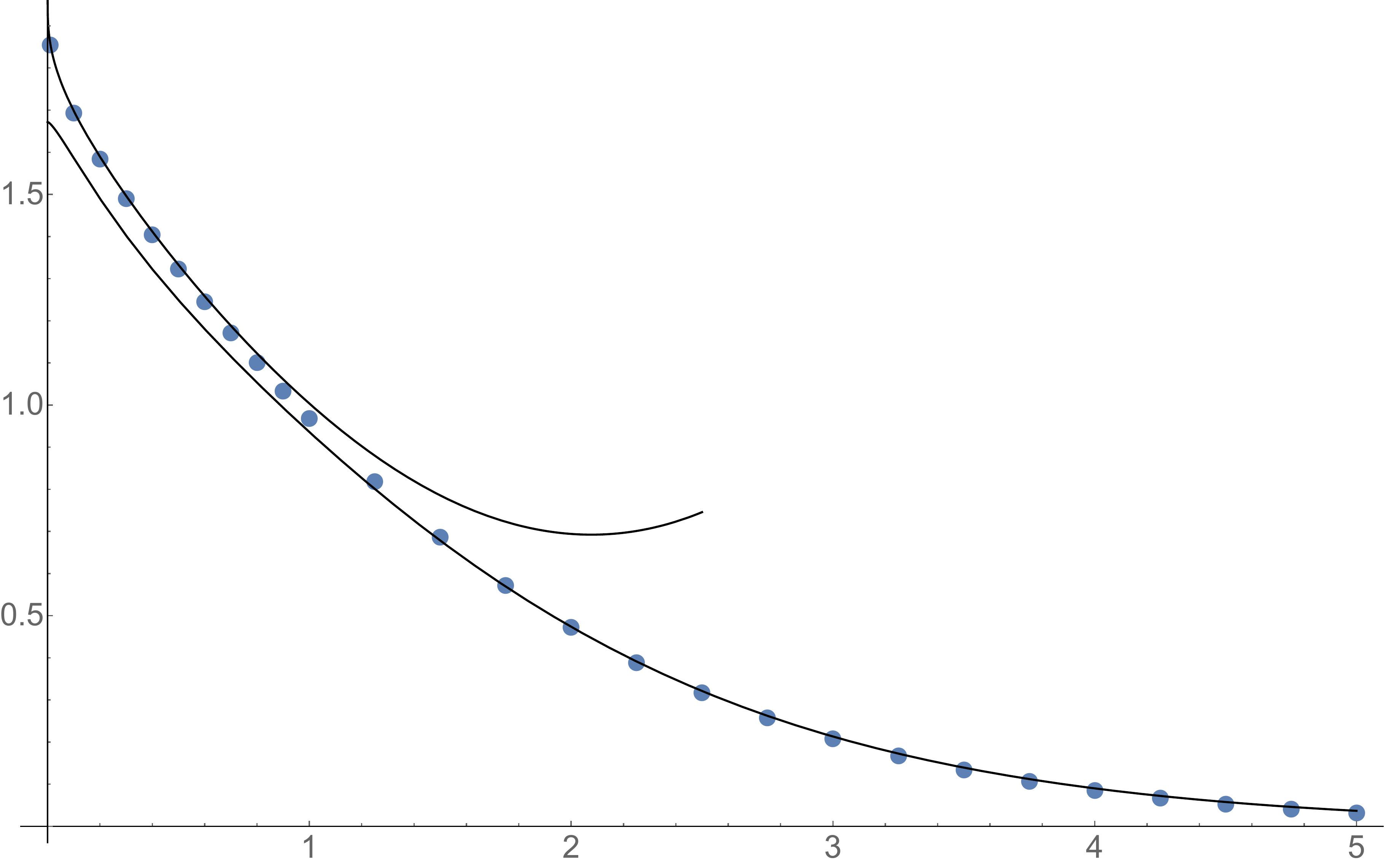}};
\node at (0.2,10.8) {\Large$c_{\rm eff}$};
\node at (13,9) {\Large$k=0$};
\node at (17,0.15) {\Large $r$};
\draw[thick,->] (8,7) -- (6.5,4);
\node at (8,7.4) {\Large small-$r$};
\draw[thick,->] (3,2) -- (4.2,3.8);
\node at (3,1.6) {\Large large-$r$};
\end{tikzpicture}
}
\vspace{1cm}

\resizebox{0.92\textwidth}{!}{
\begin{tikzpicture}
%\draw[help lines] (0,0) grid (16,10);
\node at (8,5) {\includegraphics[width=17cm]{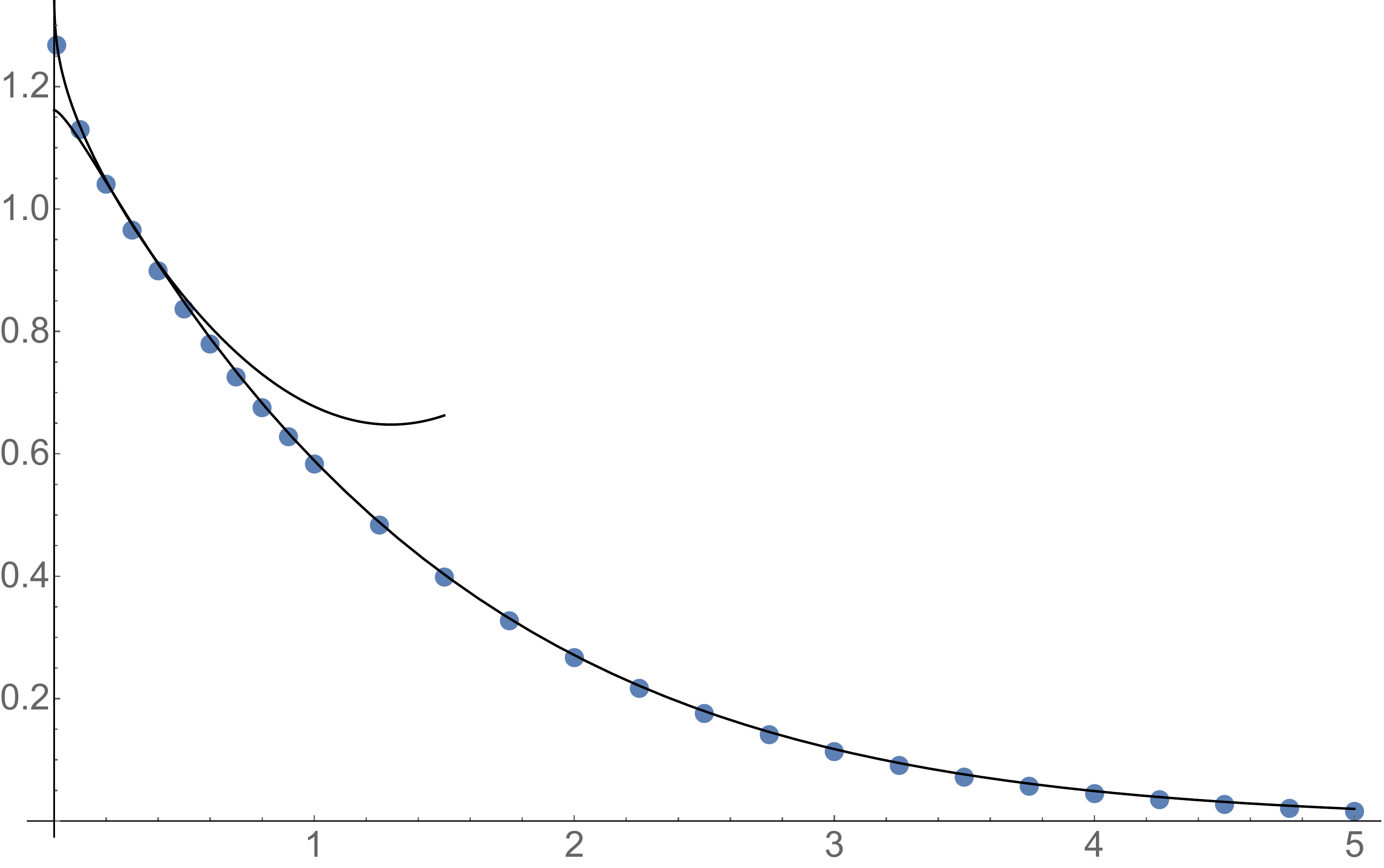}};
\node at (13,9) {\Large$k=0.2$};
\node at (0.2,10.8) {\Large$c_{\rm eff}$};
\node at (17,0.15) {\Large $r$};
\draw[thick,->] (6,7) -- (4.5,5.2);
\node at (6.3,7.5) {\Large small-$r$ };
\draw[thick,->] (3.3,2.2) -- (4.25,3.4);
\node at (3,1.7) {\Large large-$r$ };
\end{tikzpicture}
}
\end{center}
\caption{$c_{\rm eff}\,\equiv\,-6\,\mathfrak{F}(r,k)$  \eqref{kjsdfoiqwec}  plotted as a function of $r=mR$ for $n\,=\,0.5$ with $k\,=\,0$ and $k=0.2$. 
The dots were obtained from the numerical solution of the NLIE \eqref{aatast},\,\eqref{hasstsa}. The small-$r$ asymptotic 
 comes from \eqref{jashsaasy}
 while  ``large-$r$'' represents \eqref{aosiaosioa}. For the corresponding numerical data see tab.\,\ref{tabsdf}.}
 \label{fig1u} 
\end{figure}

 Much more effort is needed to derive directly from  eqs.\eqref{aatast},\,\eqref{hasstsa}
 the asymptotic formulae \eqref{jashsaasy}-\eqref{aosiaosioa}  describing the behaviour of the $k$-vacuum energy at $r\to 0$ and $r\to +\infty$.
It is, in fact,  possible to do this analytically, but here we only present some  evidence obtained through
the numerical solution of the NLIE system \eqref{aatast} (see fig.\,\ref{fig1u} and tab.\,\ref{tabsdf}).

\begin{table}
\centering
\begin{tabular}{|c||c|c|c|}
\hline
\multicolumn{4}{|c|}{$k=0$} \\ \hline
$r=mR$ & $c_{\rm eff}$ & small-$r$ & large-$r$ \\ \hline
$10^{-5}$ & $1.963104810570$ & $1.963104810585$ & $*** $ \\ \hline
$10^{-4}$ & $1.947406984923$ & $1.947406984949$ & $ ***$ \\ \hline
$10^{-3}$ & $1.919080208689$ & $1.919080211060$ & $*** $ \\ \hline
$0.01     $  & $1.859823363074$ & $1.859823782193$ & $ *** $ \\ \hline
$0.10       $  & $1.698224051017$ & $1.698316219415$ & $ ***$ \\ \hline
$0.2	0    $ & $1.589016773023$ & $1.589515588521$ & $ ***$ \\ \hline
%$0.30	    $ & $1.495267335056$ & $1.496631739040$ & $*** $ \\ \hline
$0.40	    $ & $1.409038525289$ & $1.411851755076$ & $*** $ \\ \hline
%$0.50	    $ & $1.327781544191$ & $1.332745087272$ & $1.2490889 $ \\ \hline
$0.6	0    $ & $1.250428169768$ & $1.258361320834$&$1.1804081$ \\ \hline
%$0.70	    $ & $1.176474302310$ & $1.188317075333$ & $1.1153583 $ \\ \hline
$0.8	0    $ & $1.105667568636$ & $1.122485892386$&$1.0531828 $ \\ \hline
%$0.9	0    $ & $1.037877117738$ & $1.060869630730$&$0.9934185$ \\ \hline
$1.00	    $ & $0.973032006280$ & $1.003537933054$ & $0.9358090 $ \\ \hline
%$1.25	    $ & $0.823548865527$ & $0.879697661766$ & $0.8006354 $ \\ \hline
$1.5	0    $ & $0.691697724451$ & $0.785383466436$&$0.6782200 $ \\ \hline
%$1.75	    $ & $0.576785372435$ & $***$ &    $0.5691250$ \\ \hline
$2.0	0    $ & $0.477804027757$ & $***$ &  $0.4735652 $ \\ \hline
%$2.25	    $ & $0.393469853094$ & $ *** $ &  $0.3911742 $ \\ \hline
$2.5	 0   $ & $0.322319977959$ & $ *** $ &$0.3210983 $ \\ \hline
%$2.75	    $ & $0.262815733062$ & $ *** $ & $0.2621749 $ \\ \hline
$3.00	    $ & $0.213430553126$ & $ *** $ & $0.2130984 $ \\ \hline
%$3.25	    $ & $0.172715049929$ & $ *** $ & $0.1725446 $ \\ \hline
$3.5	0    $ & $0.139339032762$ & $ *** $&$0.1392523 $ \\ \hline
%$3.75	    $ & $0.112113814605$ & $ *** $ & $0.1120700 $ \\ \hline
$4.0    0   $ & $0.089999431197$ & $ *** $&$0.0899774$ \\ \hline
%$4.25     $ & $0.072101368981$ & $ *** $ & $0.0720904$ \\ \hline
$4.5	 0   $ & $0.057660716365$ & $ *** $&$0.0576552$ \\ \hline
%$4.75	    $ & $0.046040754900$ & $ *** $ & $0.0460380$ \\ \hline
$5.00	    $ & $0.036712137455$ & $ *** $ & $0.0367108$ \\ \hline
\hline
\multicolumn{4}{|c|}{$k=0.2$} \\ \hline
%$r=mR$ & $c_{\rm eff}$ & $c_{\rm eff}^{(UV)} $ & $ c_{\rm eff}^{(IR)} $ \\ \hline
$10^{-5}$ & $1.364830335405$ & $1.364830335417$ & $*** $ \\ \hline
$10^{-4}$ & $1.350312382232$ & $1.350312382295$ & $ ***$ \\ \hline
$10^{-3}$ & $1.324531940186$ & $1.324531948818$ & $*** $ \\ \hline
$0.01     $  & $1.271882443291$ & $1.271883920213$ & $ *** $ \\ \hline
$0.10      $  & $1.134046123560$ & $1.134350188655$ & $1.11141992 $ \\ \hline
$0.20	    $ & $1.044536780739$ & $1.046137755580$ & $1.04399668 $ \\ \hline
%$0.30	    $ & $0.969842342230$ & $0.974147856348$ & $0.97572173 $ \\ \hline
$0.40	    $ & $0.902842376206$ & $0.911610323133$ & $0.90993729 $ \\ \hline
%$0.50	    $ & $0.841160905676$ & $0.856473595623$ & $0.84762573 $ \\ \hline
$0.6	0    $ & $0.783720297274$ & $0.807969967819$ & $0.78900239 $ \\ \hline
%$0.7	0    $ & $0.729938242332$ & $0.765817508094$ & $0.73400100 $ \\ \hline
$0.8	0    $ & $0.679456178786$ & $0.729950921532$ & $0.68245438 $ \\ \hline
%$0.9	0    $ & $0.632027315664$ & $***$ & $0.63416713 $ \\ \hline
$1.0	0    $ & $0.587464397842$ & $ *** $ & $0.58894471 $ \\ \hline
%$1.25	    $ & $0.487628701005$ & $ *** $ & $0.48813080 $ \\ \hline
$1.5	   0 $ & $0.402767285740$ & $ *** $ & $0.40287319 $ \\ \hline
%$1.75	    $ & $0.331122649244$ & $ *** $ & $0.33109713$ \\ \hline
$2	.00    $ & $0.271053581853$ & $***  $ & $0.27100091$ \\ \hline
%$2.25	    $ & $0.221018297286$ & $ *** $ & $0.22097243$ \\ \hline
$2.50	    $ & $0.179589431194$ & $ *** $ & $0.17955720$ \\ \hline
%$2.75	    $ & $0.145470593576$ & $ *** $ & $0.14545014$ \\ \hline
$3.00	    $ & $0.117506056343$ & $ *** $ & $0.11749384$ \\ \hline
%$3.25	    $ & $0.094682491140$ & $ *** $ & $0.09467548$ \\ \hline
$3.5	0    $ & $0.076124137247$ & $ *** $ & $0.07612023$ \\ \hline
%$3.75	    $ & $0.061083393894$ & $ *** $ & $0.06108127$ \\ \hline
$4.0 0      $ & $0.048928716276$ & $ *** $ & $0.04892758$ \\ \hline
%$4.25     $ & $0.039131323073$ & $ *** $ & $0.03913072$ \\ \hline
$4.50	    $ & $0.031251806913$ & $ *** $ & $0.03125149$ \\ \hline
%$4.75	    $ & $0.024927368178$ & $ *** $ & $0.02492721$ \\ \hline
$5.0	0    $ & $0.019860097547$ & $ *** $ & $0.01986001$ \\ \hline
\end{tabular}
\vspace{0.5cm}
\caption{The numerical data for $c_{\rm eff}\,\equiv\,-6\,\mathfrak{F}(r,k)$  with $n\,=\,0.5$, $k\,=\,0$ and $0.2$. 
The small-$r$ asymptotic was obtained by \eqref{jashsaasy} whereas the large-$r$ asymptotic comes from \eqref{aosiaosioa}. \label{tabsdf}} 
\end{table}
%\FloatBarrier\noindent

The remarkable feature of the formulae \eqref{aatast},\,\eqref{hasstsa} is that they do not depend explicitly on $n$. Hence, they can be applied to the case with
$n$ formally set to infinity, i.e., to the $O(3)$ sigma model. Nothing particularly special happens to the kernels \eqref{hasatast2}; as $n\to\infty$ they  just become the rational functions
 \bea\label{jasssayas}
G(\theta)
%&=&
%\frac{\sin(\frac{2\pi}{n+2})}{(n+2)\sinh(\frac{\theta+\ri\pi}{n+2})\sinh(\frac{\theta-\ri\pi}{n+2})}
&=&
\frac{2\pi}{(\theta+\ri\pi)(\theta-\ri\pi)}\nonumber\\
G_1(\theta)&=&
\frac{4\pi^2\,\theta}{(\theta+\frac{\ri\pi}{2})(\theta-\frac{\ri\pi}{2})(\theta+\frac{3\ri\pi}{2})(\theta-\frac{3\ri\pi}{2})}
\\
G_2(\theta)&=&G(\theta)-\frac{2\pi}{(\theta+2\ri\pi)(\theta-2\ri\pi)}\ .
\nonumber
\eea
Also the asymptotic formula  \eqref{aosiaosioa} describing the large-$r$ behaviour is, in  the $O(3)$ limit,
 \bea\label{aosioa}
 &&{\mathfrak F}(r,k)=-\frac{r}{\pi^2 }\, \big(\, 2\, c(2k)+1\big)\ K_1(r)+\frac{r}{2\pi^2}\ \big(\,2\,c(2k)+1\,\big)^2\   K_1(2r)-\\
 &&\ \ \ \ \frac{2r}{\pi^3}\ \big(\,1+c(2k)\,\big)\ \int_{-\infty}^\infty\rd\nu\ K_{1-\ri\nu}(r)K_{\ri\nu}(r)\ 
\Big(\,2\, c(2k) \,\re^{-\pi |\nu|}-  \re^{-2\pi |\nu|}\Big)
+O\big(\re^{-3r}\big)\ .\nonumber
\eea
%\FloatBarrier\noindent

The situation is much more subtle for the  small-$r$  asymptotic.
Let us recall that for finite $n$ the asymptotic formula \eqref{jashsaasy} can be applied only for  $|k|<\frac{1}{n+2}$.
This implies that in the limit $n\to\infty$ the applicability of this formula is restricted to the case $k=0$, and the only 
information it provides  is that $\lim\limits_{r\to 0}{\mathfrak F}(r,0)=-\frac{1}{3}$. \FloatBarrier

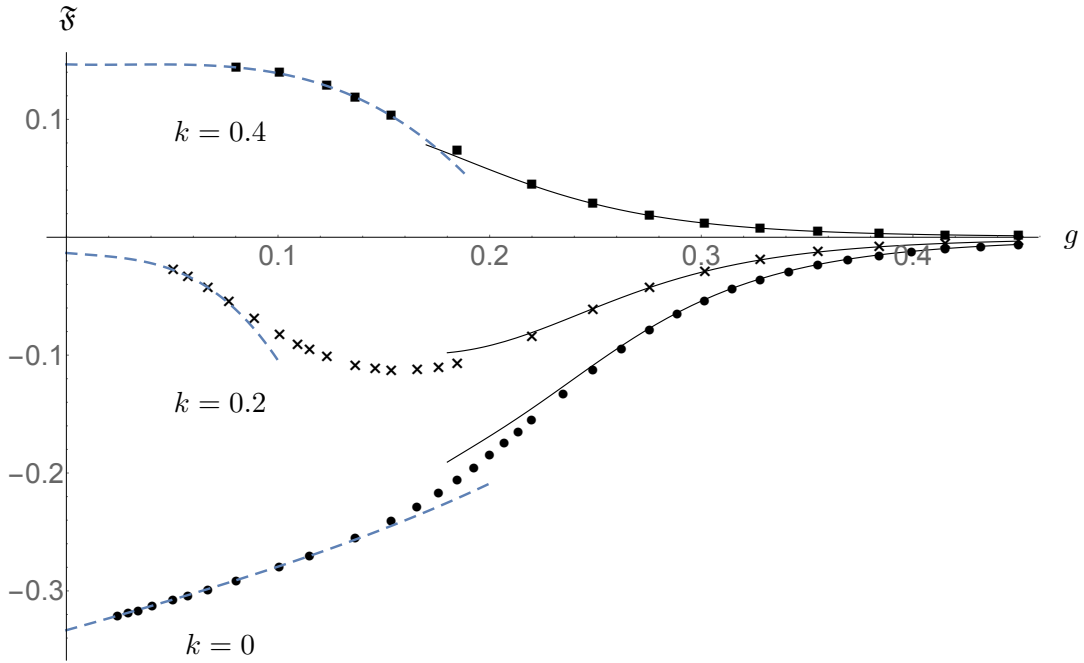
\begin{figure}
\centering
\scalebox{1}{
\begin{tikzpicture}
%\draw[step=1cm,gray,very thin] (-8,-5) grid (8,5);
\node at (-6.,4.5) {$\mathfrak{F}$};
\node at (7.2,1.55) {$g$};
\node at (-4,-3.8) {$k=0$};
\node at (-4,-0.6) {$k=0.2$};
\node at (-4,3) {$k=0.4$};
\node at (0,0) {\includegraphics[width=13.6cm]{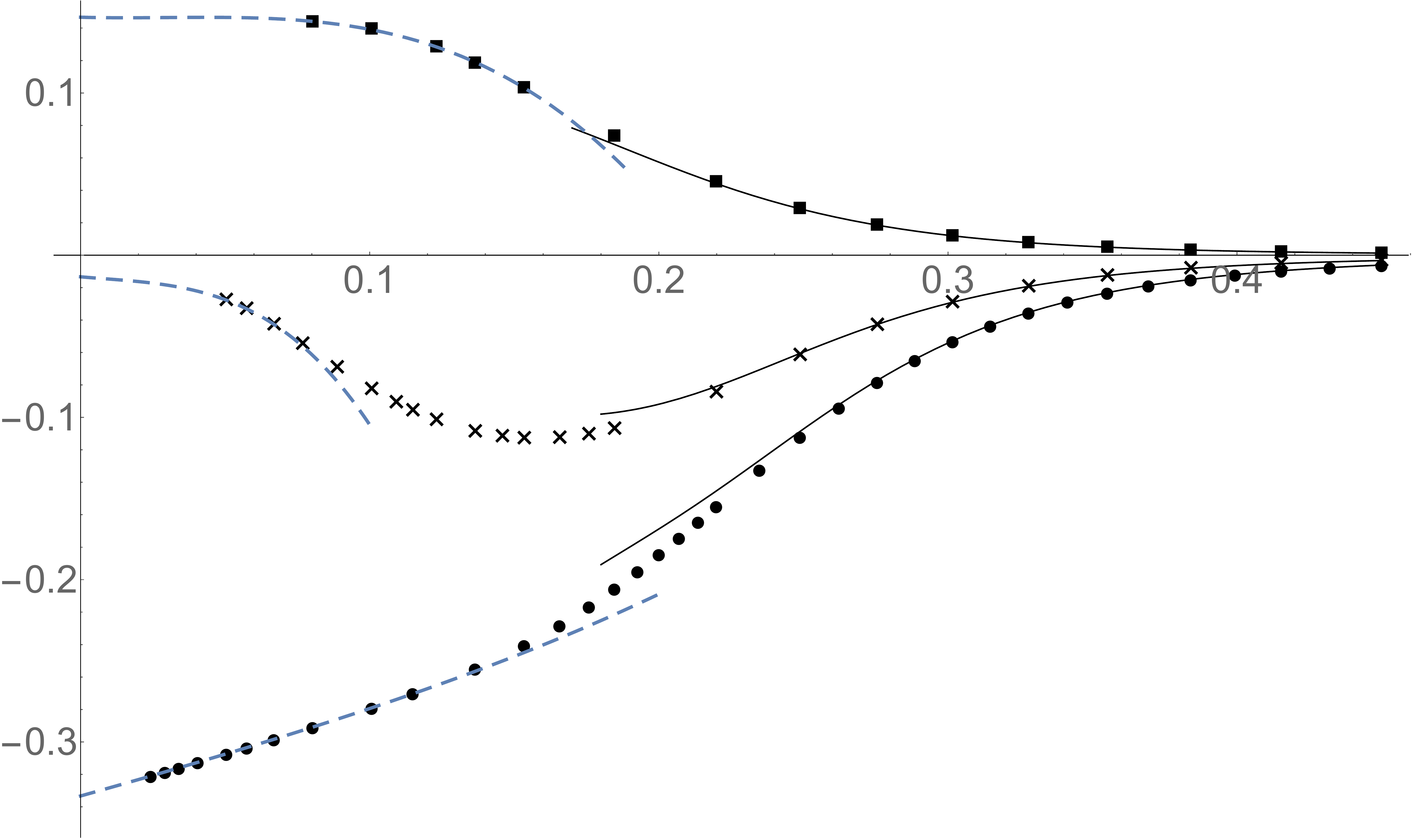}};
\end{tikzpicture}
}
\vspace{0.5cm}
\caption{A plot of\  $\mathfrak{F}(r,k)$ with $k=0,\,0.2$ and $0.4$ as a function of the running coupling constant $g(r)$ \eqref{eruidmnas} for the $O(3)$ sigma model.
The large-$r$ asymptotics, depicted by the black curves,
follow from eq.\,\eqref{aosioa}. For small $r$ and $k=0.2, \,0.4$,  the numerical data was fitted by a cubic polynomial of the form \eqref{hasast} with 
$a_0$ given by \eqref{aasjsaua}. The result of the fit is represented by the dashed line.
%Our rough  numerical estimates for the coefficients $ (a_0,a_1)$ are given by $\approx (-0.02,0.5)$ and $\approx (0.11,0.9)$ for $k=0.2$ and $k=0.4$ %respectively.
 For $k=0$ a quartic fit was used \eqref{sdfoiqwnm} 
and the coefficients were found to be $(a_1,a_2,a_3)=(0.5,0.25,1.0)$.
Note that the smallest value of the running coupling that we reached
is $g=0.0242\,\ldots$ (for $k=0$), whereas the largest value is $g=0.449\ldots\,$. These correspond to $r=10^{-15}$ and $r=5$, respectively.
\label{figkjqw}}
\end{figure}

As it follows from general  perturbative arguments, ${\mathfrak F}(r,0)$ should admit the power series expansion in terms
of the running coupling constant for the $O(3)$ NLSM.  It is convenient to  choose the RG scheme in which the running coupling $g=g(r)$  satisfies the RG flow equation \cite{Polyakov:1975rr,Hikami:1977vr}
\bea
r\,\frac{\rd g}{\rd r}=\frac{g^2}{1-g}=g^2+g^3+\ldots\  .
\eea
The solution to this equation which we will use is
\bea\label{eruidmnas}
g^{-1}\ \re^{-\frac{1}{g}}={\textstyle \frac{1}{32\pi}\ \re^{\gamma_E+1}}\ r\ .
\eea
The funny  constant  $\frac{1}{32\pi}\, \re^{\gamma_E+1}=0.048\ldots$ is chosen following the convention  
from the works \cite{Hasenfratz:1990zz,Luscher:1991wu,Shin:1996gi}.
 With this choice the gap between the vacuum and the first excited state energies in the $k=0$ sector, $\Delta E_0$, admits the perturbative expansion where the term $\propto g^2$ is absent:
$R\Delta E_0/(2\pi)=g+ g^3+1.19\, g^4+O(g^5)$.
The small-$r$  behaviour of ${\mathfrak F}(r,0)$ should admit the
%\FloatBarrier\noindent  
 asymptotic expansion of the form
\bea\label{sdfoiqwnm}
{\mathfrak F}(r,0)\asymp -{\textstyle \frac{1}{3}}+a_1\, g(r)+ a_2\, g^2(r)+a_3\,g^3(r)+a_4\ g^4(r)+ \ldots\ .
\eea 
The first coefficient in this series is known $a_1=\half$ \cite{Fateev:1992tk}. 
All others can, in principle, be calculated within the renormalized perturbation theory for the $O(3)$ NLSM.
Instead of doing so, we estimated their value by fitting the data obtained from the numerical solution of the NLIE. The fitting suggests that, in all likelihood,
$a_2=\frac{1}{4}$ and $a_3\approx 1$. Also, our numerical results for $k=0$ are in a full agreement with the numerical data quoted
 in ref.\,\cite{Balog:2009ze}.
To the best of our knowledge, the vacuum energies with $0<|k|\leq\half$ have not been discussed in the literature.\footnote{The case $k=\frac{1}{2}$ is of special interest 
for the application of resurgence theory to the problem of instanton summation in the ${\mathbb C}{\mathbb P}^{N-1}$ NLSM \cite{Dunne:2012ae}.}
One can  expect that for non-zero $k$
\bea\label{hasast}
{\mathfrak F}(r,k)\asymp a_0(k)+a_1(k)\, g(r)+ a_2(k)\,g^2(r)+a_3(k)\,g^3(r)+\ldots
\eea
with
\bea\label{aasjsaua}
a_0(k)=-{\textstyle{ \frac{1}{3}}}+2\, |k|\, \big(\,1-|k|\,\big)\ \ \ \  \ \ \ \ \ {\rm for}\ \ \ |k|\leq \half\ .
\eea
The last formula can be understood as follows. In the ultraviolet limit the effect of the target space curvature
becomes negligible and  $(-\frac{1}{3})$ here represents  the contribution of two massless Goldstones. However, with non-zero $k$,
the quasiperiodic boundary condition \eqref{aisosaio} implies the presence of  
conical singularities at the north and  south poles of the 2-sphere. The result of the work \cite{Dabholkar:1994ai}
for a string propagation  on a cone yields eq.\eqref{aasjsaua}.
Our numerical data seems to be in agreement with this prediction.
 Some of the obtained results are depicted  in fig.\,\ref{figkjqw}. 
 Note that as $k$ approaches to $\half $, the calculations for  small $r$ require a considerable amount of computational resources.

To complete this subsection, let us return to the case of finite $n$. As  has been already mentioned, the quantization condition \eqref{sasaa} admits the $R$-independent solution $p_1=0$
for $p_2=\half$. The latter corresponds to $k=\frac{1}{n+2}$. For this case, as follows from  \eqref{hassaysa}, the value of the effective central charge at $r=0$ is given by $\frac{2(n-1)}{n+2}$.
For integer $n\geq 2$ this coincides with the central charge  $c_n$  \eqref{sisias} of the ${\mathbb Z}_n$ parafermions CFT.
Based on the results of the work  \cite{Fateev:1991bv},
one can expect that the $k$-vacuum energy  with $k=\frac{1}{n+2}$ and $n=2,\,3,\ldots$ coincides with the ground state energy 
of the non-critical model  referred to as $H_n^{(0)}$ in \cite{Fateev:1991bv}.  The model can be described by means of the Euclidean action
\bea
{\cal A}_{H_n^{(0)}}={\cal A}_{{\mathbb Z}_n}-\lambda\ \int\rd^2 x\, \big(\psi^+{\bar \psi}^++\psi^-{\bar \psi}^-\big)\ , 
\eea
which  is  the critical action of the ${\mathbb Z}_n$ parafermions CFT  perturbed by  the relevant operator of the scale dimension $d=2-\frac{2}{n}$.
According to the work \cite{Fateev:1991bv}, the small-$r$ expansion  for the scaling function ${\mathfrak F}$  in this case  reads as follows
\bea
{\mathfrak  F}(r,{\textstyle\frac{1}{n+2}}\big)=-{\textstyle \frac{1}{6}}\ c_n+2\sum_{j=2}^\infty F_j\ r^{\frac{2j}{n}}+2\, F_{\rm ( \log)}\ \big({\textstyle \frac{r}{2\pi}}\big)^2\ \log (r)\ ,
\eea
with
\bea
F_{\rm (\log)}=\begin{cases}
-\frac{n-1}{2 n}\ \ \ \  \ &{\rm for}\ \ \ n\ \ \ {\rm  odd}\\
-\frac{1}{2}\ \ \ \  \ &{\rm for}\ \ \ n\ \ \ {\rm even}
\end{cases}\ .
\eea
For $n=2$  the model  $H_n^{(0)}$ coincides with the free theory of a massive Majorana fermion, and therefore
 \bea
{\mathfrak F}(r,{\textstyle\frac{1}{4}}\big)=-\frac{r}{2\pi^2}\ \int_{-\infty}^\infty\rd\theta\ \cosh(\theta)\ \log\Big(1+\re^{-r\cosh(\theta)}\Big)\ .
\eea
This was checked from the numerical solution of the NLIE \eqref{aatast}.

 \subsection{${\mathbb A}$, ${\mathbb B}$ and ${\mathbb T}$ \label{sec53}}

We are now ready to discuss  the general integrable structures in the quantum sausage model.
In fact, they are   almost identical to those from the cigar/sine-Liouville CFT. We
 will place a special emphasis on the aspects of the integrable structures which are related to the presence of the finite correlation length in the theory.

\bigskip

Recall  that the  sine-Liouville model possesses an infinite set of  involutive local IM.
At the   end of sec.\,\ref{sec20} we mentioned only the integrals  of the left chirality. Of course,  there are also the ``right'' local IM, so that
the full  commuting set is $\big\{{\mathfrak i}_{2j-1}, {\bar {\mathfrak i}}_{2j-1}\big\}_{j=1}^\infty$.
Remarkably 
(see, e.g., \cite{Lukyan} for details), 
all the local  IM are  invariant under the reflection $\varphi\mapsto-\varphi$, and therefore  they can be interpreted as the local IM for   both
theories  \eqref{asusauas} and \eqref{asuas}. 
This  observation suggests that the quantum sausage NLSM
 possesses  the infinite set of  local  IM 
$\big\{{\mathbb  I}_{2j-1}, {\bar  {\mathbb I}}_{2j-1}\big\}_{j=1}^\infty$ which can be thought of, in a certain sense, as a deformation of the conformal one \cite{Fateev:1996ea}.
In particular 
\bea\label{iasisaaaasu}
{{\mathbb  I}}_{2j-1}= \int_0^R\frac{\rd x}{2\pi}\ \Big(\,   \sum_{l+m=j } C^{(j)}_{lm}\ (\partial_+\varphi)^{2 l}(\partial_+ \vartheta)^{2m }+\ldots\Big)
%{\bar {\mathfrak  I}}_{2j-1}&=& \int_0^R\frac{\rd x}{2\pi}\ \Big(\,   \sum_{l+m=j } C^{(j)}_{lm}\ (\partial_-\varphi)^{2 l}(\partial_- \vartheta)^{2m }+\ldots\Big)\ ,\nonumber
%\nonumber
%\\
%{\hat {\mathbb I}}^{(-)}_{2j-1}&=& \int_0^R\frac{\rd x_1}{2\pi}\ 
%\Big(\,   \sum_{l+m=j } C^{(j)}_{lm}\ ({\bar \partial}\varphi)^{2 l}({\bar \partial}
%\vartheta)^{2m }+\ldots\Big)\ ,
\eea
and similar for ${\bar {\mathbb  I}}_{2j-1}$ with $\partial_+$ replaced by $\partial_-$.
Here we use  the  light cone variables $x^{\pm}=x^0\pm x^1$, the constants  $C^{(j)}_{lm}$ are the same as in eqs.\,\eqref{jasssay}-\eqref{kasausau},
and
the dots stand for monomials which include higher derivatives of 
$ \varphi$ and  $ \vartheta$, as well as  terms proportional to powers of $ \mathpzc{M}$.
It should be emphasized that the $\varphi$ and $\vartheta$ in  \eqref{iasisaaaasu} are \emph{local} fields whose dynamics are governed by the  dual action \eqref{sasa} and,
if considering the neutral sector of the model,
the periodic boundary conditions \eqref{aaisoa}.
A special r${\hat{\rm o}}$le belongs to the integrals
\bea
{{\mathbb  I}}_{1}&=& \int_0^R\frac{\rd x}{2\pi}\ \Big(\,  (\partial_+\varphi)^{2 }+(\partial_+\vartheta)^{2 }-
4 \mathpzc{M}\,
\cosh\big(\sqrt{n}\varphi\big) \cos\big(\sqrt{n+2}\, \vartheta\big)\,\Big)\nonumber\\[0.4cm]
{\bar {\mathbb  I}}_{1}&=&\big(\,  \partial_+\mapsto \partial_-\, \big)\ ,
%\\
%{\bar {\mathfrak  I}}_{1}&=& \int_0^R\frac{\rd x}{2\pi}\ \Big(\,  (\partial_-\varphi)^{2 }+(\partial_-\vartheta)^{2 }-
%4 \mathpzc{M}\, 
%\cosh\big(\sqrt{n}\varphi\big) \cos\big(\sqrt{n+2}\, \vartheta\big)\nonumber
%\ \Big)\nonumber\\
%{\hat {\mathbb I}}^{(-)}_{1}&=&- \int_0^R\frac{\rd x_1}{2\pi}\ \Big(\, 
 % ({\bar \partial}\varphi)^{2 }+({\bar \partial}\vartheta)^{2 }+
%4\mu\, 
%\cosh\big(\sqrt{n}\varphi\big) \cos\big(\sqrt{n+2}\, \vartheta\big)
% \Big) \ ,
\eea
whose sum, ${{\mathbb H}}_R={ {\mathbb I}}_{1}+{\bar {\mathbb I}}_{1}$, and difference, ${ {\mathbb P}}_R={ {\mathbb  I}}_{1}-{\bar {\mathbb I}}_{1}$,
coincide with the  Hamiltonian and  the total
momentum, respectively.
%The Hamiltonian of the theory, ${\mathbb H}_R$,  and the total
%momentum, ${\mathbb P}_R$, are given by
%${{\mathbb H}}_R={ {\mathfrak  I}}_{1}+{\bar {\mathfrak I}}_{1}$ and 
%${ {\mathbb P}}_R={ {\mathfrak  I}}_{1}-{\bar {\mathfrak I}}_{1}$ with
It is expected that
the common 
eigenvectors  of $\big\{{{\mathbb  I}}_{2j-1},{\bar {\mathbb I}}_{2j-1}\big\}_{j=1}^\infty$ form a basis
 in each invariant subspace ${\cal H}_{k}^{(K)}$ of the  Hilbert space of the sausage NLSM. Let's denote the corresponding  $k$-vacuum   eigenvalues by
  $\big\{{ \mathfrak  I}_{2j-1},{\bar {\mathfrak  I}}_{2j-1}\big\}_{j=1}^\infty$.  For the  $k$-vacuum the total momentum is zero, so 
  \bea
  {\mathfrak   I}_{1}={\bar {\mathfrak  I}}_{1}=\half\ {\cal E} R+{\textstyle \frac{\pi}{2 R}}\ {\mathfrak  F}(r,k)\ .
  \eea
  This relation together  with \eqref{hasstsa},  allows one to express the vacuum eigenvalues of  
  ${\mathbb   I}_{1}$ and  ${  {\bar {\mathbb I}}}_{1}$ in terms of the solution to
  the NLIE \eqref{aatast}.
  It is not difficult to  find  similar expressions for the other local IM (to be compared with formulae \eqref{jasshaa} from Appendix \ref{appC1}):
  \vspace{0.2cm}
  \bea\label{dsfnbmnassq}
  {\mathfrak I}_{2j-1}=\Big(\frac{m}{4}\Big)^{2j-1}\!\!\!\int\limits_{-\infty}^{\infty}\frac{\rd\theta}{\pi}
 \bigg( \! (-1)^{j}\,\re^{+ (2j-1)\theta}\, \log\big(1+\re^{-\omega(\theta)}\big)+ 2\,\Im m\Big[   \re^{+(2j-1)(\theta-\ri\gamma)}\, L(\theta-\ri \gamma)\Big]
 \!\bigg)\nonumber\\
 \\
{\bar{\mathfrak  I}}_{2j-1}=\Big(\frac{m}{4}\Big)^{2j-1}\!\!\!\int\limits_{-\infty}^{\infty}\frac{\rd\theta}{\pi}
 \bigg( \!(-1)^{j}\,\re^{- (2j-1)\theta}\, \log\big(1+\re^{-\omega(\theta)}\big)- 2\, \Im m\Big[   \re^{- (2j-1)(\theta-\ri\gamma)}\, L(\theta-\ri \gamma)\Big]
 \!\bigg)\nonumber
 \eea
For $r\ll 1$, similar to formula \eqref{jashsaasy} for  $\mathfrak{F}(r,k)$,  the vacuum eigenvalues of the higher spin local IM  can be approximated by
\bea\label{jsassays}
{\mathfrak   I}_{2j-1}={ \bar  {\mathfrak  I}}_{2j-1}\approx \big({\textstyle \frac{2\pi}{R}}\big)^{2j-1}\   i_{2j-1}\big(p(r),\half\,(n+2)\,k\,\big)\ .
\eea
Here, $i_{2j-1}(p_1,p_2)$ are the vacuum eigenvalues of the chiral local IM ${\mathfrak i}_{2j-1}$ and $p=p(r)$ is the solution of  eq.\,\eqref{jsasasa}.
Tab.\,\ref{tabwe} demonstrates the  quality of this approximation for the first few local IM.
\begin{table}
\centering
\begin{tabular}{|c||c|c|}
\hline
\multirow{2}{*}{$r=mR$} & \multirow{2}{*}{$\big(\frac{R}{2\pi}\big)^3\ {\mathfrak I}_3$} & \multirow{2}{*}{$i_3(p(r), p_2)$ } \\[-0.1cm]
 &  &  \\ \hline
& & \\[-0.45cm]
$10^{-1}$ & $3.45716396595\times 10^{-4}$ & $3.45832599760\times 10^{-4}$ \\ \hline
& & \\[-0.45cm]
$10^{-3}$ & $3.81476584833\times 10^{-4}$ & $3.81476594313\times 10^{-4}$ \\ \hline 
& & \\[-0.45cm]
$10^{-5}$ & $3.92737566343\times 10^{-4}$ & $3.92737566334\times 10^{-4}$ \\
\hline\hline
\multirow{2}{*}{$r=mR$} & \multirow{2}{*}{$\big(\frac{R}{2\pi}\big)^5\ {\mathfrak I}_5$} & \multirow{2}{*}{$i_5(p(r), p_2)$ } \\[-0.1cm]
 &  &  \\ \hline
& & \\[-0.45cm]
$10^{-1}$ & $-2.6148731\times 10^{-5}$&$-2.6151868\times 10^{-5}$ \\ \hline
& & \\[-0.45cm]
$10^{-3}$ &$-2.8189874\times 10^{-5}$ & $-2.8189869\times 10^{-5}$ \\ \hline
& & \\[-0.45cm]
$10^{-5}$ & $-2.8833109\times 10^{-5}$ & $-2.8833124\times 10^{-5}$ \\ \hline
\end{tabular}
\vspace{1cm}
\caption{The vacuum eigenvalues of the first two higher spin local IM in the sausage model for $n=1$ and $p_2=\frac{3}{2}\,k=\frac{5}{13}$.
The numerical values were calculated from the solution of the NLIE and formula \eqref{dsfnbmnassq}.
The last column gives the vacuum eigenvalues of the chiral local IM, $i_3(p(r),p_2)$ and $i_5(p(r),p_2)$,
where $p(r)$ is the solution to the quantization condition \eqref{jsasasa}. The limiting values at $r=0$ are given by $i_3(0,p_2)= \frac{39031}{95964960} =4.067\ldots \times 10^{-4}, $
$i_5(0,p_2)= -\frac{137442779}{4638370376640}= -2.963 \ldots\times 10^{-5}$. Explicit expressions for $i_3$ and $i_5$ can be found in ref.\,\cite{Lukyanov:2003nj}.}
\label{tabwe}
\vspace{1.5cm}
\end{table}
\bigskip

Note that  ${\mathfrak   I}_{2j-1}={\bar {\mathfrak  I}}_{2j-1}$ for any $j=1,\,2,\ldots\,$. These  relations can be easily understood since the model  \eqref{sasa}
is $\mathlarger{ \mathlarger{\mathpzc{P}}}$-invariant and that  under the parity transformation
\bea
 {\mathlarger{\mathpzc{P}}}\,  {{\mathbb  I}}_{2j-1}\,   \mathlarger{\mathpzc{P}}={\bar {\mathbb  I}}_{2j-1}\ .
\eea
Another important global symmetry is  $ \mathlarger{ \mathlarger{\mathpzc{C}}}$-invariance. Acting on the local fields it flips the sign of $\vartheta$ while keeping $\varphi$ unchanged.
All the local IM are $\mathpzc{C}$-invariant operators, i.e.,
\bea\label{ajsyas}
 \mathlarger{\mathpzc{C}}\,  {{\mathbb  I}}_{2j-1}\,  \mathlarger{\mathpzc{C}}={{\mathbb  I}}_{2j-1}\ ,\ \ \ \  \ \  \mathlarger{\mathpzc{C}}\  
 {\bar {\mathbb  I}}_{2j-1}\   \mathlarger{\mathpzc{C}}={\bar {\mathbb  I}}_{2j-1}\ .
 \eea
Since $\mathlarger{\mathlarger{\mathpzc{C}}}\, |\,{\rm vac}\,\rangle_k= |\,{\rm vac}\,\rangle_{-k} $,  this explains the fact that the vacuum eigenvalues  
${ \mathfrak  I}_{2j-1}$ are even functions of $k$.
The last discrete  symmetry that we shall consider is   the invariance of the dual action  \eqref{sasa} w.r.t. the transformation
\bea
\vartheta(r,t)\,\ \mapsto\,\ {\mathbb U}\,\vartheta(t,x)\, {\mathbb U}^{-1}\,=\,\vartheta(t,x)+\frac{2\pi}{\sqrt{n+2}}\ ,
\eea
where  the  unitary operator  ${\mathbb U}$ is the Flouquet-Bloch operator which is just a constant  phase factor
when it acts on the subspace ${\cal H}_k^{(K)}$:
\bea
{\mathbb U}\ {\cal H}_k^{(K)}=\re^{2\pi\ri k}\ {\cal H}_k^{(K)}\ .
\eea
Of course, $[{\mathbb U}, {{\mathbb  I}}_{2j-1}]=[{\mathbb U}, {\bar {\mathbb  I}}_{2j-1}]=0$.
\bigskip

\begin{table}
\centering
\begin{tabular}{|c||c|c||c|c||c|c|}
\hline
\multirow{2}{*}{$r=mR$} & \multirow{2}{*}{$\big(\frac{R}{2\pi}\big)^{\frac{2}{n+2}}\ {\mathfrak S}_1$} &
\multirow{2}{*}{$\tilde{s}_1(p(r),p_2)$} & \multirow{2}{*}{$\big(\frac{R}{2\pi}\big)^{\frac{2}{n+2}}\ \bar{{\mathfrak S}}_1$} &
\multirow{2}{*}{$\tilde{s}_1(p(r),-p_2)$} \\
& & & & \\ \hline
& & & & \\[-0.45cm]
$10^{-1}$ & $0.01936579$ & $0.01877110$ & $0.13129934$&$0.12726735$ \\ \hline
& & & & \\[-0.45cm]
$10^{-3}$ & $0.03374979$ & $0.03374058$ & $0.22882227$ & $0.22875988$ \\ \hline
& & & & \\[-0.45cm]
$10^{-5}$ & $0.04003149$ & $0.04003134$ & $0.27141198$ & $0.27141097$  \\
\hline
\end{tabular}

\vspace{0.3cm}
\bigskip
\begin{tabular}{|c||c|c|}\hline
\multirow{2}{*}{$r=mR$} & \multirow{2}{*}{$\big(\frac{r}{8\pi}\big)^{2k}\ {\mathfrak{S}}$} &
\multirow{2}{*}{$S\big(p_2|\,\ri p(r)\big)$} \\
& & \\ \cline{1-3}
$10^{-1}$ & $0.35644580$ & $0.35731884$ \\ \cline{1-3}
$10^{-3}$ & $0.24462688$ & $0.24463961$ \\ \cline{1-3}
$10^{-5}$ & $0.19673087$ & $0.19673105$ \\ \cline{1-3}
\end{tabular}
\vspace{0.5cm}
\caption{The vacuum eigenvalues of the dual nonlocal IM ${\mathbb  S}_{1}$, $\bar{\mathbb  S}_{1}$ and 
${\mathbb  S}$ for $n=\frac{9}{2}$ and  $p_2=\frac{13}{4}\,k=\frac{5}{13}$. \label{kjhsdfjkh}
Eq.\,\eqref{sdfjqwas} was used to find the numerical values of ${\mathfrak S}_1$,
$\bar{\mathfrak S}_1$
from the solution to the NLIE,
whereas the corresponding formula for ${\mathfrak{S}}$ is \eqref{vacljkhadfd}.
 The vacuum eigenvalues $\tilde{s}_1(p_1,p_2)$ of the chiral dual nonlocal IM are given by \eqref{oqwekjaxfd}
 and the expression for $S\big(p_2|\, q\big)$ is found in \eqref{pqowisjmdf}.
  Finally, $p(r)$ is the solution of the quantization condition \eqref{jsasasa}.
\label{asksaasuas}}
  \vspace{0.5cm}
\end{table}

$\!\!\!\!\!$
Together with the local IM, the sausage model possesses the set of  dual nonlocal IM
 $\big\{{{\mathbb S}}_{j},{\bar {\mathbb S}}_{j}\big\}_{j=1}^\infty$, which again can be
understood as  a  deformation of the corresponding conformal  set.  In contrast to the local IM, they are not   $ \mathlarger{\mathlarger{\mathpzc{C}}}$-invariant operators.
 Instead they satisfy the relations
 \bea
 \mathlarger{ \mathpzc{C}\mathpzc{P}}\  {{\mathbb  S}}_{j}\   \mathlarger{\mathpzc{C}\mathpzc{P}}={\bar {\mathbb  S}}_{j}\ .
  \eea

  \noindent This implies that the analog of eq.\eqref{jsassays} for the set $\big\{ {\mathfrak S}_{j},{\bar {\mathfrak  S}}_{j}\big\}_{j=1}^\infty$ of $k$-vacuum eigenvalues 
 of the
  dual nonlocal IM reads as follows
  \bea\label{jsassaysrter}\nonumber \\
{  \mathfrak S}_{j}\approx \big({\textstyle \frac{2\pi}{R}}\big)^{\frac{2j}{n+2}}\   \tilde{s}_{j}\big(p(r),+\half\,(n+2)\,k\,\big)\ , \ \ \ \
{ \bar  {\mathfrak  S}}_{j}\approx \big({\textstyle \frac{2\pi}{R}}\big)^{\frac{2j}{n+2}}\   \tilde{s}_{j}\big(p(r),-\half\,(n+2)\,k\,\big)\, .
\eea
\smallskip

\noindent
Similarly as for the local IM, the vacuum eigenvalues of the dual nonlocal IM can be expressed through the solution of the NLIE:
\bea\label{sdfjqwas}\nonumber \\
\!\!{\mathfrak S}_j={ \frac{2}{n+2}} \Big(\frac{m}{4}\Big)^{\frac{2j}{n+2}}\!\!\!
 \int\limits_{-\infty}^{\infty}\frac{\rd\theta}{\pi} \,\bigg(\!\!
\sin\big({\textstyle\frac{\pi j}{n+2}}\big)\ \re^{+\frac{2j \theta}{n+2}}\, \log\big(1+\re^{-\omega(\theta)}\big)
\! -\!
 \Im m\Big[  \re^{+\frac{2j( \theta-\ri\gamma)}{n+2}}\, L(\theta-\ri \gamma)\,\Big]\!\bigg)
 \nonumber\\[0.2cm]
\\[0.2cm]
\!\!{\bar {\mathfrak S}}_j={ \frac{2}{n+2}} \Big(\frac{m}{4}\Big)^{\frac{2j}{n+2}}\!\!\!
 \int\limits_{-\infty}^{\infty}\frac{\rd\theta}{\pi} \,\bigg(\!\!
\sin\big({\textstyle\frac{\pi j}{n+2}}\big)\ \re^{-\frac{2j \theta}{n+2}}\, \log\big(1+\re^{-\omega(\theta)}\big)
\! +\!
  \Im m\Big[  \re^{-\frac{2j( \theta-\ri\gamma)}{n+2}}\, L(\theta-\ri \gamma)\,\Big]
\!\bigg)\nonumber\\ \nonumber
\eea
In tab.\,\ref{kjhsdfjkh} we present  numerical data illustrating formulae \eqref{jsassaysrter} for ${\mathfrak S}_1$ and $\bar{\mathfrak S}_1$.

\medskip

We  are  now able to synthesize our study of the quantum sausage model in the form of the following conjecture. It is  expected  that the theory possesses  the operators ${\mathbb A}(\theta)$, ${\mathbb B}(\theta)$ and ${\mathbb T}(\theta)$ satisfying
the set of conditions listed on the following page.
\vfill
\pagebreak
\begin{enumerate}[label=(\roman*)]

\item\label{i1}  {\bf Commutativity:} \hspace{0.8cm} $\big[{\mathbb A}(\theta),{\mathbb A}(\theta')\big]=
\big[{\mathbb B}(\theta),{\mathbb B}(\theta')\big]=\big[{\mathbb A}(\theta),{\mathbb B}(\theta')\big]$
\smallskip

\item  {\bf  Analyticity:}  The operators ${\mathbb A}(\theta)$, ${\mathbb B}(\theta)$ and ${\mathbb T}(\theta)$ are entire functions of $\theta$.
\smallskip

\item  {\bf Global symmetries:}
\vspace{-0.8cm}
\bea
\hspace{1.85cm}&& \mathlarger{\mathpzc{C}\mathpzc{P}}\,  {\mathbb A}(\theta)\,  \mathlarger{\mathpzc{C}\mathpzc{P}}=  {\mathbb A}(-\theta)\ ,\ \ \ \ \  \mathlarger{\mathpzc{C}\mathpzc{P}}\,  {\mathbb B}(\theta)\, \mathlarger{ \mathpzc{C}\mathpzc{P}}={\mathbb B}(-\theta)\nonumber\\[0.1cm]
&& \mathlarger{\mathpzc{P}}\,  {\mathbb T}(\theta)\, \mathlarger{ \mathpzc{P}}=  
{\mathbb T}(-\theta)\ ,\ \ \ \  \ \  \ \, \ \mathlarger{\mathpzc{C}}\, {\mathbb  T}(\theta)\,  \mathlarger{\mathpzc{C}}=  {\mathbb  T}(\theta)\nonumber\\[0.1cm]
&&[ {\mathbb {U}},\,  {\mathbb A}(\theta)]=
[ {\mathbb {U}},\,  {\mathbb B}(\theta)]=
[{\mathbb {U}},\,  {\mathbb T}(\theta)]=0\nonumber
\eea

\vspace{-0.2cm}
\item  {\bf (Quasi)periodicity:}\hspace{0.4cm}
${\mathbb B}(\theta+\ri\pi)={\mathbb U}\ {\mathbb B}(\theta-\ri\pi)$,\ \ ${\mathbb T}(\theta+\ri\pi n)={\mathbb T}(\theta)$
\smallskip

\item {\bf Hermiticity:} \hspace{1.7cm}${\mathbb A}^\dagger(\theta)={\mathbb A}(\theta^*)\ ,\ \ 
 {\mathbb B}^\dagger(\theta)=\mathbb{B}(\theta^*)\ ,\ \ {\mathbb  T}^\dagger(\theta)={\mathbb T}(\theta^*)$
\smallskip

\item {\bf Functional relation:}
\bea\
\hspace{-1cm}{\mathbb A}\big(\theta\!-\!{\textstyle \frac{\ri\pi (n+1)}{2}}\big)\,{\mathbb  A}\big(\theta\!+\!{\textstyle \frac{\ri\pi (n+1)}{2}}\big)\!
-\!{\mathbb A}\big(\theta\!-\!{\textstyle \frac{\ri\pi (n-1)}{2}}\big)\,  {\mathbb A}\big(\theta\!+\!{\textstyle \frac{\ri\pi (n-1)}{2}}\big)=
 {\mathbb B}\big(\theta\!-\!{\textstyle \frac{\ri\pi}{2}}\big)\, {\mathbb B}\big(\theta\!+\!{\textstyle \frac{\ri\pi}{2}}\big)\nonumber
\eea
\item\label{i7}{\bf $T-Q$ relation:}\hspace{1.15cm} ${\mathbb T}\big(\theta+{\textstyle \frac{\ri\pi n}{2}}\big)\,{\mathbb A}(\theta)=
{\mathbb U}^{-\frac{1}{2}}\ {\mathbb A}(\theta+\ri\pi)+{\mathbb U}^{+\frac{1}{2}}\ {\mathbb A}(\theta-\ri\pi)$

\item\label{i8} {\bf Asymptotic behaviour of ${\mathbb A}(\theta)$:}
\bea
\hspace{-1cm}{\mathbb  A}(\theta)\,\asymp\,  {\mathbb S}^{\pm \frac{1}{2}}\ \exp\bigg(-\frac{r\,\cosh(\theta)}{4\sin(\frac{\pi n}{2})}\bigg)\ 
\exp\big(-a^{(\pm)}(\theta)\big)\ \ \ \ \ \ \ \ {\rm as}\ \ \ \  \Re  e (\theta)\to\pm \infty\nonumber
\eea

with $|\Im m (\theta)|<{\textstyle \frac{\pi}{2}}\, (n+2)$, and
\bea
\hspace{-0.5cm}a^{(+)}(\theta)=
\sum_{j=1}^\infty  {\mathbb S}_j\  \big({\textstyle \frac{m}{4}}\,\re^{+\theta}\big)^{-\frac{2j}{n+2}}\ ,\ \ \ \ \ 
a^{(-)}(\theta)= \sum_{j=1}^\infty  {\bar {\mathbb S}}_j\  \big({\textstyle \frac{m}{4}}\,\re^{-\theta}\big)^{-\frac{2j}{n+2}}
 \nonumber
\eea
\item\label{i9} {\bf Asymptotic behaviour of ${\mathbb  B}(\theta)$:}
\bea
\hspace{-1cm}{\mathbb  B}(\theta)\,\asymp\,  {\mathbb S}^{\pm \frac{1}{2}}\ \exp\bigg(-\frac{r\theta\sinh(\theta)}{2\pi}\, \bigg)\ 
\exp\big( -b^{(\pm)}(\theta)\big)\ \ \ \ \ \ \ \ {\rm as}\ \ \ \  \Re  e (\theta)\to\pm \infty\nonumber
\eea
with $|\Im m (\theta)|<\pi$, and
\bea
\hspace{-1cm}b^{(+)}(\theta)&=&\big({\mathbb I}_{1}-\half\,{\cal E}R\,\big)\ {\textstyle \frac{4}{m}}\ \re^{-\theta}+
\sum_{j=1}^\infty\Big(\,{\mathbb I}_{2j+1}\, \big({\textstyle \frac{m }{4}}\,\re^{+\theta}\big)^{-1-2j}+   {\mathbb S}_{j}\  \big({\textstyle \frac{m }{4}}\,\re^{+\theta}\big)^{-\frac{2j}{n+2}}
 \,\Big)\nonumber \\
 \hspace{-1cm}b^{(-)}(\theta)&=&\big({\bar {\mathbb I}}_{1}-\half\,{\cal E}R\,\big)\ {\textstyle \frac{4}{m}}\ \re^{+\theta}+
\sum_{j=1}^\infty\Big(\,{\bar {\mathbb I}}_{2j+1}\, \big({\textstyle \frac{m }{4}}\,\re^{-\theta}\big)^{-1-2j}+   {\bar {\mathbb S}}_{j}\  \big({\textstyle \frac{m}{4}}\,\re^{-\theta}\big)^{-\frac{2j}{n+2}}
 \,\Big)\nonumber
\eea
%and $C$ stands for some  $\theta$ and $R$  independent $c$-number.
\item\label{i10} {\bf Zeroes}: Let ${\mathfrak A}^{(\psi)}(\theta)$, ${\mathfrak B}^{(\psi)}(\theta)$, $\re^{2\pi\ri k}$ be the
eigenvalues of the operators ${\mathbb A}(\theta)$, ${\mathbb B}(\theta)$, ${\mathbb U}$, respectively, corresponding
to a common eigenvector $|\,\psi\,\rangle$. If  $\theta_j$ is  a zero of ${\mathfrak B}^{(\psi)}(\theta)$, then
\bea
\exp\big( -{\textstyle\frac{\ri}{2}}\,  r\sinh(\theta_j)-2\pi\ri k\,\big)\  \ 
\frac{{\mathfrak A}^{(\psi)}\big(\theta_j-\ri\pi -{\textstyle \frac{\ri\pi n}{2}}\big)}{
{\mathfrak A}^{(\psi)}\big(\theta_j-\ri\pi +{\textstyle \frac{\ri\pi n}{2}}\big)}=-1\ .\nonumber
\eea
All zeroes of ${\mathfrak B}^{(\psi)}(\theta)$ are simple and accumulate towards  infinity along the lines $\Im m(\theta)=\pi \pmod{2\pi}$.
%\pm\pi,\,\pm 3\pi,\ldots$\ .

\end{enumerate}

\bigskip
\noindent
Clearly, the conjectured properties of   ${\mathbb  A}(\theta)$, ${\mathbb B}(\theta)$ and  
${\mathbb T}(\theta)$  are  inspired   by  those of  their chiral counterparts $\alpha_+(\theta)$, $\beta_+(\theta)$ and  $\tau(\lambda)$ and the 
global symmetries of the model.
Unlike the chiral case, the subscript
was not included in the notation of  operators ${\mathbb A}$ and ${\mathbb B}$. 
It can be restored by setting ${\mathbb A}_+\equiv {\mathbb A}$ and 
${\mathbb B}_+\equiv{\mathbb B}$.  The properties of the $ \mathlarger{\mathlarger{\mathpzc{C}}}$-conjugated operators  ${\mathbb A}_-\equiv
\mathlarger{\mathlarger{\mathpzc{C}}}\,{\mathbb A}\,\mathlarger{\mathlarger{\mathpzc{C}}}$,
${\mathbb B}_-\equiv\mathlarger{\mathlarger{\mathpzc{C}}}\,{\mathbb B}\,\mathlarger{\mathlarger{\mathpzc{C}}}$ can be easily deduced from \ref{i1}-\ref{i10}.
Perhaps only the $\theta$-independent operator ${\mathbb S}$, which appears in the large-$\theta$  asymptotic expansions  \ref{i8} and  \ref{i9}, requires some elucidations.
Before presenting  them, let us first  discuss  the vacuum eigenvalues of  ${\mathbb A}(\theta)$ and ${\mathbb B}(\theta)$.
The obvious counterparts   to the formulae \eqref{skdjfhsdfhk1},\, \eqref{skdjfhsdfhk2} from Appendix \ref{appC1} read as
\bea\label{ajassay}
    \log {\mathfrak A}(\theta)&=& -\frac{r\,\cosh(\theta)}{4\sin(\frac{\pi n}{2})}
 +\int_{-\infty}^\infty\frac{\rd\theta'}{2\pi\ri }\ \Big[\,F_1(\theta-\theta'+\ri\gamma) \ L(\theta'-\ri \gamma)\\[0.2cm]
 &-&
F_1(\theta-\theta'-\ri\gamma)\,   \Big(L(\theta'-\ri \gamma)\Big)^* \ \Big]
 +\int_{-\infty}^\infty\frac{\rd\theta'}{\pi}\,
 F_2(\theta-\theta')\ \log\Big(1+\re^{-\omega(\theta')}\Big)\,\nonumber
  \eea
  where $|\Im m(\theta)|<\frac{\pi}{2}\,(n+2)-\gamma$, and
  \bea\label{jssys}
 \log {\mathfrak B}(\theta)&=&   -\frac{r\theta\sinh(\theta)}{2\pi}
+
\int_{-\infty}^\infty\frac{\rd\theta'}{2\pi\ri}\ \Big[\,
 F_3(\theta-\theta'+\ri\gamma)\ L(\theta'-\ri \gamma)\\[0.2cm]
 &-&F_3(\theta-\theta'-\ri\gamma)\ \Big(L(\theta'-\ri \gamma)\Big)^*\ \Big]
- \int_{-\infty}^\infty\frac{\rd\theta'}{\pi}\ 
 F_4(\theta-\theta')\ \log\Big(1+\re^{-\omega(\theta')}\Big)\ ,\nonumber
  \eea
 with $|\Im m (\theta)|<\pi-\gamma$. Now $\varepsilon(\theta)$ and  $\omega(\theta)$  solve  the massive NLIE 
 \eqref{aatast},\,\eqref{hasatast2} and the explicit form  of the functions $F_i(\theta)$ are given in \eqref{hasgsaf}. 
It is easy to see that these formulae combined with the asymptotics \ref{i8} and  \ref{i9},  yield the expressions \eqref{dsfnbmnassq} and \eqref{sdfjqwas} for the  vacuum eigenvalues of the local and dual nonlocal IM.
%Eq.\,\eqref{jssys} implies also a certain choice   for the  constant $C$ in \ref{i9}. As a matter of fact,
%$C$ is a  non-universal constant and its value can be chosen at will by means of the substitution  $B(\theta)\mapsto \re^{c r\cosh(\theta)}\ B(\theta)$.
%Such a substitution will not affect  the set of conditions   \ref{i1}-\ref{i9}  except for the shift
%$C\mapsto C+c$. Of course,  this adds the extra term ``$cr\cosh(\theta)$''  to the r.h.s. of \eqref{jssys}.
Notice that
the term $(-k\theta)$ is absent in \eqref{ajassay}, \eqref{jssys} compared with the analogous formulae  \eqref{skdjfhsdfhk1},\,\eqref{skdjfhsdfhk2}.
This is consistent with the absence of the factor $\re^{-k\theta}$ in the asymptotics  \ref{i8} and \ref{i9} compared with 
 the corresponding
eqs.\,\eqref{jassysay},\,\eqref{ashsaysa}
and \eqref{qpwoeier} for the chiral case. In connection with this, note that the operator $\hat{k}$ is ill defined and only its
exponent $\mathbb{U}=\exp({2\pi\ri\hat{k}})$ makes sense in the massive theory.

In the next  subsection we will point out that  the eigenvalues of the operator ${\mathbb  S}$  
play a special r$\hat{\rm o}$le in the ODE/IQFT correspondence. 
Eqs.\,\eqref{ajassay},\,\eqref{jssys}   predict that in  the case of   the  $k$-vacuum states, its eigenvalue is given by
\bea\label{vacljkhadfd}
{\mathfrak S}=\exp\bigg(\frac{2}{n+2}\ \int_{-\infty}^{\infty}\frac{\rd\theta}{\pi} \ 
  \Im m\big(   L(\theta-\ri \gamma)\,\big)\,\bigg)\ .
\eea
The small-$r$ behaviour of ${\mathfrak S}$  is given by a formula similar to \eqref{jsassays},\,\eqref{jsassaysrter} (see tab.\,\ref{asksaasuas}):
\bea
{\mathfrak S}\approx \Big(\frac{8\pi}{r}\Big)^{2 k}\  S\big(\half \,(n+2)\, k\, |\, \ri p(r)\big)\ ,
\eea
where
\bea\label{pqowisjmdf}
S(p_2|\,q)=\big(n+2\big)^{\frac{4p_2}{n+2}}\ \ \frac{\Gamma(\frac{1}{2}+p_2+q) \Gamma(\frac{1}{2}+p_2-q)}
{\Gamma(\frac{1}{2}-p_2+q) \Gamma(\frac{1}{2}-p_2-q)}
 \frac{\Gamma(1-2 p_2)}{\Gamma(1+2 p_2)}
\frac{\Gamma(1+\frac{2 p_2}{n+2})}{\Gamma(1-\frac{2 p_2}{n+2})}\ .
\eea
Notice that   $S(p_2|\,\ri p_1)$ can be  expressed in terms of the vacuum eigenvalues  \eqref{aashsat} of the operators $b_\pm$ defined in \eqref{jasaasy}:
$S(p_2|\,\ri p_1)=b^{({\rm vac})}_-(p_1,p_2)/b^{({\rm vac})}_+(p_1,p_2)$. 
%In a sector of given  $k$, for  $r\to 0$     $\log{\mathfrak S}$ diverges as $(-2k)\log r $.
%This clarifies the appearance
%of the  extra  factor 
%$\re^{-k\theta}$ in the asymptotic formulae \eqref{jassysay},\,\eqref{ashsaysa},\,\eqref{qpwoeier}
%for   the chiral operators $\alpha_+(\theta)$, $\beta_+(\theta)$ which is absent  in the analogous formulae for $A(\theta)$, $B(\theta)$.

The operator ${\mathbb T}(\theta)$ is the transfer-matrix in the sausage model -- the quantum counterpart of the Wilson loop \eqref{aoispspa} which was
the starting point of this paper. By means of the $T-Q$ equation \ref{i7} it is expressed in terms of the operator ${\mathbb A}(\theta)$ and, of course,
commutes with both ${\mathbb A}$ and ${\mathbb B}$ 
for any values of the  spectral parameter $\theta$. We did not include the formula which described its large-$\theta$
asymptotic in the list \ref{i1}-\ref{i10} since it is an immediate consequence of the $T-Q$ equation and  the  asymptotic \ref{i8} for 
${\mathbb A(\theta)}$. Notice that unlike for the  Toda-type theory,
the transfer-matrix in the sausage model does not generate  the local IM through its asymptotic expansion.

\bigskip
Finally we can turn to the case of  the $O(3)$  NLSM. There is no reason to expect that the $n\to\infty$ limit is problematic for the operator ${\mathbb B}(\theta)$.
Introduce the notation
\bea
{\mathbb B}_\infty(\theta)=\lim\limits_{n\to\infty}{\mathbb B}(\theta)\ .
 \eea
Using eqs.\eqref{jssys},\,\eqref{hasgsaf}, one finds   the relation which expresses its vacuum eigenvalue in terms of the 
solution to the NLIE \eqref{aatast} with the kernels \eqref{jasssayas}:
 \bea\label{jssygftftfs}
 \log \mathfrak{B}_\infty(\theta)&=&   -\frac{r\theta\sinh(\theta)}{2\pi}
+
\int_{-\infty}^\infty\frac{\rd\theta'}{2\pi\ri}\ \Big[\,
 f_3(\theta-\theta'+\ri\gamma)\ L(\theta'-\ri \gamma)\\[0.2cm]
 &-&f_3(\theta-\theta'-\ri\gamma)\ \Big(L(\theta'-\ri \gamma)\Big)^*\ \Big]
- \int_{-\infty}^\infty\frac{\rd\theta'}{\pi}\ 
 f_4(\theta-\theta')\ \log\Big(1+\re^{-\omega(\theta')}\Big)\ \nonumber
  \eea
where $|\Im m(\theta)|<\pi-\gamma$, and
 \bea
 {  f}_3(\theta)={ \frac{1}{\theta}}-{ \frac{1}{\sinh(\theta)}}\ ,\ \ \ \ 
  f_4(\theta)=
  \frac{\pi}{2(\theta+\frac{\ri\pi}{2})(\theta-\frac{\ri\pi}{2})}-\frac{1}{2\cosh(\theta)}\ .
  \eea
The situation with the operators ${\mathbb A}(\theta)$ and ${\mathbb T}(\theta)$  is slightly more delicate. In this case, one can expect that the following limits exist
\bea
{\mathbb A}_\infty(\theta)&=&\lim_{n\to\infty}{\mathbb  A}\big(\theta-{\textstyle \frac{\ri\pi n}{2}}\big)
\exp\Big(\,{\textstyle\frac{1}{4}}\ r\cot\big({\textstyle\frac{\pi n}{2}}\big)\, \cosh(\theta)\,\Big)\nonumber\\
\mathbb{T}_\infty(\theta)&=&\lim_{n\to\infty} \mathbb{T}(\theta)\, 
 \exp\Big(-
  {\textstyle\frac{1}{2}}\  r\cot\big({\textstyle\frac{\pi n}{2}}\big)\, \cosh(\theta)\,\Big)
   \eea
and the limiting operators satisfy the $T-Q$ equation in the form
\bea
{\mathbb T}_\infty(\theta)\,{\mathbb  A}_\infty(\theta)={\mathbb U}^{-\frac{1}{2}}\ \ {\mathbb A}_\infty(\theta+\ri\pi)+
{\mathbb U}^{+\frac{1}{2}}\ {\mathbb A}_\infty(\theta-\ri\pi)
\eea
(recall that in the sector ${\cal H}_{k}^{(K)}$ the operator ${\mathbb U}^{+\frac{1}{2}}$ becomes just a phase factor $(-1)^{K-1}\,\re^{\ri \pi k}$).
With eqs.\,\eqref{ajassay},\,\eqref{hasgsaf}, it is easy to see that
\bea\label{ajassaasdsfay}
    \log \mathfrak{A}_\infty(\theta)&=&  {\textstyle \frac{\ri}{4}}\, r\sinh(\theta)
 +\int_{-\infty}^\infty\frac{\rd\theta'}{2\pi\ri }\ \Big[\,f_1(\theta-\theta'+\ri\gamma) \ L(\theta'-\ri \gamma)\\[0.2cm]
 &-&
f_1(\theta-\theta'-\ri\gamma)\,   \Big(L(\theta'-\ri \gamma)\Big)^* \ \Big]
 +\int_{-\infty}^\infty\frac{\rd\theta'}{\pi}\,
 f_2(\theta-\theta')\ \log\Big(1+\re^{-\omega(\theta')}\Big)\,\nonumber
  \eea
with $\Im m(\theta)<\pi-\gamma$, 
\bea
 {  f}_1(\theta)= \frac{1}{\theta+\ri\pi}\ ,\ \ \ \ \ 
 f_2(\theta)=-\frac{\pi}{2(\theta+\frac{3\ri\pi}{2}) (\theta+\frac{\ri\pi}{2})}\,,
 \eea
and $\varepsilon(\theta)$, $\omega(\theta)$ are defined through the solution of the NLIE \eqref{aatast},\,\eqref{jasssayas}.

   \subsection{ODE/IQFT for the  sausage model}
In sec.\,\ref{sec4.3} we  briefly discussed the  ODE/IM correspondence for the cigar NLSM.
Recall that the correspondence relates the eigenvalues of the chiral transfer-matrices to 
the connection coefficients for the  family of second order differential equations   ${\cal D}(\theta)\,\psi=0$ 
with the operators ${\cal D}(\theta)$ of the form \eqref{aoisaisa}.
 The generalization of the construction to the sausage model is based on the ideas from the work \cite{Lukyanov:2010rn} 
 and goes along the following line.
 \bigskip

As far as our attention was confined to the CFT, 
there was no need to separately consider the antiholomorphic operators, 
${\overline{\cal D}}{({ \bar \theta})}=
-\partial_{\bar z}^2+{\overline T}_{\bar L}({\bar z})
+\re^{2{\bar  \theta}}\ { \overline{\cal P}}({\bar z})$,
since there was only a nomenclature difference 
between the holomorphic and antiholomorphic cases.
In the massive QFT,
following  \cite{Lukyanov:2010rn},
one should substitute the pair
$\big({\cal D}{(\theta_0+\theta)},{\overline
  {\cal D}}(\theta_0-{ \theta})\big)$ 
by a pair of $(2\times 2)$-matrix valued differential operators
\bea\label{asopssaopasopq}
{\boldsymbol { D}}{(\theta)}=\partial_z-{\boldsymbol
  A}_z\ ,\ \ \ \ \  \overline{{\boldsymbol { D}}}{(\theta)}= 
\partial_{\bar z}-{ {\boldsymbol A}}_{\bar z}
\eea
with
\beq\label{ystopsso}
\begin{array}{rcl}
%{\boldsymbol { D}}{(\lambda)}&=&\partial_z-{\boldsymbol A}_z\ :\ \ \ \ \ 
{\boldsymbol
  A}_z&=&-{\textstyle\frac{1}{2}}\ \partial_z\eta\,\sigma_3+ \sigma_+\,\re^{+\eta}+ 
\sigma_-\,\re^{2\theta_0+2\theta}\, {\cal P}(z)\, \re^{-\eta}\\[.3cm]
%{\boldsymbol {\bar  D}}^{({\bar \lambda})}&=&\partial_{\bar z}-{
%{\boldsymbol A}}_{\bar z}\ :\ \ \ \ \  
{ {\boldsymbol A}}_{\bar z}&=&+
{\textstyle\frac{1}{2}}\ \partial_{\bar
  z}\eta\,\sigma_3+ \sigma_-\, 
\re^{+\eta}+\sigma_+\, \re^{2\theta_0-2\theta} \ {\overline{\cal P}}({\bar z})\,\re^{-\eta}\ ,
\end{array}
\eeq
where $\sigma_3,\sigma_\pm=(\sigma_1\pm \ri \sigma_2)/2$ are the
standard Pauli matrices and ${\cal P}(z)$ is given by \eqref{oasioaq}. 
In fact, $({\boldsymbol A}_z,\,  {\boldsymbol A}_{\bar z})$ 
form a
$\mathfrak{sl}(2)$  connection whose flatness 
is a necessary condition for the existence of a solution to the  linear problem 
\bea\label{apsosaospa}
%\big(\partial_z-{\boldsymbol A}_z\big)
{\boldsymbol { D}}{(\theta)}\, {\boldsymbol \Psi}=0\ ,\ \ \  \ \ \ 
%\big(\partial_{\bar z}-{\boldsymbol A}_{\bar z}\big)
 {\boldsymbol {\overline D}}{({\theta})}\,{\boldsymbol \Psi}=0\ .
\eea
The zero-curvature relation leads to 
the Modified Sinh-Gordon (MShG) equation:
 \bea\label{asospsosap}
\partial_z\partial_{\bar z}\eta-\re^{2\eta}+ \rho^4\ |{\cal P}(z)|^2\  \re^{-2\eta}=0\ ,\qquad 
\rho=\re^{\theta_0}\ .
\eea
In refs.\,\cite{Lukyanov:2013wra,Bazhanov:2013cua}, a  class of singular solutions to this partial differential  equation 
distinguished by special monodromy properties of the associated linear
problem \eqref{apsosaospa} was introduced.  Together with the  singularities at $z=z_1,\,z_2,\,z_3$,  the solutions   are allowed to have 
the  so-called  apparent  singularities, which do not  affect  the monodromy properties of the auxiliary linear problem \eqref{apsosaospa}.
In the limit $\theta_0\to -\infty$ with $\theta_+=\theta_0+\theta$ kept fixed, the system \eqref{apsosaospa} 
can be reduced to
${\cal D}(\theta_+)\,\psi=0,\ \partial_{\bar{z}}\psi=0$ and the apparent singularities manifest themselves as the monodromy free singularities
for the operator ${\cal D}(\theta_+)$ of the form  \eqref{aoisaisa}. Parallel to this, the limit $\theta_0\to -\infty$ with $\theta_-=\theta_0-\theta$
kept fixed can be considered, which leads to the corresponding antiholomorphic 
equations ${ \overline{\cal D}}(\theta_-)\,\psi=0$ and $\partial_{z}\psi=0$.

In the same works \cite{Lukyanov:2013wra,Bazhanov:2013cua},
 evidence was presented that the linear problem  \eqref{apsosaospa}  built from 
the special singular solutions of the MShG equation makes up the ODE part for  the 
ODE/IQFT correspondence where the  IQFT  counterpart is the so-called Fateev model \cite{Fateev:1996ea}.
The latter  is governed by
the    Lagrangian
%\bea\label{aposoasio}
%{\cal L}&=& \frac{1}{16\pi}\ \sum_{i=1}^3
%\big(\, (\partial_t\varphi_i)^2-(\partial_x\varphi_i)^2\,\big)\\
%&+&2\mathpzc{M} \
%\big(\, \re^{\ri\, \alpha_3\varphi_3}\ \cos(\alpha_1\varphi_1+\alpha_2\varphi_2)+\re^{-\ri \alpha_3\varphi_3}\
%\cos(\alpha_1\varphi_1-\alpha_2\varphi_2)\,\big)
%\nonumber
%\eea
\begin{equation}\label{aposoasio}
{\cal L}={\textstyle  \frac{1}{16\pi}}\, \sum_{i=1}^3
(\partial_\mu\varphi_i)^2
+2\mathpzc{M} \,
\big(\, \re^{\ri\, \alpha_3\varphi_3}\ \cos(\alpha_1\varphi_1+\alpha_2\varphi_2)+\re^{-\ri \alpha_3\varphi_3}\
\cos(\alpha_1\varphi_1-\alpha_2\varphi_2)\,\big)
\end{equation}
for the three scalar fields $\varphi_i=\varphi_i(t,x)$ which satisfy the 
 periodic boundary conditions
\bea\label{sissiaosai}
\varphi_i(t,x+R)=\varphi_i(t,x)\ .
\eea
It is important that the  dimensionless coupling constants $\alpha_i$   satisfy the linear constraint
\bea\label{aposapoas}
\alpha_1^2+\alpha_2^2+\alpha_3^2=\half\ ,
\eea
so that the coupling $\mathpzc{M}$ in the renormalized Lagrangian   \eqref{aposoasio} has the dimensions of mass, $\mathpzc{M} \sim [\,{\rm mass}\,]$.
Within the ODE/IQFT correspondence the
parameters are identified as follows 
\beq\label{sssaopsa}
a_i=4\,\alpha_i^2
\,,\ \ \ \ \ \ \  \qquad
(i=1,2,3)\ ,
\eeq
whereas the relation between the dimensionless parameter $ \mathpzc{M}\!R$ and $\rho$ from \eqref{asospsosap} is  given by 
\bea\label{saossaops}
\rho=\half\  \mathpzc{M}R\ .
\eea
Although the original considerations of  refs. \cite{Lukyanov:2013wra,Bazhanov:2013cua}
 were focused on the ODE/IQFT correspondence with
 all three  parameters $a_1$, $a_2$, $a_3$ positive, in the subsequent work  \cite{Bazhanov:2014joa} evidence was presented 
that the correspondence remains valid with minimum modifications to the case $a_1,\,a_2>0$ and $a_3<0$.
 In  the recent works \cite{Bazhanov:2016glt,Bazhanov:2017xky}, the same conclusion was reached for  $a_1,\,a_2>0$ and $a_3=0$.
 Among the tasks of the current paper is to argue that the  ODE/IQFT correspondence remains valid for
  \bea\label{aisssau}
  a_1=-n,\ \ \ \ \ \ a_2=n+2\ ,\ \ \ \  \ \ \ a_3=0 \ \ \ \ \ {\rm with}\ \  \ n>0\,.
  \eea
  In this case, the coupling $\alpha_3$ in the Lagrangian  \eqref{aposoasio} vanishes and
 the field $\varphi_3$ is decoupled.  The interaction part turns out to be the Lagrangian for the dual action of the sausage model  \eqref{sasa}
provided the identifications $\varphi_1=2\varphi$, $\varphi_2=2\vartheta$ are made.
 Notice that with the $m$\,-\,$\mathpzc{M}$ relation for the sausage model \eqref {jassasaytast}, formula \eqref{saossaops} can be re-written as
 \bea
 \rho=\frac{r}{8\pi}\ .
 \eea

 The ODE/IQFT correspondence suggests   that for any common eigenvector $|\,\psi\,\rangle\in {\cal H}_k^{(1)}$ of the commuting family of  operators
${\mathbb A}(\theta)$ and ${\mathbb B}(\theta)$,   there exists a singular  solution of the MShG equation \eqref{asospsosap} with
 ${\cal P}(z)$  given by \eqref{oasioaq} and the  parameters $a_i$  as in \eqref{aisssau}.
  The solution should be such that 
  $\re^{-\eta}$ is a {\it  smooth, single valued
  complex function without zeroes}
on the punctured Riemann sphere.
% At the punctures the solution behaves in such a way that
%Apart from the apparent singularities, $\re^{-\eta}$ is singular  
In the  vicinity of $z=z_1,\,z_3$, the leading behaviour is described by
%so that
 \bea\label{ajsssay}
\re^{-\eta}\sim |{\cal P}(z)|^{-\frac{1}{2}}\  \ \  \ \ {\rm as}\ \ \ \ \ \ |z-z_i|\to 0\ ,
\eea
 %and
 % also
 whereas  
 in the neighbourhood of the second puncture
\bea\label{sospsaosap}
\re^{-\eta}\sim |z-z_2|^{1-(n+2)|k|}\ \ \ \ \ \ \  {\rm
  as}\ \ \ \ \ \  \ \ |z-z_2|\to  0
\eea
with $0<|k|<\half$.\footnote{At $|k|=0,\,\half$  the  leading asymptotic
\eqref{sospsaosap} involves logarithms. Here we ignore such
subtleties.} The description of the  apparent singularities involves some technical details that are 
completely analogous to those  discussed in ref.\,\cite{Bazhanov:2013cua}.
 In the case of the vacuum state, the apparent singularities are absent and the solution $\eta$ is real.
Notice that the point
$z=\infty$ on the sphere is assumed to be  regular, so that
\bea
\label{sospsaosapyst}
\re^{-\eta}\sim |z|^{2}\ \ \ \ \ \ \ \  {\rm as}\ \ \ \ \ \  \ \   |z|\to  \infty\ .
\eea

As it was mentioned in the previous subsection, the eigenvalue of the operator ${\mathbb S}$ which appears in the 
 large $\theta$-asymptotic formulae  \ref{i8} and \ref{i9} is of special interest. Let us introduce the
``regularized'' value of the solution at the puncture $z=z_2$ as
\bea
\eta=\big(\, (n+2)\, |k|-1\,\big)\ \log|z-z_2|+\eta^{(\rm reg)}+o(1)\,.
\eea
Then for the solution corresponding to the eigenvector  $|\,\psi\,\rangle\in {\cal H}_k^{(1)}$ with $0<k<\half $,   the following formula holds:
\bea\label{spssaps}
{{\mathfrak S}^{(\psi)}}=\bigg(\frac{\rho}{n+2}\bigg)^{-2k}\
\frac{\Gamma(k)}{\Gamma(1-k)}\ \frac{\exp(\eta^{(\rm reg)})}{(n+2)}\
\bigg|\frac{z_{13}}{z_{12}z_{23}}\bigg|^{-(n+2) k}\ ,
\eea
where  we use the shortcut notation $z_{ij}=z_i-z_j$.
\bigskip

We can now describe, in precise terms,  the ODE/IQFT correspondence for the sausage model. Consider  the  auxiliary linear  problem \eqref{apsosaospa} 
associated with the singular solution of the MShG equation.
 The puncture $z=z_2$ is a regular singular point for this system of ODE. 
 In the vicinity of this point, assuming that  $0<k<\half$,  one can introduce
 the basis solutions by means of the following asymptotic formulae as $z\to z_2$:
\bea
{\boldsymbol  \Theta}_{-  }(z,{\bar z}\,|\,\theta\,)&\to& \frac{\re^{+\ri \beta_2}}{\sqrt{\sin(2\pi k)}}\ 
\re^{-k(\theta-\frac{\ri\pi n}{2})}
 \bigg(
 %\re^{\frac{4}{n+2}(\theta-\frac{\ri\pi n}{2})}\
%\frac{{\bar z}_{i+1}-{\bar z}_i} 
%{{ z}_{i+1}-{ z}_i}\ 
\frac{{ z}-{ z}_2} {{\bar z}-{\bar z}_2}\,\bigg)^{+\frac{1}{4}\, (1-k(n+2))}  \begin{pmatrix}
1\\
0
\end{pmatrix}\nonumber\\[-0.2cm]
&&
\\[-0.2cm]
 {\boldsymbol  \Theta}_{+  }(z,{\bar z}\,|\,\theta\,)&\to&\frac{\re^{-\ri \beta_2}}{\sqrt{\sin(2\pi k)}}\
 \re^{+k(\theta-\frac{\ri\pi n}{2})}\
 \bigg(
 %\re^{\frac{4}{n+2}(\theta-\frac{\ri\pi n}{2})}\
%\frac{{\bar z}_{i+1}-{\bar z}_i} 
%{{ z}_{i+1}-{ z}_i}\ 
\frac{{ z}-{ z}_2} {{\bar z}-{\bar z}_2}\,\bigg)^{-\frac{1}{4}\, (1-k(n+2))}\begin{pmatrix}
0\\
1
\end{pmatrix}
\nonumber
\eea
where, for convenience, the constant phase factor is set to be
 \bea
 \re^{\ri\beta_2}=\bigg(\frac{z_{12}z_{23}}{z_{13}}\ \frac{{\bar z}_{13}}{{\bar z}_{12}{\bar z}_{23}}\bigg)^{\frac{k}{4}(n+2)}\ .
 \eea
Unlike $z=z_2$, the puncture at $z=z_1$ is an irregular singular point for the auxiliary linear  problem.
In its neighbourhood, and for   $\frac{\pi}{2} \,(n-1)\leq \Im m(\theta)\leq \frac{\pi}{2}\,(n+1)$, another  solution can be uniquely defined using the WKB asymptotic condition:
 \bea
 {\boldsymbol  \Xi}(z,{\bar z}|\,\theta\,)&\to& |{\cal P}(z)|^{\frac{1}{4}}\ \exp\bigg(-\rho\re^\theta\int^z_{z_2}\rd z\  \sqrt {{\cal P}(z)}-
 \rho\re^{-\theta}\int^{\bar z}_{{\bar z}_2}\rd {\bar z}\ \sqrt{ \overline{ {\cal P}}({\bar z})}\bigg)\nonumber\\
 &\times&
  \begin{pmatrix}
+
\re^{-\frac{\theta}{2}}\,
\big({\cal P}(z)\big)^{-\frac{1}{4}}\\
-
\re^{+\frac{\theta}{2}}\,
\big(\overline{{\cal P}}({\bar z})\big)^{-\frac{1}{4}}
\end{pmatrix}\ \ \ \ \ \ \  \ {\rm as}\ \ \ \ z\to z_1\ .
 \eea
  There must be a linear relation between these three solutions and hence,
 \bea
 {\boldsymbol  \Xi}\big(z,{\bar z}|\,\theta+{\textstyle\frac{\ri\pi n}{2}} \,\big)={\mathfrak A}_+^{(\psi)}(\theta)\ 
 {\boldsymbol  \Theta}_{-  }\big(z,{\bar z}|\,\theta+{\textstyle\frac{\ri\pi n}{2}} \,\big)+{\mathfrak A}_-^{(\psi)}(\theta)\ 
 {\boldsymbol  \Theta}_{+  }\big(z,{\bar z}|\,\theta+{\textstyle\frac{\ri\pi n}{2}} \,\big)\ .
 \eea
The ODE/IM correspondence  states that the connection coefficients
${\mathfrak A}_+^{(\psi)}(\theta)$ and ${\mathfrak A}_-^{(\psi)}(\theta)$ coincide with the eigenvalues of the operators ${\mathbb A}(\theta)$ and
$\mathlarger{\mathlarger{\mathpzc{C}}}\,{\mathbb A}(\theta)\,\mathlarger{\mathlarger{\mathpzc{C}}}$, for the common eigenvector $|\psi\rangle\in {\cal H}_k^{(1)}$ associated
 with the singular solution of the MShG equation. The eigenvalues of the transfer-matrices $\mathbb{T}_{\frac{1}{2}}\equiv \mathbb{T}$, and more
  generally $\mathbb{T}_j$ with $j=\frac{1}{2},1,\ldots\;$,  can be obtained by the formulae
  similar to eqs.\,\eqref{aksasuusa}-\eqref{aksasuusa2}:
\bea
{\boldsymbol \Xi}(z,{\bar z}\,|\,\theta+\ri\pi(2j+\half )\big)&=&
{\mathfrak T}^{(\psi)}_j(\theta+\ri\pi j)  \ 
{\boldsymbol \Xi}(z,{\bar z}\,|\,\theta+{\textstyle\frac{\ri\pi}{2}}\big) \\[0.2cm]
&-&{\mathfrak T}^{(\psi)}_{j-\frac{1}{2}}(\theta+\ri\pi (j+{\textstyle\frac{1}{2}})) \ {\boldsymbol \Xi}(z,{\bar z}\,|\,\theta-{\textstyle\frac{\ri\pi}{2}}\big)\,.\nonumber 
\eea
Finally, the eigenvalues of the operators 
$\mathbb{B}(\theta)$ and $\mathlarger{\mathlarger{\mathpzc{C}}}\,{\mathbb B}(\theta)\,\mathlarger{\mathlarger{\mathpzc{C}}}$
can also be expressed in terms of  certain connection coefficients of the ODE system \eqref{apsosaospa}.
For this purpose, one needs to introduce suitable basis solutions in the vicinity of the third puncture $z=z_3$. The corresponding formulae are simple generalizations of \eqref{assasay}
and we do not present them here.

 \section{Discussion}

In this work we considered  the problem of the quantization of the sausage NLSM.
To conclude, let's summarize and discuss the
 key results of the paper.
%\begin{itemize}

%\item
We demonstrated that the flat connection for  the classical sausage model admits 
the ultralocal gauge and
thus, ``Hamiltonian Methods in the Theory of Solitons'' \cite{Faddeev:1987ph} can be applied 
without modifications. In connection with this, we believe that the problem with ultralocality  for 
 %more general 
 other integrable NLSM  should
be revisited.
It would be interesting to see how the
 integrable canonical structures of ref.\,\cite{Maillet:1985ek}
fit in with the results of this work.

%\item
Since the classical integrable structures  in the sausage model turn out to be similar to
those from the integrable KdV/sine-Gordon hierarchy, in our study of the quantum
model we closely followed the ideas of the works \cite{Bazhanov:1994ft,Bazhanov:1996dr,Bazhanov:1998dq}. We paid special attention 
to the integrable structures of the cigar NLSM --- the CFT governing the ultraviolet
behaviour of the quantum sausage. 
In particular we constructed the BLZ type representation for the chiral transfer-matrices in the quantum cigar.

%\item 
The chiral transfer-matrices 
depend on a number of parameters and can be considered 
in the parameter domain where they are not directly related to the cigar NLSM.
In this case, they are still of physical interest since they
can be interpreted as the transfer-matrices for the minimal $\mathbb{Z}_n$ parafermionic models from ref.\,\cite{Fateev:1985mm}.
The situation here resembles the interplay between the quantum Liouville theory and the BPZ minimal models. 
We constructed  lattice transfer-matrices and presented numerical evidence that in the scaling limit
they become the chiral transfer-matrices in the parafermionic regime. We believe 
that it may hint as to how to proceed with the lattice formulation of the cigar and sausage models.
To go further in this direction the most promising approach is, perhaps, the 
 method of separation of variables  \cite{Sklyanin:1995bm} which was 
successfully applied to a similar problem appearing in the quantization of the   sinh-Gordon model
 \cite{Smirnov:1998kv,Lukyanov:2000jp,Zamolodchikov:2000kt,AlZ,
Bytsko:2006ut,Teschner:2007ng,Borot:2014dea}.
Another interesting possibility 
 is related to the work \cite{Ikhlef:2011ay}, where some  spectral properties of
the cigar NLSM were observed to appear in the scaling limit of a certain inhomogeneous version of the 6-vertex model.

%\item
One of the most effective methods for the calculation of the spectrum of commuting families of operators
including the transfer-matrices in integrable
quantum field theory is based
on the ODE/IQFT correspondence. 
From our study of the parafermionic transfer-matrix, we proposed the ODE counterpart
in the correspondence for the cigar NLSM. 
It turns out to be identical to that which was  introduced  earlier in the context of the so-called paperclip model 
in the  works \cite{Lukyanov:2003nj,Lukyanov:2005nr}. 
Based on the results of these papers, we derived non-linear integral equations for determining
the vacuum eigenvalues of the chiral transfer-matrix which work both for the cigar and the parafermionic regimes.
We believe that this might be a good starting point for applying   the 
powerful fermionic methods  \cite{Boos:2006mq,Boos:2008rh,Jimbo:2008kn,Boos:2009fs,Jimbo:2010jv}
to the sausage/$O(3)$ NLSM.

%\item
In refs.\,\cite{Lukyanov:2010rn,Bazhanov:2013cua}, 
a conceptual explanation was given of how the ODE/IQFT correspondence for integrable conformal field theory
can be generalized to the massive IQFT. Following this route, we extended the ODE/IQFT correspondence from the cigar to the sausage NLSM.
With the correspondence one can uncover the basic integrable structures by studying the properties of the 
connection coefficients of the ordinary differential equations. The main result of this paper is the list of properties \ref{i1}-\ref{i10}
from sec.\,\ref{sec53} for the commuting families of operators in the sausage model which includes the quantum transfer-matrix.
The technical result that deserves to be mentioned is the system of NLIE which describes the vacuum eigenvalues 
of the commuting families of operators. Among other things, it allows one to calculate the $k$-vacuum energies of the sausage/$O(3)$ NLSM.

There are many results in the literature concerning the energy spectrum of the $O(3)$ sigma model in the sector with $
k=0$ \cite{Wiegmann:1984pw,Balog:2009ze,Gromov:2008gj,Balog:2003yr}. 
%In particular,
 In ref.\,\cite{Fateev:1992tk} a
 system of TBA equations was proposed which allows one to calculate the ground state energy  for $k=0$
and integer values of the dimensionless coupling $n\ge 3$ of the sausage model.
Recently  Ahn,  Balog and Ravanini   \cite{BalogN} transformed this system of TBA to a system of three non linear integral equations
which, it is affirmed, works for any real positive $n$.
Their main assumption is a periodicity condition for the $Q$-function given by eq.\,(3.16) from that paper. 
In our investigations, we did not find any trace
of a $Q$-function satisfying such a strong periodic condition. Nevertheless, the numerical results 
presented in fig.\,2 
from that
paper
seem to be in agreement with the data obtained from the 
solution of our NLIE \eqref{aatast},\,\eqref{hasstsa} with $k=0$ and $n=1$.
This situation needs to be clarified.

\smallskip
Let us briefly touch on  some  problems  which have  not been  discussed in the text
but are  directly related to the subject of this work.
We  did not make any mention of the  sausage model with the topological term
equal to $\pi$ which is also  expected to be an integrable  QFT \cite{Zamolodchikov:1992zr,Fateev:1992tk}.
Another closely related model is the so-called
3D sausage model introduced by Fateev in \cite{Fateev:1996ea}.
In the work  \cite{Lukyanov:2012zt}, classical integrability was established
for a four-parameter family of NLSM  with
torsion which includes the 3D sausage as a two parameter subfamily.  
We believe that extending the ODE/IQFT approach to these  models will be 
useful, both as a  step in the development of the method, and in terms of
applications. Some results  in this direction have already been obtained in ref.\,\cite{Bazhanov:2014joa}.
There are also the remarkable works \cite{Feigin:2015raa,Alfimov:2014qua,muxin} on 
toroidal algebras, which are deeply  connected to this field.

All the  models mentioned above   are based on  the  $\mathfrak{
sl}(2)$-algebra and its associated integrable structures.
Since the work  of Klimcik \cite{Klimcik:2014bta} 
there has been increasing interest in  ``deformed'' integrable NLSM associated with  higher rank Lie 
algebras \cite{Hoare:2014pna,Delduc:2014uaa,Sfetsos:2015nya, Klimcik:2016rov}.
The first principle quantization of such theories  seems to be a very interesting problem.
In the recent works \cite{Litvinov:2016mgi, Litvinov}, an important step in this direction was taken where
a one parameter deformation  was found of the set    
of ``circular brane'' local integrals of motion  introduced in ref.\,\cite{Lukyanov:2003rt}. This 
offers the possibility for the quantization of  the
 deformed $O(N)$ NLSM  along the lines of this work.

Perhaps the main  motivation for studying NLSM is based on the fact that
certain types  of SUSY sigma models
are at the heart of the celebrated AdS/CFT correspondence, and integrability
is an important possibility. In particular, the NLSM associated with the AdS side of
the correspondence for ${\cal N}=4$ SUSY Yang Mills theory was argued to
be integrable \cite{Bena:2003wd,Beisert:2010jr}.  As briefly discussed in the introduction, the study of the 
first principles quantization of the NLSM
by traditional techniques has proven to be difficult. A similar
situation exists with  sigma models on supergroups and superspaces, which
are expected to provide theoretical descriptions  of  condensed matter systems with disorder
\cite{ef}. That is where one is most tempted to try the power of the ODE/IQFT
approach.

 \section*{Acknowledgment}

S.\,L.'s gratitude goes to A.\,B.~Zamolodchikov for sharing his insights and support.

\smallskip
\noindent
The research of G.\,K. and  S.\,L. is supported by the NSF under grant number
NSF-PHY-1404056. 
%\cite{Beisert:2010jr}

%%%\newpage

 \appendix

 \section{Appendix \label{askjhdaA}}

  \begin{table}
 %[b]
\begin{center}
\begin{tabular}{| r | c|| r| c| }
\hline
& & &\\[-0.4cm]
$ n$  &  $ \tilde{t}_2$ & $n$ & $\tilde{t}_2$   \\
\hline \hline
1   & 0     & 6 &   $0.0658731$                          \\
2   & ${\sqrt{2}\over 48}$   & 7 & $0.0613178$ \\
3   & $0.0546105$ & 8 &   $0.0561029$ \\
4   & $0.0661040$  & 9 & $0.0509101$ \\
5   & $0.0683646$ & 10 &  $0.0460445$ \\
\hline\hline
\end{tabular}
\end{center}
\caption{ Numerical values of ${\tilde t}_2$
for $ {\mathfrak j}={\mathfrak m}=0$ (from ref.\,\cite{Lukyanov:2005nr}).}
\label{tab1}
\end{table}
To investigate the scaling behaviour of ${\cal T}^{(N)}(\mu)$  \eqref{jasasysa}-\eqref{asiissiaias},
 we conducted numerical work for integer $n$ 
  when the discretized operator is a  finite dimensional matrix that can be
   diagonalized  by means of  the Bethe ansatz (see Appendix \ref{p_b} for details).
 We focused only on the vacuum eigenvalue  in the sector
     ${\cal H}_{{\mathfrak j}-\frac{\mathfrak m}{2}}^{(N)}$
 and considered the cases with $n=2,\,3,\ldots, \,6$ and all admissible values of
$\mathfrak{j},\mathfrak{m}$ \eqref{siaisaiais}. 
Let $\tau^{(\rm vac)}(\lambda)$ be the vacuum eigenvalue of the chiral 
transfer-matrix in the parafermionic subspace  ${\cal V}_{{\mathfrak j}}^{(\mathfrak m)}$. 
We expect that it can be obtained from the vacuum eigenvalue of ${\cal T}^{(N)}(\mu)$
by using the formula \eqref{jashyas} which explicitly describes the scaling limit of the discretized operator. 
To estimate  numerical values of $\tau^{(\rm vac)}(\lambda)$ we used data obtained
for a set of finite $N$ and then performed a certain interpolation procedure to $N = \infty$.
The results were compared with predictions coming from  the properties of   $\tau^{(\rm vac)}(\lambda)$
 discussed in sec.\,\ref{sec22}, 
 specialized to the values $p_1=\frac{\ri}{2}\,\mathfrak{m}$ and $p_2=\mathfrak{j}+\half$. Agreement was found in all cases considered. In this appendix, some of our numerical work is presented.
\bigskip

 \begin{table}
\centering
\begin{tabular}{|c||ccccc||c|}
\hline
 & & & & & &  \\[-0.5cm]
root $\#$& $N = 501$ & $N = 1001$ & $N = 1500$ & $N=2600$ & $N=\infty$\, & $2u_l^{\frac{n}{2}}$\\
\hline
$1$ & $0.4818860$ & $0.4818829$ &  $0.4818820$ & $0.4818814$ & $0.4818809$ &$0.47349$\\
$2$ & $1.4891566$ & $1.4891424$ &  $1.4891392$ & $1.4891372$ & $1.4891359$&$1.48725$ \\
$3$ & $2.4919329$ & $2.4918863$ &  $2.4918769$ & $2.4918715$ & $2.491868$&$2.49093$ \\
$4$ & $3.4935044$ & $3.4933890$ &  $3.4933666$ & $3.4933541$ & $3.493348$&$3.49276$ \\
$5$ & $4.4946127$ & $4.4943769$ &  $4.4943321$ & $4.4943074$ & $4.494295$&$4.49387$ \\
$6$ & $5.4955294$ & $5.4951073$ &  $5.4950277$ & $5.4949844$ & $5.494962$&$5.49464$ \\
$7$ & $6.4963870$ & $6.4956974$ &  $6.4955682$ & $6.4954981$ & $6.495463$&$6.49521$ \\
$8$ & $7.4972634$ & $7.4962107$ &   $7.4960140$ & $7.4959077$ & $7.495854$ &$7.49564$\\
$9$ & $8.4982121$ & $8.4966857$ &  $8.4964010$ & $8.4962476$ & $8.496171$&$8.49599$ \\
$10$&$9.4992734$ & $9.4971480$ &  $9.4967523$ & $9.4965392$ & $9.496432$&$9.49628$ \\
$11$ &$10.500481$ & $10.497616$ &  $  10.497084$ & $10.496797$ & $10.49665$&$10.49652$ \\
$12$ &$11.501864$ & $11.498105$ &  $  11.497406$ & $11.497031$ & $11.49684$&$11.49672$ \\
$13$ &$12.503448$ & $12.498625$ &  $  12.497729$ & $12.497248$ & $12.49701$ &$12.49690$\\
$14$ &$13.505258$ & $13.499187$ &  $  13.498060$ & $13.497455$ & $13.49715$&$13.49705$ \\
$15$& $14.507318$ & $14.499799$ &  $  14.498404$ & $14.497655$ & $14.49728$&$14.49719$ \\
%$16$& $15.5096504$ & $15.5004689$ &  $15.4987656$ & $15.497852$ & $  15.4973959$&$15.49731$ \\
\hline
\end{tabular}
\caption{Numerical values of $\frac{2N}{\pi}\big[\mu_l^{(N)}\big]^n$, where
%$2\,\big(u^{(N)}_l\big)^{\frac{n}{2}}$, with $u^{(N)}_l=\big(\frac{N}{\pi}\big)^{\frac{2}{n}}\,\big[\mu^{(N)}_l\big]^2$.
 $\mu^{(N)}_l>0$ are the roots  of the 
vacuum eigenvalue  of the discretized operator  $\mathcal{T}^{(N)}(\mu)$ for
  $n=4$, $\mathfrak{j}=\mathfrak{m}=0$.  The column ``$N=\infty$''  was obtained by interpolating the results for finite $N$.
  The entries in the last column were calculated by using  the
asymptotic formula \eqref{safwrwa} truncated at the first non-zero 
term in the series.}\label{tab2}
\end{table}
Let $\{u_l\}^\infty_{l=1}$ be the set of zeroes of $\tau^{(\rm vac)}(\lambda)$
  considered as a function of $\lambda^2$. From the  numerical data it was found that all the zeroes are simple, real, positive, and accumulate towards $\lambda^2\,=\,\infty$ with the leading asymptotic behaviour
\bea
\quad u_l\sim \, \big({\textstyle \frac{1}{2}}\big)^{\frac{2}{n}}\times
\begin{cases}\label{rtcase}
 \big(l-\half\,\big)^{\frac{2}{n}}\ \ \ \ &{\rm for}\ \ \ \ 0\leq 2{\mathfrak j}<\frac{n}{2}\\
 \big(l-\half +\frac{n}{n+2}\,\big)^{\frac{2}{n}}\ \ \ \ &{\rm for}\ \ \ \ 2{\mathfrak j}
 =\frac{n}{2}\ \ \;\;\;\; (n-{\rm even})
\end{cases}
\eea
For $0\leq 2{\mathfrak j}<\frac{n}{2}$, this is consistent with the asymptotically exact
formula,
\bea\label{safwrwa}
 u_l^{\frac{n}{2}}+{\textstyle \frac{1}{2\pi}}\  \sum_{m=1}^\infty{\tilde  {g}}_{m}\big({\textstyle\frac{\ri}{2}}\, {\mathfrak m}, 
 {\mathfrak j}+\half \big)\ 
\sin\big({\textstyle\frac{2\pi  m}{n+2}}\big)\ u_l^{-\frac{n m}{n +2}}\asymp
{\textstyle\frac{1}{2}}\,\big(l-\half\,\big)\ ,
\eea
which can be easily derived from eqs.\,\eqref{tseriega}-\eqref{ieurweiukj}. Knowledge of the coefficients  $\tilde{g}_m$ allows us to compute systematic corrections to the leading asymptotic behaviour \eqref{rtcase}. As it follows from eq.\,\eqref{jsasuausa}, the first coefficient is
\begin{equation}\label{uierert}
{\tilde  g}_{1}(p_1,p_2)=\frac{{\tilde {t}}_{1}(p_1,p_2)}{2\, \cos\big({\textstyle \frac{2\pi p_2 }{n+2}}\big)}\,, 
\end{equation}
 with $\tilde {t}_{1}(p_1,p_2)$ -- vacuum eigenvalue 
of  
 $\tilde {\mathfrak t}_{1}$ -- given by eq. \eqref{asuausai}. Notice that for $p_2=\mathfrak{j}+\half=\frac{n+2}{4}$ ($n$-even), the denominator in \eqref{uierert} is zero so that \eqref{safwrwa} is no longer valid. Also when $\,\mathfrak{j}=\mathfrak{m}\,=\,0$, $\tilde{g}_1$ vanishes, but for this case the second term in the sum in \eqref{safwrwa} is known, since
\begin{equation}
{\tilde g}_2(0,{\textstyle \frac{1}{2}})\,=\,\frac{ {\tilde  t}_2(0,\half)}{2\, \cos\big({\textstyle \frac{\pi}{n+2}}\big)}\ \nonumber
\end{equation}
 and numerical values of ${\tilde  t}_2(0,\half)$ were calculated  in ref.\,\cite{Lukyanov:2005nr} and are reproduced in tab.\,\ref{tab1}.
 Truncating 
 the  series in \eqref{safwrwa} at the first non vanishing term, we calculated the  corrections to the leading asymptotic \eqref{rtcase}.
% From \eqref{jashyas}, 
This was compared to the zeroes of the 
vacuum eigenvalue  of  $\mathcal{T}^{(N)}(\mu)$ for increasing $N$. In all cases good agreement was observed. As an example,
in tab.\,\ref{tab2} the results for $n=4$, $\mathfrak{j}=\mathfrak{m}=0$ are shown.

\bigskip
 
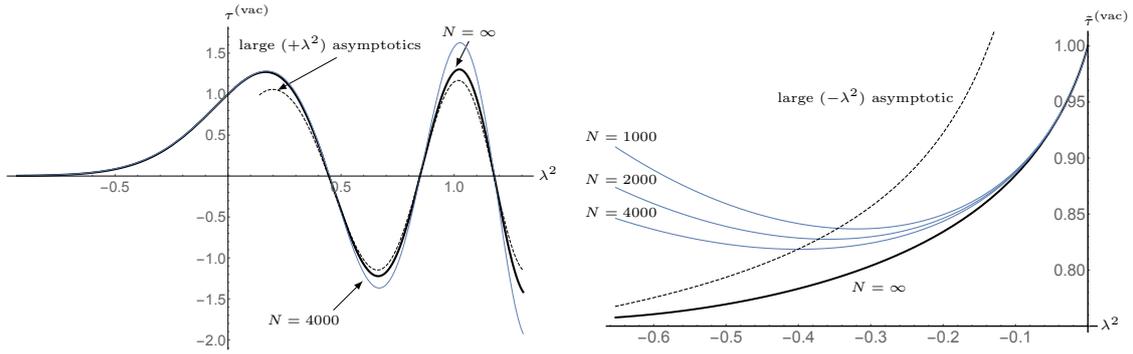
\begin{figure}
\scalebox{0.87}{
\begin{tikzpicture}
\node at (12.15,2.6) {\tiny$\tilde{\tau}^{(\rm vac)}$};
\node at (-0.85,2.7) {\tiny$\tau^{(\rm vac)}$};
\node at (12.2,-2.05) {\tiny$\lambda^2$};
\node at (3.7,0.24) {\tiny$\lambda^2$};
\node at (4.8,0.8) {\tiny $N=1000$};
\node at (4.8,0.15) {\tiny $N=2000$};
\node at (4.8,-0.35) {\tiny $N=4000$};
\node at (8.7,-1.5) {\tiny $N=\infty$};
\node at (8.5,1.4) {\tiny large $(-\lambda^2)$ asymptotic};
\node at (2.5,2.4) {\tiny $N=\infty$};
\draw [->] (2.5,2.25) -- (2.3,1.85);
\node at (0,-2.) {\tiny $N=4000$};
\draw [->] (0.45,-1.85) -- (0.9,-1.4);
\node at (0.4,2.2) {\tiny large $(+\lambda^2)$ asymptotics};
\draw [->] (0.5,2) -- (-0.4,1.5);
\node at (8.3,0) {\includegraphics[width=7.48cm]{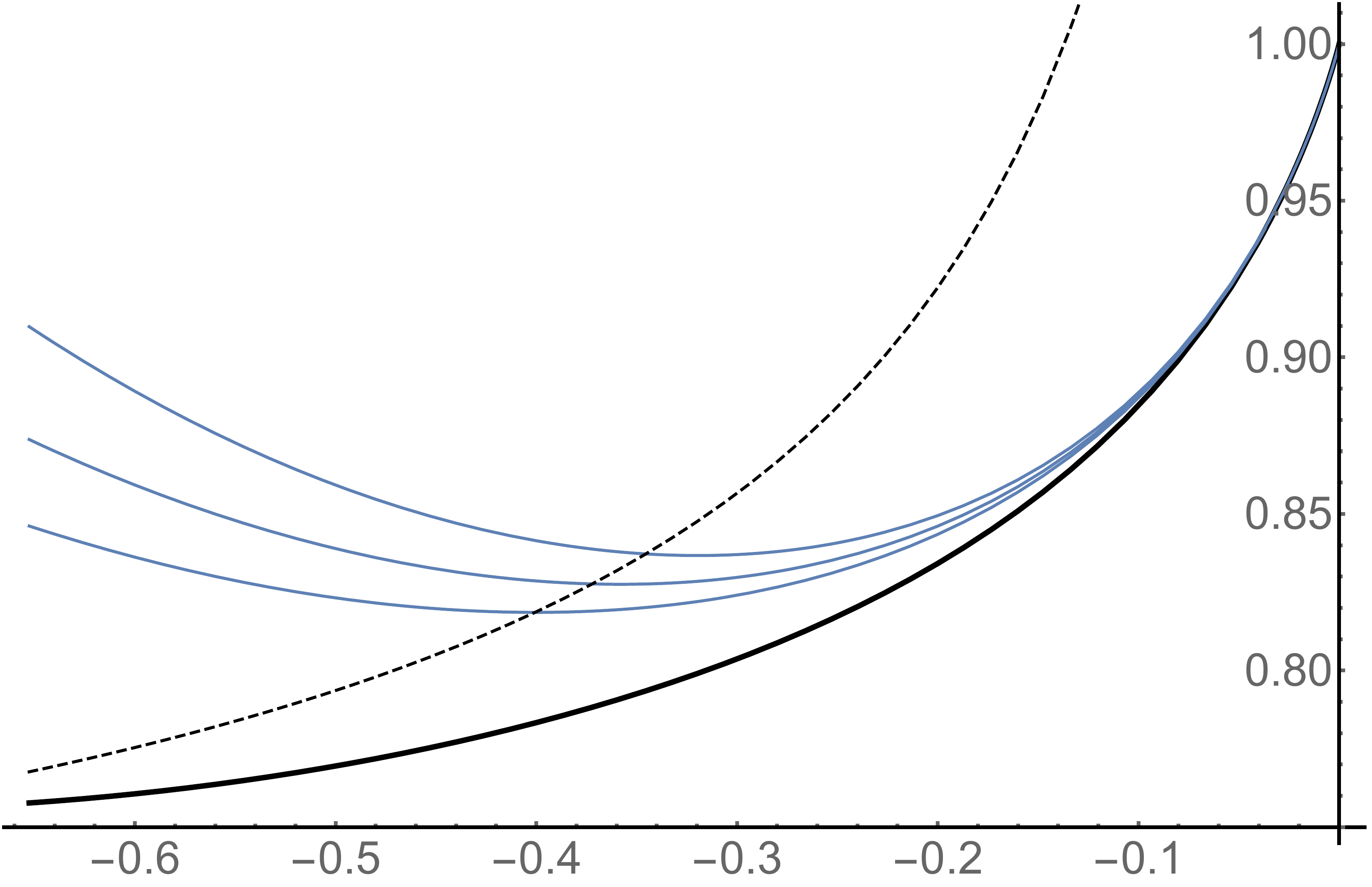}};
\node at (-0.5,0) {\includegraphics[width=7.99cm]{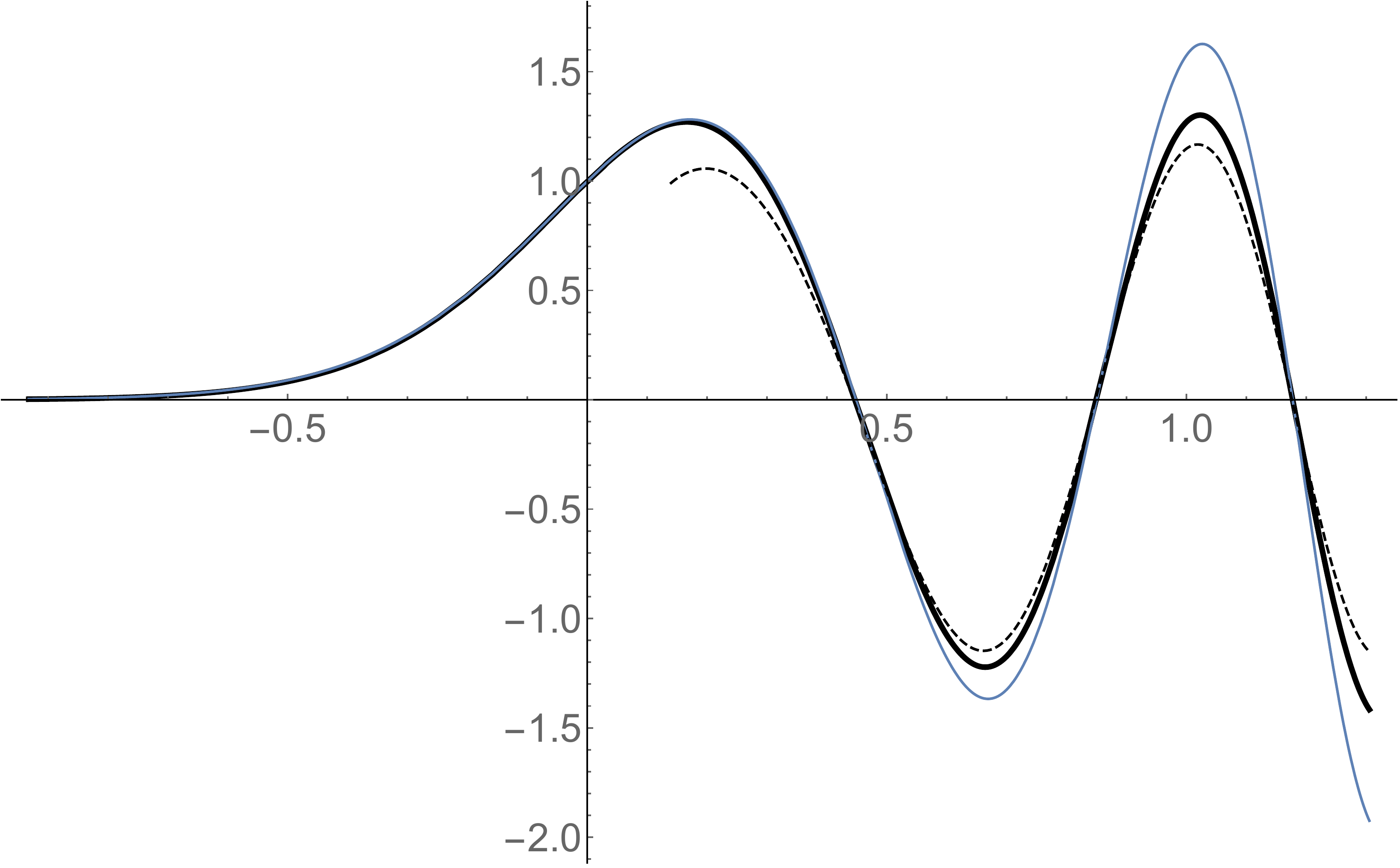}};
\end{tikzpicture}
}
\caption{On the left panel, a plot of $\tau^{(\rm vac)}$ for $n=3$, $2\mathfrak{j}=\mathfrak{m} = 1$  compared to its
large $(+\lambda^2)$ asymptotic following from eq.\,\eqref{tseriega}. On the right panel, $\tilde{\tau}^{(\rm vac)}\,=\,\tau^{(\rm vac)}\exp\big(2\pi\, (-\lambda^2)^{\frac{3}{2}}\big)$ is plotted and compared with the large $(-\lambda^2)$ asymptotic derived from eqs.\,\eqref{asasisao},\,\eqref{jasusau}.
 The scaling function was numerically estimated by interpolating to $N=\infty$  the data for
  $N=500,1000,2000,4000$. }
\label{figA1}
\end{figure}

As $\lambda^2\to-\infty$, the asymptotic behaviour of   $\tau^{(\rm vac)}$
is dictated by eqs.\,\eqref{asasisao},\,\eqref{jasusau}. 
Truncating the  sum in \eqref{jasusau}  at the first non-zero term and   substituting
${\tilde {\mathfrak t}}_j$  by its vacuum eigenvalue, we  compared  this to
the results of the $N=\infty$  interpolation.
The agreement was good considering that the interpolation procedure becomes rapidly less efficient for increasing
values of  $(-\lambda^2)$. Fig.\,\ref{figA1} shows a plot of the estimated scaling function 
versus the asymptotics for $n=3$ and $2\mathfrak{j}=\mathfrak{m}=1$.

\bigskip

 Another check that can be made is to consider the  Taylor expansion of $\tau^{(\rm vac)}(\lambda)$ at zero
 following from   formulae \eqref{saosapassap} and \eqref{jssusau}. The coefficient 
 $t_1(p_1,p_2)$  $(p_1={\textstyle\frac{\ri}{2}}\, {\mathfrak m}, 
p_2= {\mathfrak j}+\half )$
 can be compared to the corresponding term in  the vacuum eigenvalue  of the discretized operator:
\begin{equation}
\mathcal{T}^{(N,{\rm vac})}(\mu)=2\cos\big(\textstyle{\frac{\mathfrak{m}\pi}{n}}\big)+t_1^{(N)}\,\mu^2+O(\mu^4) \,.\nonumber
\end{equation}
Note that ${t}_1^{(N)}$
 is a divergent quantity for large $N$ and must be regularized.  According to eq.\,\eqref{jashyas}, for $n>2$,  the following limit
 exists and converges to $t_1$:
\begin{equation}\label{fdsfda}
t_1\big({\textstyle\frac{\ri}{2}}\, {\mathfrak m}, {\mathfrak j}+\half \big)=\lim_{N\to\infty}\,{t}_1^{(N, {\rm reg})}\ ,\ \  \  \  \ \ \ 
 {t}_1^{(N, {\rm reg})}=\big({\textstyle\frac{\pi}{N}}\big)^{\frac{2}{n}}\, \Big(\,{t}_1^{(N)}+
2N\ {\textstyle \frac{\cos(\frac{\mathfrak{m}\pi}{n})}{\cos(\frac{\pi}{n})}}\,\Big)\  .
\end{equation}
We compared the value of $t_1\big({\textstyle\frac{\ri}{2}}\, {\mathfrak m}, {\mathfrak j}+\half \big)$ given by eq.\,\eqref{jssusau} to the
numerical values of 
${t}_1^{(N, {\rm reg})}$ and found good agreement for $n=3,4\ldots,6$ and all the allowed values of $\mathfrak{j},\mathfrak{m}$.
A few cases are presented in tab.\,\ref{tabA2}.

Finally, let us mention that for $n=2$,
 analytic expressions exist for both $\tau^{\rm (vac)}$ and the vacuum eigenvalue of $\mathcal{T}^{(N)}$. In the case $\mathfrak{j}=\mathfrak{m}=0$, 
\begin{equation}
 \mathcal{T}^{(N,{\rm vac})}(\mu)=2\, \prod\limits_{m=1}^N\Big(1-\mu^2\cot\big({\textstyle\frac{\pi}{2N}\, (m-\half)}\big)\Big)\ ,\nonumber
\end{equation}
and using the formula \eqref{jashyas}, the scaling limit can be taken explicitly to yield
\begin{equation}
\tau^{\rm (vac)}(\lambda)=\Big( \frac{\re}{2}\Big)^{2\lambda^2}\,\frac{2\sqrt{\pi}}{\Gamma\big(\frac{1}{2}-
2 \lambda^2\big)}\, .\nonumber
\end{equation}
It is easy to verify that  this is consistent with the properties of the chiral transfer-matrix discussed in sec.\,\ref{sec22}.
For $n=2$ and  $2\mathfrak{j}=\mathfrak{m}=1$, the discretized operator turns out to be zero for any $N$ and hence, $\tau(\lambda)=0$.

\begin{table}
%[b]
\centering
\begin{tabular}{|c|c|c|c|c|c||c|}
\hline
$n=6$ & $N=100$ & $N=200$ & $N=400$ & $N=800$ & $N=\infty$ & eq.\,\eqref{jssusau} \\
\hline
$2\mathfrak{j}=\mathfrak{m}=0$& $0.54474$ &  $0.54519$ & $0.54542$ & $0.54553$ & $0.5456440$ & $0.5456445$ \\
\hline
$2\mathfrak{j}=\mathfrak{m}=2$ & $0.43807$ & $0.44710$ & $0.45357$ & $0.45818$ & $0.469649$ & $0.469446$ \\ \hline
\hline
$n=5$ & $N=100$ & $N=200$ & $N=400$ & $N=750$ & $N=\infty$ & eq.\,\eqref{jssusau}\\
\hline
$2\mathfrak{j}=\mathfrak{m}=0$ & $0.86236$ & $0.86271$ & $0.86287$ & $0.86294$ & $0.8630048$ & $0.8630049$ \\ \hline
$2\mathfrak{j}=\mathfrak{m}=2$& $0.40173$ &  $0.40751$ & $0.41144$ & $0.41390$ & $0.419808$ & $0.419632$ \\
\hline
\end{tabular}
\caption{The regularized  value ${t}_1^{(N, {\rm reg})}$ \eqref{fdsfda}
for a variety of cases and increasing $N$ compared 
to the expression for $t_1\big({\textstyle\frac{\ri}{2}}\, {\mathfrak m}, {\mathfrak j}+\half \big)$ given by eq.\,\eqref{jssusau}. The column ``$N=\infty$'' was
obtained by interpolation.}\label{tabA2}
\end{table}

%\newpage

  \section{Appendix\label{p_b} }

In this appendix we will consider the vacuum eigenvalue of the matrices $\mathcal{Z}_\pm(\mu)$ in the space
$\mathcal{H}^{(N)}_{\mathfrak{j}-\frac{\mathfrak{m}}{2}}$.
Recall that $\mathcal{H}^{(N)}_{\mathfrak{j}-\frac{\mathfrak{m}}{2}}$ denotes the eigenspace  of the matrix  
$\pi_{{\cal H}^{(N)}}({\tt Z})$ \eqref{jasasasuy},\,\eqref{jasasauyas}
having eigenvalue 
$\omega^{{\mathfrak j}-\frac{\mathfrak m}{2}}$, where $\mathfrak{j}$ and $\mathfrak{m}$ are restricted as in \eqref{siaisaiais}.
Our considerations are entirely based on the  properties of $\mathcal{Z}_\pm(\mu)$ \ref{reli}-\ref{relv} listed in sec.\,\ref{sec4.1}.
\bigskip

Let $\mathcal{Z}^{(\psi)}_\pm(\mu)$  be the eigenvalue corresponding to a common eigenvector $|\psi\rangle$ of the commuting family $\mathcal{Z}_\pm(\mu)$.
Using the analytical conditions \ref{reliv} and $\mu\to-\mu$ symmetry \ref{relv}, it can be written in the form,
\bea\label{nvfxcvbv}
 \mathcal{Z}^{(\psi)}_\pm(\mu)&=&B^{(N,\psi)}\ \mu^\mathfrak{m}\ \prod\limits_{i=1}^{(n-1)N-2\mathfrak{j}-\mathfrak{m}}\,\Big(1\mp \frac{\mu}{\mu_i}\Big) \qquad 
 (n - {\rm odd})
\eea
and
\bea
\label{nvfxcvbv2}
 \cal{Z}^{(\psi)}_+(\mu)&=&B^{(N,\psi)}\ \mu^\mathfrak{m}\ \prod\limits_{i=1}^{\frac{nN}{2}- \mathfrak{j}-\frac{\mathfrak{m}}{2}}\,
\bigg(1-\frac{\mu^2}{v_i}\bigg)\nonumber\\
&&\hspace{6.4cm} ( n- {\rm \ even})\\[-0.2cm]
 %\\[0.2cm]
\mathcal{Z}^{(\psi)}_-(\mu)&=&B^{(N,\psi)}\ \mu^\mathfrak{m}\ \prod\limits_{i=1}^{\frac{(n-2)N}{2}-\mathfrak{j}-\frac{\mathfrak{m}}{2}}\,\bigg(1-\frac{\mu^2}{w_i}\bigg)\, \nonumber
\eea
From the $T-Q$ type relations  \ref{reliii}, it follows that the overall coefficient $B^{(N,\psi)}$ (depending on the state $|\psi\rangle$)
is the same for both $\mathcal{Z}_+^{(\psi)}$ and $\mathcal{Z}_-^{(\psi)}$.
Another consequence of this relation is that
the  roots
satisfy the following Bethe ansatz equations:
\begin{equation}\label{ksdjfh1}
 \prod\limits_{i=1}^{(n-1)N-2\mathfrak{j}-\mathfrak{m}}\frac{\mu_i+q^{-1}\,\mu_l}{\mu_i+q^{+1}\,\mu_l}\,=\,-q^{2\mathfrak{m}}\,
\bigg(\frac{1-q^{+\frac{1}{2}}\,\mu_l}{1-q^{-\frac{1}{2}}\,\mu_l}\bigg)^{2N}\hspace{1cm} (n - {\rm odd})
\end{equation}
and
\begin{eqnarray}\label{ksdjfh2}
&&\prod\limits_{i=1}^{\frac{nN}{2}- \mathfrak{j}-\frac{\mathfrak{m}}{2}}\frac{v_i-q^{-2}\,w_l}{v_i-q^{+2}\,w_l}\,=\,-q^{2\mathfrak{m}}\nonumber\\[0.2cm]
&& \hspace{9cm} ( n - {\rm even})\\[-0.2cm]
&&\!\!\!\prod\limits_{i=1}^{\frac{(n-2)N}{2}-\mathfrak{j}-\frac{\mathfrak{m}}{2}}\frac{w_i-q^{-2}\,v_l}{w_i-q^{+2}\,v_l}
\,=\,-q^{2\mathfrak{m}}\,\bigg(\frac{1-q^{+1}\,v_l}{1-q^{-1}\,v_l}\bigg)^{2N}\nonumber
\end{eqnarray}
Similar equations
for the Fateev-Zamolodchikov spin chain \eqref{sisisai} with periodic boundary
conditions were   previously derived
in the works  \cite{Albertini} and \cite{Ray}  for odd and even $n$, respectively.
Notice that the constant $B^{(N,\psi)}$  in  \eqref{nvfxcvbv},\,\eqref{nvfxcvbv2} is  determined (up to an overall sign) by  the quantum Wronskian type relations \ref{relii}.

The Bethe ansatz equations are valid for all integer $n\ge 2$ and $\mathfrak{j}, \mathfrak{m}$ restricted to \eqref{siaisaiais}, except for 
$2\mathfrak{j}=\mathfrak{m}=\frac{n}{2}$ ($n$ even) which requires special attention. 
In this case, for certain sectors of $\mathcal{H}^{(N)}_{0}$ a significant simplification occurs; $\mathcal{Z}_-^{(\psi)}$ vanishes so that
 the $T-Q$  type relations \ref{reliii} become trivial and the quantum Wronskian  type relations \ref{relii} can be used to obtain
much simpler equations for the roots. For instance, for the vacuum eigenvalue, 
$\mathcal{Z}_-^{(\rm vac)}(\mu)=0$ and
$\mathcal{Z}_+^{({\rm vac})}$ is given explicitly by
\begin{equation}\label{faslkdere}
{\mathcal Z}_+^{(\rm vac)}(\mu)=2\, \sqrt{N}\ \mu^{\frac{n}{2}}\,\prod\limits_{l=1}^{N-1}\Big(1+\mu^n\,\cot\big({\textstyle\frac{\pi l}{2N}}\big)\Big)\,, 
 \hspace{0.5cm}\ \ \big(\, 2\mathfrak{j}=\mathfrak{m}={\textstyle\frac{n}{2}}\,,\ 
\ \ n-{\rm  \ even}\,\big)\, .
\end{equation}
Recall that the vacuum is defined as the lowest energy state
 of the Fateev-Zamolodchikov spin chain Hamiltonian \eqref{sisisai}, \eqref{sisisai2}, 
 which commutes
with  both $\mathcal{Z}_+(\mu)$ and $\mathcal{Z}_-(\mu)$ for any $\mu$.

\begin{figure}
\centering
\scalebox{0.92}{
\begin{tikzpicture}
\node at (0,0) {\includegraphics[width=7.65cm]{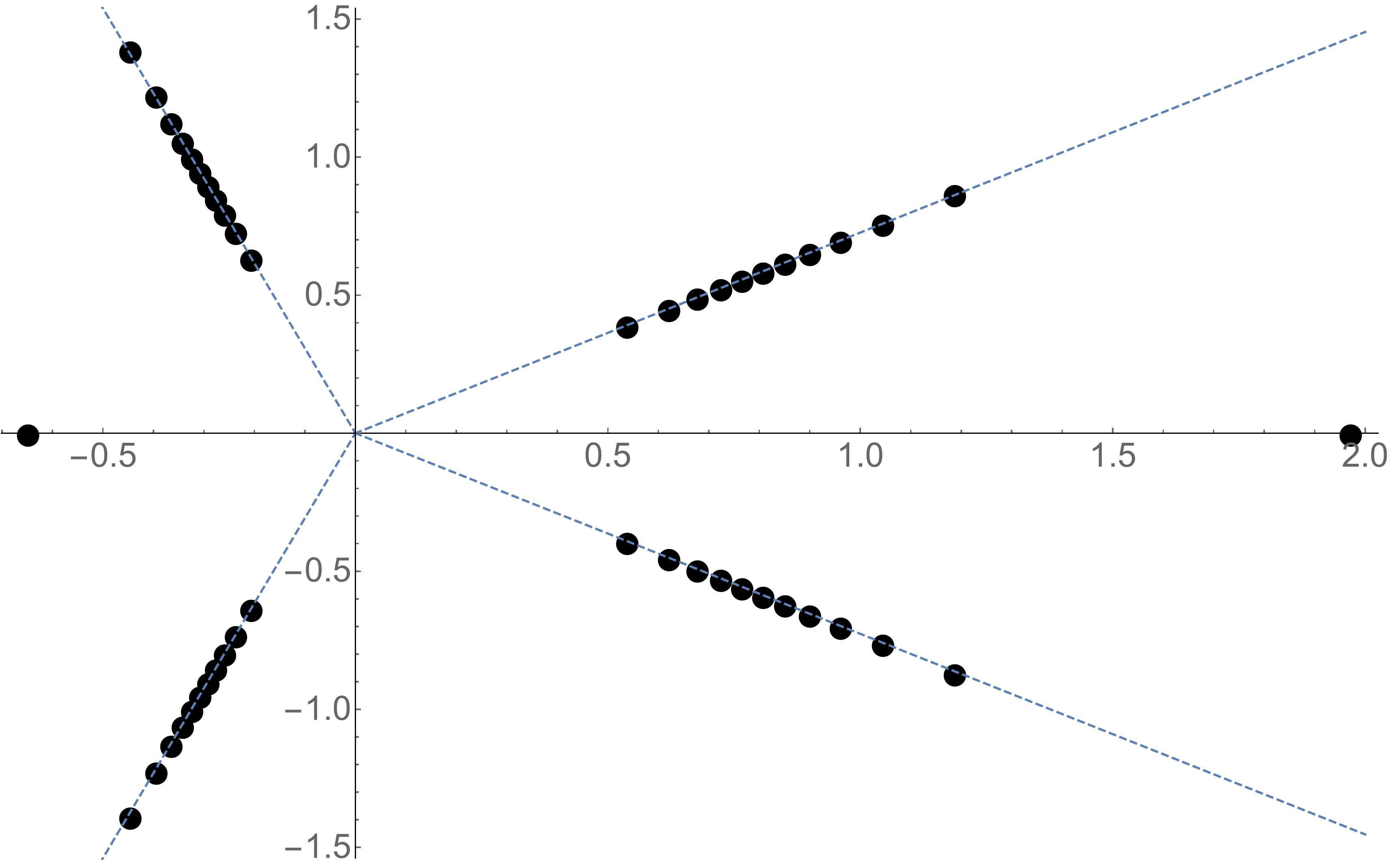}};
\node at (1,2) {\small$\mu$};
\draw (1,2.03) circle (0.3cm);
%\node at (12,1) {\small$\mu^2$};
%\draw (12,1.03) circle (0.3cm);
\node at (10.5,2) {\small$\mu^2$};
\draw (10.5,2.03) circle (0.3cm);
\node at (8.5,0) {\includegraphics[width=7.65cm]{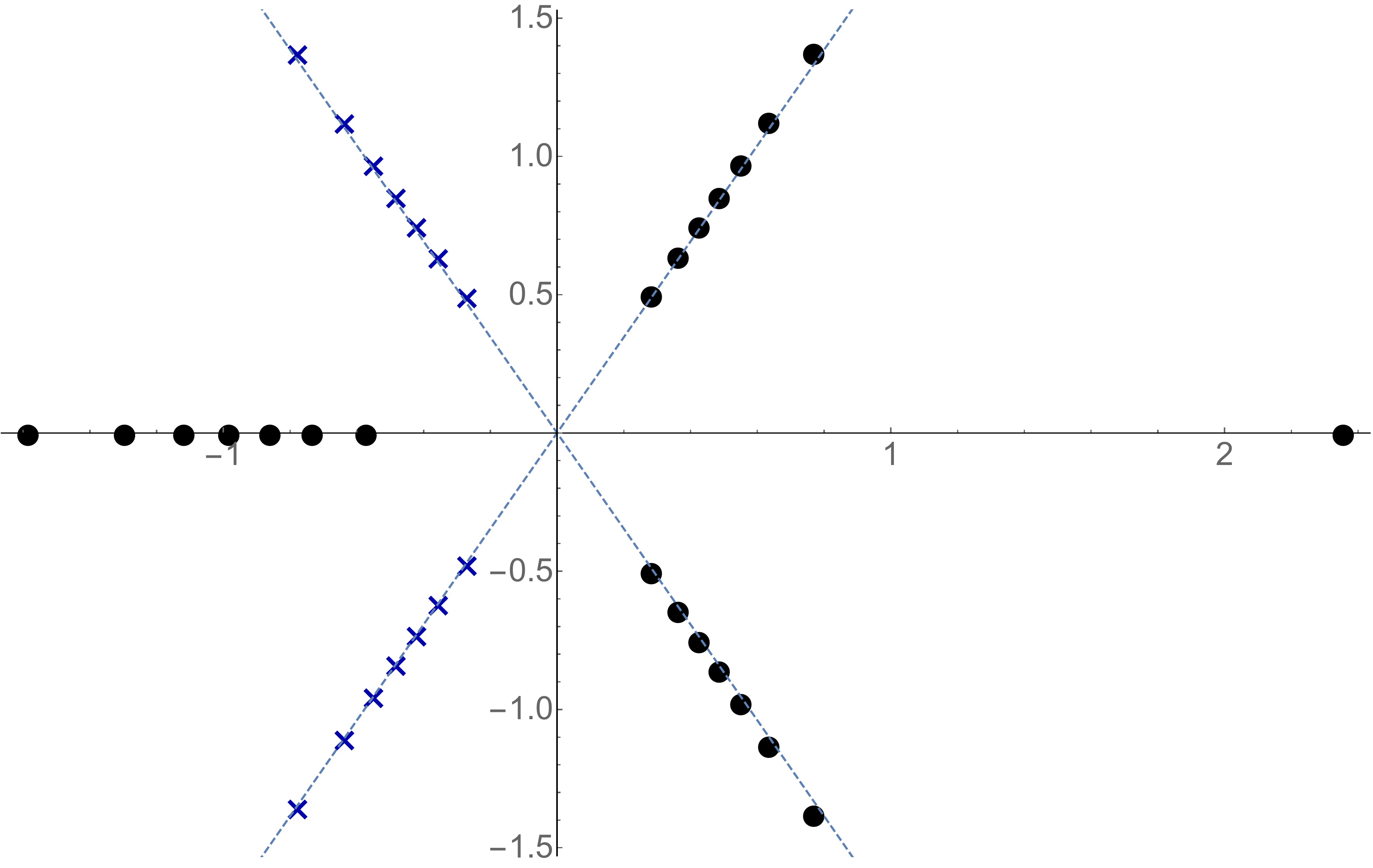}};
\end{tikzpicture}
}
\caption{On the left panel, the roots of  ${\cal Z}^{\rm (vac)}_+(\mu)$ are depicted  in the complex plane for $n=5$, $\mathfrak{j}=1, \ \mathfrak{m}=0$ and
$N=12$. On the right panel, the roots of $\mathcal{Z}_+^{(\rm vac)}$ (circles) and  $\mathcal{Z}_-^{(\rm vac)}$ (crosses) as functions of $\mu^2$ are shown
for  $n=6$, $2\mathfrak{j}=3,\  \mathfrak{m}=1$ and
$N=8$. 
%The dashed lines are the rays on which the roots accumulate, given by eqs.\,\eqref{fshgfdq1}-\eqref{fshgfdq2}.
\label{figB1}}

\end{figure}
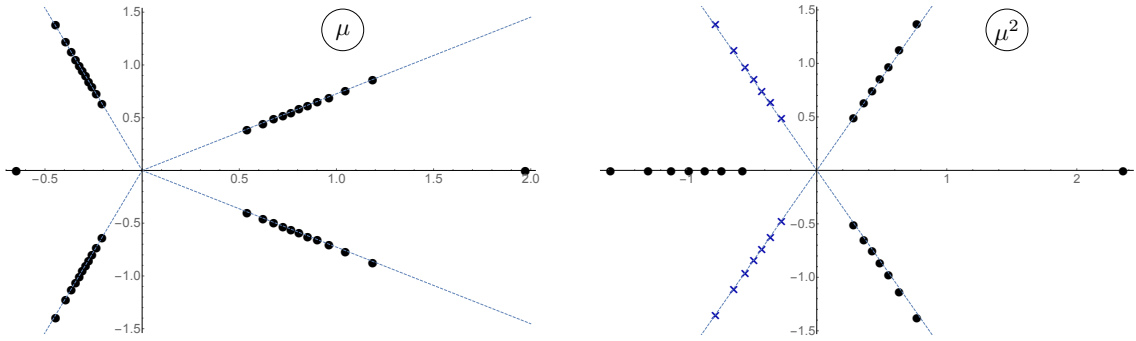

\bigskip
We studied the solutions to the Bethe ansatz equations corresponding to the low energy states $|\,\psi\,\rangle$  of the Fateev-Zamolodchikov spin chain.
It was found that the roots accumulate along the rays given by (see fig.\,\ref{figB1})
\vspace{0.1cm}
\begin{equation}
\arraycolsep=0.8cm
\begin{array}{lll}
\arg(\mu)\;\hspace{0.1em}=\pm{\textstyle \frac{\pi}{n}}\, p\,,& p=1,3,\ldots,n-2 & (\mu_i-{\rm roots}) \\[0.4cm]
\arg(\mu^2)={\textstyle \frac{2\pi}{n}}p\,,&  p=1,3,\ldots,\,n-1
&(v_i - {\rm roots})\\[0.4cm]
\arg(\mu^2)= {\textstyle \frac{2\pi}{n}}\,p\,,& p=2,4,\ldots,n-2  & (w_i - {\rm roots})\, \nonumber
\end{array}
\end{equation}
\vspace{0.1cm}

\noindent
\noindent
In the scaling limit most of the roots become densely packed along the rays. However we observed that at the edges of the distribution,
 the roots exhibit a certain scaling behaviour. 
In particular, at the edge next to zero  of the locus labeled by the integer $p$,  with index $i$ enumerating the roots ordered by increasing absolute value,
 the following limits exist
\begin{equation}
\arraycolsep=0.8cm
\begin{array}{ccc}
\lim\limits_{N\to \infty\atop i-{\rm fixed}}N^\frac{1}{n}\,\mu_{i,p}^{(N,\psi)}\, ,
 & \lim\limits_{N\to \infty\atop i-{\rm fixed}}N^\frac{2}{n}\,v_{i,p}^{(N,\psi)}\, , 
&\lim\limits_{N\to \infty\atop i-{\rm fixed}}N^\frac{2}{n}\,w_{i,p}^{(N,\psi)}\, .
\end{array}\nonumber
\end{equation}
Here we temporarily exhibit the dependence of the roots on $N$ and the state $|\,\psi\,\rangle$.
Also, the scaling limit can be defined for the coefficient $B^{(N,\psi)}$ in formulae\ \eqref{nvfxcvbv}, \eqref{nvfxcvbv2}:
\begin{equation}\label{sdkjfhasfd}
B^{(\psi)}={\rm s}\!\!\!\lim\limits_{N\rightarrow\infty}\, (\pi/N)^{\frac{\mathfrak{m}}{n}}\,B^{(N,\psi)}\ .
%\hspace{1.9cm} 
%\big(s=\mathfrak{j}-\textstyle{\frac{1}{2}}\,\mathfrak{m}\, \big)\, .
\end{equation}

Keeping $N$ finite, consider the logarithm of the r.h.s of eqs.\,\eqref{ZOD} and \eqref{ZEV} for a given eigenvalue.
% using the form $\mathcal{Z}_\pm^{(\psi)}$ \eqref{nvfxcvbv}, \eqref{nvfxcvbv2}.
%and keeping $N$ finite.
% Take the logarithm and then expand at $\lambda=0$.
With $\mathcal{Z}_\pm^{(\psi)}$ of the form \eqref{nvfxcvbv}, \eqref{nvfxcvbv2} it
 is straightforward to find their Taylor series at $\lambda=0$. In the case of odd $n$, the expansion coefficients are given by
\begin{eqnarray}\label{coefkjsdfh1}
M^{(N)}_{m}&=&\frac{1}{m}\  \Big(\frac{\pi}{N}\Big)^{\frac{m}{n}}\ \bigg(\ \sum\limits_i\,  \mu_i^{-m}\ +\ \frac{(-1)^m\ N}{\cos\big(\frac{\pi m}{2n}\big)}\ \bigg)\,
\ \, \hspace{0.8cm}(m<n)\nonumber\\[0.3cm]
M^{(N)}_n&=&\frac{\pi }{n N}\ \sum\limits_i \,\mu_i^{-n}\ 
+\ \frac{2}{ n}\ (n-1)\ \log\Big( {\frac{N\re}{\pi}}\Big) \\[0.3cm]
M^{(N)}_m&=&\frac{1}{m}\  \Big(\frac{\pi}{N}\Big)^{\frac{m}{n}}\  
 \sum\limits_i \, \mu_i^{-m}\, \hspace{4.3cm} (m>n)\   \nonumber
\end{eqnarray}
For even $n$,
\begin{eqnarray}\label{coefkjsdfh2}
&&\hspace{-1.2cm}\!\!\!\arraycolsep=0.1cm\begin{array}{ll}V^{(N)}_{m}\,=\,\dfrac{1}{m}\ \Big(\dfrac{\pi}{N}\Big)^{\frac{2m}{n}}\ \
 {\displaystyle\sum\limits_i}\, v_i^{-m}\,
 &
\end{array}\nonumber\hspace{6.55cm}\, \,\,\big(2m<n\big)  \\[0.5cm]
&&\hspace{-1.2cm}\!\!\!\arraycolsep=0.1cm\begin{array}{ll}
W^{(N)}_{m}\,=\,\dfrac{1}{m}\ \Big(\dfrac{\pi}{N}\Big)^{\frac{2m}{n}}\ \bigg(\,{\displaystyle\sum\limits_i}\, w_i^{-m}+
\dfrac{N}{\cos\big(\frac{\pi m}{n}\big)}\,\bigg)\, &
\end{array}
\nonumber\hspace{3.9cm} \,\,\big(2m<n\big)  \\[0.5cm]
&&\hspace{-1.2cm}\!\!\!\arraycolsep=0.2cm\begin{array}{ll}
V^{(N)}_{\frac{n}{2}}\,=\,\dfrac{2\pi}{nN}\ {\displaystyle\sum\limits_i}\, v_i^{-\frac{n}{2}}\,+\,2\log\Big(\dfrac{N\re}{\pi}\Big)\,&
\end{array}\\[0.5cm]
&&\hspace{-1.2cm}\!\!\!\arraycolsep=0.2cm\begin{array}{ll}
W^{(N)}_{\frac{n}{2}}\,=\,\dfrac{2\pi}{nN}\ {\displaystyle\sum\limits_i}\, w_i^{-\frac{n}{2}} \,-\,\dfrac{2}{n}\ (n-2)\ \log\Big(\dfrac{N\re}{\pi}\Big) &
\end{array}\nonumber\\[0.5cm]
&&\hspace{-1.2cm}\arraycolsep=0.1cm\begin{array}{ll}
V^{(N)}_{m}\,=\,\dfrac{1}{m}\ \Big(\dfrac{\pi}{N}\Big)^{\frac{2m}{n}}\ {\displaystyle\sum\limits_i} \,v_i^{-m}\, , &
\hspace{0.8cm} W^{(N)}_{m}\,=\,\dfrac{1}{m}\ \Big(\dfrac{\pi}{N}\Big)^{\frac{2m}{n}}\ {\displaystyle\sum\limits_i}\, w_i^{-m}\, 
\end{array} \nonumber \hspace{0.8cm} \big(2m>n\big)
\end{eqnarray} 
It is expected that the following limits exist,
\begin{equation}\label{nmvcxkj}
\arraycolsep=0.4cm
\begin{array}{lll}
M^{(\psi)}_m ={\rm s}\!\!\!\lim\limits_{N\to \infty} M_m^{(N)}\,  & & (n-{\rm odd}) \\[0.6cm]
V^{(\psi)}_m= {\rm s}\!\!\!\lim\limits_{N\to \infty} V_m^{(N)}\, ,&
W^{(\psi)}_m= {\rm s}\!\!\!\lim\limits_{N\to \infty} W_m^{(N)}\,      &    (n-{\rm even} )
\end{array}
\end{equation}
and coincide with the expansion coefficients in $\lambda\equiv \re^{\frac{\theta}{n}}$ of the CFT eigenvalues of $\log\zeta_\pm$:
\begin{equation}
\log\zeta_\pm^{(\psi)}(\theta) =\log B^{(\psi)}\,+{\textstyle\frac{\mathfrak{m}}{n}}\,\theta
\,\mp{\textstyle \frac{2}{n}}\, \theta\, \re^\theta-\sum\limits_{m=1}^\infty\,(\pm1)^m\  M^{(\psi)}_m\  \re^{\frac{m \theta}{n}}
 \hspace{0.5cm} (n -{\rm odd})
\nonumber\\[0.2cm]
\end{equation}
\begin{eqnarray}
&&\hspace{-0.07cm}\log\zeta_+^{(\psi)}(\theta)=\log B^{(\psi)}\,+{\textstyle \frac{\mathfrak{m}}{n}}\, \theta-
\sum\limits_{l=1}^\infty\, V^{(\psi)}_{m}\  \re^{\frac{2m \theta}{n}}
\hspace{2.3cm}\ \  \ \ \ \ \, \,\, (n-{\rm even})\nonumber \\[0.2cm]
&&\hspace{-0.07cm}\log\zeta_-^{(\psi)}(\theta) =\log B^{(\psi)}\,+{\textstyle \frac{\mathfrak{m}}{n}}\, \theta+
{\textstyle \frac{4}{n}}\, \theta\, \re^\theta-\sum\limits_{m=1}^\infty\, W^{(\psi)}_{m}\  \re^{\frac{2m \theta}{n}}
\, . 
\nonumber\end{eqnarray}

Recall that the symbol ``${\rm slim}$'' stands for the scaling limit which is applied for  low energy  eigenstates only. For  numerical checks,
we focused only on the vacuum of the Fateev-Zamolodchikov spin chain  \eqref{sisisai}, \eqref{sisisai2}. Our numerical work confirmed the existence of 
the limits \eqref{nmvcxkj} for $n=3,\,4,\,\ldots, 6$ and all admissible values of $\mathfrak{j}$ and $\mathfrak{m}$ \eqref{siaisaiais}. 
Since a few  of the expansion coefficients in \eqref{hasasasy} 
are available in explicit form, we have the following analytical predictions for some of the limits in \eqref{nmvcxkj}.

\begin{table}
% n = 4, j = 1, m = 0
\centering
\scalebox{0.9}{
\begin{tabular}{|c||c|c|c|c|c||c|}
\hline
$n=4$, $\mathfrak{j}=1$, $\mathfrak{m}=0$& $N=101$ & $N=201$ & $N=400$ & $N=1001$ & $N=\infty$ & exact \\ \hline
$V_1^{(\rm vac)}$ &2.4301852  &2.4299253 &2.4298202 &2.4297709 & 2.4297498 &2.4297502 \\ \hline
$V_{2}^{(\rm vac)}-W_{2}^{(\rm vac)}$& 3.1094496&3.1094648 & 3.1094686&3.1094697 & 3.1094699& 3.1094699 \\ \hline
\hline
$n=4$, $2\mathfrak{j}=\mathfrak{m}=1$ & $N=201$ & $N=401$ & $N=1001$ & $N=1500$ & $N=\infty$ & exact \\ \hline
$V_1^{(\rm vac)}$ &-0.970065 &-0.962059 &-0.955825 &-0.954074 &-0.948453 &-0.948425 \\ \hline
$W_1^{(\rm vac)}$ &2.020549 &2.016718 &2.013751 &2.012921 &2.010289 &2.010250 \\ \hline
\hline
$n=5$, $2\mathfrak{j}=\mathfrak{m}=2$ & $N=100$ & $N=200$ & $N=400$ & $N=750$ & $N=\infty$ & exact \\ \hline
$M_1^{(\rm vac)}$  &-1.09540 &-1.09018 & -1.08669&-1.08453 & -1.07962&-1.07956 \\ \hline
$M_2^{(\rm vac)}$ &0.77960 &0.77649 &0.77445 &0.77320 &0.77054 &0.77039 \\ \hline
\end{tabular}
}
\caption{Numerical values of the coefficients \eqref{coefkjsdfh1},\,\eqref{coefkjsdfh2} for the vacuum of the Fateev-Zamolodchikov spin chain
\eqref{sisisai}, \eqref{sisisai2}.
 The column $N=\infty$ was
obtained  by interpolating the finite-$N$ data. The last column lists the exact predictions
given in \eqref{sdfkjha1},\,\eqref{sdfkjha2}.\label{tabB2}}
\end{table}

Let $f_{0,1}=f_{0,1}(p_1,p_2)$  be  defined by eq.\,\eqref{jsasusau} and
$\gamma(x)\equiv\Gamma(x)/\Gamma(1-x)$. Then for $n>2$ one has (here the superscript ``{\footnotesize{(vac)}}'' in the notation for the coefficients \eqref{nmvcxkj}
 is  omitted):
\begin{eqnarray}\label{sdfkjha1}
&&M_m(\mathfrak{m},\mathfrak{j})=0 \hspace{7.8cm} (m=1,3,\ldots,n-2-2\mathfrak{m}\,) \nonumber\\[0.4cm] 
&&M_1\big({\textstyle\frac{n-1}{2}},{\textstyle\frac{n-1}{4}}\big)=
%-n^{-\frac{1}{n}}\,\frac{\Gamma\big(\frac{1}{2}-\frac{1}{2n}\big)}{\Gamma\big(\frac{1}{2}+\frac{1}{2n}\big)} \nonumber\\[0.4cm]
-n^{-\frac{1}{n}}\ \gamma\big(\half-{\textstyle\frac{1}{2n}}\big)\nonumber\\[0.4cm]
&&M_2(\mathfrak{m},\mathfrak{j})=-f_{0,1}\big(\,{\textstyle-\frac{\ri\,\mathfrak{m}}{2}},\mathfrak{j}+{\textstyle\frac{1}{2}}\,\big)\, 
\hspace{5cm} 
\big(\mathfrak{m} =0,1,2,\ldots,{\textstyle{\frac{n-3}{2}}}\big)\nonumber\\[0.4cm]
&&M_2\big({\textstyle\frac{n-1}{2}},{\textstyle\frac{n-1}{4}}\big)=-f_{0,1}\big(\,{\textstyle-\frac{\ri(n-1)}{4}},\mathfrak{j}+{\textstyle\frac{1}{2}}\,\big)
+ {\textstyle \frac{1}{2}}\,n^{-\frac{2}{n}}\,
\ \gamma^2\big(\half-{\textstyle\frac{1}{2n}}\big)\\[0.4cm]
&&M_n(0,\mathfrak{j})=2\log\big({\textstyle\frac{\re}{2}}\big)+
{\textstyle \frac{2}{n}}\, \big(\gamma_E-\log(n)\big)+4\psi(1+\mathfrak{j})-
\psi(1+{\textstyle\frac{n}{2}})+\gamma_E \nonumber \\[0.8cm]
&&V_m(\mathfrak{m},\mathfrak{j})=W_m(\mathfrak{m},\mathfrak{j})\hspace{6.7cm}  \big(m=1,2,\ldots,{\textstyle \frac{n-2}{2}}-\mathfrak{m}\,\big)\nonumber\, \nonumber\\[0.4cm]
&&V_1(\mathfrak{m},\mathfrak{j})=W_1(\mathfrak{m},\mathfrak{j})=-f_{0,1}\big(\,{\textstyle-\frac{\ri\,\mathfrak{m}}{2}},\mathfrak{j}+{\textstyle\frac{1}{2}}\,\big)\, 
\hspace{3.3cm} \big(\mathfrak{m} =0,1,2,\ldots,{\textstyle{\frac{n}{2}-2}}\big)\nonumber \\[0.4cm]
&&V_1\big({\textstyle\frac{n-2}{2}},\mathfrak{j}\big)=-f_{0,1}\big(\,{\textstyle-\frac{\ri(n-2)}{4}},\mathfrak{j}+{\textstyle\frac{1}{2}}\,\big)
-n^{-\frac{2}{n}}
\ \gamma\big(\half-{\textstyle\frac{1}{n}}\big)\label{sdfkjha2}\\[0.4cm]
&&W_1\big({\textstyle\frac{n-2}{2}},\mathfrak{j}\big)=-f_{0,1}\big(\,{\textstyle-\frac{\ri(n-2)}{4}},\mathfrak{j}+{\textstyle\frac{1}{2}}\,\big)
+n^{-\frac{2}{n}} \ \gamma\big(\half-{\textstyle\frac{1}{n}}\big)\nonumber\\[0.4cm]
&&V_{\frac{n}{2}}(0,\mathfrak{j})-W_{\frac{n}{2}}(0,\mathfrak{j})=4\log\big({\textstyle\frac{\re}{2}}\big)+
{\textstyle \frac{4}{n}}\, \big(\gamma_E-\log(n)\big)+8\,\psi(1+\mathfrak{j})-
2\,\psi(1+{\textstyle\frac{n}{2}})+2\gamma_E\, \nonumber 
\end{eqnarray}
%where $\gamma_E$ is the Euler constant and
%$\psi(z)=\partial_z\log\Gamma(z)$.

\noindent
The  numerical data  agreed with these explicit formulae. This is shown, for a few cases, in tab.\,\ref{tabB2}. 

As was already mentioned,  the constant $\big(B^{(N,\psi)}\big)^2$ can be found using the quantum Wronskian type relations \ref{relii}  from sec.\,\ref{sec4.1}.
The r.h.s. of these relations  is proportional to  the lattice  shift operator $\mathbb{P}^{(N)}$ \eqref{isisaai} whose eigenvalues are pure phases \eqref{hsaasyas}.
By explicit diagonalization of $\mathcal{Z}_\pm$ for small $N$ we found  that
\bea
\frac{1}{\pi}\ \arg\big( B^{(N,\psi)}\,\big) ={ \frac{1}{nN}}\ \big(\,(2\mathfrak{j}-s)\,s+n\,(L-\bar{L})\,\big)+s\ \ \ \ \ \ \ \ ({\rm mod}\, 2)\ ,
\eea
where $s=\mathfrak{j}-\frac{1}{2}\, \mathfrak{m}$ and $L$, $\bar{L}$ are non-negative integers  depending on the state $|\psi\rangle$. For the
vacuum state $L={\bar L}=0$, and the overall sign of the limit $B^{({\rm vac})}$ \eqref{sdkjfhasfd} is $(-1)^s$. This coincides with
the sign factor   in   $B_s(\mathfrak{m})$  \eqref{iusdfkjasd}. For large values of $N$, when direct  diagonalization becomes impossible,
we verified by means of the Bethe ansatz that the absolute value of $(\pi/N)^{\frac{\mathfrak{m}}{n}}\,B^{(N, {\rm vac})}$ 
 converges to $(-1)^s\, B_s(\mathfrak{m})$
(see tab.\,\ref{tabB3}).

%We also checked that
 %$(\pi/ N)^{\frac{\mathfrak{m}}{n}}\,B^{(N, \psi)}$
 %corresponding to  the vacuum state  approaches the analytical prediction  \eqref{iusdfkjasd} as $N\to \infty$
 %(see tab.\,\ref{tabB3}).

Recall that $2\mathfrak{j}=\mathfrak{m}=\frac{n}{2}$ with even $n$ is a special case.  Using eq.\,\eqref{faslkdere} the scaling functions can 
be found explicitly,
\begin{equation}
\zeta_+^{(\rm vac)}(\theta)=\frac{2\sqrt{\pi}\ \re^{\frac{\theta}{2}}}{\Gamma\big(1+2\re^\theta\big)}\, ,
\hspace{1.5cm}\zeta_-^{(\rm vac)}(\theta)=0 \, .\nonumber
\end{equation}
This formula can be applied for $n=2$.  For the remaining $n=2$  case,  $\mathfrak{j}=\mathfrak{m}=0$,  it is easy to show that for finite $N$
\begin{equation}
\mathcal{Z}^{(\rm vac)}_+(\mu)=\sqrt{2}\, \prod\limits_{m=1}^N\Big(1-\mu^2\,\cot\big(\textstyle{\frac{\pi}{4N}\,(2m-1)}\big)\Big)\,,\hspace{1.5cm}
\mathcal{Z}^{(\rm vac)}_-(\mu)=\sqrt{2}\, , \nonumber
\end{equation}
so that the scaling functions are given by
\begin{equation}
\zeta_+^{(\rm vac)}(\theta)=\frac{\sqrt{2\pi}}{\Gamma\big(\frac{1}{2}+2\re^{\theta}\big)}\ 
\Big(\,\frac{2}{\re}\,\Big)^{2\re^{\theta}}\, ,\hspace{1.5cm} 
\zeta_-^{(\rm vac)}(\theta)=\sqrt{2}\ \exp\big(2\theta\,\re^{\theta}\,\big)\nonumber\,.
\end{equation}

\begin{table}
\scalebox{0.93}{
\begin{tabular}{|c||c|c|c|c|c||c|}
\cline{7-7}
\multicolumn{5}{}{} & &  \multirow{2}{*}{$(-1)^s\, B_s({\mathfrak m})$} \\
\multicolumn{5}{}{} & &   \\
\cline{1-7}
\multirow{3}{*}{$n=3$, $2\mathfrak{j}=\mathfrak{m}=1$} & $N=101$ & $N=201$ & $N=401$ & $N=1001$ & $N=\infty$ & \multirow{3}{*}{1.626210} \\ \cline{2-6}
& & & & & & \\[-0.4cm]
  &1.621564 &1.623528 &1.624666 &1.625467 &1.626213&\\ \hline
\hline
\multirow{3}{*}{$n=6$, $2\mathfrak{j}=\mathfrak{m}=2$}  & $N=100$ & $N=200$ & $N=400$ & $N=800$ & $N=\infty$ &\multirow{3}{*}{ 1.82536}  \\ 
\cline{2-6}
& & & & & & \\[-0.4cm]
  &1.79398 &1.80320 &1.80971 &1.81430 &1.82531 & \\ \hline
\hline
\multirow{3}{*}{$n=6$, $2\mathfrak{j}=3$, $\mathfrak{m}=1$}  & $N=2$ & $N=4$ & $N=6$ & $N=8$ & $N=\infty$ &\multirow{3}{*}{4.10} \\ 
\cline{2-6}
& & & & & & \\[-0.4cm]
 &4.178 &4.148 &4.135 & 4.127 & 4.08 &\\ \hline
\end{tabular}
}
\caption{The absolute value of $(\pi/N)^{\frac{\mathfrak{m}}{n}}\,B^{(N, {\rm vac})}$  corresponding to the vacuum state of
 the Fateev-Zamolodchikov spin chain \eqref{sisisai},\,\eqref{sisisai2}.
 The column ``$N=\infty$'' contains the results  of numerical interpolation  from the finite $N$ data.
 The analytical expression for $B_s(\mathfrak{m})$  is  given by  \eqref{iusdfkjasd}.  \label{tabB3}}
 \end{table}

\newpage

\section{Appendix\label{appC1}}

In this appendix we sketch some technical details in the derivation of the system of  NLIE \eqref{hasatast}.

Suppose $\theta$, $p_1^2$ and $p_2$ are real, then  eqs.\,\eqref{exna}, Hermiticity conditions
\eqref{sassao},\,\eqref{iuywqesadj},\ \eqref{jssasu} and the  periodicity \eqref{ashsasya}  imply that
\bea\label{exnall}
&&\re^{-\ri\pi \re^{\theta}}\  \Big[\,\re^{-\ri\pi p_2}\ \beta_-(\theta)\, 
{\alpha}_{+}\big(\theta-{\textstyle \frac{\ri\pi n}{2}}\big)-
\,\re^{+\ri\pi p_2}\,  \beta_+(\theta)\ {\alpha}_{-}\big(\theta-{\textstyle \frac{\ri\pi n}{2}}\big)\, \Big]=\nonumber\\[0.2cm]
 &&\re^{+\ri\pi \re^{\theta}}\  \Big[\,\re^{-\ri\pi p_2}\,  \beta_+(\theta)\ {\alpha}_{-}\big(\theta+{\textstyle \frac{\ri\pi n}{2}}\big)-
 \re^{+\ri\pi p_2}\ \beta_-(\theta)\, 
{\alpha}_{+}\big(\theta+{\textstyle \frac{\ri\pi n}{2}}\big)
\, \Big]\ ,\\[0.4cm]
&&\re^{+\ri\pi \re^{\theta}}\   \Big[\, \re^{-\ri\pi p_2}\ \beta_-(\theta+\ri\pi)\,  
{\alpha}_{+}\big(\theta-{\textstyle \frac{\ri\pi n}{2}}\big)-\re^{+\ri\pi p_2}\ 
\beta_+(\theta+\ri\pi)\,  {\alpha}_{-}\big(\theta-{\textstyle \frac{\ri\pi n}{2}}\big)\, \Big]=\nonumber\\[0.2cm]
&&\re^{-\ri\pi \re^{\theta}}\   \Big[\, \re^{-\ri\pi p_2}\ 
\beta_+(\theta+\ri\pi)\,  {\alpha}_{-}\big(\theta+{\textstyle \frac{\ri\pi n}{2}}\big)-
\re^{+\ri\pi p_2}\ \beta_-(\theta+\ri\pi)\,  
{\alpha}_{+}\big(\theta+{\textstyle \frac{\ri\pi n}{2}}\big)\,\Big]
\, .\nonumber
\eea
Due to the analyticity of the operators  $\alpha_\pm(\theta)$ and $\beta_\pm(\theta)$, these relations should  be satisfied  for any complex
$\theta$.
Let us introduce the shortcut notations
\bea
B_0=\frac{\beta_+(\theta)}{\beta_-(\theta)}\ ,\ \ B_1=\frac{\beta_+(\theta+\ri\pi)}{\beta_-(\theta+\ri\pi)}\ ,\ \ 
U=\re^{2\pi\ri p_2}\ \frac{\alpha_{+}(\theta+{\textstyle \frac{\ri\pi n}{2}})}{\alpha_{-}(\theta+{\textstyle \frac{\ri\pi n}{2}})}\ ,\ \ 
A_\pm=\re^{\mp 2\pi\ri p_2}\ \frac{\alpha_{\pm}(\theta-{\textstyle \frac{\ri\pi n}{2}})}{\alpha_{\pm}(\theta+{\textstyle \frac{\ri\pi n}{2}})}\nonumber
\eea
and
$\Lambda=\exp\big(2\pi\ri\, \re^\theta\,\big)$.
Then \eqref{exnall} can  be rewritten as
\bea\label{hsssay}
B_0=U\ \frac{1+\Lambda^{-1}\, A_+}{1+\Lambda^{-1}\, A_-}\ ,\ \ \ \ \ \ \ \ \ \ \ \ B_1=U\ \frac{1+\Lambda\,A_+}{1+\Lambda\,A_-}\  .\nonumber
\eea
Solving these equations w.r.t. $A_+$ and $A_-$, one finds
\bea
A_+=-\half\, \big(\Lambda+\Lambda^{-1}\,\big)+\frac{\Lambda-\Lambda^{-1}}{B_1-B_0}\  \big(\,B_0 B_1\,U^{-1}-\half\,B_0-\half\, B_1\, \big)\nonumber
\eea
and similar for $A_-$. 
This formula, combined with the quantum Wronskian relation \eqref{ajsasy} written in the form
\bea
\frac{\Lambda-\Lambda^{-1}}{B_1-B_0}=-
\frac{ \beta_-(\theta)\,\beta_-(\theta+\ri\pi)}{2 \ri\, \sin\big(\frac{2\pi p_2}{n+2}\big)}\ ,\nonumber
\eea
leads to 
\bea
A_+(\theta)=-\cos\big(2\pi\re^{\theta}\big)
-
\frac{\beta_+(\theta)\,\beta_+ (\theta+\ri\pi)}{2\ri\, U(\theta)\sin\big(\frac{2\pi p_2}{n+2}\big)}+
\frac{
   \beta_+(\theta)\,\beta_-(\theta+\ri\pi)+\beta_-(\theta)\,\beta_+(\theta+\ri\pi)}{4\ri\ \sin\big(\frac{2\pi p_2}{n+2}\big)}\ .\nonumber
\eea
Together  with   the periodicity condition $\beta_\pm(\theta+\ri\pi)=\beta_\pm(\theta-\ri\pi)$  the last equation   implies
\bea
A_+\big(\theta-{\textstyle\frac{\ri\pi}{2}}\big)-A_+\big(\theta+{\textstyle\frac{\ri\pi}{2}}\big)=
\frac{\beta_+\big(\theta+{\textstyle\frac{\ri\pi}{2}}\big)
\beta_-\big(\theta-{\textstyle\frac{\ri\pi}{2}}\big)}
{2\ri\,\sin\big(\frac{2\pi p_2}{n+2}\big)}
\ 
\Big(\,U^{-1}\big(\theta+{\textstyle\frac{\ri\pi}{2}}\big)- U^{-1}\big(\theta-{\textstyle\frac{\ri\pi}{2}}\big)\Big)\ .\nonumber
\eea
As it follows from  the quantum Wronskian relation \eqref{hsassat}:
\bea
U^{-1}\big(\theta+{\textstyle\frac{\ri\pi}{2}}\big)- U^{-1}\big(\theta-{\textstyle\frac{\ri\pi}{2}}\big)=
\frac{2\ri\  \re^{-2\pi\ri p_2}\ \sin\big(\frac{2\pi p_2}{n+2}\big)}{\alpha_+(\theta+{\textstyle\frac{\ri\pi (n-1)}{2}}\big)
\alpha_+\big(\theta+{\textstyle\frac{\ri\pi (n+1)}{2}}\big)}\ .\nonumber
\eea
This can be substituted into the previous formula, yielding eq.\,\eqref{jassay} with the subscript ``$+$''. Of course the 
formula is valid  for the ``$-$'' case also.
%\bea
%\frac{\alpha_{+}(\theta-{\textstyle \frac{\ri\pi (n+1)}{2}})}{\alpha_{+}(\theta+{\textstyle \frac{\ri\pi (n-1)}{2}})}-
%\frac{\alpha_{+}(\theta-{\textstyle \frac{\ri\pi (n-1)}{2}})}{\alpha_{+}(\theta+{\textstyle \frac{\ri\pi( n+1)}{2}})}
%=
%\frac{  \beta_+(\theta)\,\beta_+ (\theta+\ri\pi)}{\alpha_+(\theta+{\textstyle\frac{\ri\pi (n-1)}{2}}\big)
%\alpha_+\big(\theta+{\textstyle\frac{\ri\pi (n+1)}{2}}\big)}\ .\nonumber
%\eea
%The result  is equivalent to eq.\,\eqref{jashsatg} with the notation  $\varepsilon$ given in  \eqref{hasssay}.

\bigskip
Let us now  take a closer look  at the second equation in  \eqref{exnall} specialized to the eigenvalues corresponding to a common eigenvector $|\,\psi\,\rangle$.
Suppose $\theta_j$ is a zero of $\beta_+^{(\psi)}(\theta)$. As follows from the quantum Wronskian relation \eqref{ajsasy}, 
$\beta^{(\psi)}_-(\theta_j)\not=0$, and therefore
we  conclude that
\bea
\re^{-\ri\pi (\re^{\theta_j}+p_2)}\  
{\alpha}^{(\psi)}_{+}\big(\theta_j-\ri\pi -{\textstyle \frac{\ri\pi n}{2}}\big)
=-\re^{\ri\pi (\re^{\theta_j}+p_2)}\ 
{\alpha}^{(\psi)}_{+}\big(\theta_j-\ri\pi +{\textstyle \frac{\ri\pi n}{2}}\big)\ ,\nonumber
\eea
which can be equivalently written in the form  \eqref{jassaya}.

\bigskip
As  was mentioned in the main body of the text, the zeroes of the entire periodic function
$ \beta^{({\rm vac})}_+(\theta)=\beta^{({\rm vac})}_+(\theta+\ri\pi)$ are simple,   located on the lines
$\Im m (\theta)=\pi\, (2 m+1),\, m\in{\mathbb Z}$, and accumulate toward $\Re e (\theta)\to +\infty$.
Also, assuming that the parameters $p_1$ and $p_2$ are restricted as in cases \ref{sdkjfhewqa}, \ref{sdkjfhewqb} from sec.\,\ref{sec45},
 it is expected that the entire function $ \alpha^{({\rm vac})}_+(\theta)$ does not have any zeroes within the strip
$|\Im  m(\theta)|<\frac{\pi}{2}\, (n+2)$. Therefore, as follows from the definition \eqref{hasssay},  $\varepsilon^{({\rm vac})}(\theta)$
is an analytic function for $|\Im m (\theta)|<\pi$ where it has the leading asymptotic behaviour \eqref{ahsassat}  at $\Re e (\theta)\to+\infty$.
Combining this analytic information with the ``quantization condition'' \eqref{aisisi} for the zeroes of  $ \beta^{({\rm vac})}_+(\theta)$ 
and the asymptotic behaviour \big(see eq.\,\eqref{qpwoeier}\big)
\bea
\log\beta^{(\rm vac)}_+(\theta)=-2\theta\,\re^{\theta}-k\,\theta+o(1)\ \ \ \  \ \ {\rm as}\ \ \ \ \ \ \Re e(\theta)\to+\infty\ \ \ \&\  \ \big|\Im m(\theta)\big|<\pi\ ,\nonumber
\eea
with $k=\frac{2p_2}{n+2}$, it is a simple exercise (see however appendix \ref{apDD}) to derive a dispersion-type relation
\bea\label{asdjhdsfe1}
&&\log\Big( \beta^{({\rm vac})}_+\big(\theta-{\textstyle \frac{\ri\pi}{2}}\big)\, \beta^{({\rm vac})}_+\big(\theta+{\textstyle \frac{\ri\pi}{2}}\big)\Big)
=2\pi\,\re^{\theta}-2k\,\theta-\\
&& \Im m\bigg[ \int_{-\infty}^{\infty}\frac{\rd\theta'}{\pi}\, 
\frac{1}{1+\re^{2\theta-2\theta'+2\ri\gamma}}\Big(2L^{({\rm vac})}(\theta'-\ri \gamma)+
\ri \big(\varepsilon^{({\rm vac})}(\theta'-\ri\gamma)-4\pi\,\re^{\theta'-\ri\gamma}+2\pi k\, \big)\Big) \bigg].\nonumber
\eea
Here $\gamma\in (0,{\textstyle\frac{\pi}{2}})$ is an arbitrary constant and the notation
\bea\label{ajsasuas}
L^{({\rm vac})}(\theta)=\log\Big(1+\exp\big(-\ri \varepsilon^{\rm( vac)}(\theta)\,\big)\Big)
\eea
is used. 

\bigskip
The next important property   employed in the derivation of the 
 system of integral equations \eqref{hasatast}-\eqref{hasatast2} is that  $\varepsilon^{(\rm vac)}(\theta)$ can be written
 in terms of the Fourier integral 
 \bea\label{ajsuauy}
 {\varepsilon}^{(\rm vac)}(\theta)=4\pi\,\re^\theta-2\pi k+
\int\limits_{{\mathbb  R}+\ri 0}\frac{\rd\nu}{2\pi}\ \re^{\ri\nu\theta}\ {\tilde  \varepsilon}(\nu)\ .
 \eea
Notice that the existence of the Fourier transform   is ensured  by the asymptotic behaviour   \eqref{ahsassat} at $\theta\to+\infty$, and
formulae \eqref{aisisosa},\,\eqref{hsssaty} for $\theta\to-\infty$.  
One can expect that the function ${\tilde \varepsilon}(\nu)$ decays sufficiently fast
as
$\nu\to\pm\infty$, so that
 the integral in   \eqref{ajsuauy} converges for any $\theta$ in the
strip  of analyticity $|\Im m(\theta)|<\pi$.
It is not difficult to see now that
 \begin{eqnarray}\label{asdjhdsfe2}
 \log\Big[\alpha_+^{(\rm vac)}\big(\theta-{\textstyle \frac{\ri\pi(n+1)}{2}}\big)\alpha_+^{(\rm vac)}\big(\theta+{\textstyle \frac{\ri\pi(n+1)}{2}}\big)\Big]\!=\!
 2\pi\re^{\theta}\!-\!2k\theta\!+\!\ri \!\!\!\int\limits_{{\mathbb  R}+\ri 0}\frac{\rd\nu}{2\pi}\,\re^{\ri\nu\theta}\
 \frac{\cosh(\frac{\pi (n+1)\nu}{2})}{\sinh(\frac{\pi n\nu}{2}) }
\  {\tilde  \varepsilon}(\nu) \nonumber \\[0.2cm]
\end{eqnarray}
 and also   that the imaginary part of the function \eqref{ajsasuas} with $\theta$ having infinitesimally small negative imaginary part, can be represented
 by the convergent  integral
 \bea
 \Im m\big( L^{(\rm vac)}(\theta-\ri 0)\big)=\int\limits_{{\mathbb R}+\ri 0}\frac{\rd\nu}{2\pi}\ \re^{\ri\nu\theta}\ {\tilde  L}(\nu)\ .\nonumber
 \eea
 Similarly for the function $\omega(\theta)$ \eqref{ajshsahyt} with $\theta$ real, one has
\bea
{\omega}^{(\rm vac)}(\theta)=4\pi\,\re^\theta+
\int\limits_{{\mathbb  R}+\ri 0}\frac{\rd\nu}{2\pi}\ \re^{\ri\nu\theta}\ {\tilde  \omega}(\nu)\,,\qquad \ \ \ \ 
\log\big(1+\re^{-\omega^{(\rm vac)}(\theta)}\,\big)=\int\limits_{{\mathbb R}+\ri 0}\frac{\rd\nu}{2\pi}\ \re^{\ri\nu\theta}\ {\tilde M}(\nu)\ .\nonumber
 \eea
 The remaining part of the derivation of the NLIE consists of straightforward manipulations with the Fourier images
 ${\tilde  \varepsilon}$, ${\tilde  L}$, $ {\tilde  \omega}$, ${\tilde M}$. Finally, going back to functions of the variable $\theta$, one derives the system of integral
 equations \eqref{hasatast}-\eqref{hasatast2}.
 
\bigskip

Knowing the functions $\varepsilon^{(\rm vac)}(\theta)$, $\omega^{(\rm vac)}(\theta)$ from the solution of the NLIE, 
and the asymptotic formulae \eqref{jassysay},\,\eqref{qpwoeier}, 
one can recover the vacuum eigenvalues of the operators $\alpha_+(\theta)$ and $\beta_+(\theta)$ from \eqref{asdjhdsfe1},\,\eqref{asdjhdsfe2}.
The corresponding explicit relations are given below, where we drop the superscript ``{\footnotesize (vac)}'' like in the NLIE\, \eqref{hasatast}-\eqref{hasatast2}:
\bea\label{skdjfhsdfhk1}
    \log{\alpha}_+(\theta)&=& -{ \frac{\pi}{\sin(\frac{\pi n}{2})}}\   \re^{\theta}-k\theta
 +\int_{-\infty}^\infty\frac{\rd\theta'}{2\pi\ri }\ \Big[\,F^{\rm (CFT)}_1(\theta-\theta'+\ri\gamma) \ L(\theta'-\ri \gamma)\\[0.2cm]
 &-&
F^{\rm (CFT)}_1(\theta-\theta'-\ri\gamma)\,   \Big(L(\theta'-\ri \gamma)\Big)^* \ \Big]
 +\int_{-\infty}^\infty\frac{\rd\theta'}{\pi}\,
 F_2(\theta-\theta')\ \log\Big(1+\re^{-\omega(\theta')}\Big)\,\nonumber
  \eea
  valid for $| \Im m( \theta)|< \frac{\pi}{2}\,(n+2)-\gamma$, and
  \bea\label{skdjfhsdfhk2}
 \log \beta_+(\theta)&=& 
  -2\theta\,\re^\theta -k\theta+
\int_{-\infty}^\infty\frac{\rd\theta'}{2\pi\ri}\ \Big[\,
 F^{\rm (CFT)}_3(\theta-\theta'+\ri\gamma)\ L(\theta'-\ri \gamma)\\[0.2cm]
 &-&F^{\rm (CFT)}_3(\theta-\theta'-\ri\gamma)\ \Big(L(\theta'-\ri \gamma)\Big)^*\ \Big]
- \int_{-\infty}^\infty\frac{\rd\theta'}{\pi}\ 
 F_4(\theta-\theta')\ \log\Big(1+\re^{-\omega(\theta')}\Big)\nonumber
  \eea
for $| \Im m( \theta)|< \pi-\gamma$.  Here the kernels are given by
$ F^{\rm (CFT)}_1(\theta)= F_1(\theta)-\frac{1}{n+2}$,  $\qquad$ $ F^{\rm (CFT)}_3(\theta)= F_3(\theta)-\frac{1}{n+2}$ with
 \bea\label{hasgsaf}
  F_1(\theta)&=&{ \frac{1}{n+2}}\ \tanh\Big({\frac{\theta}{n+2}}\Big)\ ,\ \ \ \ 
  %\nonumber \\[0.2cm]
 F_2(\theta)=\frac{\sin(\frac{\pi}{n+2})}{2(n+2)\cosh\big(\frac{\theta+\frac{\ri\pi}{2}}{n+2}\big)
 \cosh\big(\frac{\theta-\frac{\ri\pi}{2}}{n+2}\big)}\nonumber\\[0.3cm]
 F_3(\theta)&=& { \frac{1}{n+2}}\ 
 \coth\Big({\frac{\theta}{n+2}}\Big)-{ \frac{1}{\sinh(\theta)}}
\\[0.3cm]
  F_4(\theta)&=&
\frac{\sin(\frac{\pi}{n+2})}{2(n+2)\sinh(\frac{\theta+\frac{\ri\pi}{2}}{n+2})
 \sinh(\frac{\theta-\frac{\ri\pi}{2}}{n+2})}-\frac{1}{2\cosh(\theta)}
  \ .\nonumber
  \eea
The vacuum eigenvalues of the chiral transfer-matrix can be obtained using the $T-Q$ relation
\bea\label{kas}
\tau^{(\rm vac)}(\ri\lambda)=\frac{\alpha_+(\theta+\ri\pi )}{ \alpha_+(\theta)}+
\frac{\alpha_+(\theta-\ri\pi )}{ \alpha_+(\theta)}  \qquad {\rm with} \qquad \lambda=\re^{\frac{\theta}{n}}\,.\nonumber
\eea
Combining  \eqref{skdjfhsdfhk1},\,\eqref{skdjfhsdfhk2} with the general asymptotic expansions at $\Re e(\theta)\to+\infty$ found in \eqref{jassysay},\,\eqref{qpwoeier},
the expressions for the local and dual nonlocal integrals of motion follow
 \bea\label{jasshaa}
&& i_{2m-1}(p_1,p_2)=\int_{-\infty}^{\infty}\frac{\rd\theta}{\pi}
 \bigg( 2\, \Im m\Big[  \re^{(2m-1)(\theta-\ri\gamma)} L(\theta-\ri \gamma)\Big]
 +(-1)^{m} \re^{(2m-1)\theta} \log\big(1+\re^{-\omega(\theta)}\big)\!\bigg)\nonumber\\[0.2cm]
 &&\\[0.2cm]
&&{\tilde s}_m(p_1,p_2)=-{ \frac{2}{n+2}} 
 \int_{-\infty}^{\infty}\frac{\rd\theta}{\pi}\bigg( \Im m\Big[  \re^{\frac{2m( \theta-\ri\gamma)}{n+2}} L(\theta-\ri \gamma)\,\Big]
 \!-\!
\sin\big({\textstyle\frac{\pi m}{n+2}}\big)\, \re^{\frac{2m \theta}{n+2}} \,\log\big(1+\re^{-\omega(\theta)}\big)\!\bigg).\nonumber
\eea
%Also, using the formula \eqref{skdjfhsdfhk2} one can express
%the vacuum eigenvalue of the operator $b_+\equiv\lim\limits_{\theta\to-\infty}\beta_{+}(\theta)$ in terms of the solution to the NLIE
%\bea
%\log b_+^{(\rm vac)}(p_1,p_2)&=&\ldots\nonumber \\
%\eea

\section{Appendix\label{apDD}}

Here we discuss the modifications to the integral equations \eqref{hasatast}-\eqref{hasatast2} for the
case of real $p_1\ne 0$ and $p_2>-\frac{1}{2}$, when the asymptotics of the functions
$\varepsilon^{\rm(vac)}(\theta)$ and $\omega^{(\rm{vac})}(\theta)$ 
  at $\theta\to-\infty$  oscillate \eqref{dksjfhsdf1}.

The first important difference in this case is that 
 $\alpha_+^{(\rm vac)}(\theta)$ has a set of zeroes $\{\theta_m^{(\alpha)}\}_{m=1}^{\infty}$ in the strip 
$|\Im  m(\theta)|<\frac{\pi}{2}\, (n+2)$ whose asymptotic behaviour is given by relation \,\eqref{oiuqwmnaspx}.
Secondly,
in the derivation of  \eqref{hasatast} presented in the previous appendix, 
%takes the
%advantage of working with Fourier transformations on the infinite
%line. 
%In this derivation 
 we implicitly assumed that all values $\theta_*$ on the real axis, 
such that  $\varepsilon^{({\rm vac})}(\theta_*)=\pi \pmod{2\pi}$
%\beq\label{thetastar}
%\varepsilon^{({\rm vac})}(\theta_*)=\pi \pmod{2\pi}\,,
%\eeq
arise from the quantization condition \eqref{aisisi}, 
i.e., $\theta_*=\theta^{(\rm vac)}_j-\ri\pi$ for
some $j=1,2,\ldots$ (recall that $\Im m(\theta_j)=\pi$).
In other words all 
such $\theta_*$ are related to the zeroes of 
$\beta_+^{\rm (vac)}(\theta)$ and, therefore, form an increasing 
semi-infinite sequence extending towards $+\infty$ on the real axis
(see \eqref{hsassatas}). For the oscillating asymptotics  \eqref{dksjfhsdf1} this is no longer
true. 
Indeed, it is easy to check from  \eqref{dksjfhsdf1} that 
the condition
\bea
\varepsilon^{({\rm vac})}({\tilde \theta}_m)=-\pi\,(2m-1)  \ \ \ \ \ \ {\rm with}\ \ \ m=1,\,2,\ldots\nonumber
\eea
%\eqref{thetastar}
 is  satisfied for an infinite set of values 
$\big\{{\tilde \theta}_m\big\}_{m=1}^\infty$  which extend towards $-\infty$ such that
% with exponential accuracy,
%\beq
%{\tilde \theta}_m=
%\theta_0-\frac{n}{4 \ri p_1}\, \log\bigg[\frac{\cos\big(\pi(p_2+\ri
%  p_1)\big)}{\cos\big(\pi(p_2-\ri p_1)\big)}\bigg]
%-\frac{\pi n}{2p_1}\  m+o\big( (m/p_1)^{-\infty}\big)
%\nonumber
%\eeq
\beq
{\tilde \theta}_m
=-\mathlarger{\textstyle\frac{n}{2p_1}}\, \Big(\pi m-\mathlarger{{\textstyle\frac{1}{2}}}\, 
\delta(p_1,p_2)\Big)+o\big( (m/p_1)^{-\infty}\big)\, ,
\nonumber
\eeq
valid up to an exponentially small correction. Here $$\delta(p_1,p_2)=
4p_1\theta_0/n+\ri\, \log\big[\cos(\pi(p_2+\ri
  p_1))/\cos(\pi(p_2-\ri p_1)\big]$$ coincides with the scattering phase defined by eq.\,\eqref{ksjshs}.
%\bea
%\Theta_1=\Theta_0-\frac{n\pi}{2 p_1}-\frac{n}{4 \ri p_1}\, \log\bigg[\frac{\cos\big(\pi(p_2+\ri
 % p_1)\big)}{\cos\big(\pi(p_2-\ri p_1)\big)}\bigg]\ ,\nonumber
%\eea
 %$\theta_0$ is defined in \eqref{dksjfhsdf1}.
 % and the phase of the logarithm
%is set to be  zero at $p_2=0$ and $p_1\not=0$.  
In the terminology of the Bethe ansatz 
we have an infinite number of ``holes'' where the phase passes a
resonant value  without a corresponding zero $\theta_j$.  Therefore  the integrals in the r.h.s. of   \eqref{hasatast}
contain spurious contributions from non-existent roots. 
To exclude these unwanted contributions one needs to add extra source terms to
the r.h.s. of eqs. \eqref{hasatast}.

Introduce the notation
\bea\label{eq-with-cdd0}
J^{(\varepsilon)}(\theta)&=&-\ri\, \sum\limits_{m=1}^\infty \log\bigg[
\frac{S
(\theta-{\tilde \theta}_m)}{S
\big(\theta+{\textstyle \frac{\ri \pi }{2}}\, (n+2)-\theta_m^{(\alpha)}\,\big)}\bigg]\nonumber \\[0.5cm]
&& \\[-0.4cm]
J^{(\omega)}(\theta)&=&-\sum\limits_{m=1}^\infty \log\bigg[
\frac{t
\big(\theta+{\textstyle \frac{\ri\pi}{2}}-{\tilde \theta}_m\big)}{t
\big(\theta+{\textstyle \frac{\ri \pi }{2}}+{\textstyle \frac{\ri \pi }{2}}\, (n+2)-\theta_m^{(\alpha)}\,\big)}\bigg] \nonumber
%\nonumber
\eea
where $S(\theta)$ and $t(\theta)$ are defined in \eqref{jassasa} below.
%\beq
%\begin{array}{rclrcl}
%S(\theta)=\displaystyle 
% \re^{-\frac{2\ri\pi  }{n+2}}\ \ 
%\ds 
%-\frac{\sinh\big(\frac{\ri\pi-\theta}{n+2}\big)}
%{\sinh\big(\frac{\ri\pi+\theta}{n+2}\big)}\ ,
%\quad\quad\quad
%&\widetilde{C}^{(\varepsilon)}(\theta)&=&\ds \re^{-\frac{2\ri \pi}{n+2}}\ \ 
%\frac{\cosh\big(\frac{\theta+\ri\pi}{n+2}\big)}
%{\cosh\big(\frac{\theta-\ri\pi}{n+2}\big)}
%\\[.6cm]
%S^{(\omega)}(\theta)=\ds
%\frac{\sinh\big(\frac{\frac{\ri\pi}{2}+\theta}{ n +2}\big)
%\sinh\big(\frac{\frac{\ri\pi}{2}-\theta}{ n +2}\big)}{
%\sinh\big(\frac{\theta+\frac{3\ri\pi}{2}}{ n +2}\big)
%\sinh\big(\frac{\frac{3\ri\pi}{2}-\theta}{ n +2}\big)}=t\big(\theta+{\textstyle\frac{\ri\pi }{2}}\big)
%\quad
%&\widetilde{C}^{(\omega)}(\theta)&=&\ds
%\frac{\cosh\big(\frac{\theta+\frac{\ri \pi}{2}}{ n +2}\big)
%\cosh\big(\frac{\theta-\frac{\ri \pi}{2}}{ n +2}\big)
%}
%{\cosh\big(\frac{\theta+\frac{3\ri \pi}{2}}{ n +2}\big)
%\cosh\big(\frac{\theta-\frac{3\ri \pi}{2}}{ n +2}\big)}
%\end{array}\ .\nonumber
%\eeq
Then the modified equations   \eqref{hasatast}  can be written as 
\bea\label{eq-with-cdd}
\varepsilon(\theta-\ri\gamma)&= &4\pi\,\re^{\theta-\ri\gamma}-2\pi k +J^{(\varepsilon)}(\theta-\ri\gamma)
+
\int_{-\infty}^\infty\frac{\rd \theta'}{2\pi\ri }\
\Big[G(\theta-\theta'-2\ri\gamma)\  \big(L(\theta'-\ri \gamma)\big)^*
\nonumber\\[.4cm]
&-& G(\theta-\theta')\  L(\theta'-\ri \gamma)\Big]+
\int_{-\infty}^\infty\frac{\rd \theta'}{2\pi}\ G_1(\theta-\theta'-\ri\gamma)\  \log\big(1+\re^{-\omega(\theta')}\big)\nonumber
\\[0.4cm]
\omega(\theta)&=&4\pi\, \re^{\theta}+J^{(\omega)}(\theta)
+\Im m\bigg[\int_{-\infty}^\infty\frac{\rd\theta'}{\pi}\,G_1(\theta-\theta'+\ri\gamma)\  L(\theta'-\ri \gamma)
\,\bigg]\\[0.4cm]
&-&
\int_{-\infty}^\infty\frac{\rd\theta'}{\pi}\, G_2(\theta-\theta')\, \log\big(1+\re^{-\omega(\theta')}\big)\nonumber
\\[0.4cm]
L(\theta)&=&\log\big(1+\re^{-\ri \varepsilon(\theta)}\,\big)\ .\nonumber
\eea
One can check that the leading terms in the asymptotics  \eqref{dksjfhsdf1} 
%with an arbitrary value of $\theta_0$ 
solves these equations at $\theta\to -\infty$, i.e., when
the exponential terms  proportional to $\re^\theta$ in the r.h.s. are omitted.

\section{Appendix\label{appD}}
Here, we present the explicit form of  the two particle scattering amplitudes for the sausage model \cite{Fateev:1992tk}. 
The $S$-matrix satisfies the Yang-Baxter equation
and was originally introduced as the Boltzmann weights of the so-called $19$-vertex model \cite{Zamolodchikov:1980ku}.
\bea\label{jassasa}
 S(\theta)&=&S_{++}^{++}(\theta)=S_{--}^{--}(\theta)=-\frac{\sinh(\frac{\ri\pi-\theta}{n+2})}
{\sinh(\frac{\ri\pi+\theta}{n+2})}\nonumber\\[0.4cm]
T(\theta)&=&S_{+-}^{+-}(\theta)=S_{-+}^{-+}(\theta)= S(\ri\pi-\theta)\nonumber\\[0.4cm]
t(\theta)&=&S_{+0}^{+0}(\theta)=S_{0+}^{0+}(\theta)=S_{-0}^{-0}(\theta)=S_{0-}^{0-}(\theta)=
\frac{\sinh(\frac{\theta}{n+2})\, \sinh(\frac{\ri\pi-\theta}{n+2})} 
{\sinh(\frac{2\ri \pi-\theta}{n+2})\, \sinh(\frac{\ri\pi+\theta}{n+2})} \nonumber\\[0.4cm]
r(\theta)&=&a(\ri\pi-\theta)=S_{+0}^{0+}(\theta)=
S_{0+}^{+0}(\theta)=S_{-0}^{0-}(\theta)=S_{0-}^{-0}(\theta)=S^{+-}_{00}(\ri\pi-\theta)=S^{-+}_{00}(\ri\pi-\theta)
\nonumber\\[0.4cm]
&=&
S_{+-}^{00}(\ri\pi-\theta)=S_{-+}^{00}(\ri\pi-\theta)
=
-\ri\ \frac{\sin(\frac{2\pi}{n+2})\, \sinh(\frac{\ri\pi-\theta}{n+2})} 
{\sinh(\frac{2\ri \pi-\theta}{n+2})\, \sinh(\frac{\ri\pi+\theta}{n+2})}
\\[0.4cm]
 R(\theta)&=& S_{-+}^{+-}(\theta)=S_{+-}^{-+}(\theta)=\frac{\sin(\frac{\pi}{n+2}) \sin(\frac{2\pi}{n+2})}
 { \sinh(\frac{2\ri\pi-\theta}{n+2}) \sinh(\frac{\ri\pi+\theta}{n+2})} \nonumber\\[0.4cm]
  \sigma(\theta)&=&  S_{00}^{00}(\theta)=S_{+0}^{+0}(\theta)+S_{-+}^{+-}(\theta) \nonumber
 \eea
As a $9\times 9$ matrix ${ \boldsymbol  S}^{\rm (2\mapsto 2)}$ satisfies the conditions
\bea
&&\big(\,{ \boldsymbol  S}^{\rm (2\mapsto 2)}\,\big)^\dag\;{ \boldsymbol  S}^{\rm (2\mapsto 2)}={\boldsymbol I}^{(2)} \ \ \ \ \ \  {\rm for} \ \  \Im m (\theta)=0\,\nonumber \\[0.4cm]
&&{ \det \boldsymbol  S}^{\rm (2\mapsto 2)}(\theta)=\bigg[\,\frac{\sinh^2(\frac{\ri\pi-\theta}{n+2})}
{\sinh^2(\frac{\ri\pi+\theta}{n+2})}\ \frac{\sinh(\frac{2\ri\pi+\theta}{n+2})}
{\sinh(\frac{2\ri\pi-\theta}{n+2})}\,\bigg]^4\,.
\eea


\begin{thebibliography}{99}





%\cite{Faddeev:1979gh} 
\bibitem{Faddeev:1979gh} 
  L.~D.~Faddeev, E.~K.~Sklyanin and L.~A.~Takhtajan,
  \emph{The quantum inverse problem method. 1},
  \emph{Theor.\ Math.\ Phys.}\  {\bf 40} (1980) 688
  [\emph{Teor.\ Mat.\ Fiz.}\  {\bf 40} (1979) 194].
  %%CITATION = TMPHA,40,688;%%
  %381 citations counted in INSPIRE as of 21 Jun 2017

%\cite{Sklyanin:1980ij}
%\bibitem{Sklyanin:1980ij} 
 % E.~K.~Sklyanin,
  %``Quantum version of the method of inverse scattering problem,''
 % J.\ Sov.\ Math.\  {\bf 19}, 1546 (1982)
%  [Zap.\ Nauchn.\ Semin.\  {\bf 95}, 55 (1980)].
%  doi:10.1007/BF01091462
  %%CITATION = doi:10.1007/BF01091462;%%
  %76 citations counted in INSPIRE as of 21 Jun 2017
  
% \bibitem{Baxter:1971sam}
%R.~J. Baxter,
% ``Generalized ferroelectric model on a square lattice'',
%{ Stud. Appl. Math.} {\bf 1} 51 (1971).
%51--69.
%%CITATION = SAPMB,1,51;%%.

  
  %\cite{Baxter:1972wg}
\bibitem{Baxter:1972wg} 
R.~J. Baxter, \emph{Partition function of the eight-vertex lattice model}, \emph{Ann. Phys.}\ {\bf 70} (1972) 193.
 % R.~J.~Baxter,
 % Stud. Appl. Math. {\bf 1}, 51 (1971);
  %``Eight vertex model in lattice statistics and one-dimensional anisotropic Heisenberg chain. 1. Some fundamental eigenvectors,''
 % Annals Phys.\  {\bf 76}, 1 (1973);
%  doi:10.1016/0003-4916(73)90439-9; 
%Annals Phys.\  {\bf 76}, 25 (1973).
%  doi:10.1016/0003-4916(73)90440-5
  %%CITATION = doi:10.1016/0003-4916(73)90439-9;%%
  %165 citations counted in INSPIRE as of 22 Jun 2017

%\cite{Baxter:1972wf}
%\bibitem{Baxter:1972wf} 
%  R.~J.~Baxter,
  %``Eight vertex model in lattice statistics and one-dimensional anisotropic Heisenberg chain. 2. Equivalence to a generalized ice-type lattice model,''
 % Annals Phys.\  {\bf 76}, 25 (1973).
 % doi:10.1016/0003-4916(73)90440-5
  %%CITATION = doi:10.1016/0003-4916(73)90440-5;%%
  %146 citations counted in INSPIRE as of 22 Jun 2017




  %\cite{Baxter:1982zz}
\bibitem{Baxter:1982zz} 
  R.~J.~Baxter,
  \emph{Exactly solved models in statistical mechanics}, Academic Press, London U.K. (1982).
  %%CITATION = INSPIRE-1120339;%%
  %133 citations counted in INSPIRE as of 22 Jun 2017

%\url{https://physics.anu.edu.au/theophys/_files/Exactly.pdf}


%\cite{Destri:1987ze}
\bibitem{Destri:1987ze} 
  C.~Destri and H.~J.~de Vega,
  \emph{Light cone lattice approach to fermionic theories in 2-$D$: the massive Thirring model},
  \emph{Nucl.\ Phys.}\  {\bf B 290} (1987) 363.
  %doi:10.1016/0550-3213(87)90193-3
  %%CITATION = doi:10.1016/0550-3213(87)90193-3;%%
  %108 citations counted in INSPIRE as of 23 Jun 2017







%\cite{Pohlmeyer:1975nb}
\bibitem{Pohlmeyer:1975nb} 
  K.~Pohlmeyer,
  \emph{Integrable Hamiltonian systems and interactions through quadratic constraints},
 \emph{Commun.\ Math.\ Phys.}\ {\bf 46} (1976) 207.
%  doi:10.1007/BF01609119
  %%CITATION = doi:10.1007/BF01609119;%%
  %563 citations counted in INSPIRE as of 23 Jun 2017


  %\cite{Zakharov:1973pp}
\bibitem{Zakharov:1973pp} 
  V.~E.~Zakharov and A.~V.~Mikhailov,
 \emph{Relativistically invariant two-dimensional models in field theory integrable by the inverse problem technique},
 \emph{Sov.\ Phys.\ JETP} {\bf 47} (1978) 1017
  [\emph{Zh.\ Eksp.\ Teor.\ Fiz.}\  {\bf 74} (1978) 1953].
  %%CITATION = SPHJA,47,1017;%%
  %343 citations counted in INSPIRE as of 23 Jun 2017

\bibitem{Polyakov:1983tt}
A.~M. Polyakov and P.~B. Wiegmann,
\emph{Theory of nonabelian goldstone bosons in two dimensions},
\emph{Phys. Lett.}\
  {\bf B 131} (1983) 121.
%%CITATION = PHLTA,B131,121;%%.

\bibitem{Faddeev:1985qu}
L.~D. Faddeev and N.~Yu. Reshetikhin,
\emph{Integrability of the
 principal chiral field model in (1+1)-dimension},
\emph{Ann. Phys.}\
 {\bf 167} (1986) 227.
%%CITATION = APNYA,167,227;%%.


\bibitem{ZZ79}
A.~B. Zamolodchikov and { Al}.~B. Zamolodchikov,
\emph{Factorized S-matrices in two dimensions as the exact solutions of 
certain relativistic quantum field theory models}, \emph{Ann. Phys.}\
{\bf 120} (1979)  253.

%\cite{Bytsko:1994ae}
\bibitem{Bytsko:1994ae}
  A.~G.~Bytsko,
  \emph{The zero-curvature representation for the nonlinear $O(3)$ sigma model},
\emph{Zapiski Nauchnykh Seminarov POMI} {\bf 215} (1994) 100, English translation in 
  \emph{J.\ Math.\ Sci.}\  {\bf 85} (1997) 1619
%  doi:10.1007/BF02355322
  [\href{http://arxiv.org/abs/hep-th/9403101}{\ttfamily hep-th/9403101}].
  %%CITATION = doi:10.1007/BF02355322;%%
  %3 citations counted in INSPIRE as of 26 Sep 2017

\bibitem{Bazhanov:1994ft}
  V.~V.~Bazhanov, S.~L.~Lukyanov and A.~B.~Zamolodchikov,
 \emph{Integrable structure of conformal field theory, quantum KdV theory and
  thermodynamic Bethe ansatz},
  \emph{Commun.\ Math.\ Phys.}\  {\bf 177} (1996) 381 
[\href{http://arxiv.org/abs/hep-th/9412229}{\ttfamily hep-th/9412229}].
%%CITATION = CMPHA,177,381;%%


%\cite{Fioravanti:2001bx}
\bibitem{Fioravanti:2001bx} 
  D.~Fioravanti and M.~Rossi,
\emph{A Braided Yang-Baxter algebra in a theory of two coupled lattice quantum KdV: Algebraic properties and ABA representations},
\emph{J.\ Phys.\ {\bf A}} {\bf 35} (2002) 3647 
%  doi:10.1088/0305-4470/35/16/306
  [\href{http://arxiv.org/abs/hep-th/0104002}{\ttfamily  hep-th/0104002}].
  %%CITATION = doi:10.1088/0305-4470/35/16/306;%%
  %25 citations counted in INSPIRE as of 01 Dec 2017


%\cite{Fioravanti:2002sq}
\bibitem{Fioravanti:2002sq} 
  D.~Fioravanti and M.~Rossi,
\emph{Exact conserved quantities on the cylinder 1: Conformal case},
  \emph{JHEP} {\bf 0307} (2003) 031 
%  doi:10.1088/1126-6708/2003/07/031
  [\href{http://arxiv.org/abs/hep-th/0211094}{\ttfamily hep-th/0211094}].
  %%CITATION = doi:10.1088/1126-6708/2003/07/031;%%
  %13 citations counted in INSPIRE as of 01 Dec 2017

%\cite{Fateev:1992tk}
\bibitem{Fateev:1992tk} 
  V.~A.~Fateev, E.~Onofri and A.~B.~Zamolodchikov,
 \emph{Integrable deformations of the $O(3)$ sigma model. The sausage model},
 \emph{Nucl.\ Phys.}\ {\bf B 406} (1993) 521.
  %doi:10.1016/0550-3213(93)90001-6
  %%CITATION = doi:10.1016/0550-3213(93)90001-6;%%
  %87 citations counted in INSPIRE as of 22 Jun 2017

%\cite{Lukyanov:2012zt}
\bibitem{Lukyanov:2012zt} 
  S.~L.~Lukyanov,
 \emph{The integrable harmonic map problem versus Ricci flow},
  \emph{Nucl.\ Phys.}\ {\bf B 865} (2012) 308 [\href{http://arxiv.org/abs/1205.3201}{\ttfamily hep-th/1205.3201}].
  %doi:10.1016/j.nuclphysb.2012.08.002
    %\href{http://arxiv.org/abs/1205.3201}{{\ttfamily }}.
 % [arXiv:hep-th/1205.3201].
  %%CITATION = doi:10.1016/j.nuclphysb.2012.08.002;%%
  %14 citations counted in INSPIRE as of 23 Jun 2017

 %\cite{Sklyanin:1979gh}
\bibitem{Sklyanin:1979gh}
E.~K. Sklyanin, \emph{On the complete integrability of the Landau-Lifshitz equation},
LOMI E-79-3 (1980).

%\url{https://www.researchgate.net/publication/247031108_On_the_complete_integrability_of_the_Landau-%Lifshitz_equation}%\cite{Faddeev:1987ph}
\bibitem{Faddeev:1987ph} 
  L.~D.~Faddeev and L.~A.~Takhtajan,
  \emph{Hamiltonian methods in the theory of solitons}, Springer-Verlag,
  Berlin (1987).
  % (Springer series in Soviet mathematics)
   %47 citations counted in INSPIRE as of 23 Jun 2017
%  \cite{Bazhanov:1994ft,Bazhanov:1996dr,Bazhanov:1998dq}
%\cite{Bazhanov:1994ft}

%\cite{Bazhanov:1996dr}
\bibitem{Bazhanov:1996dr}
  V.~V.~Bazhanov, S.~L.~Lukyanov and A.~B.~Zamolodchikov,
 \emph{Integrable structure of conformal field theory II. Q-operator and DDV equation},
  \emph{Commun.\ Math.\ Phys.}\  {\bf 190} (1997) 247
 % \href{http://arxiv.org/abs/hep-th/9604044}{
  [\href{http://arxiv.org/abs/hep-th/9604044}{\ttfamily hep-th/9604044}].
%}

  %[arXiv:hep-th/9604044].





%\cite{Bazhanov:1998dq}
\bibitem{Bazhanov:1998dq}
  V.~V.~Bazhanov, S.~L.~Lukyanov and A.~B.~Zamolodchikov,
 \emph{Integrable structure of conformal field theory III. The Yang-Baxter relation},
  \emph{Commun.\ Math.\ Phys.}\  {\bf 200} (1999) 297
 %  \href{http://arxiv.org/abs/hep-th/9805008}{
[\href{http://arxiv.org/abs/hep-th/9805008}{\ttfamily hep-th/9805008}].
%}
  %[arXiv:hep-th/9805008].
  
 %\cite{Voros:1992}
\bibitem{Voros:1992}
A.~Voros, \emph{Spectral Zeta Functions},
\emph{Adv.\ Stud.\ Pure\ Math.}\ {\bf 21} (1992) 327.



 %\cite{Voros:1999}
\bibitem{Voros:1999}
A.~Voros, \emph{An exact solution method for 1D polynomial Schr\"{o}dinger equations},
 \emph{J.\ Phys.}\ {\bf A 32} (1999) 5993 [\href{http://arxiv.org/abs/math-ph/9902016}{\ttfamily math-ph/9902016}].
%   \href{http://arxiv.org/abs/math-ph/9902016}{}.
  % [arXiv:math-ph/9902016]









%\cite{Voros:1999bz}
%\bibitem{Voros:1999bz}
%  A.~Voros,
  %``Exact resolution method for general 1D polynomial Schr\"odinger equation,''
  %J.\ Phys.\ A  {\bf 32}, 5993 (1999) [math-ph/9902016].
  %%CITATION = JPAGB,A32,5993;%%
  
  %\cite{Dorey:1998pt}
\bibitem{Dorey:1998pt}
  P.~Dorey and R.~Tateo,
 \emph{Anharmonic oscillators, the thermodynamic Bethe ansatz, and nonlinear
  integral equations},
  \emph{J.\ Phys.}\  {\bf A 32} (1999) L419 [\href{http://arxiv.org/abs/hep-th/9812211}{\ttfamily hep-th/9812211}].
 % \href{http://arxiv.org/abs/hep-th/9812211}{}.
  %arXiv:hep-th/9812211].
  %%CITATION = JPAGB,A32,L419;%%
  
 
  
  \bibitem{Bazhanov:1998wj}
  V.~V.~Bazhanov, S.~L.~Lukyanov and A.~B.~Zamolodchikov,
  \emph{Spectral determinants for Schroedinger equation and Q-operators of
  conformal field theory},
  \emph{J.\ Stat.\ Phys.}\ {\bf 102} (2001) 567 [\href{http://arxiv.org/abs/hep-th/9812247}{\ttfamily hep-th/9812247}].
   %\href{http://arxiv.org/abs/hep-th/9812247}{{\ttfamily [arXiv:hep-th/9812247]}}.
  %[arXiv:hep-th/9812247].

%\cite{Suzuki:1999hu}
\bibitem{Suzuki:1999hu} 
  J.~Suzuki,
  \emph{Functional relations in Stokes multipliers and solvable models related to $U_q(A^{(1)}_n)$},
 \emph{J.\ Phys.}\ {\bf A 33} (2000) 3507 [\href{http://arxiv.org/abs/hep-th/9910215}{\ttfamily hep-th/9910215}].
 % doi:10.1088/0305-4470/33/17/308
% \href{http://arxiv.org/abs/hep-th/9910215}{{\ttfamily [arXiv:hep-th/9910215]}}.
 % [arXiv:hep-th/9910215].
  %%CITATION = doi:10.1088/0305-4470/33/17/308;%%
  %25 citations counted in INSPIRE as of 27 Jun 2017
%\cite{Bazhanov:2003ni}

\bibitem{Bazhanov:2003ni}
  V.~V.~Bazhanov, S.~L.~Lukyanov and A.~B.~Zamolodchikov,
  \emph{Higher-level eigenvalues of Q-operators and Schroedinger equation},
 \emph{Adv.\ Theor.\ Math.\ Phys.}\  {\bf 7} (2003) 711 [\href{http://arxiv.org/abs/hep-th/0307108}{\ttfamily hep-th/0307108}].
  %\href{http://arxiv.org/abs/hep-th/0307108}{{\ttfamily [arXiv:hep-th/0307108]}}.
 % [arXiv:hep-th/0307108].

% \cite{AlZ}
\bibitem{AlZ}
Al.~B.~Zamolodchikov,
\emph{Generalized Mathieu equation and Liouville TBA}
in
 \emph{Quantum Field Theories in Two Dimensions, Collected works of Alexei Zamolodchikov vol.\,2},
edited by  A. Belavin, Ya. Pugai and A. Zamolodchikov,
World Scientific (2012).

%\cite{Lukyanov:2010rn,Lukyanov:2013wra}}
\bibitem{Lukyanov:2010rn}
  S.~L.~Lukyanov and A.~B.~Zamolodchikov,
 \emph{Quantum sine(h)-Gordon model and classical integrable equations},
  \emph{JHEP} {\bf 1007} (2010) 008 [\href{http://arxiv.org/abs/1003.5333}{\ttfamily arXiv:1003.5333}].
   %[arXiv:hep-th/1003.5333].
  %\href{http://arxiv.org/abs/1003.5333}{{\ttfamily [arXiv:1003.5333]}}.




%\cite{Dorey:2007zx}
\bibitem{Dorey:2007zx} 
  P. Dorey, C. Dunning and R. Tateo,
  \emph{The ODE/IM correspondence},
  \emph{J.\ Phys.}\ {\bf A\,40} (2007) R205 
[\href{http://arxiv.org/abs/hep-th/0703066}{\ttfamily hep-th/0703066}].
 % doi:10.1088/1751-8113/40/32/R01
 % [arXiv:hep-th/0703066].
 % \href{http://arxiv.org/abs/hep-th/0703066}{{\ttfamily [arXiv:hep-th/0703066]}}.
  %%CITATION = doi:10.1088/1751-8113/40/32/R01;%%
  %121 citations counted in INSPIRE as of 27 Jun 2017


%\cite{Dorey:2000ma}
%\bibitem{Dorey:2000ma} 
%  P.~Dorey, C.~Dunning and R.~Tateo,
  %``Differential equations for general SU(n) Bethe ansatz systems,''
 % J.\ Phys.\ A {\bf 33}, 8427 (2000)
 % doi:10.1088/0305-4470/33/47/308
%  [hep-th/0008039].
  %%CITATION = doi:10.1088/0305-4470/33/47/308;%%
  %39 citations counted in INSPIRE as of 27 Jun 2017


%\cite{Dorey:2012bx}
\bibitem{Dorey:2012bx} 
  P.~Dorey, S.~Faldella, S.~Negro and R.~Tateo,
  \emph{The Bethe ansatz and the Tzitzeica-Bullough-Dodd equation},
  \emph{Phil.\ Trans.\ Roy.\ Soc.\ Lond.}\  {\bf A 371} (2013) 20120052 [\href{http://arxiv.org/abs/1209.5517}{\ttfamily arXiv:1209.5517}].
%  doi:10.1098/rsta.2012.0052
%\href{http://arxiv.org/abs/1209.5517}{{\ttfamily [arXiv:1209.5517]}}.
 % [arXiv:1209.5517 [math-ph]].
  %%CITATION = doi:10.1098/rsta.2012.0052;%%
  %14 citations counted in INSPIRE as of 26 Jun 2017
%\cite{Masoero:2015rcz}

%\cite{Adamopoulou:2014fca}
\bibitem{Adamopoulou:2014fca} 
  P.~Adamopoulou and C.~Dunning,
  \emph{Bethe ansatz equations for the classical $A_n^{(1)}$ affine Toda field theories},
  \emph{J.\ Phys.}\  {\bf A 47} (2014) 205205  [\href{http://arxiv.org/abs/1401.1187}{\ttfamily arXiv:1401.1187}].
 % doi:10.1088/1751-8113/47/20/205205
 % [arXiv:1401.1187 [math-ph]].
 % \href{http://arxiv.org/abs/1401.1187}{{\ttfamily [arXiv:1401.1187]}}.
  %%CITATION = doi:10.1088/1751-8113/47/20/205205;%%
  %9 citations counted in INSPIRE as of 27 Jun 2017

%\cite{Ito:2015nla}
\bibitem{Ito:2015nla}
K.~Ito and C.~Locke, 
\emph{ODE/IM correspondence and Bethe ansatz for affine Toda
 field equations}, \emph{Nucl. Phys.}\ {\bf B 896} (2015) 763 [\href{http://arxiv.org/abs/1502.00906}{\tt arXiv:1502.00906}].
 % \href{http://dx.doi.org/10.1016/j.nuclphysb.2015.05.016}{{ Nucl. Phys.}
 % {\bfseries B 896}, 763  (2015) }
  %763--778
%\href{http://arxiv.org/abs/1502.00906}{{\ttfamily [arXiv:1502.00906]}}.
%%CITATION = ARXIV:1502.00906;%%.

\bibitem{Masoero:2015rcz1} 
  D.~Masoero, A.~Raimondo and D.~Valeri,
  \emph{Bethe ansatz and the spectral theory of affine lie algebra-valued connections I. The simply-laced case},
  \emph{Commun.\ Math.\ Phys.}\  {\bf 344} (2016) 719 [\href{http://arxiv.org/abs/1501.07421}{\tt arXiv:1501.07421}].

\bibitem{Masoero:2015rcz2} 
  D.~Masoero, A.~Raimondo and D.~Valeri,
  \emph{Bethe ansatz and the spectral theory of affine lie algebra-valued connections II: The non simply-laced case},
 \emph{Commun.\ Math.\ Phys.}\  {\bf 349} (2017) 1063 [\href{http://arxiv.org/abs/1511.00895}{\tt arXiv:1511.00895}].
 % doi:10.1007/s00220-016-2744-2
  %\href{http://arxiv.org/abs/1511.00895}{{\ttfamily [arXiv:1511.00895]}}.
  %[arXiv:1511.00895 [math-ph]].
  %%CITATION = doi:10.1007/s00220-016-2744-2;%%
  %5 citations counted in INSPIRE as of 27 Jun 2017
  
%\cite{Ito:2016qzt}
\bibitem{Ito:2016qzt} 
  K.~Ito and H.~Shu,
 \emph{ODE/IM correspondence for modified $B_2^{(1)}$ affine Toda field equation},
  \emph{Nucl.\ Phys.}\  {\bf B 916} (2017) 414 [\href{http://arxiv.org/abs/1605.04668}{\tt arXiv:1605.04668}].
 % doi:10.1016/j.nuclphysb.2017.01.009
  % \href{http://arxiv.org/abs/1605.04668}{{\ttfamily [arXiv:1605.04668]}}.
 % [arXiv:1605.04668 [hep-th]].
  %%CITATION = doi:10.1016/j.nuclphysb.2017.01.009;%%
  %1 citations counted in INSPIRE as of 27 Jun 2017
  
  %\cite{Babenko:2017fmu}
\bibitem{Babenko:2017fmu} 
  C.~Babenko and F.~Smirnov,
  \emph{Suzuki equations and integrals of motion for supersymmetric CFT},  \href{http://arxiv.org/abs/1706.03349}{\tt arXiv:1706.03349}.
%    \href{http://arxiv.org/abs/1706.03349}{{\ttfamily [arXiv:1706.03349]}}.
 % arXiv:1706.03349 [hep-th].
  %%CITATION = ARXIV:1706.03349;%%
  
%\cite{Lukyanov:2010rn,Dorey:2012bx,Lukyanov:2013wra,Bazhanov:2013cua}

%\cite{Hamilton}
\bibitem{Hamilton}
R. S.  Hamilton, 
\emph{The Ricci flow on surfaces},
Mathematics and general relativity, Contemporary Mathematics vol. 71,
 Amer. Math. Soc., Providence RI, (1988) 237.
 %, pp. 237-262.
%Mathematics and general relativity (Santa Cruz, CA, 1986) Contemp. Math., vol. 71,
 %Amer. Math. Soc., Providence, RI, (1988) 237.
 %, pp. 237-262.

  %\cite{Elitzur:1991cb}
\bibitem{Elitzur:1991cb} 
  S.~Elitzur, A.~Forge and E.~Rabinovici,
  \emph{Some global aspects of string compactifications},
 \emph{Nucl.\ Phys.}\  {\bf B 359} (1991) 581.
  %doi:10.1016/0550-3213(91)90073-7
  %%CITATION = doi:10.1016/0550-3213(91)90073-7;%%
  %230 citations counted in INSPIRE as of 30 May 2016


%\cite{Witten:1991yr}
\bibitem{Witten:1991yr} 
  E.~Witten,
  \emph{On string theory and black holes},
  \emph{Phys.\ Rev.}\ {\bf D 44} (1991) 314.
  %doi:10.1103/PhysRevD.44.314
  %%CITATION = doi:10.1103/PhysRevD.44.314;%%
  %1059 citations counted in INSPIRE as of 19 May 2016
  


%\cite{Str79,KR87a}
\bibitem{Str79}
Y.~G. Stroganov, 
\emph{A new calculation method for partition functions in some
  lattice models}, 
  \emph{ Phys. Lett.} {\bf A 74} (1979) 116. 
  %116--118.
   \bibitem{BP82}
R.~J. Baxter and P.~A. Pearce, 
\emph{Hard hexagons: interfacial tension and         
 correlation length}, 
 \emph{J. Phys.} {\bf A 15} (1982) 897.
  % 897--910.
  \bibitem{KR87a}
A.~N. Kirillov and N.~Yu. Reshetikhin,
\emph{Exact solution of the integrable
 {$XXZ$} Heisenberg model with arbitrary spin. {I}. The ground state and
 the excitation spectrum}, 
 \emph{J. Phys.}\ {\bfseries A 20} (1987) 1565.
  %1565--1585.
  
  %\cite{Lukyanov:2003nj}
\bibitem{Lukyanov:2003nj} 
  S.~L.~Lukyanov, E.~S.~Vitchev and A.~B.~Zamolodchikov,
  \emph{Integrable model of boundary interaction: the paperclip},
  \emph{Nucl.\ Phys.}\  {\bf B 683} (2004) 423 [\href{http://arxiv.org/abs/hep-th/0312168}{\tt hep-th/0312168}].
  % \href{http://arxiv.org/abs/hep-th/0312168}{{\ttfamily [arXiv:hep-th/0312168]}}.
 % doi:10.1016/j.nuclphysb.2004.02.010
  %[hep-th/0312168].
  %%CITATION = doi:10.1016/j.nuclphysb.2004.02.010;%%
  %47 citations counted in INSPIRE as of 09 Jun 2017




%\cite{Lukyanov:2005nr}
\bibitem{Lukyanov:2005nr} 
  S.~L.~Lukyanov, A.~M.~Tsvelik and A.~B.~Zamolodchikov,
  \emph{Paperclip at $\theta = \pi$},
  \emph{Nucl.\ Phys.}\  {\bf B 719} (2005) 103 [\href{http://arxiv.org/abs/hep-th/0501155}{\tt hep-th/0501155}].
  %doi:10.1016/j.nuclphysb.2005.04.040
   %\href{http://arxiv.org/abs/hep-th/0501155}{{\ttfamily [hep-th/0501155]}}.
  %[hep-th/0501155].
  %%CITATION = doi:10.1016/j.nuclphysb.2005.04.040;%%
  %7 citations counted in INSPIRE as of 09 Jun 2017
  
  
  
  %\cite{Lukyan}
  \bibitem{Lukyan}
  S.~ L.~Lukyanov and A.~B.~Zamolodchikov,  \emph{Integrability in 2D field theory/sigma models},
 Lecture notes  for the 2016 Les Houches school, to appear.
 
 %\cite{Bakas:1991fs}
\bibitem{Bakas:1991fs} 
  I.~Bakas and E.~Kiritsis,
  \emph{Beyond the large N limit: non-linear $W_\infty$ as symmetry of the SL(2,R) / U(1) coset model},
 \emph{Int.\ J.\ Mod.\ Phys.}\ {\bf A 7S1A} (1992) 55 [\href{http://arxiv.org/abs/hep-th/9109029}{\tt hep-th/9109029}].
  %[Int.\ J.\ Mod.\ Phys.\ A {\bf 7}, 55 (1992)]
%  doi:10.1142/S0217751X92003720
 %\href{http://arxiv.org/abs/hep-th/9109029}{{\ttfamily [arXiv:hep-th/9109029]}}.
 % [arXiv:hep-th/9109029].
  %%CITATION = doi:10.1142/S0217751X92003720;%%
  %63 citations counted in INSPIRE as of 19 May 2016
  \bibitem{Dijkgraaf:1991ba} 
  R.~Dijkgraaf, H.~L.~Verlinde and E.~P.~Verlinde,
  \emph{String propagation in a black hole geometry},
  \emph{Nucl.\ Phys.}\  {\bf B 371} (1992) 269.
 % doi:10.1016/0550-3213(92)90237-6
  %%CITATION = doi:10.1016/0550-3213(92)90237-6;%%
  %394 citations counted in INSPIRE as of 29 May 2016
  
  %\cite{ZAM}
\bibitem{ZAM}
A.~B.~Zamolodchikov and Al.~B.~Zamolodchikov, unpublished notes (1995).\footnote{Although Zamolodchikov's notes
have never been published, they were broadly distributed  within the scientific community.}
  %\cite{Dijkgraaf:1991ba}

  
  

  


  
 %\cite{Hernandez:1989zj}
%\bibitem{Hernandez:1989zj} 
 % O.~F.~Hernandez,
  %``Feigin-fuchs Bosonization of Lykken Parafermions and SU(1,1) {Kac-Moody} Algebras,''
 % Phys.\ Lett.\ B {\bf 233}, 355 (1989).
 % doi:10.1016/0370-2693(89)91322-1
  %%CITATION = doi:10.1016/0370-2693(89)91322-1;%%
  %16 citations counted in INSPIRE as of 28 Jun 2017
  
  
  
  
  
  
  
  
  

  
   %\cite{Gerasimov:1989mz}
\bibitem{Gerasimov:1989mz} 
  A.~Gerasimov, A.~Marshakov and A.~Morozov,
  \emph{Free field representation of parafermions and related coset models},
  \emph{Nucl.\ Phys.}\ {\bf B 328} (1989) 664.
  %[Theor.\ Math.\ Phys.\  {\bf 83}, 466 (1990)]
 % [Teor.\ Mat.\ Fiz.\  {\bf 83}, 186 (1990)].
 % doi:10.1016/0550-3213(89)90224-1
  %%CITATION = doi:10.1016/0550-3213(89)90224-1;%%
  %58 citations counted in INSPIRE as of 28 Jun 2017
 
  %\cite{Fateev:1985mm}
\bibitem{Fateev:1985mm} 
  V.~A.~Fateev and A.~B.~Zamolodchikov,
  \emph{Nonlocal (parafermion) currents in two-dimensional conformal quantum field theory and self-dual critical points in $Z_N$-symmetric statistical systems},
  \emph{Sov.\ Phys.\ JETP} {\bf 62} (1985) 215
  [\emph{Zh.\ Eksp.\ Teor.\ Fiz.}\  {\bf 89} (1985) 380].
  %%CITATION = SPHJA,62,215;%%
  %660 citations counted in INSPIRE as of 22 Jun 2017
  

%\cite{Fateev:1982wi}
\bibitem{Fateev:1982wi} 
  V.~A.~Fateev and A.~B.~Zamolodchikov,
 \emph{Self-dual solutions of the star triangle relations in $Z_N$-models},
 \emph{Phys.\ Lett.}\ {\bf A 92} (1982) 37.
 % doi:10.1016/0375-9601(82)90736-8
  %%CITATION = doi:10.1016/0375-9601(82)90736-8;%%
  %118 citations counted in INSPIRE as of 22 Jun 2017
  
  
  %\cite{Felder:1988zp}
\bibitem{Felder:1988zp} 
  G.~Felder,
  \emph{BRST approach to minimal models},
  \emph{Nucl.\ Phys.}\  {\bf B 317} (1989) 215
 [\emph{Erratum ibid} \emph{Nucl.\ Phys.}\  {\bf B 324} (1989) 548].
 % doi:10.1016/0550-3213(89)90481-1, 10.1016/0550-3213(89)90568-3
  %%CITATION = doi:10.1016/0550-3213(89)90481-1, 10.1016/0550-3213(89)90568-3;%%
  %358 citations counted in INSPIRE as of 28 Jun 2017
  
%\cite{Lukyanov:2003nj,Lukyanov:2005nr}






%\bibitem{Iz}
%A. G. Izergin and V.  E. Korepin, ``Lattice regularization of two-dimensional quantum field theory models,'' (in Russian) 
%Zap. Nauchn. Sem. LOMI, 120, pp.75-91,
%``Nauka'', Leningrad, 1982. 
%\cite{Izergin:1981mc}
\bibitem{Izergin:1981mc} 
  A.~G.~Izergin and V.~E.~Korepin,
  \emph{The lattice quantum {sine-Gordon} model},
 \emph{Lett.\ Math.\ Phys.}\  {\bf 5} (1981) 199.
  %doi:10.1007/BF00420699
  %%CITATION = doi:10.1007/BF00420699;%%
  %36 citations counted in INSPIRE as of 28 Jun 2017
  

%\bibitem{Skl83}
%E.~K. Sklyanin, 
%``Some algebraic structures connected with the {Y}ang-{B}axter
  %equation. {R}epresentations of a quantum algebra'',  
%  Func. Anal. Appl.
% {\bfseries 17} no.~4, (1983) 273--284.

%\cite{Sklyanin:1983ig}
\bibitem{Sklyanin:1983ig} 
  E.~K.~Sklyanin,
  \emph{Some algebraic structures connected with the Yang-Baxter equation. Representations of quantum algebras},
  \emph{Funct.\ Anal.\ Appl.}\  {\bf 17} (1983) 273
  [\emph{Funkt.\ Anal.\ Pril.}\  {\bf 17} (1983) 34].
%  doi:10.1007/BF01076718
  %%CITATION = doi:10.1007/BF01076718;%%
  %140 citations counted in INSPIRE as of 28 Jun 2017
  


%\cite{Bazhanov:1989nc}
\bibitem{Bazhanov:1989nc} 
  V. V. Bazhanov and Y. G. Stroganov,
  \emph{Chiral Potts model as a descendant of the six vertex model},
 \emph{J.\ Stat.\ Phys.}\  {\bf 59} (1990) 799.
 % doi:10.1007/BF01025851
  %%CITATION = doi:10.1007/BF01025851;%%
  %130 citations counted in INSPIRE as of 22 Jun 2017


%\cite{Baxter:1999mn}
\bibitem{Baxter:1999mn} 
  R. J. Baxter, V. V. Bazhanov and J. H. H. Perk,
  \emph{Functional relations for transfer matrices of the chiral Potts model},
  \emph{Int.\ J.\ Mod.\ Phys.}\  {\bf B 4} (1990) 803.
  %doi:10.1142/S0217979290000395
  %%CITATION = doi:10.1142/S0217979290000395;%%
  %79 citations counted in INSPIRE as of 22 Jun 2017



%\cite{Sklyanin:1995bm}
\bibitem{Sklyanin:1995bm} 
  E.~K.~Sklyanin,
  \emph{Separation of variables --- new trends},
 \emph{Prog.\ Theor.\ Phys.\ Suppl.}\  {\bf 118} (1995) 35 [\href{http://arxiv.org/abs/solv-int/9504001}{\tt solv-int/9504001}].
  %doi:10.1143/PTPS.118.35
  % \href{http://arxiv.org/abs/solv-int/9504001}{{\ttfamily [arXiv:solv-int/9504001]}}.
  %[solv-int/9504001].
  %%CITATION = doi:10.1143/PTPS.118.35;%%
  %157 citations counted in INSPIRE as of 21 Jun 2017
  
 %\cite{Lukyanov:2013wra,Bazhanov:2013cua}
  %\cite{Lukyanov:2013wra}
\bibitem{Lukyanov:2013wra} 
  S.~L.~Lukyanov,
 \emph{ODE/IM correspondence for the Fateev model},
  \emph{JHEP} {\bf 1312} (2013) 012 [\href{http://arxiv.org/abs/1303.2566}{\tt arXiv:1303.2566}].
 % doi:10.1007/JHEP12(2013)012
  %\href{http://arxiv.org/abs/1303.2566}{{\ttfamily [arXiv:1303.2566]}}.
 % [arXiv:1303.2566 [hep-th]].
  %%CITATION = doi:10.1007/JHEP12(2013)012;%%
  %11 citations counted in INSPIRE as of 09 Jun 2017

%\cite{Bazhanov:2013cua}
\bibitem{Bazhanov:2013cua} 
  V.~V.~Bazhanov and S.~L.~Lukyanov,
  \emph{Integrable structure of quantum field theory: classical flat connections versus quantum stationary states},
  \emph{JHEP} {\bf 1409} (2014) 147 [\href{http://arxiv.org/abs/1310.4390}{\tt arXiv:1310.4390}].
 %  \href{http://arxiv.org/abs/1310.4390}{{\ttfamily [arXiv:1310.4390]}}.
 % doi:10.1007/JHEP09(2014)147
 % [arXiv:1310.4390 [hep-th]].
  %%CITATION = doi:10.1007/JHEP09(2014)147;%%
  %11 citations counted in INSPIRE as of 09 Jun 2017

 
 %\cite{Fioravanti:2004cz}
\bibitem{Fioravanti:2004cz} 
  D.~Fioravanti,
  \emph{Geometrical loci and CFTs via the Virasoro symmetry of the mKdV-SG hierarchy: An excursus},
 \emph{Phys.\ Lett.}\ B {\bf 609} (2005) 173 [\href{http://arxiv.org/abs/hep-th/0408079}{\tt hep-th/0408079}].
  % \href{http://arxiv.org/abs/hep-th/0408079}{{\ttfamily [arXiv:hep-th/0408079]}}.
  %[arXiv:hep-th/0408079].
  %%CITATION = HEP-TH/0408079;%%

  
 %\cite{Feigin:2007mr}
\bibitem{Feigin:2007mr} 
  B.~Feigin and E.~Frenkel, 
\emph{Quantization of soliton systems and Langlands duality},
\emph{Adv. Stud. Pure Math.}\ {\bf 61} (2011) 185 [\href{http://arxiv.org/abs/0705.2486}{\tt arXiv:0705.2486}].
%  \href{http://arxiv.org/abs/0705.2486}{{\ttfamily [arXiv:0705.2486]}}.
 % [arXiv:math.QA/0705.2486].
  %%CITATION = ARXIV:0705.2486;%%
  %6 citations counted in INSPIRE as of 01 Sep 2013
  
  
 % \cite{Klummpe:1991vs,Destri:1994bv}
  %\cite{Klummpe:1991vs}
\bibitem{Klummpe:1991vs} 
  A.~Klumper, M.~T.~Batchelor and P.~A.~Pearce,
  \emph{Central charges of the 6- and 19- vertex models with twisted
  boundary conditions},
  \emph{J.\ Phys.}\ {\bf A 24} (1991) 3111.
  %%CITATION = JPHGB,A24,3111;%%
  %91 citations counted in INSPIRE as of 28 Aug 2014
  
  %\cite{Destri:1992ey}
%\bibitem{Destri:1992ey} 
  %C.~Destri and H.~J.~de Vega,
  %``New approach to thermal Bethe ansatz,''
 % hep-th/9203064.
  %%CITATION = HEP-TH/9203064;%%
  %14 citations counted in INSPIRE as of 28 Jun 2017
  
  %\cite{Destri:1994bv}
\bibitem{Destri:1994bv1} 
  C.~Destri and H.~J.~De Vega, 
\emph{New approach to thermal Bethe ansatz},
\emph{Phys. Rev. Lett.}\ {\bf 69} (1992) 2313
[\href{http://arxiv.org/abs/hep-th/9203064}{\tt hep-th/9203064}].
%  \href{http://arxiv.org/abs/hep-th/9203064}{{\ttfamily [arXiv:hep-th/9203064]}}; 
  %%CITATION = HEP-TH/9407117;%%
  %132 citations counted in INSPIRE as of 28 Aug 2014
  
  %\cite{Destri:1994bv}
\bibitem{Destri:1994bv2} 
  C.~Destri and H.~J.~De Vega, 
  \emph{Unified approach to thermodynamic Bethe ansatz and finite size
  corrections for lattice models and field theories},
  \emph{Nucl.\ Phys.}\  {\bf B 438} (1995) 413 [\href{http://arxiv.org/abs/hep-th/9407117}{\tt hep-th/9407117}].
   % \href{http://arxiv.org/abs/hep-th/9407117}{{\ttfamily [arXiv:hep-th/9407117]}}.
 % [hep-th/9407117].
  %%CITATION = HEP-TH/9407117;%%
  %132 citations counted in INSPIRE as of 28 Aug 2014



  %\cite{Zamolodchikov:1995aa}
\bibitem{Zamolodchikov:1995aa} 
  A.~B.~Zamolodchikov and Al.~B.~Zamolodchikov,
  \emph{Structure constants and conformal bootstrap in Liouville field theory},
  \emph{Nucl.\ Phys.}\  {\bf B 477} (1996) 577 [\href{http://arxiv.org/abs/hep-th/9506136}{\tt hep-th/9506136}].
 % \href{http://arxiv.org/abs/hep-th/9506136}{{\ttfamily [arXiv:hep-th/9506136]}}.
  %doi:10.1016/0550-3213(96)00351-3
 % [hep-th/9506136].
  %%CITATION = doi:10.1016/0550-3213(96)00351-3;%%
  %495 citations counted in INSPIRE as of 28 Jun 2017

%\cite{Fateev:1979dc,Bukhvostov:1980sn,Luscher:1981tq}
%\cite{Fateev:1979dc}
\bibitem{Fateev:1979dc} 
  V.~A.~Fateev, I.~V.~Frolov and A.~S.~Schwarz,
  \emph{Quantum fluctuations of instantons in the nonlinear $\sigma$ model},
  \emph{Nucl.\ Phys.}\  {\bf B 154} (1979) 01.
  %doi:10.1016/0550-3213(79)90367-5
  %%CITATION = doi:10.1016/0550-3213(79)90367-5;%%
  %173 citations counted in INSPIRE as of 28 Jun 2017

%\cite{Bukhvostov:1980sn}
\bibitem{Bukhvostov:1980sn} 
  A.~P.~Bukhvostov and L.~N.~Lipatov,
  \emph{Instanton --- anti-instanton interaction in the $O(3)$ nonlinear $\Sigma$ model and an exactly soluble fermion theory},
  \emph{Nucl.\ Phys.}\  {\bf B 180} (1981) 116.
  %[Pisma Zh.\ Eksp.\ Teor.\ Fiz.\  {\bf 31}, 138 (1980)].
 % doi:10.1016/0550-3213(81)90157-7
  %%CITATION = doi:10.1016/0550-3213(81)90157-7;%%
  %36 citations counted in INSPIRE as of 28 Jun 2017
  
  %\cite{Luscher:1981tq}
\bibitem{Luscher:1981tq} 
  M.~ L${\rm \ddot u}$scher,
  \emph{Does the topological susceptibility in lattice sigma models scale according to the perturbative renormalization group?},
 \emph{Nucl.\ Phys.}\ {\bf B 200} (1982) 61.
  %doi:10.1016/0550-3213(82)90058-X
  %%CITATION = doi:10.1016/0550-3213(82)90058-X;%%
  %131 citations counted in INSPIRE as of 28 Jun 2017

%\cite{Luscher:1985dn}
\bibitem{Luscher:1985dn1} 
  M.~L${\rm \ddot u}$scher,
  \emph{Volume dependence of the energy spectrum in massive quantum field theories. I. Stable particle states},
  \emph{Commun.\ Math.\ Phys.}\  {\bf 104} (1986) 177.
  %doi:10.1007/BF01211589
  %%CITATION = doi:10.1007/BF01211589;%%
  %655 citations counted in INSPIRE as of 28 Jun 2017

\bibitem{Luscher:1985dn2} 
  M.~L${\rm \ddot u}$scher,
  \emph{Volume dependence of the energy spectrum in massive quantum field theories. II. Scattering states},
  \emph{Commun.\ Math.\ Phys.}\ {\bf 105} (1986) 153.
  %doi:10.1007/BF01211589
  %%CITATION = doi:10.1007/BF01211589;%%
  %655 citations counted in INSPIRE as of 28 Jun 2017



  %\cite{BalogN}
  \bibitem{BalogN}
  C.~Ahn, J.~Balog and F.~Ravanini,
  \emph{NLIE for the sausage model}, \href{http://arxiv.org/abs/1701.08933}{\tt arXiv:1701.08933}.
 % \href{http://arxiv.org/abs/1701.08933}{{\ttfamily [arXiv:1701.08933]}}.
 % preprint 2017  [arXiv:hep-th/1701.08933]
  
  %\cite{Polyakov:1975rr}
\bibitem{Polyakov:1975rr} 
  A.~M.~Polyakov,
  \emph{Interaction of goldstone particles in two-dimensions. Applications to ferromagnets and massive Yang-Mills fields},
  \emph{Phys.\ Lett.}\  {\bf  B 59} (1975) 79.
  %doi:10.1016/0370-2693(75)90161-6
  %%CITATION = doi:10.1016/0370-2693(75)90161-6;%%
  %591 citations counted in INSPIRE as of 28 Jun 2017
  
  %\cite{Hikami:1977vr}
\bibitem{Hikami:1977vr} 
  S.~Hikami and E.~Brezin,
  \emph{Three loop calculations in the two-dimensional nonlinear sigma model},
  \emph{J.\ Phys.}\ {\bf A 11} (1978) 1141.
  %doi:10.1088/0305-4470/11/6/015
  %%CITATION = doi:10.1088/0305-4470/11/6/015;%%
  %86 citations counted in INSPIRE as of 28 Jun 2017


%\cite{Hasenfratz:1990zz}
\bibitem{Hasenfratz:1990zz} 
  P.~Hasenfratz, M.~Maggiore and F.~Niedermayer,
  \emph{The exact mass gap of the $O(3)$ and $O(4)$ nonlinear sigma models in $d=2$},
 \emph{Phys.\ Lett.}\  {\bf B 245} (1990) 522.
%  doi:10.1016/0370-2693(90)90685-Y
  %%CITATION = doi:10.1016/0370-2693(90)90685-Y;%%
  %190 citations counted in INSPIRE as of 27 Jun 2017

%\cite{Luscher:1991wu}
\bibitem{Luscher:1991wu} 
  M.~L$\ddot{\rm u}$scher, P.~Weisz and U.~Wolff,
 \emph{A numerical method to compute the running coupling in asymptotically free theories},
 \emph{Nucl.\ Phys.}\  {\bf B 359} (1991) 221.
 % doi:10.1016/0550-3213(91)90298-C
  %%CITATION = doi:10.1016/0550-3213(91)90298-C;%%
  %290 citations counted in INSPIRE as of 27 Jun 2017

%\cite{Shin:1996gi}
\bibitem{Shin:1996gi} 
  D.~S.~Shin,
  \emph{A determination of the mass gap in the $O(N)$ sigma model},
  \emph{Nucl.\ Phys.}\ {\bf B 496} (1997) 408 [\href{http://arxiv.org/abs/hep-lat/9611006}{\tt hep-lat/9611006}].
  %doi:10.1016/S0550-3213(97)00197-1
  %\href{http://arxiv.org/abs/hep-lat/9611006}{{\ttfamily [arXiv:hep-lat/9611006]}}.
  %[hep-lat/9611006].
  %%CITATION = doi:10.1016/S0550-3213(97)00197-1;%%
  %18 citations counted in INSPIRE as of 27 Jun 2017



%\cite{Balog:2009ze}
\bibitem{Balog:2009ze} 
  J.~Balog and A.~Hegedus,
  \emph{The finite size spectrum of the 2-dimensional $O(3)$ nonlinear $\sigma$-model},
  \emph{Nucl.\ Phys.}\ {\bf B 829} (2010) 425 [\href{http://arxiv.org/abs/0907.1759}{\tt arXiv:0907.1759}].
  %doi:10.1016/j.nuclphysb.2009.11.010
  %   \href{http://arxiv.org/abs/0907.1759}{{\ttfamily [arXiv:0907.1759]}}.
  %[arXiv:0907.1759 [hep-th]].
  %%CITATION = doi:10.1016/j.nuclphysb.2009.11.010;%%
  %14 citations counted in INSPIRE as of 26 Jun 2017

  
  %\cite{Dunne:2012ae}
\bibitem{Dunne:2012ae} 
  G.~V.~Dunne and M.~Unsal,
  \emph{Resurgence and trans-series in quantum field theory: the $\mathbb{CP}^{N-1}$ model},
  \emph{JHEP} {\bf 1211} (2012) 170 [\href{http://arxiv.org/abs/1210.2423}{\tt arXiv:1210.2423}].
 % doi:10.1007/JHEP11(2012)170
 % \href{http://arxiv.org/abs/1210.2423}{{\ttfamily [arXiv:1210.2423]}}.
  %[arXiv:1210.2423 [hep-th]].
  %%CITATION = doi:10.1007/JHEP11(2012)170;%%
  %106 citations counted in INSPIRE as of 28 Jun 2017
 
  
  
%\cite{Dabholkar:1994ai}
\bibitem{Dabholkar:1994ai} 
  A.~Dabholkar,
  \emph{Strings on a cone and black hole entropy},
  \emph{Nucl.\ Phys.}\ {\bf B 439} (1995) 650 [\href{http://arxiv.org/abs/hep-th/9408098}{\tt hep-th/9408098}].
  %doi:10.1016/0550-3213(95)00050-3
  %   \href{http://arxiv.org/abs/hep-th/9408098}{{\ttfamily [arXiv:hep-th/9408098]}}.
%  [hep-th/9408098].
  %%CITATION = doi:10.1016/0550-3213(95)00050-3;%%
  %54 citations counted in INSPIRE as of 21 Jun 2017
  
   %\cite{Fateev:1991bv}
\bibitem{Fateev:1991bv} 
  V.~A.~Fateev and A.~B.~Zamolodchikov,
  \emph{Integrable perturbations of $Z_N$ parafermion models and the $O(3)$ sigma model},
  \emph{Phys.\ Lett.}\ {\bf B 271} (1991) 91.
 % doi:10.1016/0370-2693(91)91283-2
  %%CITATION = doi:10.1016/0370-2693(91)91283-2;%%
  %92 citations counted in INSPIRE as of 23 Jun 2017
  
  %\cite{Fateev:1996ea}
\bibitem{Fateev:1996ea}
  V.~A.~Fateev,
  \emph{The sigma model (dual) representation for a two-parameter family of integrable quantum field theories},
  \emph{Nucl.\ Phys.}\ B {\bf 473} (1996) 509.
  %%CITATION = NUPHA,B473,509;%%
  
  %\cite{Bazhanov:2014joa}
\bibitem{Bazhanov:2014joa} 
  V.~V.~Bazhanov, G.~A.~Kotousov and S.~L.~Lukyanov,
  \emph{Winding vacuum energies in a deformed $O(4)$ sigma model},
  \emph{Nucl.\ Phys.}\ {\bf B 889} (2014) 817 [\href{http://arxiv.org/abs/1409.0449}{\tt arXiv:1409.0449}].
  %doi:10.1016/j.nuclphysb.2014.11.005
   %  \href{http://arxiv.org/abs/1409.0449}{{\ttfamily [arXiv:1409.0449]}}.
 % [arXiv:1409.0449 [hep-th]].
  %%CITATION = doi:10.1016/j.nuclphysb.2014.11.005;%%
  %5 citations counted in INSPIRE as of 26 Jun 2017
  
%\cite{Bazhanov:2016glt}
\bibitem{Bazhanov:2016glt} 
  V.~V.~Bazhanov, S.~L.~Lukyanov and B.~A.~Runov,
  \emph{Vacuum energy of the Bukhvostov-Lipatov model},
  \emph{Nucl.\ Phys.}\  {\bf B 911} (2016) 863 [\href{http://arxiv.org/abs/1607.04839}{\tt arXiv:1607.04839}].
 % doi:10.1016/j.nuclphysb.2016.08.031
  %[arXiv:1607.04839 [hep-th]];
%%CITATION = doi:10.1016/j.nuclphysb.2016.08.031;%%
  
%\cite{Bazhanov:2017xky}
\bibitem{Bazhanov:2017xky} 
  V.~V.~Bazhanov, S.~L.~Lukyanov and B.~A.~Runov,
\emph{Bukhvostov-Lipatov model and quantum-classical duality},
[\href{http://arxiv.org/abs/1711.09021}{\ttfamily arXiv:1711.09021}].
%%CITATION = ARXIV:1711.09021;%%


%\cite{Maillet:1985ek}
\bibitem{Maillet:1985ek} 
  J.~M.~Maillet,
  \emph{New integrable canonical structures in two-dimensional models},
  \emph{Nucl.\ Phys.} {\bf B 269} (1986) 54.
  %doi:10.1016/0550-3213(86)90365-2
  %%CITATION = doi:10.1016/0550-3213(86)90365-2;%%
  %116 citations counted in INSPIRE as of 26 Jun 2017




%%\cite{Smirnov:1998kv}
\bibitem{Smirnov:1998kv} 
  F. A. Smirnov, \emph{Quasiclassical study of form-factors in finite volume}, \href{http://arxiv.org/abs/hep-th/9802132}{\tt hep-th/9802132}.
  % \href{http://arxiv.org/abs/hep-th/9802132}{{\ttfamily [arXiv:hep-th/9802132]}}.
%  hep-th/9802132.
  %%CITATION = HEP-TH/9802132;%%
  %41 citations counted in INSPIRE as of 26 Jun 2017\cite{Lukyanov:2000jp}
\bibitem{Lukyanov:2000jp} 
  S.~L.~Lukyanov,
  \emph{Finite temperature expectation values of local fields in the sinh-Gordon model},
  \emph{Nucl.\ Phys.}\  {\bf B 612} (2001) 391 [\href{http://arxiv.org/abs/hep-th/0005027}{\tt hep-th/0005027}].
  %doi:10.1016/S0550-3213(01)00365-0
   %\href{http://arxiv.org/abs/hep-th/0005027}{{\ttfamily [arXiv:hep-th/0005027]}}.
  %[hep-th/0005027].
  %%CITATION = doi:10.1016/S0550-3213(01)00365-0;%%
  %34 citations counted in INSPIRE as of 26 Jun 2017
%\cite{Lukyanov:2000jp,Teschner:2007ng}
%\cite{Teschner:2007ng}
%\cite{Zamolodchikov:2000kt}
\bibitem{Zamolodchikov:2000kt} 
  A.~B.~Zamolodchikov,
  \emph{On the thermodynamic Bethe ansatz equation in the sinh-Gordon model},
  \emph{J.\ Phys.}\ {\bf A 39} (2006) 12863 [\href{http://arxiv.org/abs/hep-th/0005181}{\tt hep-th/0005181}].
  %doi:10.1088/0305-4470/39/41/S09
 % \href{http://arxiv.org/abs/hep-th/0005181}{{\ttfamily [arXiv:hep-th/0005181]}}.
 % [hep-th/0005181].
  %%CITATION = doi:10.1088/0305-4470/39/41/S09;%%
  %27 citations counted in INSPIRE as of 26 Jun 2017
  %\cite{Bytsko:2006ut}
\bibitem{Bytsko:2006ut} 
  A.~G.~Bytsko and J.~Teschner,
  \emph{Quantization of models with non-compact quantum group symmetry: modular XXZ magnet and lattice sinh-Gordon model},
  \emph{J.\ Phys.}\  {\bf A 39} (2006) 12927 [\href{http://arxiv.org/abs/hep-th/0602093}{\tt hep-th/0602093}].
  %doi:10.1088/0305-4470/39/41/S11
 % \href{http://arxiv.org/abs/hep-th/0602093}{{\ttfamily [arXiv:hep-th/0602093]}}.
 % [hep-th/0602093].
  %%CITATION = doi:10.1088/0305-4470/39/41/S11;%%
  %83 citations counted in INSPIRE as of 26 Jun 2017
\bibitem{Teschner:2007ng} 
  J.~Teschner,
 \emph{On the spectrum of the sinh-Gordon model in finite volume},
  \emph{Nucl.\ Phys.}\  {\bf B 799} (2008) 403 [\href{http://arxiv.org/abs/hep-th/0702214}{\tt hep-th/0702214}].
  %doi:10.1016/j.nuclphysb.2008.01.021
  % \href{http://arxiv.org/abs/hep-th/0702214}{{\ttfamily [arXiv:hep-th/0702214]}}.
 % [hep-th/0702214].
  %%CITATION = doi:10.1016/j.nuclphysb.2008.01.021;%%
  %43 citations counted in INSPIRE as of 26 Jun 2017
%\cite{Borot:2014dea}
\bibitem{Borot:2014dea} 
  G.~Borot, A.~Guionnet and K.~K.~Kozlowski,
  \emph{Asymptotic expansion of a partition function related to the sinh-model},  \href{http://arxiv.org/abs/1412.7721}{\tt arXiv:1412.7721}.
  %doi:10.1007/978-3-319-33379-3
  % \href{http://arxiv.org/abs/1412.7721}{{\ttfamily [arXiv:1412.7721]}}.
 % arXiv:1412.7721 [math-ph].
  %%CITATION = doi:10.1007/978-3-319-33379-3;%%
  
  %\cite{Ikhlef:2011ay}
\bibitem{Ikhlef:2011ay} 
  Y.~Ikhlef, J.~L.~Jacobsen and H.~Saleur,
  \emph{An integrable spin chain for the $SL(2,R)/U(1)$ black hole sigma model},
  \emph{Phys.\ Rev.\ Lett.}\  {\bf 108} (2012) 081601 [\href{http://arxiv.org/abs/1109.1119}{\tt arXiv:1109.1119}].
  %doi:10.1103/PhysRevLett.108.081601
 %  \href{http://arxiv.org/abs/1109.1119}{{\ttfamily [arXiv:1109.1119]}}.
  %[arXiv:1109.1119 [hep-th]].
  %%CITATION = doi:10.1103/PhysRevLett.108.081601;%%
  %13 citations counted in INSPIRE as of 26 Jun 2017


%\cite{Boos:2006mq,Boos:2008rh,Jimbo:2008kn,Boos:2009fs,Jimbo:2010jv}
  %\cite{Boos:2006mq}
\bibitem{Boos:2006mq} 
  H.~Boos, M.~Jimbo, T.~Miwa, F. A. Smirnov and Y.~Takeyama,
  \emph{Hidden Grassmann structure in the XXZ model},
  \emph{Commun.\ Math.\ Phys.}\  {\bf 272} (2007) 263 [\href{http://arxiv.org/abs/hep-th/0606280}{\tt hep-th/0606280}].
 % doi:10.1007/s00220-007-0202-x
 %\href{http://arxiv.org/abs/hep-th/0606280}{{\ttfamily [arXiv:hep-th/0606280]}}.
 % [hep-th/0606280].
  %%CITATION = doi:10.1007/s00220-007-0202-x;%%
  %56 citations counted in INSPIRE as of 28 Jun 2017
  
  %\cite{Boos:2008rh}
\bibitem{Boos:2008rh} 
  H.~Boos, M.~Jimbo, T.~Miwa, F. A. Smirnov and Y.~Takeyama,
 \emph{Hidden Grassmann structure in the XXZ model II: creation operators},
 \emph{Commun.\ Math.\ Phys.}\  {\bf 286} (2009) 875 [\href{http://arxiv.org/abs/0801.1176}{\tt arXiv:0801.1176}].
  %doi:10.1007/s00220-008-0617-z
  %\href{http://arxiv.org/abs/0801.1176}{{\ttfamily [arXiv:0801.1176]}}.
 % [arXiv:0801.1176 [hep-th]].
  %%CITATION = doi:10.1007/s00220-008-0617-z;%%
  %47 citations counted in INSPIRE as of 28 Jun 2017
  
  %\cite{Jimbo:2008kn}
\bibitem{Jimbo:2008kn} 
  M.~Jimbo, T.~Miwa and F. A. Smirnov,
  \emph{Hidden Grassmann structure in the XXZ model III: introducing the Matsubara direction},
 \emph{J.\ Phys.}\ {\bf A 42} (2009) 304018 [\href{http://arxiv.org/abs/0811.0439}{\tt arXiv:0811.0439}].
   %\href{http://arxiv.org/abs/0811.0439}{{\ttfamily [arXiv:0811.0439]}}.
  %doi:10.1088/1751-8113/42/30/304018
  %[arXiv:0811.0439 [math-ph]].
  %%CITATION = doi:10.1088/1751-8113/42/30/304018;%%
  %39 citations counted in INSPIRE as of 28 Jun 2017
  
  %\cite{Boos:2009fs}
\bibitem{Boos:2009fs} 
  H.~Boos, M.~Jimbo, T.~Miwa and F. A. Smirnov,
  \emph{Hidden Grassmann structure in the XXZ model IV: CFT limit},
\emph{Commun.\ Math.\ Phys.}\  {\bf 299} (2010) 825 [\href{http://arxiv.org/abs/0911.3731}{\tt arXiv:0911.3731}].
 %  \href{http://arxiv.org/abs/0911.3731}{{\ttfamily [arxiv:0911.3731]}}.
 % doi:10.1007/s00220-010-1051-6
  %[arXiv:0911.3731 [hep-th]].
  %%CITATION = doi:10.1007/s00220-010-1051-6;%%
  %37 citations counted in INSPIRE as of 28 Jun 2017


%\cite{Jimbo:2010jv}
\bibitem{Jimbo:2010jv} 
  M.~Jimbo, T.~Miwa and F. A. Smirnov,
  \emph{Hidden Grassmann structure in the XXZ model V: sine-Gordon model},
 \emph{Lett.\ Math.\ Phys.}\  {\bf 96} (2011) 325 [\href{http://arxiv.org/abs/1007.0556}{\tt arXiv:1007.0556}].
 %  \href{http://arxiv.org/abs/1007.0556}{{\ttfamily [arxiv:1007.0556]}}.
  %doi:10.1007/s11005-010-0438-9
  %[arXiv:1007.0556 [hep-th]].
  %%CITATION = doi:10.1007/s11005-010-0438-9;%%
  %29 citations counted in INSPIRE as of 28 Jun 2017

  
  
  
  
  
  %\cite{Wiegmann:1984pw}
\bibitem{Wiegmann:1984pw} 
 P.~B.~Wiegmann,
\emph{On the theory of nonabelian goldstone bosons in two-dimensions; exact solution of the $SU(N)\otimes SU(N)$ nonlinear $\sigma$ model},
  \emph{Phys.\ Lett.}\  {\bf B 141} (1984) 217.
  %doi:10.1016/0370-2693(84)90205-3
  %%CITATION = doi:10.1016/0370-2693(84)90205-3;%%
  %103 citations counted in INSPIRE as of 23 Jun 2017
  
    %\cite{Balog:2003yr}
\bibitem{Balog:2003yr}
  J.~Balog and A.~Hegedus,
  \emph{TBA equations for excited states in the $O(3)$ and $O(4)$ nonlinear sigma model},
  \emph{J.\ Phys.}\  {\bf A 37} (2004) 1881 [\href{http://arxiv.org/abs/hep-th/0309009}{\tt hep-th/0309009}].
 %\href{http://arxiv.org/abs/hep-th/0309009}{{\ttfamily [arXiv:hep-th/0309009]}}. 
 % [arXiv:hep-th/0309009].
  %%CITATION = HEP-TH/0309009;%%
  %32 citations counted in INSPIRE as of 16 Apr 2014
  
 % \cite{Gromov:2008gj}
\bibitem{Gromov:2008gj} 
 N.~Gromov, V.~Kazakov and P.~Vieira,
  \emph{Finite volume spectrum of 2D field theories from Hirota dynamics},
  \emph{JHEP} {\bf 0912} (2009) 060 [\href{http://arxiv.org/abs/0812.5091}{\tt arXiv:0812.5091}].
 %   \href{http://arxiv.org/abs/0812.5091}{{\ttfamily [arXiv:0812.5091]}}.
  %[arXiv:hep-th/0812.5091].
  %%CITATION = ARXIV:0812.5091;%%
  %53 citations counted in INSPIRE as of 16 Apr 2014
  
 %\cite{Zamolodchikov:1992zr}
\bibitem{Zamolodchikov:1992zr} 
  A.~B.~Zamolodchikov and Al.~B.~Zamolodchikov,
  \emph{Massless factorized scattering and sigma models with topological terms},
  \emph{Nucl.\ Phys.}\ {\bf  B 379} (1992) 602.
  %doi:10.1016/0550-3213(92)90136-Y
  %%CITATION = doi:10.1016/0550-3213(92)90136-Y;%%
  %150 citations counted in INSPIRE as of 28 Jun 2017




 
%\cite{Feigin:2015raa}
\bibitem{Feigin:2015raa} 
  B.~Feigin, M.~Jimbo, T.~Miwa and E.~Mukhin,
  \emph{Quantum toroidal $\mathfrak{g}{{\mathfrak{l}}_{1}}$ and Bethe ansatz},
 \emph{J.\ Phys.}\ {\bf  A 48} (2015) 244001 [\href{http://arxiv.org/abs/1502.07194}{\tt arXiv:1502.07194}].
  %doi:10.1088/1751-8113/48/24/244001
  % \href{http://arxiv.org/abs/1502.07194}{{\ttfamily [arXiv:1502.07194]}}.
  %[arXiv:1502.07194 [math.QA]].
  %%CITATION = doi:10.1088/1751-8113/48/24/244001;%%
  %16 citations counted in INSPIRE as of 27 Jun 2017  
  
  %\cite{Alfimov:2014qua}
\bibitem{Alfimov:2014qua} 
  M.~N.~Alfimov and A.~V.~Litvinov,
  \emph{On spectrum of ILW hierarchy in conformal field theory II: coset CFT's},
  \emph{JHEP} {\bf 1502} (2015) 150 [\href{http://arxiv.org/abs/1411.3313}{\tt arXiv:1411.3313}].
%  doi:10.1007/JHEP02(2015)150
 %\href{http://arxiv.org/abs/1411.3313}{{\ttfamily [arXiv:1411.3313]}}.
  %[arXiv:1411.3313 [hep-th]].
  %%CITATION = doi:10.1007/JHEP02(2015)150;%%
  %10 citations counted in INSPIRE as of 27 Jun 2017

%\cite{Feigin:2015raa,Alfimov:2014qua,muxin}
%\cite{muxin}
\bibitem{muxin}
B.~Feigin, M.~Jimbo, E.~Mukhin,
\emph{Integrals of motion from quantum toroidal algebras}, \href{http://arxiv.org/abs/1705.07984}{\tt arXiv:1705.07984}.
% \href{http://arxiv.org/abs/1705.07984}{{\ttfamily [arXiv:1705.07984]}}.
%arXiv:1705.07984 [math.QA]




 	


 	
 	
  %\cite{Klimcik:2014bta}
\bibitem{Klimcik:2014bta} 
  C.~Klim\v{c}\'{i}k,
  \emph{Integrability of the bi-Yang-Baxter sigma-model},
 \emph{Lett.\ Math.\ Phys.}\  {\bf 104} (2014) 1095 [\href{http://arxiv.org/abs/1402.2105}{\tt arXiv:1402.2105}].
%  \href{http://arxiv.org/abs/1402.2105}{{\ttfamily [arXiv:1402.2105]}}.
  %doi:10.1007/s11005-014-0709-y
 % [arXiv:1402.2105 [math-ph]].
  %%CITATION = doi:10.1007/s11005-014-0709-y;%%
  %50 citations counted in INSPIRE as of 26 Jun 2017


%\cite{Hoare:2014pna}
\bibitem{Hoare:2014pna} 
  B.~Hoare, R.~Roiban and A.~A.~Tseytlin,
  \emph{On deformations of $AdS_n \times S^n$ supercosets},
  \emph{JHEP} {\bf 1406} (2014) 002 [\href{http://arxiv.org/abs/1403.5517}{\tt arXiv:1403.5517}].
  %doi:10.1007/JHEP06(2014)002
 % \href{http://arxiv.org/abs/1403.5517}{{\ttfamily [arXiv:1403.5517]}}.
 % [arXiv:1403.5517 [hep-th]].
  %%CITATION = doi:10.1007/JHEP06(2014)002;%%
  %79 citations counted in INSPIRE as of 26 Jun 2017

 %\cite{Delduc:2014uaa}
\bibitem{Delduc:2014uaa} 
  F.~Delduc, M.~Magro and B.~Vicedo,
  \emph{Integrable double deformation of the principal chiral model},
  \emph{Nucl.\ Phys.}\  {\bf B 891} (2015) 312 [\href{http://arxiv.org/abs/1410.8066}{\tt arXiv:1410.8066}].
 % doi:10.1016/j.nuclphysb.2014.12.018
%  \href{http://arxiv.org/abs/1410.8066}{{\ttfamily [arXiv:1410.8066]}}.
 % [arXiv:1410.8066 [hep-th]].
  %%CITATION = doi:10.1016/j.nuclphysb.2014.12.018;%%
  %17 citations counted in INSPIRE as of 26 Jun 2017
  
%\cite{Sfetsos:2015nya}
\bibitem{Sfetsos:2015nya} 
  K.~Sfetsos, K.~Siampos and D.~C.~Thompson,
  \emph{Generalised integrable $\lambda$- and $\eta$- deformations and their relation},
  \emph{Nucl.\ Phys.}\  {\bf B 899} (2015) 489 [\href{http://arxiv.org/abs/1506.05784}{\tt arXiv:1506.05784}].
  % \href{http://arxiv.org/abs/1506.05784}{{\ttfamily [arXiv:1506.05784]}}.
 % doi:10.1016/j.nuclphysb.2015.08.015
%  [arXiv:1506.05784 [hep-th]].
  %%CITATION = doi:10.1016/j.nuclphysb.2015.08.015;%%
  %31 citations counted in INSPIRE as of 26 Jun 2017

%\cite{Klimcik:2016rov}
\bibitem{Klimcik:2016rov} 
  C.~Klim\v{c}\'{i}k,
\emph{Poisson-Lie T-duals of the bi-Yang-Baxter models},
  \emph{Phys.\ Lett.}\  {\bf B 760} (2016) 345 [\href{http://arxiv.org/abs/1606.03016}{\tt arXiv:1606.03016}].
  %doi:10.1016/j.physletb.2016.06.077
    %\href{http://arxiv.org/abs/1606.03016}{{\ttfamily [arXiv:1606.03016]}}.
  %[arXiv:1606.03016 [hep-th]].
  %%CITATION = doi:10.1016/j.physletb.2016.06.077;%%
  %8 citations counted in INSPIRE as of 26 Jun 2017


%\cite{Litvinov:2016mgi}
\bibitem{Litvinov:2016mgi} 
  A.~Litvinov and L.~Spodyneiko,
  \emph{On W algebras commuting with a set of screenings},
  \emph{JHEP} {\bf 1611} (2016) 138 [\href{http://arxiv.org/abs/1609.06271}{\tt arXiv:1609.06271}].
 % \href{http://arxiv.org/abs/1609.06271}{{\ttfamily [arXiv:1609.06271]}};
 % doi:10.1007/JHEP11(2016)138
  %[arXiv:1609.06271 [hep-th]];
  %A.~Litvinov and L.~Spodyneiko, 
%  private communication (to be published).
  %%CITATION = doi:10.1007/JHEP11(2016)138;%%

%\cite{Litvinov:2016mgi}
\bibitem{Litvinov} 
V.~Fateev,  A.~Litvinov and L.~Spodyneiko, private communications, to be published.
 
 
 
 %\cite{Lukyanov:2003rt}
\bibitem{Lukyanov:2003rt}
  S. L. Lukyanov and A. B. Zamolodchikov,
  \emph{Integrable circular brane model and Coulomb charging at large conduction}, \emph{J.\,\,Stat.\,\,Mech.} {\bf 0405} (2004) P05003
 [\href{http://arxiv.org/abs/hep-th/0306188}{\tt hep-th/0306188}].
   %\href{http://arxiv.org/abs/hep-th/0306188}{{\ttfamily [arXiv:hep-th/0306188]}}.
 % [arXiv:hep-th/0306188].
  %%CITATION = HEP-TH/0306188;%%
  %9 citations counted in INSPIRE as of 05 Mar 2013
 
 
 
%\cite{Bena:2003wd}
\bibitem{Bena:2003wd}
  I. Bena, J. Polchinski and R. Roiban,
  \emph{Hidden symmetries of the $AdS_5\times S^5$ superstring},
\emph{Phys. Rev.} {\bf D 69} (2004) 046002 [\href{http://arxiv.org/abs/hep-th/0305116}{\tt hep-th/0305116}].
  %Phys.\,\,Rev.\,\,{\bf D 69},\,\,046002\,\,(2004)
% \href{http://arxiv.org/abs/hep-th/0305116}{{\ttfamily [arXiv:hep-th/0305116]}}.
 % [hep-th/0305116].
  %%CITATION = HEP-TH/0305116;%%
  %602 citations counted in INSPIRE as of 07 Oct 2013
  

  
  %\cite{Beisert:2010jr}
\bibitem{Beisert:2010jr}
  N.~Beisert, C.~Ahn, L.~F.~Alday, Z.~Bajnok, J.~M.~Drummond, L.~Freyhult, N.~Gromov and R.~A.~Janik et al.,
  \emph{Review of AdS/CFT integrability: an overview},
  \emph{Lett.\ Math.\ Phys.}\  {\bf 99}  (2012) 03 [\href{http://arxiv.org/abs/1012.3982}{\tt arXiv:1012.3982}].
  %\href{http://arxiv.org/abs/1012.3982}{{\ttfamily [arXiv:1012.3982]}}.
  %[arXiv/hep-th:1012.3982].
  %%CITATION = ARXIV:1012.3982;%%



\bibitem{ef}
K. B. Efetov, \emph{Supersymmetry
in disorder and chaos}, Cambridge University
Press, New York (1997).

%\bibitem{Zirnbauer}  
%M.R. Zirnbauer, ``Conformal
%field theory of the integer quantum Hall plateau transition''~
%[arXiv/hep-th:9905054].

  
 % \cite{Albertini} 
\bibitem{Albertini}  
 G. Albertini, \emph{Bethe-ansatz type equations for the Fateev-Zamolodchikov spin model}, \emph{J. Phys.} {\bf A 25} (1992) 1799.
  
  %\cite{Ray}
  \bibitem{Ray}
 S. Ray, \emph{Bethe ansatz study for ground state of Fateev Zamolodchikov model}, \emph{J.  Math. Phys.} {\bf 38} (1997) 1524.
  
  %\cite{Zamolodchikov:1980ku}
\bibitem{Zamolodchikov:1980ku} 
V.~A.~Fateev and  A.~B.~Zamolodchikov,
  \emph{Model factorized S matrix and an integrable Heisenberg chain with spin 1} (in Russian),
  \emph{Sov.\ J.\ Nucl.\ Phys.}\  {\bf 32} (1980) 298.
 % [\emph{Yad.\ Fiz.}\  {\bf 32} (1980) 581].
  %%CITATION = SJNCA,32,298;%%
  %142 citations counted in INSPIRE as of 28 Jun 2017
  
\end{thebibliography}
\end{document}